\documentclass[twoside,a4paper,10pt,notitlepage]{book}

\usepackage{amsmath}
\usepackage{amssymb}
\usepackage{graphicx}
\usepackage{fancyhdr}              	
\usepackage{color}                          
\usepackage{subfig}
\usepackage{hyperref}
\usepackage{etoolbox}
\usepackage[toc]{multitoc}

\font\msytw=msbm9 scaled\magstep1
\apptocmd{\sloppy}{\hbadness 10000\relax}{}{}

\hypersetup{ 
colorlinks=true, 
breaklinks=true, 
urlcolor= blue, 
linkcolor= red, 
citecolor=blue, 
pdftitle={Glassy and jammed states of harmonic spheres}, 
pdfauthor={Hugo Jacquin}, 
pdfsubject={Ph.D. Thesis}
}

\graphicspath{{graphics/}}                  

\makeatletter
\newcommand*{\toccontents}{\@starttoc{toc}}
\makeatother

\let\a=\alpha \let\b=\beta  \let\g=\gamma  \let\d=\delta \let\e=\varepsilon
\let\z=\zeta     \let\th=\theta  \let\lam=\lambda
\let\m=\mu    \let\n=\nu         \let\p=\pi    \let\r=\rho
\let\s=\sigma \let\t=\tau   \let\f=\varphi \let\ph=\phi 
   \let\o=\omega
\let\G=\Gamma \let\D=\Delta   
\let\Si=\Sigma     \let\P=\Psi
\let\O=\Omega

\def\OO{{\cal{O}}} \def\PP{{\cal{P}}} \def\DD{{\cal{D}}} \def\FF{{\cal{F}}} \def\MM{{\cal{M}}} \def\NN{{\cal{N}}}
\def\TT{{\cal{T}}} \def\LL{{\cal{L}}} \def\CC{{\cal{C}}} \def\SS{{\cal{S}}}

\def\ha{{\hat{a}}} \def\hac{{\hat{a}^{+}}} \def\hn{{\hat{n}}} \def\hW{{\hat{W}}} \def\hWs{{\hat{W}_{sym}}}

 \def\ketPsi{{| \Psi  \rangle}}   
\def\bra0{{\langle  0|}} \def\ket0{{|0 \rangle}} \def\ketC{{| \CC  \rangle}}  
 \def\braP{{\langle P |}} \def\braC{{\langle  \CC |}} 

\let\io=\infty
\def\to{\rightarrow}
\def\la{\left\langle}
\def\ra{\right\rangle}

\def\wt{\widetilde}
\def\Tr{{\rm Tr}\,}
\def\erf{{\text{erf}}}
\def\ie{{i.e. }}\def\eg{{e.g. }}

\def\olr{{\overline{\r}}} 
\def\olth{{\overline{\theta}}} 
\def\bfr{{\boldsymbol{\rho}}}
\def\bfn{{\boldsymbol{\nu}}}
\def\ola{{\overline{a}}}
\def\bfh{{{\bf h}}}

\def\bfw{{{\bf w}}}
\def\hr{{\hat{\r}}}
\def\th{{\tilde{h}}}
\def\tc{{\tilde{c}}}
\def\hr{{\hat{\rho}}} 

\def\bx{\overline{x}} \def\by{\overline{y}}  \def\bff{\overline{f}}
\def\RRR{\hbox{\msytw R}}

\newcommand{\beq}{\begin{equation}}
\newcommand{\eeq}{\end{equation}}
\newcommand{\ba}{\begin{align}}
\newcommand{\ea}{\end{align}}

\begin{document}

\begin{titlepage}
\thispagestyle{empty}

\author{Hugo Jacquin}
\title{Glass and jamming transition of simple liquids:\\static and dynamic theory}
\maketitle

We study the glass and jamming transition of finite-dimensional models of simple liquids: hard-spheres,
harmonic spheres and more generally bounded pair potentials that modelize frictionless spheres in interaction. 
At finite temperature, we study their glassy dynamics via field-theoretic methods by resorting 
to a mapping towards an effective quantum mechanical evolution, and show that such an approach resolves 
several technical problems encountered with previous attempts. We then study the static, mean-field version
of their glass transition via replica theory, and set up an expansion in terms of the corresponding static order parameter.
Thanks to this expansion, we are able to make a direct and exact comparison between historical Mode-Coupling results
and replica theory. Finally we study these models at zero temperature within the hypotheses of the random-first-order-transition
theory, and are able to derive a quantitative mean-field theory of the jamming transition.

The theoretic methods of field theory and liquid theory used in this work are presented in a mostly 
self-contained, and hopefully pedagogical, way. This manuscript is a corrected version of my PhD thesis,
defended in June, 29th, under the advisorship of Fr\'ed\'eric van Wijland, and also contains the result of collaborations
with Ludovic Berthier and Francesco Zamponi. The PhD work was funded by a CFM-JP Aguilar grant, and conducted 
in the Laboratory MSC at Universit\'e Denis Diderot --Paris 7, France.

\vfill

\toccontents

\end{titlepage}

\newpage
\pagestyle{fancy}

\chapter{Introduction}
\label{chap:intro}

\section{Theory of amorphous solids}

This thesis is devoted to the theoretical analysis of the properties of amorphous solids. Amorphous solids are ubiquitous
in daily life: window glass, optical fibers, metallic glasses, plastics, ... all of those are amorphous solids, i.e. materials 
that have a disordered microscopic structure similar to that of dense liquids, while being solid on macroscopic scales. 
Despite many years of research efforts by theoreticians and experimentalists, the understanding of this kind of materials 
is still largely empirical, and no comprehensive physical theory has been devised yet. Many 
theoretical scenarios exist \cite{BB11}, but for the moment a coherent and quantitative theory of amorphous solids that 
would start from the microscopic scale, aiming at deriving the existence and properties of an amorphous solid phase, is still 
lacking. 

This state of affairs poorly compares with the situation in liquid theory \cite{hansen} or solid-state theory 
\cite{ashcroft_mermin}, where accurate quantitative theories allow for first-principles predictions for virtually any model.
In liquid state theory, the validity of the ergodic hypothesis, i.e. the assumption that the system is able to visit all its 
possible configurations in a ``short" time (when compared to the typical time of an experiment) allows for the use of 
equilibrium statistical mechanics, greatly reducing the theoretical difficulty. In the case of the solid state, the existence of a 
fixed lattice on which particles are attached allow for the treatment of quantum fluctuations thanks to invariance properties 
of the lattice, and to the localized classical trajectories of the particles.

In the case of amorphous solids, the disordered, liquid-like, positions of the particles require a description with the level of 
complexity of the liquid state theory, at least for what concerns static properties, but the ergodic hypothesis is not verified:
amorphous materials are generically out of equilibrium, either because their relaxation time is comparable to the duration
of an experiment or because microscopic configurations are forbidden due to mechanical constraints. Additionally, no
underlying lattice symmetry is present to simplify the problem. As a consequence, the theoretical treatment of amorphous
solids is mostly based on tools borrowed from liquid state theory, and a short introduction to its formalism is presented in 
the second part of chapter \ref{chap:formalism} of this thesis. 

Many of the theoretical developments in the field of the glass transition have been concentrated on the study of idealized
models \cite{MPV87}, that aim at suppressing the complexity of the problem while keeping the essential ingredients 
needed to observe a phenomenology akin to that of amorphous solids. In this thesis we will systematically start from 
realistic
finite-dimensional models, trying to derive from first-principles, and in a controlled way, the existence and properties of an 
amorphous phase, and to understand the relations between different existing theoretical scenarios.
Because of the difficulty of dealing with finite-dimensional, non-idealized systems, the range of questions 
addressed by such an approach is limited when compared to what can be learnt from numerical experiments or by
heuristic arguments, but we believe that the quantitative implementation of the vast amount of theoretical ideas that 
emerged in the last decades in realistic models will give a firm basis for future developments of the field, and is 
complementary to other approaches.

\subsubsection{Glassy and jammed systems}

Amorphous materials can be separated into two classes: the ones that are composed of molecules or atoms of size of the 
order of the Angstr\"om, and are formed at finite temperature, that are called structural glasses \cite{BK05}, and the ones 
composed of 
a large number of particles of sizes ranging from the micrometer to the centimeter, insensitive to thermal 
fluctuations \cite{Rogers}.

Structural glasses can be formed starting from virtually any stable liquid, by many procedures \cite{An95}, the most
common one being the quench: the 
liquid has to be sufficiently rapidly cooled down below its melting temperature, avoiding the formation of crystalline states 
\cite{Ca09}. This is always 
possible thanks to the nature of the transition from liquid to solid: the formation of the crystalline structure requires the 
nucleation of a droplet of crystal inside the liquid, which is locally unfavorable when compared to the liquid structure, even 
though it is thermodynamically more stable. Thus cooling down at a sufficiently large rate will not open up the possibility of such 
nucleation, at least on experimental time scales. When the liquid has been placed below its melting temperature, it is said 
to be supercooled. The physical properties of supercooled liquids are at first indistinguishable from their liquid 
counterparts and can be deduced from the standard theory of liquids. However, when the system is further cooled down, 
its dynamics start to slow down very rapidly, and its viscosity increases by many orders of magnitude. The system then falls 
out of equilibrium and becomes an amorphous solid. 

Athermal amorphous solids are formed by starting from a dilute assembly of particles in a box, and either 
pouring more and more objects in the box \cite{Be72}, inflating them \cite{LS90} at constant rate while allowing the 
particles to move, or performing a sequence of inflations and minimization of the energy of the system \cite{OSLN03}, all
procedures having the effect of increasing the density of objects, until mechanical rigidity is attained. These 
procedures are in essence non-equilibrium, free of thermal fluctuations
that govern the behavior of glasses. As a consequence, the final states attained by these procedures, sometimes called 
jammed states, strongly depend on the followed protocol. The 
question of the maximal density that can be realized by such amorphous solids and the properties of the corresponding
packings are largely open questions, and have deep connections in mathematics and computational sciences 
\cite{ConwaySloane}.

\subsection{Thermal systems: the glass transition}

\begin{figure}[htb]
\centering
\includegraphics[width=8cm]{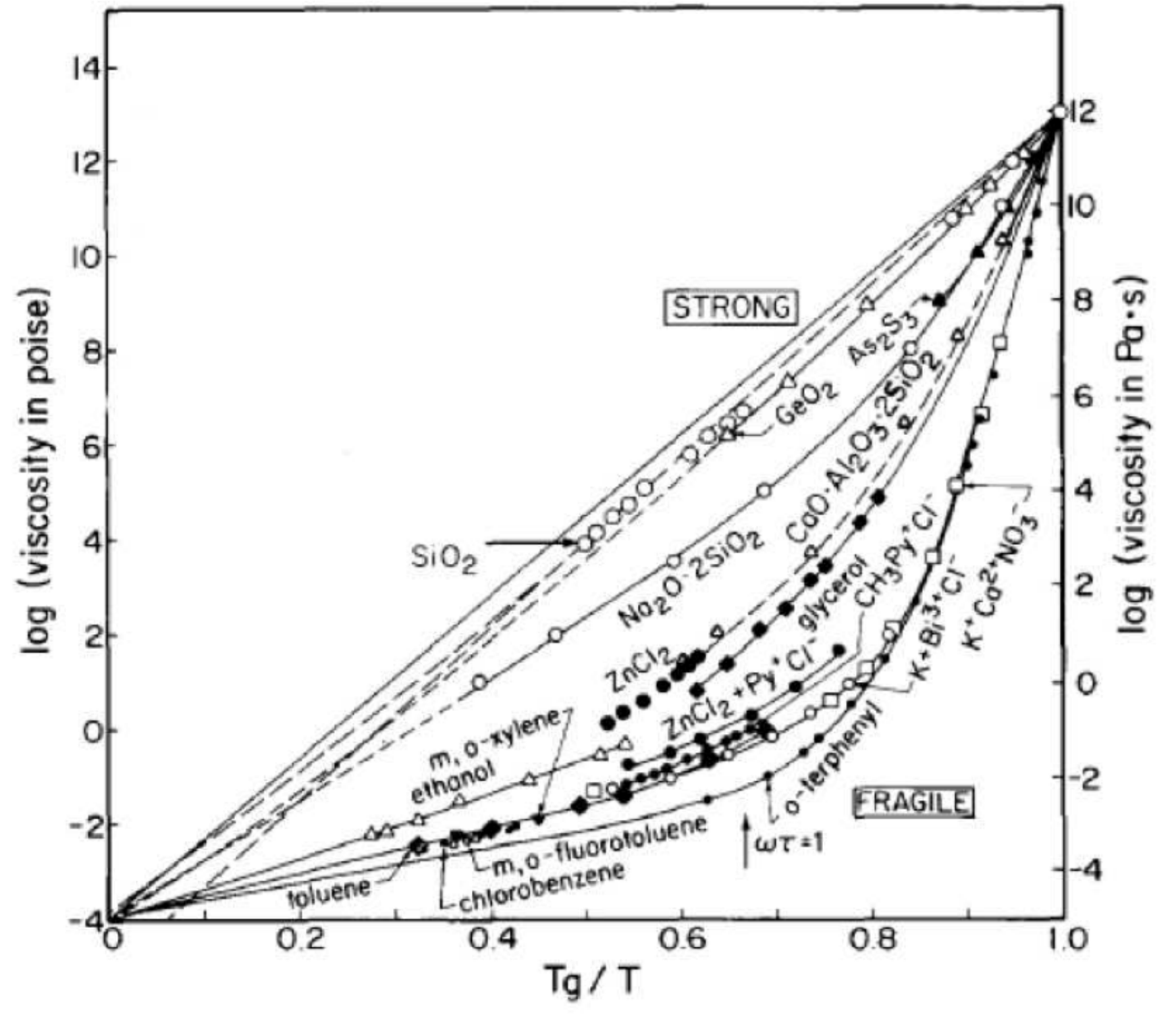}
\caption[Angell Plot for the divergence of viscosity in glass formers]{Viscosity against reduced 
temperature for various glass formers. A clear cut 
separation between two classes of glasses is observed, and corresponds to different behaviors of the 
activation energy required for relaxation. (from \cite{An88})}
\label{fig:angell-plot}
\end{figure}

The very large increase of the viscosity upon cooling liquids below their melting point is called the glass transition, suggesting the idea that
a proper thermodynamic phase transition exists from the liquid to the amorphous solid. However, experimentally no sharp
transition can be detected: for example following the evolution of the viscosity, no proper divergence can be extracted from
experiments, and the glass transition temperature $T_g$ is often arbitrarily set to the temperature at which the viscosity 
$\eta$ has reached $10^{13}$ Poise. One can alternatively follow the evolution of the relaxation time of the system, that also 
increases 
drastically upon lowering the temperature. An early important remark is that structural glasses separate into two important
classes: fragile and strong. This separation is obvious when we represent the evolution of the viscosity on an Angell plot
\cite{An88,GIXA99}, i.e. representing the logarithm of the viscosity against the temperature in units of $T_g$. An example 
of such plot, taken from \cite{An88}, is given in Figure \ref{fig:angell-plot}. By definition of $T_g$, all curves must meet at $T/T_g = 1$ for which
the logarithm of the viscosity (in base 10) must be equal to $13$. We see that two tendencies arise: strong liquids for which
the growth of the viscosity is essentially exponential, and fragile ones for which the growth is super-exponential upon
approaching the transition.
This dramatic increase of viscosity is directly linked to the relaxation time $\t_\a$ of the system through a Maxwell model 
that gives $\eta = G_{\io} \t_\a$, where $G_{\io}$ is the instantaneous shear modulus. Relaxation processes in the system
can schematically be imagined to be an accumulation of well-defined single events, such as the sudden escape of one 
particle from its local environment, that have an energy cost $E$ (the ``activation" energy). In that case the relaxation 
time will be described by an Arrhenius law:
\begin{equation}
\t_\a = \t_0 \exp \left( \frac{E}{k_B T} \right) ,
\label{relax_time}
\end{equation}
and the behavior of the viscosity in Fig. \ref{fig:angell-plot} allows us to identify the activation energy as the slope of the 
curves. For strong materials, such as Silica, the Arrhenius law is satisfied, but for fragile materials such as o-Terphenyl,
the activation energy itself depends on temperature. This is consistent with the commonly observed 
Vogel-Fulcher-Tamman law which states that the relaxation time should diverge with inverse temperature as:
\begin{align}
\t_\a = \t_0 \exp \left( \frac{D T_0}{T - T_0}  \right) ,
\label{VFT}
\end{align}
and seems to be consistent with several sets of experimental data.
This picture suggests that, for fragile liquids, the relaxation events in a supercooled liquid upon approaching its glass 
transition become more
and more energetically costly, i.e. more and more cooperative, possibly associated with a divergence at finite 
temperature $T_0$, which should thus be identified with $T_g$, the glass transition. However the range of available data
is not sufficient to provide an unambiguous fit, and other functional forms can be chosen, that support a transition
at zero temperature. The idea of a local ``cage" formed by the neighbors of each particle is however
appealing, and is thought to be the basic mechanism behind the physics of amorphous solids.

In dense liquids, a way to quantify the local environment of a single particle is to consider the radial distribution function
$g(r)$, which is the probability of finding a particle at distance $r$ of a given particle \cite{hansen}. 
To compute it, we first define the microscopic density $\hr$ as:
\begin{align}
\hr(x,t) = \sum_{i=1}^N \d(x-x_i(t)) ,
\end{align}
where $x_i(t)$ is the position of particle $i$ at time $t$. Obviously the average of this over many different experiments
will give the average number of particles in the system $\r$:
\begin{align}
\r(x,t) = \la \hr(x,t) \ra .
\label{def_micro_density_liq}
\end{align}
Not much can be learnt from this quantity since it is expected to be constant in time and space if translational and 
time-translational invariance are respected, i.e. for homogeneous liquids at equilibrium.
The radial distribution function is thus by definition related to the (normalized) equal time value of the second moment of 
this quantity:
\begin{align}
g(x-y) = \frac{\la \displaystyle \sum_{i \ne j} \d(x-x_i(0)) \d(y-x_j(0)) \ra}{\r^2} 
= 1 + \frac{\la (\hr(x,0) - \r) (\hr(y,0)-\r) \ra}{\r^2} - \frac 1 \r \d(x-y) ,
\label{def_g_intro}
\end{align}
where we made it obvious that for homogeneous systems, $g$ only depends on one variable.
The static structure factor $S(k)$ is related to the Fourier transform of $g$ by:
\begin{align}
S(k) = 1 + \r \int_x e^{i k \cdot x} \left[ g(x) - 1 \right] 
= \frac 1 \r \int_x e^{i k \cdot (x-y)} \la (\hr(x,0) - \r)(\hr(y,0) - \r) \ra .
\label{S_k_intro}
\end{align}
\begin{figure}[htb]
\centering
\includegraphics[width=14cm]{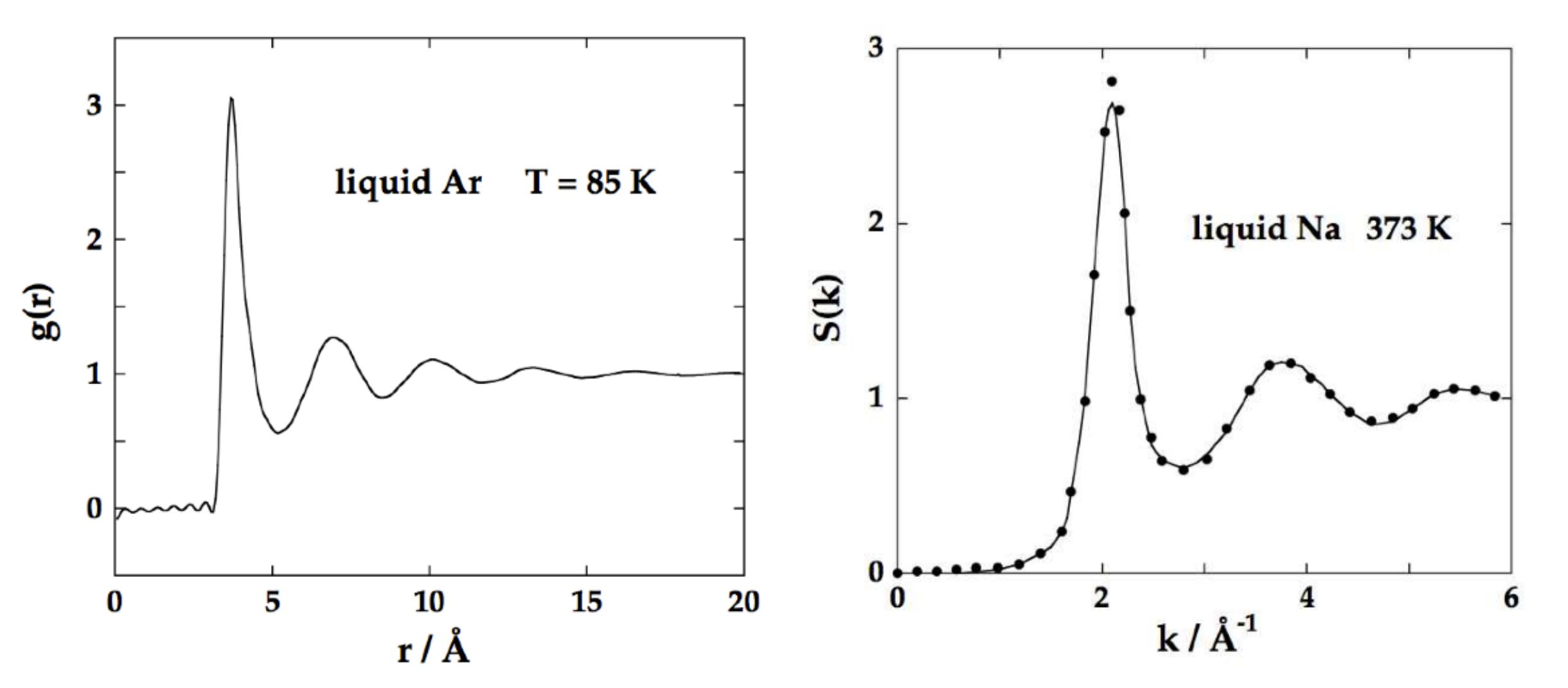}
\caption[Schematic forms of the radial distribution function and structure factor of simple liquids]
{Left: Schematic form of the radial distribution function of a simple liquid. $g(r)$ is equal to $0$ for small distances, has
marked peaks at $\s, 2 \s, \ldots$, and decays to $1$ at large distances. Right: Corresponding schematic form 
of the structure factor of a simple liquid. From \cite{hansen}}
\label{fig:g_S_liquid}
\end{figure}
The function $g$ has a characteristic 
shape in dense liquids, shown in the left frame of Fig. \ref{fig:g_S_liquid}: it is equal to zero for distances smaller than the 
particle diameters $\s$, reflecting the hard-core 
repulsion between particles, has a very strong peak at $r=\s$ reflecting the fact that a particle is surrounded by a spherical 
shell of
neighbors, then has subsequent peaks at $r=2 \s, 3 \s, \ldots$ of decreasing intensity, and finally decreases to $1$ at long
distances, reflecting the fact that no long range order exists in a liquid. The Fourier transform of $g(r)-1$, shown in the 
right frame, is called the structure factor $S(k)$, and is directly accessible in neutron diffusion experiments, or light diffusion 
experiments in the case of colloids \cite{bookDLS}.

For liquid states, the knowledge of the pair distribution function, i.e. a static two-point function is enough to 
quantitatively deduce the thermodynamics and dynamics of the system. However such observable is essentially blind to the 
presence of the glass transition, since it is observed in the experiments that the glass is essentially an arrested liquid 
configuration. A better observable has to 
be found in order to discriminate between the supercooled liquid and the amorphous solid.

\subsubsection{Order parameters for the glass transition}

This idea of a caging effect is better described by a dynamic correlation function: consider the time evolution of the position
of one particle of the fluid. If a caging effect is present, the particle will spend most of its time vibrating around its initial 
position, until it will eventually be able to escape its cage. 
Instead of computing the density-density correlation function at equal times as in Eq.(\ref{S_k_intro}), we can compute
the dynamic structure factor $S(k,t)$ with:
\begin{align}
S(k,t) = \frac 1 \r \int_x e^{ik \cdot (x-y)} \la(\hr(x,t) - \r)(\hr(y,0)-\r)\ra .
\label{def_S_k_t}
\end{align}
By definition this function reduces to the static structure factor at $t=0$.
\begin{figure}[htb]
\centering
\includegraphics[width=8cm]{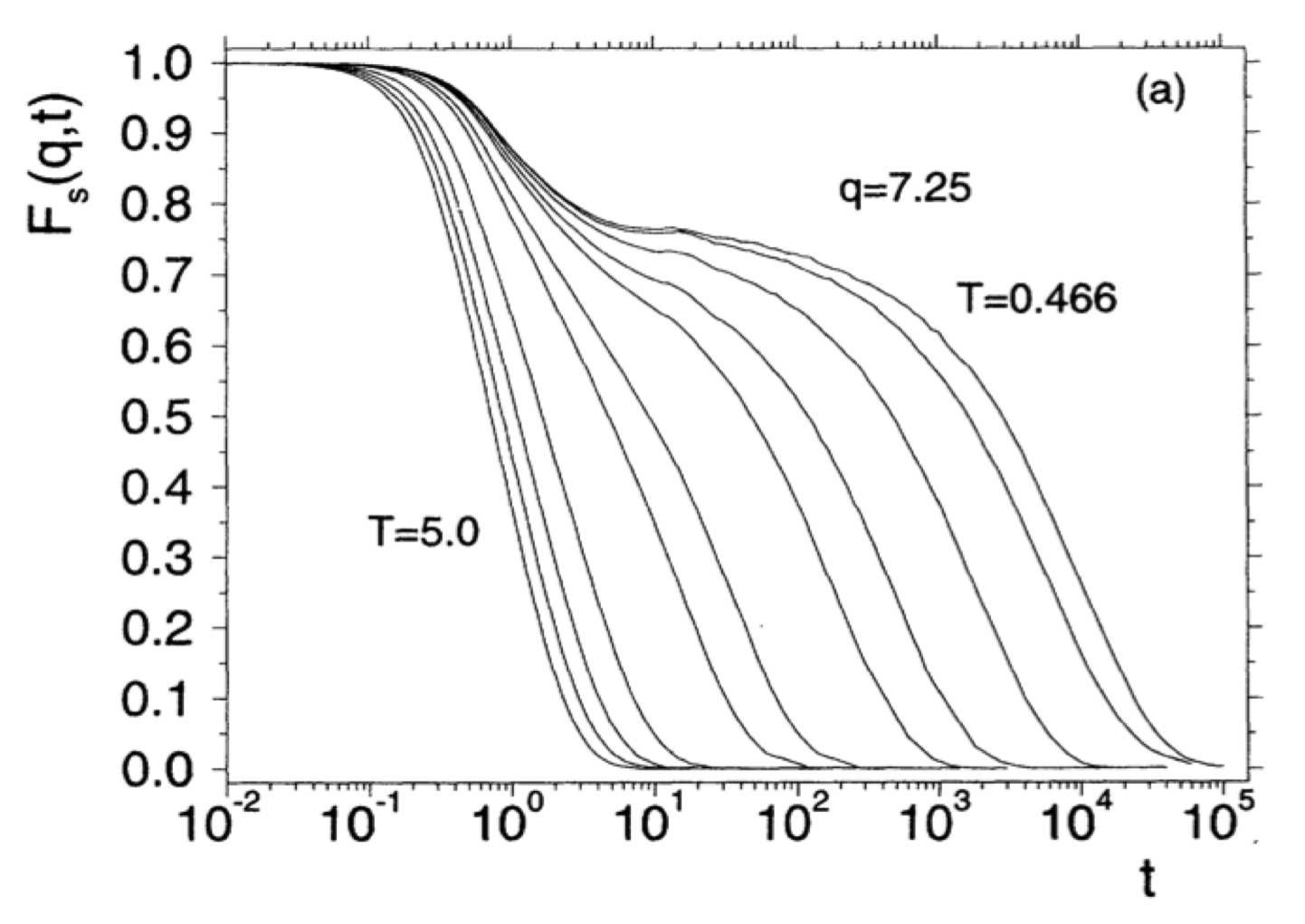}
\caption[Evolution of the dynamical structure factor upon approaching the glass transition]{Time evolution of the 
self part (i.e. restricting to $i=j$ in Eq. (\ref{def_g_intro})) of the dynamical structure factor of a model of glass-former,
for wave vector $2 \p/\s$, for several temperatures above the glass transition. (from \cite{KA95b})}
\label{fig:S_k_t}
\end{figure}
The behavior of $S(k,t)$ for a typical glass former is shown in Fig. \ref{fig:S_k_t} for one given wave
vector $k \approx 2 \p/\s$, which corresponds to a probe of the dynamics at the scale of one particle. At high enough 
temperatures, 
the relaxation is exponential just like in a liquid, particles are able to move freely in the fluid. However when the 
temperature is decreased, the function presents a two step relaxation: this is the signature of the cage effect discussed 
above. First a typical particle vibrates inside its cage, which corresponds to the initial relaxation, a process called $\b$ 
relaxation, then the relaxation saturates for a long time, while the particle is confined into its cage, finally the particle 
is able to escape its cage, and starts to explore more of its phase space, which leads eventually to its final de-correlation,
the so-called $\a$ relaxation (which explains the $\a$ subscript of the relaxation time of the
system in Eq.(\ref{relax_time})).
Below the glass transition temperature, when the system falls out of equilibrium, the plateau developped by $S(k,t)$
eventually extends to infinite times, which
reflects the fact that $\t_\a \to \io$, which implies broken ergodicity. Thus this dynamical function appears as a good order 
parameter for the glass transition: defining the non-ergodicity factor $f(k)$ as:
\begin{align}
f(k) \equiv \lim_{t \to \io} \frac{S(k,t)}{S(k)} ,
\label{def_f_k_intro}
\end{align}
$f$ jumps from $0$ in a supercooled liquid, ergodic, phase, to non-zero values below the glass transition.

The correct order parameter for the glass transition is thus a two-point quantity, and the first task of a microscopic
theory of glasses is to be able to derive from first-principles the existence of correlation functions that do not decay to 
zero at long times. This is a very different situation from that of the liquid-gas or the liquid-solid transitions: in both 
transitions the one-point density is enough to discriminate between
different phases. In the liquid-gas transition the system will separate from a gas phase with uniform density to a 
coexistence between liquid and gas, where the density takes two different values. In the liquid-solid transition, the density
switches from a uniform value in the system to a non-uniform value which presents modulations that reflect the lattice 
symmetry of the ordered phase. Thus accurate theories, such as the density functional theory \cite{Si91} described 
below, can be built by looking at the free-energy of the system as a function of density, and comparing uniform density 
profiles to non-uniform profiles. For example the Ramakrishnan-Yussouf theory of freezing \cite{RY79}, which is the 
starting point of many theories of freezing, even in the context of glasses \cite{SSW85,DV99,KM03,CKDKS05}, aims at 
finding non-uniform density profiles minima for a suitable free-energy. In this way amorphous glassy profiles can be 
obtained, as well as periodic density profiles that correspond to a crystal phase.

\subsection{A-thermal systems: the jamming transition}

\begin{figure}[htb]
\centering
\includegraphics[width=15cm]{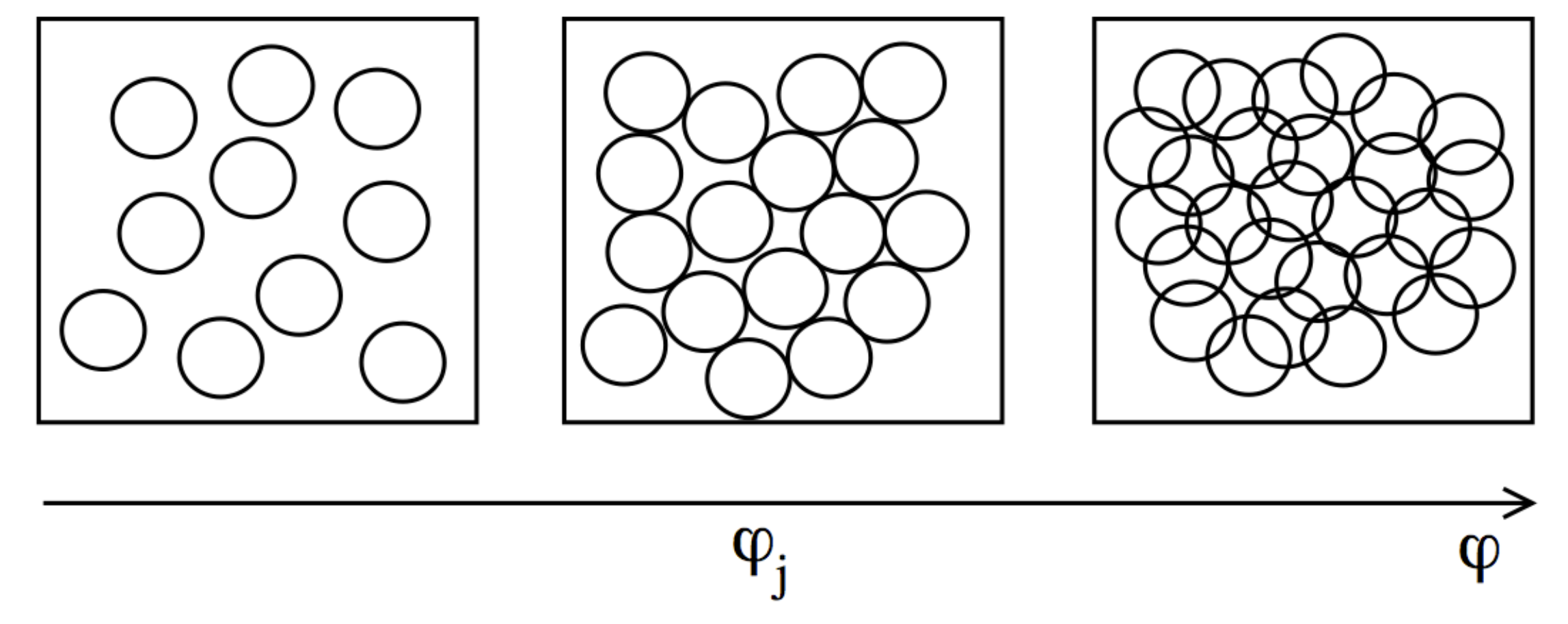}
\caption[Schematic picture of the Jamming transition]{Schematic picture of the jamming transition. Upon increasing 
$\varphi = \p \r \s^3 / 6$, the fraction of the volume of space occupied by the spheres, a transition from zero contacts states 
to states with finite number of contacts is observed.}
\label{fig:jamming_schematic}
\end{figure}
In the case of a-thermal systems such as heaps of grains, powders or foams, it has been observed experimentally 
and numerically that many physical quantities 
display critical scaling around the density at which the system acquires rigidity. A very large amount of numerical 
effort has been put since several years for the study of frictionless, spherical and deformable particles, that have a finite
range interaction, and here we restrict the discussion to such systems for simplicity.

In commonly used algorithms \cite{SDK64,OSLN03,LS90}, an assembly of such particles, initially randomly distributed, 
is gradually inflated, while minimizing its energy, or running microscopic dynamics, between each inflation step. 
Physical observables are then 
computed as averages over many realizations of one of the packing protocols described above. As long as the density is 
low, the lowest energy (amorphous) configuration is a state where there are no contacts between spheres. Of course this 
situation can not persist forever and a density exists above which no amorphous configuration without contacts can be 
found, and spheres begin to be deformed with finite energetic cost. 

This situation is
pictorially described in Figure \ref{fig:jamming_schematic}. The density $\varphi_J$ at which contacts begin to appear is 
called the Jamming point. At jamming, roughly situated around $0.64$ for 3 dimensional frictionless spheres, the average 
number of  contacts per particle $z$ jumps from $0$ to a finite value $z_c$, equal to $2 d$ for
spherical frictionless particles \cite{OLLN02}, where $d$ is the spatial dimension (2 or 3 for the systems of experimental 
interest). This value is precisely the minimal number of contacts required for a packing of spheres to be mechanically rigid: 
the packings are called isostatic \cite{Ma1864}. 
When compressing the spheres further, $z$ displays critical scaling with $\varphi - \varphi_J$, the distance to the jamming 
density, and so do thermodynamic quantities such as pressure and energy.

The radial distribution function defined above develops a diverging peak at $r=\s$ upon approaching jamming, 
reflecting the fact that particles are found at contact with probability $1$. Indeed the integral of this diverging peak counts 
the number of neighbors, and thus is equal to $2 d$ at jamming. Many more interesting scaling relations and critical behaviors, also concerning the rheology of these packings have been discovered, but the first-principles approach adopted
in our work will be limited to the aforementioned properties, and the reader is advised to refer to specific reviews
(\cite{Vh10}, \cite{JNB96}).

\begin{figure}[htb]
\centering
\includegraphics[width=7cm]{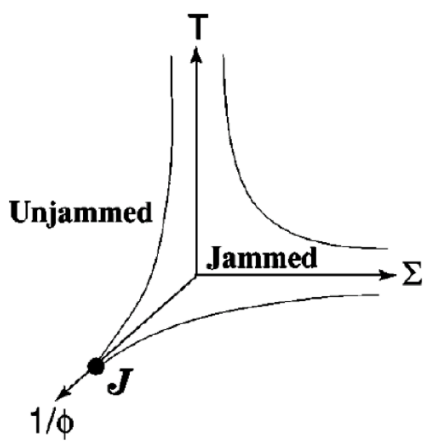}
\caption[Jamming phase diagram of Liu and Nagel]{Jamming phase diagram, as proposed by Liu and Nagel. The inverse 
volume fraction, temperature and Stress imposed on the material are the three main control parameters involved in 
amorphous solids. The temperature--density plane contains the structural and colloidal glasses, while the plane
stress--density concerns granular matter. The Jamming point is a point on the zero temperature and zero applied stress 
axis, and is postulated to influence its vicinity at $T \ne 0$. (from \cite{OSLN03})}
\label{fig:liu_diagram}
\end{figure}

The jamming transition can be seen as the extreme case of the cage effect: at jamming the particles do not have any 
space available to vibrate. It is thus tempting to associate the two phenomena by imagining that the jamming transition 
could be the real mechanism behind the glass transition. This hypothesis would support the idea of a true divergence
of the relaxation time at zero temperature.

Following this line of thought, Liu and Nagel \cite{LN98} proposed to gather all amorphous solids on a single phase 
diagram, shown in Fig. \ref{fig:liu_diagram}. In a diagram with inverse density, applied stress and temperatures as the axes,
they identify a jammed phase for high density, low temperature and low applied stress. Indeed, an amorphous solid will 
always yield under a high enough stress, or start to flow at high enough temperature, or low enough density. The 
proper jamming point lies at a high density in the zero temperature, zero applied stress axis, and is postulated to control
its vicinity. In particular the glass transition would be controlled by this zero-temperature fixed point. Although physically
appealing, such relation between jamming and glass transition is mostly hypothetical, since the studies of the jamming 
transition almost exclusively focus on the zero temperature part of this phase diagram.

\section{Current theoretical approaches}

There are currently two analytical approaches that are able to predict the divergence of the viscosity for realistic models of 
glass formers by starting from a microscopic description and performing well-defined approximations (even if they are not 
{\it a priori} justified !):
Mode-Coupling Theory (MCT) \cite{Sj80,GS87a}, which is a theory that describes the dynamics of dense liquids in terms of 
the dynamical structure factor $S(k,t)$, and the Random First Order Transition theory (RFOT) \cite{KW87a,KW87b} which 
is a theory that focuses on the long time limit of the dynamical processes, working in a static framework only. 

In the context of the jamming transition, no microscopic theory is able to predict the existence of jammed states
and deduce the critical scalings of the different physical observables that are observed numerically or in the experiments.
However, an adaptation of the RFOT to high-density states of hard spheres \cite{PZ10} has been able to identify glassy 
states of hard spheres with diverging pressure at a value close to the usual random close packing density $0.64$ in three 
dimensions, and contact number very close to the isostatic value.

\subsection{Dynamics: Mode-Coupling theory}

Starting from Hamiltonian dynamics and focusing on slowly varying collective variables such as the density, one is able to  
formally derive, using the so-called Mori-Zwanzig projection operator formalism \cite{zwanzig,Mo65}, a closed equation for 
the dynamical structure factor $S(k,t)$ defined in Eq.(\ref{def_S_k_t}) that reads:
\begin{align}
& \frac{m}{k_B T} \frac{\partial^2 S(k,t)}{\partial t^2} + \frac{k^2}{S(k)} S(k,t) + \frac \r 2 \int_0^t dt' ~ K(k,t-t') \frac{\partial S(k,t')}{\partial t'}
= 0 .
\label{mori-zwanzig_MCT}
\end{align}
All the complexity of the dynamics is now hidden in the calculation of the memory kernel $K$, which involves correlations 
between the density
and all the other hydrodynamic variables in the system. The closure approximations made within Mode-Coupling Theory (MCT)
lead to the following form for the memory kernel \cite{BGS84}:
\begin{align}
& K(k,t) =  \int_q \left[ \frac{k \cdot q}{k} c(q) + \frac{k \cdot (k-q)}{k} c(k-q) 
+ \r c^{(3)}(k,-q) \right]^2 S(q,t) S(k-q,t)
\label{kernel_MCT}
\end{align}
The third-order direct correlation function that appears in the kernel is usually neglected by resorting to the 
factorization approximation whose effect is to simply eliminate it. Furthermore, it has been shown to be negligible when 
compared to the term involving the second-order direct correlation function \cite{BGL89}.
Eq.(\ref{mori-zwanzig_MCT}) combined with Eq.(\ref{kernel_MCT}) constitutes a closed equation bearing on $S(k,t)$ that 
can be solved given the equilibrium correlations of the liquid.

Letting the time go to infinity in these equations gives a self-consistent equation for the non-ergodicity parameter $f(k)$
defined in Eq.(\ref{def_f_k_intro}):
\begin{equation}
\frac{f(k)}{1-f(k)} = \frac{\r S(k)}{2} \int_q \left[ \frac{k \cdot q}{k^2} c(q) + \frac{k \cdot(k-q)}{k^2} c(k-q) + \r c^{(3)}(q,k-q) 
\right]^2 
S(q) S(k-q) f(q) f(k-q)
\label{true_MCT}
\end{equation}
Numerically solving this self-consistent equation predicts the appearance, at constant density and below a critical 
temperature $T_{\rm{MCT}}$, of a 
non-zero solution for $f(k)$, signaling ergodicity breaking and the appearance of a glass phase.
Within MCT, the relaxation time is predicted to diverge at $T_{\rm{MCT}}$ with a power-law:
\begin{align}
\t_\a \sim \left( \frac{T-T_{\rm{MCT}}}{T_{\rm{MCT}}} \right)^{-\g} ,
\end{align}
For $T$ very close to the critical temperature, $S(k,t)$ presents a two-step behavior such as the one showed in Fig. \ref{fig:S_k_t}.  The 
approach of $S(k,t)$ to its plateau value (the $\b$-relaxation) is given by:
\begin{align}
\frac{S(k,t)}{S(k)} - f(k)  \sim t^{-a} ,    
\label{scaling_beta_MCT}
\end{align}
and the beginning of the $\a$-relaxation, i.e. the final de-correlation at long times in the ergodic phase is given by:
\begin{equation}
\frac{S(k,t)}{S(k)} - f(k) \sim - t^{b}
\label{scaling_alpha_MCT}
\end{equation}
The three exponents $a,b$ and $\g$ are not independent, but verify scaling relations \cite{BGS84,Go85}:
\begin{align}
& \lam = \frac{\G(1-a)^2}{\G(1-2a)} = \frac{\G(1-b)^2}{\G(1-2b)} , \label{lambda} \\
& \g = \frac 1{2a} + \frac 1{2b} , \label{scaling_talpha_ab_MCT}
\end{align}
so that all exponents can be deduced from the knowledge of $\lam$, which itself solely depends on the structure of the 
equation for the non-ergodicity factor Eq.(\ref{true_MCT}) \cite{Go85}.
The predictions of MCT for the power-law scalings Eq.(\ref{scaling_beta_MCT},\ref{scaling_alpha_MCT}) are well 
verified experimentally \cite{Go99}, but the prediction for the critical temperature $T_{\rm{MCT}}$ is too high, predicted 
by MCT to be higher than $T_g$, the experimental glass transition, which is at worst an upper bound for the true transition, 
if it exists at all. Thus when comparing experiments with theoretical predictions, adjustments of $\g$ and $T_{\rm{MCT}}$ are
usually made to obtain the values of $a$ and $b$ for example. Given the difficulty to obtain accurate measurements for 
very long times, the ambiguities inherent to such adjustments can not be ignored.

Recently, several experiments \cite{BEPBCPS09,EBPPSBC09} have confirmed that the scalings predicted by MCT are only 
valid when the system is not too
close to the transition, while closer to the transition the system enters another regime, where the power-law divergence
of the relaxation time is strictly ruled out.
The failure of MCT is commonly explained by the fact that MCT neglects activated events, i.e. temperature-induced 
escapes of local metastable states. MCT is thus seen as a ``mean-field" theory, even though it has been shown recently
to break down in high dimensions \cite{CIPZ11}. The approximations involved when expressing the kernel $K$ in 
Eq.(\ref{mori-zwanzig_MCT}) are thus probably ill-behaved and call for improvement \cite{Bo10}.

Careful inspection of equivalent dynamical theories have shown \cite{ABL09} under mild assumptions that the 
scaling predictions Eqs.(\ref{scaling_beta_MCT}--\ref{scaling_talpha_ab_MCT}) are in fact universal predictions for any 
dynamical theory that predicts the existence of a critical temperature at which $S(k,t)$ does not decay to zero at long 
times. Thus the quantitative failure of MCT must not hide the fact that it could provide an excellent starting point in order
find an accurate theory, able to predict the avoided singularity that seems to be observed numerically and experimentally.
Recent extensions of MCT \cite{DM86,GS87b} have claimed to predict this avoided singularity by including coupling to
currents in the theory. However, the approximations performed to obtain this theory were later shown to violate a number 
of physical requirements \cite{CR06}, and have to be rejected for now. Moreover, such currents do not exist in colloidal 
systems, and another scenario must be built to deal with that case.

\subsection{Random-First-Order Transition theory}

\begin{figure}[hbt]
\centering
\includegraphics[width=8cm]{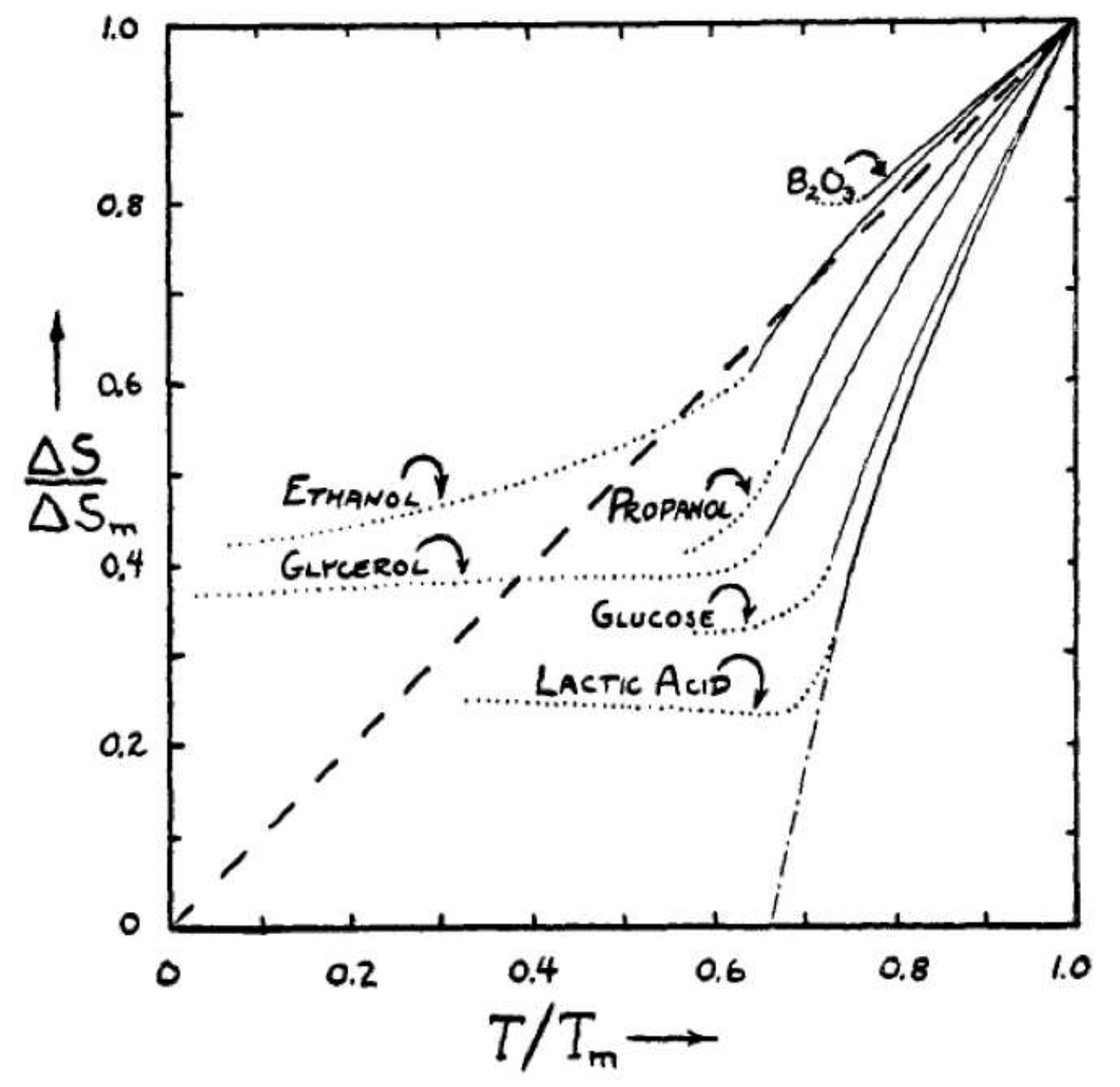}
\caption{Configurational entropy of several glass formers as a function of temperature. For several components, 
the configurational entropy seems to extrapolate down to zero at a non-zero temperature: the Kauzmann temperature.}
\label{fig:kauzmann}
\end{figure}

Kauzmann noted very early that if one separates the total entropy of a glass former into a vibrational part, that
accounts for the solid-like motions of the particles inside their cages, and a configurational part, that accounts for the 
liquid-like rearrangements that occur when cooperative movements allow the cages to reorganize, the latter is seen
to decrease upon approaching the glass transition \cite{Ka48}, as is shown in Fig. \ref{fig:kauzmann}.
The system then falls out of equilibrium and the configurational entropy saturates at a finite value. But Kauzmann noted
that one could extrapolate the decrease of the configurational entropy all the way to zero, and that this canceling would
then occur at a non-zero temperature $T_K$. Since an entropy cannot become negative, the system has to
undergo a phase transition at this point. This hypothetic transition is commonly called the Kauzmann ideal glass transition.

Adam, Gibbs and Di Marzio \cite{GD58,AG65} have interpreted this in terms of cooperatively rearranging regions, the sizes 
of which increase upon lowering the temperature. As more and more particles need to be cooperatively moved in order to 
perform a structural relaxation of the system, the relaxation time increases accordingly. The relaxation time would then 
obey:
\begin{equation}
\t_\a = A \exp \left( \frac{C}{T S_c(T)} \right) ,
\label{adam-gibbs}
\end{equation}
where $S_c$ is the configurational entropy. An entropy crisis where $S_c \to 0$ thus leads to a diverging relaxation time.
In particular if the configurational entropy was found to vanish linearly, we would obtain the VFT law of Eq.(\ref{VFT}).

This scenario has found a concrete application in the case of mean-field models of spin-glasses \cite{MPV87,CC05}, 
where it has been shown to hold exactly. The p-spin glass model \cite{De80,GM84b} is the paradigmatic model that has the 
phenomenology closest to that of structural glasses. In this model it is found that below a certain temperature $T_d$, a dynamical
transition takes place because of the appearance of an extensive number of metastable states. One is led to define the
complexity $\Si$ as the extensive part of the number $\NN$ of metastable states:
\begin{equation}
\Si = \lim_{N \to \io} \frac{1}{N} \ln \NN ,
\end{equation}
where $N$ is the number of particles in the system, and a finite $\Si$ is found to appear for $T<T_d$.

Because these are mean-field models, the free-energy barriers between these states are infinite, and the system 
dynamically gets stuck for infinite times in one state, causing ergodicity breaking. However, thermodynamically, all these 
states are equivalent and no proper transition occurs at $T_d$. Upon decreasing the temperature, the number of relevant 
metastable states diminishes, until becoming sub-extensive. At that point, a thermodynamic phase transition occurs and 
there is only one state that dominates the partition function: the ideal glass. This critical temperature is thus naturally 
associated to the Kauzmann temperature $T_K$. 

Beyond mean-field, activated events can allow for jumps between different metastable states, and one thus expects that 
the relaxation time will not diverge at $T_d$ but only at $T_K$ \cite{KW87a,BB04a}. Making the identification 
$S_c \leftrightarrow \Si$, the relaxation time should thus diverge at $T_K$ as:
\begin{equation}
\t_\a \sim \exp \left( \frac{C}{T \Si(T)} \right) ,
\label{adam-gibbs_RFOT}
\end{equation}
making contact with the Vogel-Fulcher-Tamman law and the conjecture of Kauzmann.

Furthermore, the divergence of the relaxation time was shown \cite{KW87a} to be described by equations similar to the
Mode-Coupling ones at $T_d$. The hypothetical extension of this set of predictions made on mean-field disordered spin
glasses to finite dimensional structural glasses has been named the Random-First-Order Transition theory. It is an 
elegant construction, but until recently, it was solely based on exact results found in models that can be argued to be quite
far from the reality of structural glasses.

However, a series of works of Monasson \cite{Mo95}, M\'ezard and Parisi \cite{MP96,MP99a}, and Cardenas, Franz and 
Parisi \cite{CFP99} have shown that this theoretical approach could be applied to models of structural glass formers,
with qualitative success only.

\section{Questions discussed in this work}

\subsection{Approach developed in this thesis}

In this thesis, we have studied a model of harmonic spheres. This is a system of spherical and 
frictionless particles that interact via a pair potential
\begin{equation}
v(r) = \left\{ \begin{array}{ll}
\displaystyle \varepsilon \left( 1 - \frac r \s \right)^2  \quad & \quad \text{if } r < \s \\
0 \quad & \quad \text{otherwise}
\end{array} \right. ,
\label{potential_harmS}
\end{equation}
where $\e$ is an energy scale, $r$ is the distance between two centers of spheres, and $\s$ is the diameter of the spheres.
This model has been introduced by Durian in the context of foam mechanics \cite{Du95}, and is now one of the 
paradigmatic models for studying the jamming transition \cite{OSLN03}. 

Apart from being useful in the context of jamming, this pair potential can be seen as modelling the interaction
between soft colloids in a dense regime \cite{SV99,bookmicrogel}. These colloids usually are polymer particles of poly
(methylmethacrylate) (PMMA) \cite{MMWM98,EBPPSBC09} or poly(N-isopropylacrylamide) (p-NIPAM) 
\cite{SR00,CACGHA11} immersed in a solvent. Each polymer particle undergoes Brownian motion due to the presence of
the solvent at finite temperature, and in the case where they are able to interpenetrate slightly, their mutual repulsion
can be modeled by a simple harmonic repulsion like Eq.(\ref{potential_harmS}). Their Brownian nature introduces 
thermal fluctuations, which allows for a statistical treatment of the dynamics, contrary to Hamiltonian dynamics. They have
been studied at finite temperature \cite{BW09b,BW09a,ZXCYAAHLNY09} and shown to reproduce the behavior of 
structural glass formers.

From the point of view of liquid theory, this potential has a well defined positive Fourier transform, which simplifies the 
computations. Furthermore, when the temperature decreases to zero, the finite repulsion between two particles cannot be
overcome anymore, and the particles become exactly hard-spheres. Indeed, when the temperature goes to zero, the 
energy scale $\e$ becomes infinite when compared to the thermal fluctuations (i.e. $\e / k_B T \to \io$) and the potential
becomes equivalent to an infinite repulsion for $r \le \s$, while still having zero repulsion for $r > \s$.
This choice of model allows one to investigate both the a-thermal and the thermal amorphous solids, in a simplified way.

In this thesis, we attempt, without resorting to mean-field toy models and solely by focusing on the harmonic sphere model,
to address several questions, which we summarize in the following.

\subsection{What is the theoretical status of the Mode-Coupling transition ?}

Can we find a theoretical framework where Mode-Coupling (or a corrected version of it) is well defined and easily 
generalizable, without resorting to unphysical approximations ?
This question is the object of Chapter \ref{chap:dynamics1}, where we propose a new framework that makes contact
with the field of particle physics, and that solves several difficulties inherent to the treatment of the dynamics of glasses.

In view of the recent results and the RFOT scenario, the ideal result would be to obtain a theory where the dynamical
transition is avoided, and a cross-over to a thermodynamic transition is observed.

\subsection{Does the relation between MCT and the RFOT scenario hold beyond mean-field ?}

For the moment, in non mean-field models, Mode-Coupling theory and the RFOT inspired calculations are intrinsically
different and can not be compared. Can we make bridges between the two approaches~?
This is the subject of Chapter \ref{chap:replicas}, where we compute, from replica theory, a self-consistent equation for the
non-ergodicity factor similar to Eq.(\ref{true_MCT}). We find that, in the two-mode approximation, that amounts
to expanding the static free-energy at third order in the order parameter, replica theory and MCT can
not be reconciled, although a three-body term, usually neglected within MCT, is recovered from replica theory.

As a valuable by-product, this calculation provides an expansion in 
powers of the static order parameter within replica theory. It allows to show that although the 1RSB transition found in replica
theory is qualitatively stable against addition of further corrections, these corrections are nonetheless quantitatively relevant at the 
transition. Starting from our order-parameter expansion, one should be able to unify all approximation schemes of replica 
theory in order to obtain a theory that is quantitatively efficient.

\subsection{What is the relation between jamming and the glass transition ?}

Finally, is the jamming transition at $T=0$ related to the glass transition at finite temperature ? 
Krzakala and Kurchan \cite{KK07b} then Mari, Krzakala and Kurchan \cite{MKK09} have studied 
mean field models where it has been possible to prove, using replica theory or numerical simulations, that the dynamical 
arrest of the glass transition and the jamming transition are distinct mechanisms. 
Whether this situation persists in finite dimensional models is addressed in Chapter \ref{chap:jamming}, where we show
that replica theory allows us to properly disentangle the two phenomena, and show that they are indeed different.

\chapter{Formalism of many-body systems}
\label{chap:formalism}

We have seen that the proper order parameter for the glass transition is a dynamic two-point quantity. In chapter
\ref{chap:replicas}, we will see that, in the context of the random-first-order transition theory, the glass transition can
also be characterized by a static two point quantity, via the introduction of replicas. In order to obtain theories that are
able to capture a phase transition in terms of two-point quantities, it is convenient to resort to the so-called two-particle
irreducible effective action, be it in the dynamic context or in a static context. In the dynamic case, this formalism is well
documented in the context of quantum field theory, but less so in the context of classical statistical mechanics. Quantum 
field theory textbooks give only partial accounts, or do not mention it. 
In this chapter, we present  an overview of the two-particle irreducible effective action, for a generic field theory, and in the
special case of liquid theory.

\section{Statistical field theories}
\label{sec_generic_fieldtheory}

We will generically denote by $\f$ a microstate of the system, i.e. a set of variables or
fields that completely determines its microscopic properties. In the case of equilibrium liquid theory, $\f$ can be chosen to
be the microscopic density of particles of the system. In the case of a dynamic theory, it can be for example the set of all 
time trajectories of density profiles. We will in the following denote the space and/or time variables, as well as 
internal indices (such as spin state for quantum mechanics, or replica indices in the last chapters of this manuscript), by 
a single number in subscript. Implicit summation over repeated indices is always assumed.

The macroscopic state of the system is supposed to be fully determined by a functional of $\f$, that we 
will denote by $S$. In a statistical field theory, $S$ stands for the action, in the case of equilibrium theory, $S$ would be 
replaced with the Hamiltonian of the system. The statistical weight of a particular configuration of the field, $\PP[\f]$, 
is supposed to be of the form
\beq
\PP[\f] = e^{-S[\f]} \ . 
\label{stat-weight_generic}
\eeq

We will be interested in computing the statistical average of an observable $\OO$ that depends on the field. 
This average, denoted by $\la  \cdot \ra$ in the following, is obtained by summing over all possible realizations of the
field, properly weighted by their statistical weights:
\beq
\la \OO[\f] \ra = \frac 1{Z} \Tr \OO[\f] e^{-S[\f]} \ , 
\eeq
where $\Tr \equiv \int \DD \f$ and the normalization factor is $Z$ the partition function:
\beq
Z = \Tr e^{-S[\f]} \ .
\eeq
Usually, one wants to compute averages of the field $\f$ or powers of the field. A useful way to generate all averages of
$\f$ is to artificially introduce an external field $J$ that is linearly coupled to $\f$ in the action, and consider $Z$ as a 
functional of the field $J$. Averages of $\f$ will then be generated by successively differentiating $Z$ with respect to $J$, 
and setting $J$ to zero after the calculation to come back to the original theory:
\beq\begin{split}
& \ph_1 \equiv \la \f_1 \ra = \left. \frac 1{Z} \frac{\d Z}{\d J_1} \right|_{J=0} \ ,  \\
& \la \f_1 \cdots \f_n \ra = \left. \frac 1{Z} \frac{\d^n Z}{\d J_1 \cdots \d J_n} \right|_{J=0} \ ,
\label{def-phi_generic}
\end{split} \eeq
where we defined the average field $\ph$ in equation (\ref{def-phi_generic}).
Taking the logarithm of the partition function leads to a new functional $W$
\beq
W = \ln Z \label{def_W_generic} ,
\eeq
that generates the set of all cumulants of $\f$:
\begin{align}
& \ph_1 = \frac{\d W}{\d J_1} = \left. \frac 1{Z} \frac{\d Z}{\d J_1} \right|_{J=0} \ , \label{cumulants_1}  \\
& G_{12} \equiv \la \f_1 \f_2 \ra - \la \f_1 \ra \la \f_2 \ra = \left. \frac{\d^2 W}{\d J_1 \d J_2} \right|_{J=0} \ , 
\label{def-G_generic} \\
& W^{(n)}_{1,\ldots,n} \equiv \left. \frac{\d^n W}{\d J_1 \cdots \d J_n} \right|_{J=0} \ ,
\end{align}
where we defined the propagator $G$ in Eq.(\ref{def-G_generic}) because of its specific relevance with respect to 
higher-order members of the hierarchy.
At each order, the $n$-th cumulant of $\f$ is related to all averages of $\f$ of order lesser or equal to $n$. To first order
the average coincides with the cumulant. The relation at second order is shown in Eq.(\ref{def-G_generic}) \ . We show 
here for future use the relation for three-body functions:
\beq \begin{split}
W^{(3)}_{123} = \la \f_1 \f_2 \f_3 \ra & - \la \f_1 \ra \la \f_2 \f_3 \ra + 2 \la \f_1 \ra \la \f_2 \ra \la \f_3 \ra \\
& - \la \f_2 \ra \la \f_1 \f_3 \ra \\
& - \la \f_3 \ra \la \f_1 \f_2 \ra \ ,
\end{split} \eeq
which can be rewritten as:
\beq
\la \f_1 \f_2 \f_3 \ra = W^{(3)}_{123} + \ph_1 G_{23} + \ph_2 G_{13} + \ph_3 G_{12} + \ph_1 \ph_2 \ph_3 \ .
\label{cumulants_3}
\eeq

\subsection{Expansion around a saddle-point}

The action $S$ is, in many-body problems such as the ones we will be interested in, usually too complex to be treated 
exactly, and one usually resorts to an expansion around a given approximation of the action, that is exactly solvable. In the 
case of quantum mechanics, the saddle point of the action corresponds to the classical trajectories, so that we can build 
semi-classical approximations by expanding the action around the saddle-point in inverse powers of Planck's constant.
This starting point for expansions is not always justified, depending on the statistical theory that we are considering. \\

\noindent Following the example of quantum mechanics, we evaluate the partition function at its saddle point:
\begin{align}
\left\{ \begin{array}{ll}
& \displaystyle Z \approx Z^{(0)} \equiv e^{-S[\f^*]} , \\
& \\
& \displaystyle \left. \frac{\d S[\f]}{\d \f_1} \right|_{\f^*} = 0 ,
\end{array} \right. \label{expansion_ex_1}
\end{align} 
Now we make a change of variables $\f \to \f - \f^*$ in the trace defining the partition function:
\beq\begin{split}
Z & = \Tr e^{-S[\f+ \f^*] } \\
& = Z^{(0)} ~ \Tr \exp \left(- \frac 12 \f_1 \left. \frac{\d^2 S[\f]}{\d \f_1 \d \f_2} \right|_{\f^*} \f_2 
- \sum_{N=3}^\io \left. \frac{\d^n S[\f]}{\d \f_1 \cdots \d \f_N} \right|_{\f^*} \f_1 \cdots \f_N \right) \ ,
\end{split}\eeq
where we performed an expansion of the action around $\f^*$ in powers of $\f$, and integration over repeated indices is assumed. 
The expansion around the saddle point has 
the effect to cancel the constant and linear terms in the action.
We gather all the cubic and higher order terms in a functional $S_{\rm{ng}}[\f]$ and define $V_{12}$ as the quadratic 
coefficient.
We now artificially once again add an external source $J$ coupled to $\f$ to get the following
expression of the partition function:
\beq
Z[J] = Z^{(0)} \Tr e^{- \frac 12 \f_1 V_{12} \f_2 - S_{\rm{ng}}[\f] + J_1 \f_1} \ .
\eeq
In an Ising model, the $\f$ field 
could be the local magnetization, and the source $J$ would in that case be an external magnetic field. 
In the case of liquid theory, the field $\f$ is usually the local microscopic density in the liquid, and $J$ is the chemical 
potential and a possible inhomogeneous external field.
Now calculating the partition function can be done perturbatively around the quadratic part:
\beq\begin{split}
Z[J] & = Z^{(0)} \Tr e^{-\frac 12 \f_1 V_{12} \f_2 + J_1 \f_1} e^{-S_{\rm{ng}}[\f]} \\
& = Z^{(0)} \sum_{N=0}^\io \frac{(-1)^N}{N !} \Tr {S_{\rm{ng}}[\f]}^N e^{-\frac 12 \f_1 V_{12} \f_2 + J_1 \f_1} \ .
\label{expansion_ex_2}
\end{split}\eeq
Expanding $S_{\rm{ng}}$ in powers of the field, the calculation of $Z$ is reduced to an infinite sum of averages 
of $\f$ with respect to a quadratic statistical weight. For example if $S_{\rm{ng}}$ was composed of a cubic term plus a 
quartic one:
\beq
S_{\rm{ng}}[\f] = \frac {g_3} {3!} \f_1 \f_1 \f_1 , \label{expansion_ex_3} + \frac {g_4}{4!} \f_1 \f_1 \f_1 \f_1
\eeq
the expansion of $Z$ would then become:
\beq
Z[J] = Z^{(0)} Z_{\rm{quad}}[J] \left( 1 + \sum_{N=1}^\io \frac{(-1)^N}{N !} \frac 1{Z_{\rm{quad}}} 
\Tr \left[ \frac{g_3}{3!} \f_1 \f_1 \f_1 + \frac {g_4}{4!} \f_1 \f_1 \f_1 \f_1 \right]^N e^{-\frac 12 \f_1 V_{12} \f_2 + J_1 \f_1} \right) \ , 
\label{expansion_ex_4}
\eeq
where $Z_{\rm{quad}}$ is the partition function of the quadratic theory.
This is indeed an infinite sum of averages of $\f$, under a quadratic weight. The quadratic weight is a generalization
of a Gaussian distribution and thus integrals needed to compute averages of $\f$ under such distribution are tractable,
at least on a formal level. The average of a product of fields under quadratic weight is given by Wick's theorem \cite{Wi50},
and simply expresses the fact that, for a quadratic weight, high-order averages of the field $\f$ only depend on its 
two first cumulants, that we will call $\ph^{(0)}$ and $G^{(0)}$. The calculation of these quantities is straightforward when 
looking at $Z_{\rm{quad}}$:
\beq
Z_{\rm{quad}}[J] = \Tr e^{-\frac 12 \f_1 V_{12} \f_2 + J_1 \f_1} = \rm{Cte} ~ e^{- \frac 12 \int_{12} \ln V_{12}} 
e^{\frac 12 J_1 V_{12}^{-1} J_2} 
\label{expansion_ex_5} \ , 
\eeq
hence
\begin{align}
&  \ph^{(0)}_1 = \left. \frac 1{Z} \frac{\d Z}{\d J_1} \right|_{J=0} = 0 
\label{expansion_ex_6} \ , \\
& G^{(0)}_{12} = \left. \frac{\d^2 W[J]}{\d J_1 \d J_2} \right|_{J=0} = V_{12}^{-1} \label{expansion_ex_7} \ ,
\end{align}
where we have set $J=0$ at the end of the calculations to come back to the original theory.
Thus, the full expression of $Z$ will be a sum of integrals that will contain only $G^{(0)}$ functions (often called the ``bare"
propagator in field theory). This can pictorially be written, to second order in $g_3$ and $g_4$, as (combinatorial
factors have been omitted and can be dealt with by a proper definition of the diagrams):
\beq
Z = Z^{(0)} Z_{\rm{quad}} \left( \begin{array}{ll}
1 & - ~ g_4 ~ \begin{minipage}[1,1]{1cm} \includegraphics[width=1cm]{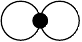} \end{minipage}
+ ~ g_3^2 ~ \begin{minipage}[1,1]{2cm} \includegraphics[width=2cm]{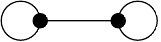} \end{minipage}
+ ~ g_3^2 ~ \begin{minipage}[1,1]{1cm} \includegraphics[width=1cm]{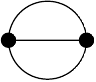} \end{minipage} \\
& + ~ g_4^2 ~ \begin{minipage}[1,1]{2cm} \includegraphics[width=2cm]{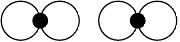} \end{minipage}
+ ~ g_4^2 ~ \begin{minipage}[1,1]{1.8cm} \includegraphics[width=1.8cm]{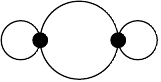} \end{minipage}
+ ~ g_4^2 ~ \begin{minipage}[1,1]{1cm} \includegraphics[width=1cm]{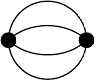} \end{minipage}
+ \ldots \end{array} \right) \ .
\label{expansion_ex_8} 
\eeq
The diagrams above are defined as follows: a black point is an integration point bearing an index, a line 
joining two black points of indices, say $1$ and $2$, is a bare propagator $V_{12}^{-1}$. Integration over repeated indices 
is understood, and numerical factors have been omitted for simplicity. Also, two diagrams standing side by side mean that
the product of the two diagrams is carried out. 

Eqs.(\ref{expansion_ex_1}--\ref{expansion_ex_8}) is only a textbook example of a starting point for expansion and its associated 
diagrammatic expansion Eq.(\ref{expansion_ex_7}). Of course 
the type of integrals, and 
thus diagrams that represent them, will depend on the particular theory considered, as well as the approximate starting 
point for the expansion. 
The diagrams of liquid theory (the Mayer diagrams \cite{Sa58,goodstein,hansen}) do not have the same 
characteristics as usual diagrams in dynamical field theories used for glasses. However, we will not need in this chapter 
to know the particular form of the diagrammatic expansion nor the starting point of calculation, and wish to stay on a more
generic level.

\subsection{The free-energy and the linked cluster theorem}

The free-energy $W$ is defined as the logarithm of the partition function in Eq.(\ref{def_W_generic}).
Taking the logarithm of the partition function has a dramatic effect on 
a diagrammatic expansion of the partition function: it systematically eliminates all diagrams that can be expressed as a
product of other diagrams (they are usually called ``connected" diagrams), and in a sense reduces the number of diagrams in 
the complete expression. As we saw in 
Eq.(\ref{def-G_generic}), the logarithm of the partition function is the generating functional of the cumulants of the field,
that have the property of clustering: they are functions that decay to zero when two coordinates are infinitely far from each 
other. This property of clustering leads to the disappearance of the disconnected diagrams in the expansion of $W$.
This is the ``linked cluster theorem", a physical demonstration of which can be found in \cite{Zinn}, while a very elegant one
in terms of replicas can be found in \cite{negele-orland}. \\

\noindent As an illustration, we show what Eq.(\ref{expansion_ex_8}) becomes upon taking its logarithm:
\beq\begin{split}
W = \ln Z =  W^{(0)} + W_{\rm{quad}} 
& - ~ g_4 ~ \begin{minipage}[1,1]{1cm} \includegraphics[width=1cm]{diag1} \end{minipage}
+ ~ g_3^2 ~ \begin{minipage}[1,1]{1.8cm} \includegraphics[width=1.8cm]{diag3} \end{minipage}
+ ~ g_3^2 ~ \begin{minipage}{1cm} \includegraphics[width=1cm]{diag4} \end{minipage}
 \label{expansion_ex_log} \\
& + ~ g_4^2 ~\begin{minipage}{1.8cm} \includegraphics[width=1.8cm]{diag5} \end{minipage}
+ ~ g_4^2 ~\begin{minipage}{1cm} \includegraphics[width=1cm]{diag6} \end{minipage}
+ \ldots
\end{split}\eeq
The diagrams that were the product of two simpler diagrams have disappeared of the expression. \\

\subsection{Reduction of diagrams: first Legendre transform}

This procedure of diagrammatic reduction can be continued by performing a Legendre transformation of $W$ with respect
to the source field $J$. Define $\G_1$, a functional of $\ph$, as:
\beq
\left\{ \begin{array}{ll}
& \G_1[\ph] = J^*_1[\ph] \ph_1 - W[J^*[\ph]] \ , \\
& \\
& \displaystyle \text{with } J^*[\ph] \text{ such that } ~ \left. \frac{\d W[J]}{\d J_1} \right|_{J^*} = \ph_1 \ .
\end{array} \right. \label{legendre_1_generic}
\eeq
We replaced one variable with its conjugate, here $J$ with $\ph$. This operation will have a dramatic effect on the 
diagrammatic expansion of $W$, resumming whole classes of diagrams.
By definition of the Legendre transform, we have:
\beq
e^{-\G_1[\ph]} = \int \DD \f ~ e^{-S[\f] + J^*[\f](\f - \ph)} \ .
\eeq
Under this form, we see why the $\G_1$ functional is often called the effective action: upon integrating the fluctuations
(performing the path integral), the microscopic action $S$ becomes a new ``effective" action $\G_1$. Expanding $\G_1$
and $S$ in Taylor series, the coupling constants in $S$ ($g_3$ and $g_4$ in our example) will be renormalized to give the
corresponding effective couplings in $\G_1$. \\

From the effective action, we obtain a new class of correlation functions, defined by the functional derivatives of the effective action:
\beq
\G^{(N)}_{1 \cdots n}[\phi] \equiv \frac{\d^N \G_1[\phi]}{\d \phi_1 \cdots \d \phi_N} \ .
\eeq
The first of these derivatives is the source field, as we can see from the definition of $\G_1$:
\beq
\frac{\d \G_1[\phi]}{\d\phi_1} = \int_2 \frac{\d J^*_2[\phi]}{\d \phi_1} \phi_2 + J^*_1[\phi] - \int_2 \frac{\d J^*_2[\phi]}{\d \phi_1} \left. \frac{\d W[J]}{\d J_2}
\right|_{J^*[\phi]} = J^*_1[\phi] \ .
\label{first_derivative_gamma1}
\eeq
The second derivative is the inverse of the propagator $G[J]$ evaluated at the value of the source $J^*[\phi]$. This can be shown by noting that:
\beq \begin{split}
\d_{12} & = \frac{\d \phi_1}{\d \phi_2} = \frac{\d }{\d \phi_2} \left( \left. \frac{\d W[J]}{\d J_1} \right|_{J^*[\phi]} \right) \\
& = \int_3 \frac{\d J^*_3[\phi]}{\d \phi_2} \left. \frac{\d^2 W[J]}{\d J_1 \d J_3} \right|_{J^*[\phi]} \ .
\end{split} \eeq
The second derivative of $W$ is the propagator, while the derivative of $J^*$ is the second derivative of $\G_1$, we thus obtain:
\beq
\d_{12} = \int_3 G_{13}[J^*[\phi]] ~ \G^{(2)}_{32}[\phi] \ .
\label{W2_G2_formal}
\eeq
Which shows that $\G^{(2)}$ is the functional inverse of $G[J^*]$:
\beq
\G^{(2)}_{12}[\ph] = G_{12}[J^*[\ph]]^{-1} \ .
\label{W2_G2_formal_bis}
\eeq

\subsubsection{Diagrammatic expansion of the effective action}

Expanding the action around $\ph$, we get:
\beq
e^{-\G_1[\ph]} = e^{-S[\ph]} \int \DD \f ~ \exp \left( - \frac 12 S^{(2)}_{12}[\phi] \f_1 \f_2 
- S_{\rm{ng}}[\f,\ph] + \int_1 \left[ J^*_1[\ph] - S^{(1)}_1[\phi] \right] \f_1 \right) \ ,
\label{action_legendre_expanded_generic}
\eeq
where $S^{(N)}_{1\cdots N}[\phi] = \left. \frac{\d^N S[\f]}{\d \f_1 \cdots \d \f_N} \right|_{\phi}$, and $S_{\rm{ng}}$ gathers all derivatives of the action of order higher than three, evaluated at $\ph$:
\beq
S_{\rm{ng}}[\f,\ph] \equiv \sum_{N=3}^\io \frac {1}{N!} \int_{1,\ldots,N} S^{(N)}_{1 \cdots N}[\phi] \f_1 \cdots \f_n \ .
\eeq
We get by taking the logarithm of Eq.(\ref{action_legendre_expanded_generic}):
\begin{align}
\G_1[\ph] = S[\ph] - \ln \left[ \int \DD \f ~ \exp \left( - \frac 12 \int_{1,2} S^{(2)}_{12}[\phi] \f_1 \f_2 
- S_{\rm{ng}}[\f,\ph] + \int_1 \left[ J^*_1[\ph] - S^{(1)}_1[\phi] \right] \f_1 \right) \right] \ .
\label{gamma1_strangesource}
\end{align}
As before, the evaluation of the path integral in Eq.(\ref{gamma1_strangesource}) can be done by expanding
around the quadratic part of the action. 
The second derivative of the action, which is the inverse of the propagator for the diagrams is now:
\beq
S^{(2)}_{12}[\ph] = \left(G_{12}^{(0)}[\phi]\right)^{-1} = \left. \frac{\d^2 S[\f]}{\d \f_1 \d \f_2} \right|_{\phi} \ .
\eeq
Before the Legendre transformation, the inverse propagator was the second derivative of the action, but evaluated at the saddle-point. \\

We explicitly extract the quadratic part of the action:
\beq
\G_1[\ph] = S[\ph] - \ln \int \DD \f ~ \exp \left( - \frac 12 \int_{1,2} S^{(2)}_{12}[\phi] \f_1 \f_2 \right) 
- \ln \la e^{-S_{\rm{ng}}[\phi,\f]} \ra_{\rm{1PI}} \ ,
\label{gamma1_expanded_around_quadratic}
\eeq
where the average is defined as:
\beq
\la \frac{}{} \bullet \frac{}{} \ra_{\rm{1PI}} \equiv \frac{\displaystyle \int\DD \f ~ 
\left( \frac{}{} \bullet \frac{}{} \right) \exp \left( - \frac 12 \int_{1,2} S^{(2)}_{12}[\phi] \f_1 \f_2 + \int_1 \left[ J^*_1[\ph] - S^{(1)}_1[\phi] \right] \f_1 \right)}
{\displaystyle \int \DD \f ~  \exp \left( - \frac 12 \int_{1,2} S^{(2)}_{12}[\phi] \f_1 \f_2 \right)} \ .
\eeq

\subsubsection{Average of the field}

The linear term here in Eq.(\ref{gamma1_strangesource}) exactly enforces that the average of $\f$ is $0$, a demonstration 
of which can be found in \cite{Ja74,calzetta}, and we reproduce it below.
We first rewrite the effective action with the change of variable $\f \to \f + \phi$ to obtain:
\beq
\G_1[\ph] = - \ln \int \DD \f ~ e^{ - S[\phi+\f] + J^*_1[\phi] \f_1 } \ .
\eeq
From this expression we can take the derivative with respect to $\phi_1$ to get:
\beq \begin{split}
\G^{(1)}_1[\phi] & = \la \frac{\d S[\phi+\f]}{\d \phi_1} - \frac{\d J^*_2[\phi]}{\d \phi_1} \f_2 \ra_{\rm{1PI}} \ , \\
& = \la  \frac{\d S[\f+\phi]}{\d \f_1}  \ra_{\rm{1PI}} - \G^{(2)}_{12}[\phi] \la \f_1 \ra_{\rm{1PI}} \ .
\end{split} \eeq
Now we also know that:
\beq \begin{split}
& \int \DD \f ~ \left[ -\frac{\d S[\f+\ph]}{\d \f_1} + J^*_1[\phi] \right] e^{-S[\f+\phi] + J^*_2[\phi]\f_2} = 0 \ , \\
\Rightarrow ~ & \la \frac{\d S[\ph+\phi]}{\d \f_1} \ra_{\rm{1PI}} = \G^{(1)}_1[\phi] \ .
\end{split} \eeq
since the integral of a derivative is zero, up to boundary terms.
Thus we get:
\beq \begin{split}
& \G^{(1)}_1[\phi] = \G^{(1)}_1[\phi] - \G^{(2)}_{12}[\phi] \la \f_1 \ra_{\rm{1PI}} \ , \\
\Rightarrow ~ & \la \f_1 \ra_{\rm{1PI}} = 0 \ .
\end{split} \eeq
The last equality is obtained because in order for the Legendre transform to be defined, the relation between $J$ and $\phi$ must be invertible, i.e.
$J^*[\phi]$ must be a monotonic function and thus $\G^{(2)}$ is invertible (in the functional sense). 
In practice this implies that we compute all diagrams with zero average field (``vacuum" diagrams).

\subsubsection{One-particle irreducibility}

We can show that the diagrams that we must calculate in order to evaluate diagrammatically the effective action are 
all ``one-particle irreducible" (1PI), which means that upon cutting one line of the diagram, they do not separate in disconnected parts.
For example in Eq.(\ref{expansion_ex_log}), the second diagram is the only one that we pictured and that is one-particle reducible:
it separates into two bubbles when cutting the central line.

A simple proof of the irreducibility of the diagrams of the effective action can be found in \cite{Zinn}, and we reproduce it here.
We start from the free energy without Legendre transformation, and add a perturbation to the action, defining:
\beq
W_\e[J] = \ln \int \DD \f ~ \exp \left( - S[\f] + J_1 \f_1 - \frac{\e}{2} \int_{1,2} \f_1 \f_2 \right) \ .
\eeq
This amounts to a shift in the inverse propagator:
\beq
\left(G^{(0)}_{12}\right)^{-1} \to \left(G^{(0)}_{12}\right)^{-1} + \e \ ,
\eeq
which gives a shift in the propagator (at first order in $\e$):
\beq
G^{(0)}_{12} \to G^{(0)}_{12} - \e \int_{1'} G^{(0)}_{11'} \int_{2'} G^{(0)}_{22'} + \OO(\e^2) \ .
\eeq
Now consider a diagram of the unperturbed free-energy. It is made of nodes and $G^{(0)}_{12}$ lines.
Now adding the perturbation doubles each line: either the line is untouched, either it is replaced by a
$\e \int G^{(0)} \int G^{(0)}$ line, which is disconnected.
At first order in $\e$, we perform only one of these replacements, and this amounts to cut open one line of the diagram.
If the resulting diagram is disconnected, this means that the original diagram was one-particle reducible.
Proving one-particle irreducibility of a given quantity thus amounts to show that the first order in $\e$ of the expansion
of this quantity is a connected function. \\

Let us look at the first order in $\e$ for the free-energy:
\beq \begin{split}
W_\e[J] & = W[J] + \e \left. \frac{d W_\e[J]}{d \e} \right|_{\e=0} + \OO(\e^2) \\
& = W[J] - \frac{\e}2 \int_{1,2} \la \f_1 \f_2 \ra + \OO(\e^2) \\
& = W[J] - \frac{\e}2 \int_{1,2} G_{12}[J] - \frac{\e}2 \int_{1} \phi_1[J] \int_2 \phi_2[J] + \OO(\e^2) \ ,
\end{split} \eeq
where we defined $G_{12}[J]$ as the propagator in presence of a source $J$, and $\phi[J]$ as the average of
the field in presence of a source.
The first order term obviously contains a disconnected part, and thus the diagrams in $W[J]$ are not 1PI. \\

We turn now to the effective action, i.e. the Legendre transform of $W_\e$. 
Its derivative with respect to the external parameter $\e$ is the 
same as that of $W$ by the properties of the Legendre transformation:
\beq \begin{split}
\frac{d \G_\e[\phi]}{d \e} & = \int_1 \frac{d J^*_1[\phi]}{d \e} \phi_1 - \left. \frac{d W_\e[J]}{d \e} \right|_{J^*[\phi]} - \int_1 \frac{d J^*_1[\phi]}{d \e} 
\left. \frac{\d W[J]}{\d J_1} \right|_{J^*[\phi]} \ , \\
& = - \left. \frac{d W_\e[J]}{d \e} \right|_{J^*[\phi]} \ .
\end{split} \eeq
Thus we obtain:
\beq
\G_\e[\phi] = \G_1[\phi] + \frac{\e}2 \int_{1,2} \phi_1 \phi_2 + \frac{\e}2 \int_{1,2} G_{12}[\phi^*[\phi]] + \OO(\e^2) \ .
\eeq
As it stands, the effective action is not 1PI either.
However, the first of the $\OO(\e)$ terms is contained in the action evaluated at the average field. Indeed
upon adding the perturbation, we made the change:
\beq
S[\f] \to S_\e[\f] = S[\f] + \frac{\e}2 \int_{1,2} \f_1 \f_2 \ .
\eeq
Thus looking at the starting point of the diagrammatic expansion of $\G$, Eq.(\ref{gamma1_strangesource}), we see that 
the first term already contains this disconnected part.
Thus we find:
\beq
\G_\e[\phi] = S_\e[\phi] - \ln \int \DD \f ~ e^{-S_\e[\phi+\f] + S_\e[\phi] + J^*_1[\phi] \f} \ ,
\eeq
and we showed that the second term has an $\OO(\e)$ part that is connected.
This proves that:
\beq \begin{split}
\G_1[\phi] = S[\phi] + \G_{\rm{1PI}}[\ph] \ ,
\end{split} \eeq
where $\G_{\rm{1PI}}$ is the sum of all ``vacuum" (i.e. with zero average field) 1PI diagrams with 
propagator $\left( S^{(2)}_{12} \right)^{-1}$.
The microscopic action is thus a resummation of the one-particle reducible (1PR) diagrams, and 
this is why the effective action is sometimes also called ``1PI effective action" or ``1PI functional".

\subsubsection{Gaussian approximation}

The calculation at lowest order (when considering that $S_{\rm{ng}}$ is negligible with respect to the quadratic part), is formally the same 
than in the expansion of $Z$ and we thus get:
\beq
- \ln \left[ \int \DD \f ~ \exp \left( - \frac 12 \left. \frac{\d^2 S[\f]}{\d \f_1 \d \f_2} \right|_{\ph} \f_1 \f_2 \right) \right] 
= \rm{Cte} + \frac 12 \int_{1,2} \ln \left( {G^{(0)}_{12}}^{-1} \right) \ .
\eeq
Finally we obtain:
\beq
\G_1[\ph] = \rm{Cte} + S[\ph] + \frac 12 \int_{1,2} \ln \left( {G^{(0)}_{12}}^{-1} \right) + \left\{ \text{1PI diagrams} \right\} \ .
\label{gamma1PI_full}
\eeq

For example, upon performing the Legendre transformation, Eq.(\ref{expansion_ex_log}) now becomes:
\beq
\G_1[\ph] = S[\ph] + \int_{12} \ln \left( {G^{(0)}_{12}}^{-1} \right)
+ ~ g_4 ~ \begin{minipage}[1,1]{1cm} \includegraphics[width=1cm]{diag1} \end{minipage}
- ~ g_3^2 ~ \begin{minipage}[1,1]{1cm} \includegraphics[width=1cm]{diag4} \end{minipage}
- ~ g_4^2 ~ \begin{minipage}[1,1]{1.8cm} \includegraphics[width=1.8cm]{diag5} \end{minipage}
- ~ g_4^2 ~ \begin{minipage}{1cm} \includegraphics[width=1cm]{diag6} \end{minipage} 
+ \ldots \label{expansion_ex_Gamma1} \ .
\eeq
Only a few diagrams are left here to this order in this expansion. The lines
in the diagrams are $G^{(0)}$.

\subsubsection{Loop expansion of the effective action}

In order to systematize the evaluation of the effective action, we can resort to the so-called loop expansion.
We introduce a parameter $\lam$ in the definition of the partition function:
\beq \begin{split}
& Z[J] = \int \DD \f ~ e^{-\frac 1{\lam} \left[ S[\f] - \int_1 J_1 \f_1 \right]} \ , \\
& W[J] = \lam \ln Z[J] \ .
\end{split} \eeq
and expand $W$ and $\G_1$, its Legendre transform, around $\lam$=0.
The lowest order term is given by the saddle-point of the functional integral:
\beq 
\left\{ \begin{array}{ll}
& W[J] = - S[\f^*[J]] + \int_1 J_1 \f^*_1[J] + \OO(\lam) \ , \\
& \\
& \displaystyle  \f^*[J] \text{ such that } \left. \frac{\d S[\f]}{\d \f_1} \right|_{\f^*[J]} = J_1 \ .
\end{array} \right.
\eeq
We can now perform the Legendre transformation.
The source $J^*[\ph]$ that selects the correct average value of the field is defined by:
\beq
\left. \frac{\d W_0[J]}{\d J_1} \right|_{J^*[\ph]} = \ph_1 \ .
\eeq
The saddle-point equation imposes that, for all values of $J$:
\beq
\frac{\d W[J]}{\d J_1} = \f^*_1[J] + \int_{1'} \frac{\d \f^*_{1'}[J]}{\d J_1} \left[ J_1 - \left. \frac{\d S[\f]}{\d \f} \right|_{\f^*[J]} \right] + \OO(\lam) = \f^*_1[J] + \OO(\lam) \ .
\eeq
Thus we have that:
\beq
\f^*_1[J^*[\ph]] = \ph_1 + \OO(\lam) \ ,
\eeq
and thus:
\beq
\G_1[\ph] = S[\ph] + \OO(\lam) \ .
\eeq
We now turn to the first order term. Returning to Eq.(\ref{gamma1_expanded_around_quadratic}), we have with the introduction of $\lam$:
\beq \begin{split}
& \G_1[\ph] = S[\ph] - \lam \ln \int \DD \f ~ \exp \left( - \frac 12 \int_{1,2} \frac{S^{(2)}_{12}[\phi]}{\lam} \f_1 \f_2 \right) 
- \lam \ln \la e^{-\frac 1{\lam} S_{\rm{ng}}[\phi,\f]} \ra_{1PI} \ , \\
& \la \frac{}{} \bullet \frac{}{} \ra_{1PI} = \frac{\displaystyle \int \DD \f ~ \bullet ~ e^{- \frac 12 \int_{1,2} \frac{S^{(2)}_{12}[\phi]}{\lam} \f_1 \f_2 + \int_1 \frac{J^*_1[\ph] - S^{(1)}_1[\phi]}{\lam} \f_1}}{\displaystyle \int \DD \f ~ e^{- \frac 12 \int_{1,2} \frac{S^{(2)}_{12}[\phi]}{\lam} \f_1 \f_2}}
\end{split} \eeq
We now perform the change of variables:
\beq
\f \to \sqrt{\lam} ~ \f 
\eeq
in the functional integrals to get:
\beq
\G_1[\ph] = \rm{Cte} + S[\ph] - \lam\ln \int \DD \f ~ \exp \left( - \frac 12 \int_{1,2} S^{(2)}_{12}[\phi] \f_1 \f_2 \right) 
- \lam\ln \la e^{-\frac 1{\lam} S_{\rm{ng}}[\phi,\sqrt{\lam} ~ \f]} \ra_{1PI} \ .
\eeq
Note that an infinite normalization factor dependant on $\lam$ comes from this change of variables and is neglected here.

First we observe that we can neglect the contribution from the non-gaussian part of the action: its lowest order term is cubic in $\f$ and thus 
gives a $\lam^{1/2}$ contribution. Thus the lowest order term coming from this functional integral is $\lam^2$ because of the $\lam$ in front of the logarithm.
We can also look at the linear term, that has a non-trivial $\lam$ dependance through the term $\G_1^{(1)} \f$. But we see also that at lowest order,
\beq
J^*_1[\ph] \equiv \G^{(1)}_1[\ph] = S^{(1)}_1[\ph] \ ,
\eeq
and the next order is $\OO(\lam)$, which means that the linear term is at least $\OO(\sqrt{\lam})$, and thus when evaluated diagrammatically, at least $\OO(\lam)$, 
and again $\OO(\lam^2)$ because of the $\lam$ in factor of the logarithm.

Finally we end up with only the Gaussian integral to compute and thus:
\beq
\G_1[\ph] = \rm{Cte} + S[\ph] + \lam\int_{1,2} \ln \left( \frac{}{} S^{(2)}_{12}[\ph] \frac{}{} \right) + \OO(\lam^2) \ .
\eeq
We can also evaluate the next order in order to see the first non-trivial diagrams.
We now have:
\beq
J^*_1[\ph] - S^{(1)}_1[\ph] = \frac{\lam}{2} \int_{23} S^{(3)}_{123} G^{(0)}_{23} + \OO(\lam^2) \ .
\eeq
An order 2 term comes from a $\OO(\lam)$ term of the expansion of the exponential.
The lowest order contribution come from the cubic and quartic terms:
\beq \begin{split}
& \int_{1\cdots 6} S^{(3)}_{123} G^{(0)}_{14} G^{(0)}_{25} G^{(0)}_{36} S^{(3)}_{456} \ , \\
& \int_{1\cdots 6} G^{(0)}_{12} S^{(3)}_{123} G^{(0)}_{34} S^{(3)}_{456} G^{(0)}_{56} \ , \\
& \int_{1\cdots 4} G^{(0)}_{12} S^{(4)}_{1234} G^{(0)}_{34} \ .
\end{split} \eeq
On the other hand the lowest order term coming form the source term is:
\beq
\int_{1\cdots 6}  G^{(0)}_{12} S^{(3)}_{123} G^{(0)}_{34} S^{(3)}_{456} G^{(0)}_{56} \ , \\
\eeq
and it cancels exactly the corresponding diagram coming from the non-gaussian part of the action,
which is 1PR. We obtain the second order expression written in Eq.(\ref{expansion_ex_Gamma1}).
At the end of the calculation, we must send back $\lam$ to $1$ to return to the original theory.
In quantum field theory, $\lam$ is equal to Planck's constant $\hbar$, which is indeed small,
and the loop expansion corresponds to a semi-classical expansion, and is thus justified.

\subsubsection{Variational principle and inverse Legendre transform}

Finally, apart from simplifying computations in terms of diagrams, the Legendre transformation provides us with a
variational 
principle to calculate both the average of the field and the free-energy. 
We recall Eq.(\ref{first_derivative_gamma1}) that the derivative of the effective action is the source field:
\beq
\frac{\d \G_1[\phi]}{\d \phi_1} = J^*_1[\phi] \ .
\eeq
Now starting from this equation, we can change again our viewpoint: instead of considering $\ph$ as a variable, and $J^*$
defined as the value of $J$ that fixes this particular value of $\la \f \ra$, we can choose for $\ph$ the physical average that 
corresponds to the initial value of $J$ (which can be zero if it had been introduced by hand), let us call it $\ph^*[J]$. 
The derivative of the effective action, evaluated at $\ph^*[J]$ is thus bound to be $J$:
\beq
\G_1^{(1)}[\ph^*[J]] = J_1 \ .
\label{stat_principle_1_generic}
\eeq
If the physical case was $J=0$, this amounts to say that $\G_1$ is extremal at the physical value of the average field.

At this particular value of $\ph$, we obtain the physical value of $W$ as:
\begin{align}
W[J] = \int_1 J_1 \ph^*_1[J] - \G_1[\ph^*[J]] . 
\end{align}
Mathematically, we have performed the inverse Legendre transform.

Eq.(\ref{stat_principle_1_generic}) is important in regards of the physical symmetries of the problem: it is used to obtain
the consequences of the symmetries of the action on the effective action (sometimes called Ward-Takahashi identities), in
order to obtain, when performing approximations, a theory that respects all physical requirements.

\noindent The program to perform approximations on the partition function is then the following:
\begin{itemize}
\item Choose an approximate starting point for the expansion of $\ln Z$
\item Write down the corresponding diagrammatic expansion
\item Perform the Legendre transform to reduce diagrams
\item Truncate the expansion by selecting a class of diagrams and obtain an approximate $\G_1[\ph]$
\item Use the variational principle Eq.(\ref{stat_principle_1_generic}) to obtain the approximate $\ph$
\item Evaluate the approximate functional $\G_1$ at this particular value of $\ph$ to get $\ln Z$
\end{itemize}
This procedure is in fact much more than a reduction: one can show that keeping only one diagram in the expansion
of $\G$ is equivalent to keeping an infinity of diagrams in $\ln Z$, thus providing better approximations.

\subsubsection{Higher-order correlation functions}

Computing a functional derivative of Eq.(\ref{W2_G2_formal}) with respect to $\ph_3$, we get formally:
\beq
0 =  \frac{\d^3 \G_1[\ph]}{\d \ph_1 \d \ph_3 \d \ph_4} \left. \frac{\d^2 W[J]}{\d J_2 \d J_4} 
\right|_{J^*}  + \frac{\d^2 \G_1[\ph]}{\d \ph_1 \d \ph_4} \frac{\d^2 \G_1[\ph]}{\d \ph_4 \d \ph_5} 
\left. \frac{\d^3 W[J]}{\d J_2 \d J_4 \d J_5} \right|_{J^*} ,
\eeq
And multiplying through by a second derivative of $\G$ and using Eq.(\ref{W2_G2_formal}), we get:
\beq \begin{split}
\G^{(3)}_{123} & = - \G^{(2)}_{11'} \G^{(2)}_{22'} \G^{(2)}_{33'} W^{(3)}_{1'2'3'} \ , \\
\Leftrightarrow ~ W^{(3)}_{123} & = - G_{11'} G_{22'} G_{33'} \G^{(3)}_{1'2'3'} \ ,
\label{W3_G3_formal}
\end{split}\eeq
where we dropped the functional dependances of the correlation functions for clarity.
This is a standard equation that can be found in any textbook on field theory, for example in \cite{Zinn}, that expresses
the relation between the so-called vertex functions and the cumulants of the field. 
We can continue this procedure, and we can in this way express all cumulants of the field as a function of only the propagator and 
the derivatives of the effective action (sometimes called vertex functions, or proper vertexes in the context of field theory).

At each order, the $n$-th order 
functional $\G^{(n)}$ involves all functionals $W^{(m)}$ of order lesser or equal to $n$. We will use such relations in 
Chapters \ref{chap:dynamics1} and \ref{chap:replicas}.
In the context of liquid theory, these functionals are related to the direct correlation functions. In the context of the
dynamics of supercooled liquids, the second order functional $\G^{(2)}$ can be identified with the memory kernel of the 
Mori-Zwanzig formalism.

In the presence of phase transitions, the propagators of the theory often develop singularities or divergences at large
wave-lengths, making the diagrams in the expansion of $W$ singular. The vertex functionals $\G^{(2)}$ being the 
inverse (in Fourier space) of the propagator, they are often free
of these divergences. For example, the direct correlation function of simple liquids develops no singularity near the liquid/gas 
transition \cite{hansen}. If the order parameter of the transition is a one-point quantity (typically $\ph$ itself), then due to 
the regularity of these functions, the phase transition can be detected with approximations of $\G_1$, whereas performing 
approximations on $W$ would lead to divergent integrals, making the analysis of the transition much harder.

Of course, when the order-parameter of the transition is a two-point quantity, which is the case in the glass transition, as 
explained in the introduction, it is necessary to go one step further in order to obtain approximations that are able to detect
the transition. This is achieved by further Legendre transforming with respect to a two-point quantity, as explained in the 
following.

\subsection{Reduction of diagrams: second Legendre transform}

As we saw in the introduction, the natural order parameter for the glass transition is a two-point correlation function.
Thus, in order to be able to detect a possible transition in terms of this order parameter, it is important to be able
to perform accurate approximations on two-point functions. This can be achieved by introducing another Legendre 
transformation \cite{LW60,DM64a,DM64b,CJT74}, with respect to a two-point quantity. 

We now add to the original partition function a source $K$ coupled to the square of the field, 
in addition to the linear source, and define:
\beq
W[J,K] = \ln \int \DD \f ~ e^{-S[\f] + J_1 \f_1 + \frac 12 K_{12} \f_1\f_2} \ ,
\eeq
The derivative of the free-energy with respect to the sources give:
\beq
\left\{ \begin{array}{ll}
& \displaystyle \frac{\d W[J,K]}{\d J_1} = \la \f_1 \ra \ , \\
& \\
& \displaystyle \frac{\d W[J,K]}{\d K_{12}} = \frac 12 \la \f_1 \f_2 \ra = \frac 12 \left( \frac{}{} G_{12}[J] + \la \f_1 \ra \la \f_2 \ra \right) \ . \\
\end{array} \right.
\eeq
We can thus perform a double Legendre transform with respect to both $J$ and $K$, which will give a functional
of $\phi$ and $G$, often called the 2PI effective action.
\beq
\left\{ \begin{array}{ll}
& \displaystyle \G_2[\phi,G] = \int_1 J^*_1[\phi,G] \phi_1 + \frac 12 \int_{1,2} K^*_{12}[\phi,G] \left( \frac{}{} \phi_1 \phi_2 + G_{12} \right) - W[J^*[\phi,G],K^*[\phi,G]] 
\ , \\ & \\
& \displaystyle J^* \text{ such that } \left. \frac{\d W[J,K]}{\d J_1} \right|_{J^*,K^*} \hspace{-0.5cm} = \phi_1 \text{ and } K^* \text{ such that } 
\left. \frac{\d W[J,K]}{\d K_{12}} \right|_{J^*,K^*} \hspace{-0.5cm} = \frac 12 \left( \frac{}{} \phi_1 \phi_2 + G_{12} \right) \ .
\end{array} \right.
\eeq

We define the first derivative of the 2PI effective action as:
\beq \left\{ \begin{array}{ll}
& \displaystyle \G^{(1,0)}_1[\phi,G] \equiv \frac{\d \G_2[\phi,G]}{\d \phi_1} \ , \\
& \\
& \displaystyle  \G^{(0,1)}_{12}[\phi,G] \equiv \frac{\d \G_2[\phi,G]}{\d G_{12}} \ . 
\end{array} \right.
\eeq
These derivatives are related to the sources by:
\beq \left\{ \begin{array}{ll}
& \displaystyle \G_1^{(1,0)}[\phi,G] = J^*_1[\phi,G] + \int_2 K^*_{12}[\phi,G] \phi_2 \ , \\
& \\
& \displaystyle \G_{12}^{(0,1)}[\phi,G] = \frac 12 K^*_{12}[\phi,G] \ ,
\end{array} \right. \eeq
or equivalently, we obtain the expression of the sources in function of $\phi$ and $G$ only:
\beq \left\{ \begin{array}{ll}
& \displaystyle J^*_1[\phi,G] = \G^{(1,0)}_1[\phi,G] - 2 \int_2 \G^{(0,1)}_{12}[\phi,G] \phi_2 \ , \\
& \\
& \displaystyle K^*_{12}[\phi,G] = 2 \G_{12}^{(0,1)}[\phi,G]  \ .
\end{array} \right. \eeq

Similarly as in the 1PI case, the second derivatives of the 2PI effective action is the inverse of a matrix of second derivatives of
the free-energy. However its explicit expression is not very useful here (although we will need it in Chapter \ref{chap:replicas}).

\subsubsection{Diagrammatic expansion of the 2PI effective action}

Similarly as before, we rewrite the effective action as a functional integral:
\beq
\G_2[\phi,G] = - \ln \int \DD \f ~ \exp \left( -S[\f] + \int_1 J^*_1 ~ (\f_1 - \phi_1) 
+ \frac 12 \int_{1,2} K^*_{12} ~ \left(\f_1 \f_2 - \phi_1 \phi_2 - G_{12} \right) \right) \ ,
\eeq
where we dropped the functional dependences of the sources.
Replacing the sources by their expression in function of the derivatives of the effective action we get:
\beq
\G_2[\phi,G] = - \ln \int \DD \f ~ \exp \left( -S[\f] + \int_1 \G^{(1,0)}_1 ~ (\f_1 - \phi_1) 
+ \int_{1,2} \G^{(0,1)}_{12} ~ \left[ \frac{}{} (\f_1- \phi_1) (\f_2 - \phi_2) - G_{12} \right] \right) \ .
\eeq
Changing variables in the functional integral we get:
\beq
\G_2[\phi,G] = - \ln \int \DD \f ~ \exp \left( -S[\phi+\f] + \int_1 \G^{(1,0)}_1 ~ \f_1 
+ \int_{1,2} \G^{(0,1)}_{12} ~ \left[ \f_1 \f_2 - G_{12} \right] \right) \ .
\label{gamma2_backgroundfield}
\eeq
Again, this can be expanded around the quadratic part of the action to generate a diagrammatic expansion of the effective action.
However the one-loop expression of the effective action is harder to extract.

\subsubsection{Average and correlations of the field}

Intuitively, we see that the particular values of the sources will be here to enforce $\la \f \ra = 0$ and $\la \f \f \ra = G$.
We can prove this with a similar procedure than the 1PI case. We take a derivative of Eq.(\ref{gamma2_backgroundfield}) with respect to $\phi$ to get:
\beq
\G^{(1,0)}_1 = - \la - \frac{\d S[\phi + \f]}{\d \phi_1} + \int_3 \G^{(2,0)}_{13} \f_3 + \int_{3,4} \G^{(1,1)}_{1,34} (\f_3\f_4 - G_{34}) \ra \ .
\eeq
Using the equation of movement again:
\beq
\la -\frac{\d S[\phi+\f]}{\d \f_1} + \G^{(1,0)} + 2 \int_{3} \G^{(0,1)}_{13} \f_{3} \ra = 0 \ ,
\eeq
to obtain:
\beq \begin{split}
\G^{(1,0)}_1 & = \la \frac{\d S[\phi + \f]}{\d \ph_1} \ra - \int_3 \G^{(2,0)}_{13} \la \f_3 \ra 
- \int_{3,4} \G^{(1,1)}_{1,34} \left( \frac{}{} \la \f_3\f_4 \ra - G_{34} \right) \ , \\
& = \G^{(1,0)}_1 + 2 \int_{1'} \G^{(0,1)}_{11'} \la \f_{1'} \ra - \int_3 \G^{(2,0)}_{13} \la \f_3 \ra 
- \int_{3,4} \G^{(1,1)}_{1,34} \left( \frac{}{} \la \f_3 \f_4 \ra - G_{34} \right) \ ,
\end{split} \eeq
and thus:
\beq
\int_3 \left[ \frac{}{} \G^{(2,0)}_{13} - 2 \G^{(0,1)}_{13} \right] \la \f_3 \ra + \int_{3,4} \G^{(1,1)}_{1,34} \left( \frac{}{} \la \f_3 \f_4 \ra - G_{34} \right) = 0 \ .
\eeq
We can also take a derivative of Eq.(\ref{gamma2_backgroundfield}) to obtain:
\beq
\int_3 \G^{(1,1)}_{3,12} \la \f_3 \ra + \int_{3,4} \G^{(0,2)}_{12,34} \left( \frac{}{} \la \f_3 \f_4 \ra - G_{34} \right) = 0 \ .
\eeq
We can combine these two last results in a matricial product:
\beq
\int_{3,4} \begin{pmatrix} 
\G^{(2,0)}_{13} - 2 \G^{(0,1)}_{13} & \G^{(1,1)}_{1,34} \\
& \\
\G^{(1,1)}_{3,12} & \G^{(0,2)}_{12,34} 
\end{pmatrix} \begin{pmatrix}
\la \f_3 \ra \d_{3,4} \\ \\ \la \f_3 \f_4 \ra - G_{34}
\end{pmatrix} = 0 \ .
\eeq
If the left hand matrix is non singular, the solution of this system is:
\beq \left\{ \begin{array}{ll}
& \displaystyle \la \f_1 \ra = 0 \ , \\
& \\
& \displaystyle \la \f_1 \f_2 \ra = G_{12} \ .
\end{array} \right. \eeq
To see that the left matrix is non-singular, we can re-express it in function of the sources to obtain:
\beq
\begin{pmatrix} 
\displaystyle \G^{(2,0)}_{13} - 2 \G^{(0,1)}_{13} & \displaystyle \G^{(1,1)}_{1,34} \\
& \\
\displaystyle \G^{(1,1)}_{3,12} & \displaystyle \G^{(0,2)}_{12,34} 
\end{pmatrix} = \begin{pmatrix}
\displaystyle \frac{\d J^*_1}{\d \phi_3} + \int_{1'} \phi_{1'} \frac{\d K^*_{11'}}{\d \phi_2} & & \displaystyle \frac 12 \frac{\d K^*_{34}}{\d \phi_1} \\
& \\
\displaystyle  \frac 12 \frac{\d K^*_{12}}{\d \phi_3} & & \displaystyle  \frac 12 \frac{\d K^*_{12}}{\d G_{34}}
\end{pmatrix} \ .
\eeq
In order for the double Legendre transform to exist, the relationship between $(J^*,K^*)$ and $(\phi,G)$ must be monotonous, thus 
the matrix above is positive definite, and thus invertible, which proves the result.

\subsubsection{Loop expansion of the 2PI effective action}

Following the systematic expansion of the 1PI effective action, we again introduce an expansion parameter $\lam$.
We define thus:
\beq \begin{split}
& Z[J,K] = \int \DD \f ~ e^{- \frac 1{\lam} \left[ S[\f] - \int_1 J_1 \f_1 - \frac 12 \int_{12} K_{12} \f_1 \f_2 \right]} \ , \\
& W[J,K] = \lam\ln Z[J,K] \ .
\end{split} \eeq
Note that this modifies the definition of the propagator:
\beq
G_{12} \equiv \frac {\la \f_1 \f_2 \ra - \la \f_1 \ra \la \f_2 \ra}{\lam} \ ,
\eeq
and thus that of the Legendre transform:
\beq
\left\{ \begin{array}{ll}
& \displaystyle \G_2[\ph,G] = \int_1 J^*_1[\ph,G] \ph_1 + \frac 12 \int_{12} K^*_{12}[\ph,G] \left( \frac{}{} \ph_1 \ph_2 + \lam G_{12} \right) 
- W[J^*[\ph,G],K^*[\ph,G]] \ , \\
& \\
& \displaystyle J^* \text{ and } K^* \text{ such that } \left. \frac{\d W[J,K]}{\d J_1} \right|_{J^*,K^*} = \phi_1 \text{ and } \left. \frac{\d W[J,K]}{\d K_{12}} \right|_{J^*,K^*} = \frac 12 \left( \frac{}{} \ph_1 \ph_2 + \lam G_{12} \right) \ . 
\end{array} \right.
\eeq
At lowest order in $\lam$, we get again the saddle equation:
\beq
\left\{ \begin{array}{ll}
& \displaystyle W[J,K] = - S[\f^*[J,K]] + \int_1 J_1 \f^*_1[J,K] + \frac 12 \int_{12} \f^*_1[J,K] K_{12} \f^*_2[J,K] \ , \\
& \\
& \displaystyle \f^*[J,K] \text{ such that } \left. \frac{\d S[\f]}{\d \f_1} \right|_{\f^*} = J_1 + \int_2 K_{12} \f^*_2[J,K] \ .
\end{array} \right.
\eeq
Performing the Legendre transformation lead us again to:
\beq
\G_2[\ph,G] = S[\ph] + \OO(\lam) \ .
\eeq
The lowest order functional is independant from the propagator, thus we are forced to explicitly evaluate the next order
in order to get the lowest order dependance on $G$.

In order to do this we expand the free-energy around the saddle point 
(we omit the dependance of $\f^*$ on the sources for compactness):
\beq \begin{split}
W[J,K] = & ~ -S[\f^*] + \int_1 J_1 \f^*_1 + \frac 12 \int_{12} K_{12} \f^*_1 \f^*_2 \\
& + \lam\ln \int \DD \f ~ \exp \left( -\frac12 \int_{12} \frac{S^{(2)}_{12}[\f^*]-K_{12}}{\lam} \f_1 \f_2 - \frac{S_{\rm{ng}}[\f^*;\f]}{\lam} \right) 
\end{split} \eeq
We perform again the change of variables:
\beq
\f \to \sqrt{\lam} \f \ ,
\eeq
to get:
\beq \begin{split}
W[J,K] = & ~ \rm{Cte} - S[\f^*] + \int_1 J_1 \f^*_1 + \frac 12 \int_{12} K_{12} \f^*_1 \f^*_2 \\
& + \lam \ln \int \DD \f ~ \exp \left( -\frac12 \int_{12} \left[ \frac{}{} S^{(2)}_{12}[\f^*] - K_{12} \right] \f_1 \f_2 - \frac{S_{\rm{ng}}[\f^*;\sqrt{\lam} ~ \f]}{\lam} \right)\ .
\end{split} \eeq
At first order we can neglect the contribution from the non-Gaussian part of the action and , and we end up with a Gaussian integral, which gives:
\beq \begin{split}
W[J,K] = & ~ \rm{Cte} - S[\f^*] + \int_1 J_1 \f^*_1 + \frac 12 \int_{12} K_{12} \f^*_1 \f^*_2 
- \frac{\lam}2 \int_{12} \ln \left( \frac{}{} S^{(2)}_{12}[\f^*] - K_{12} \right) + \OO(\lam^2) \ .
\end{split} \eeq
The derivatives of $W$ with respect to the sources are now:
\beq 
\left\{\begin{array}{ll}
\displaystyle \frac{\d W[J,K]}{\d J_1} = & \displaystyle \f^*_1 - \frac{\lam}2 \int_{345} \frac{\d \f^*_5}{\d J_1} S^{(3)}_{345}[\f^*] \left( \frac{}{} S^{(2)}_{34}[\f^*] - K_{34} \right)^{-1} + \OO(\lam^2) \ , \\
& \\
\displaystyle \frac{\d W[J,K]}{\d K_{12}} = & \displaystyle \frac 12 \f^*_1 \f^*_2 + \frac \lam 2 \left( \frac{}{} S^{(2)}_{12}[\f^*] - K_{12} \right)^{-1} \\
& \\
& \displaystyle - \frac \lam 2 \int_{345} \frac{\d \f^*_5}{\d K_{12}} S^{(3)}_{345}[\f^*] \left( \frac{}{} S^{(2)}_{34}[\f^*] - K_{34} \right)^{-1} + \OO(\lam^2) \ .
\end{array} \right.
\label{derivatives_W_oneloop}
\eeq
We also have the equation defining $\f^*$:
\beq
S^{(1)}_1[\f^*[J,K]] = J_1 + \int_2 K_{12} \f^*_2[J,K] \ ,
\eeq
which lead us to equations on the derivatives of $\f^*$:
\beq
\left\{ \begin{array}{ll}
& \displaystyle \frac{\d \f^*_1[J,K]}{\d K_{23}} = \frac{\displaystyle \left( \frac{}{} S^{(2)}_{13}[\f^*[J,K]] - K_{13} \right)^{-1} \f^*_2[J,K] 
+ \left( \frac{}{} S^{(2)}_{12}[\f^*[J,K]] - K_{12} \right)^{-1} \f^*_3[J,K]}{\displaystyle 2} \ , \\
& \\
&\quad \displaystyle \frac{\d \f^*_1[J,K]}{\d J_{2}} = \left( \frac{}{} S^{(2)}_{12}[\f^*[J,K]] - K_{12} \right)^{-1} \ .
\end{array} \right.
\eeq
We can reinsert these into the derivatives of the free-energy in Eq.(\ref{derivatives_W_oneloop}) to get their final expressions, that are 
entirely parametrized by $J$,$K$ and $\f^*$ (we omit the dependance of $\f^*$ on $J^*[\ph,G]$ and $K^*[\ph,G]$ for compactness):
\beq
\left\{ \begin{array}{ll}
& \displaystyle  \frac{\d W[J,K]}{\d J_1} = \f^* - \frac{\lam}2 \int_{345} \left( \frac{}{} S^{(2)}_{15}[\f^*] - K_{15} \right)^{-1} S^{(3)}_{345}[\f^*] 
\left( \frac{}{} S^{(2)}_{34}[\f^*] - K_{34} \right)^{-1} \ , \\
& \\
& \displaystyle  \frac{\d W[J,K]}{\d K_{12}} = \frac 12 \f^*_1 \f^*_2 + \frac \lam 2 \left( \frac{}{} S^{(2)}_{12}[\f^*] - K_{12} \right)^{-1} \\
& \\
& \displaystyle \quad \quad  - \frac{\lam}{4} \int_{345} \left( \frac{}{} S^{(2)}_{34}[\f^*] - K_{34} \right)^{-1} S^{(3)}_{345}[\f^*] 
\left[ \f^*_2 \left( \frac{}{} S^{(2)}_{15}[\f^*] - K_{15} \right)^{-1} + \f^*_1 \left( \frac{}{} S^{(2)}_{25}[\f^*] - K_{25} \right)^{-1} \right] \ .
\end{array} \right.
\label{derivatives_W_oneloop_final}
\eeq
We can now evaluate these two equations at $J^*[\ph,G]$ and $K^*[\ph,G]$. 
We know that at lowest order $\f^*$ coincides with $\ph$.
Thus in all first order terms, we can replace $\f^*[J^*,K^*]$ by $\ph$. 
By definition of $J^*$, the left hand side of the first equation of Eq.(\ref{derivatives_W_oneloop_final}) 
is $\ph$. This leads to the first order correction to the relation between $\f^*$ and $\ph$:
\beq
\f^*_1[J^*,K^*] = \ph_1 + \frac{\lam}{2} \int_{345} \left( \frac{}{} S^{(2)}_{25}[\ph] - K^*_{25} \right)^{-1} S^{(3)}_{345}[\ph] 
\left( \frac{}{} S^{(2)}_{34}[\ph] - K^*_{34} \right)^{-1} + \OO(\lam^2) \ .
\eeq
The second equation of Eq.(\ref{derivatives_W_oneloop_final}), when evaluated at $J^*$ and $K^*$ must give
\beq
\frac 12 \left( \frac{}{} \ph_1 \ph_2 + \lam G_{12} \right)
\eeq
by definition. Replacing $\f^*$ by its estimation at first order in $\lam$, we see that the complex terms cancels and we get a
simple equation on $K^*$ alone, which finally gives the desired value of $K^*$ at lowest order:
\beq
K^*_{12}[\ph,G] = S^{(2)}_{12}[\ph] - G_{12}^{-1} + \OO(\lam) \ .
\eeq
We evaluate now effective action:
\beq \begin{split}
\G_2[\ph,G] = & ~\rm{Cte} + S[\f^*] + \int_1 J^*_1 \left( \ph_1 - \f^*_1 \right) + \frac 12 \int_{12} K^*_{12} \left( \ph_1 \ph_2 + \lam G_{12} - \f^*_1 \f^*_2 \right) \\
& ~ + \frac{\lam}2 \int_{12} \ln \left( G_{12}^{-1} \right) + \OO(\lam^2) \ .
\label{gamma2_oneloop_inter}
\end{split} \eeq
Expanding the action around $\ph$, and using the fact that $\f^* - \ph$ is $\OO(\lam)$, we obtain:
\beq \begin{split}
\G_2[\ph,G] = & ~\rm{Cte} + S[\ph] + \int_1 \left( S^{(1)}_1[\ph] - J^*_1 \right) \left( \f^*_1 - \ph_1 \right) 
- \int_{12} K^*_{12} (\f^*_1 - \ph_1) \ph_2 \\
& ~ + \frac{\lam}2 \int_{12} \ln \left( G_{12}^{-1} \right) + \frac{\lam}2 \int_{12} K^*_{12} G_{12} + \OO(\lam^2) \ .
\end{split} \eeq
We know that:
\beq
J^*_1[\ph,G] = \frac{\d \G_2[\ph,G]}{\d \ph_1} - \int_2 K^*_{12}[\ph,G] \ph_2 \ .
\eeq
Thus at lowest order we have:
\beq
J^*_1[\ph,G] = S^{(1)}_1[\ph]  - \int_2 K^*_{12} \ph_2  + \OO(\lam) \ .
\eeq
Reinserting this in Eq.(\ref{gamma2_oneloop_inter}), along with the lowest order values of $J^*$ and $K^*$, 
we obtain the final result:
\beq
\G_2[\phi,G] = \rm{Cte} + S[\ph] + \frac \lam 2 \int_{12} \left( \frac{}{} S^{(2)}_{12}[\ph] - G_{12}^{-1} \right) G_{12} + \frac \lam 2 \int_{1,2} \ln \left( G_{12}^{-1} \right) + \OO(\lam^2) \ .
\eeq
Evaluating at $\lam=1$ we obtain the standard expression of the 2PI functional:
\beq
\G_2[\phi,G] = \rm{Cte} + S[\ph] + \frac 12 \int_{12} \left( \frac{}{} S^{(2)}_{12}[\ph] - G_{12}^{-1} \right) G_{12} + \frac 12 \int_{1,2} \ln \left( G_{12}^{-1} \right) + \Phi[\ph,G] \ ,
\eeq
where $\Phi$ is a sum of diagrams, with propagators $G$, average field $0$, vertices $S^{(N)}[\ph]$, and that are two-particle irreducible, 
meaning that they 
do not become disconnected when we cut two of their $G$ lines. They are thus called 2PI diagrams. The $\Phi$ functional is sometimes
called the Luttinger-Ward functional \cite{LW60,Po06}. The proof that the diagrams are 2PI is much more tedious than in the 1PI case, 
and can be found in \cite{CJT74}.
Note that the term $G^{-1} G$ has zero derivative with respect to $G$, and thus must be considered as a constant for all practical means.
For translationally invariant systems it is the integral over the reciprocal space of $1$:
\beq
\int_{12} G_{12}^{-1} G_{12} = V \int_r G(r) G^{-1}(r) = V \int_k G(k) G^{-1}(k) = V \int_k 1 \ .
\eeq

We can evaluate the lowest order values of the two sources (we have already found the value of $K^*$ but not that of $J^*$):
\beq 
\left\{ \begin{array}{ll}
& K^*_{12}[\ph,G] = S^{(2)}_{12}[\ph] - G_{12}^{-1} + \OO(\lam) \ , \\
& \\
& \displaystyle J^*_{1}[\ph,G] = S^{(1)}_1[\ph] - \int_2 G_{12}^{-1} \ph_2 + \OO(\lam) \ .
\end{array} \right.
\eeq
This can be used to evaluate the diagrams contributing to the next order. This clearly shows the role of the source terms: $K$ replaces the bare propagator
by the full one, while $J$ gets rid of the 2PR diagrams.

\noindent To continue with our example, after this Legendre transformation, Eq.(\ref{expansion_ex_Gamma2}) becomes:
\beq \begin{split}
\G_2[\ph,G] = & ~ \rm{Cte} + S[\ph] + \frac 12 \int_{12} \left( \frac{}{} S^{(2)}_{12}[\ph] - G_{12}^{-1} \right) G_{12} 
+ \frac 12 \int_{1,2} \ln \left( G_{12}^{-1}\right) \\
& ~ + g_4 ~ \begin{minipage}[1,1]{1cm} \includegraphics[width=1cm]{diag1} \end{minipage}
+ g_3^2 ~ \begin{minipage}[1,1]{1cm} \includegraphics[width=1cm]{diag4} \end{minipage}
- g_4^2 ~ \begin{minipage}[1,1]{1cm} \includegraphics[width=1cm]{diag6} \end{minipage} 
+ \ldots \label{expansion_ex_Gamma2} \ ,
\end{split} \eeq
The lines of these diagrams are now the full propagators $G$.

\subsubsection{Variational principle and inverse Legendre transformation}

Equivalently to Eq.(\ref{stat_principle_1_generic}) we have as a consequence of the Legendre transformation:
\begin{align}
\frac{\d \G_2[\ph,G]}{\d G_{12}} = \frac 12 K^*_{12}[\phi,G]
\label{stat_principle_2_generic}
\end{align}
As in the 1PI case, this equation can be seen as a variational principle that we can use to obtain
self-consistent equations on the propagator.
Performing the inverse Legendre transformation is done by setting 
the source to its original value, i.e. $0$, which means that the physical propagator extremalizes 
the 2PI functional.

Now the variational principle in Eq.(\ref{stat_principle_2_generic}) becomes:
\begin{align}
0 = S^{(2)}_{12}[\ph] - G_{12}^{-1} + 2 \frac{\d \Phi[\ph,G]}{\d G_{12}}  \ .
\end{align}
Thus we obtain an equation for the self-energy, under the form of a self-consistent equation:
\beq
\Si_{12}[\ph,G] \equiv \left( G^{(0)}_{12}[\ph,G] \right)^{-1} - G_{12}^{-1} =  2 \frac{\d \Phi[\ph,G]}{\d G_{12}} \ .
\label{stat_principle_2_sigma}
\eeq

\noindent In order to perform approximations on the correlation function, the program is then:
\begin{itemize}
\item Choose an approximate starting point for the expansion of $\ln Z$.
\item Write down the corresponding diagrammatic expansion.
\item Perform the double Legendre transform to reduce diagrams.
\item Truncate the expansion by selecting a class of diagrams and obtain an approximation of $\G_2[\ph,G]$.
\item Use the variational principles Eq.(\ref{stat_principle_2_generic}) to obtain a self consistent equation on $G$.
\item Evaluate the approximate functional $\G_2$ at these particular values of $\ph$ and $G$ to get $\ln Z$.
\end{itemize}

A crucial property of these truncation schemes is that if the Gaussian part and the non-Gaussian part of the action
are separately invariant under a linear symmetry, then the deduced self-consistent equations for the propagator will also 
possess this symmetry \cite{ABL06}. If the non-Gaussian action is invariant, and if the diagrams are evaluated at the 
physical propagator, which respects this invariance, then automatically all the 2PI diagrams will also be invariant, and so 
will be $\Si$, whatever the chosen truncation of the diagrammatic series. This is true only for a linear symmetry, for 
which the consequences on the propagators are simply relations between them, that can be enforced when evaluating the
diagrams, whereas for a non-linear symmetry, the propagators must satisfy relations that involve higher-order correlation 
functions, and building a consistent expansion would then require to keep a possibly infinite number of terms in the 
expansion.

We will exploit this property of the 2PI expansions in Chapter \ref{chap:dynamics1} in order to build self-consistent 
equations for the non-ergodicity parameter that preserves the time-reversal.

\section{Theory of liquids}

We review here the equivalent of the discussion above for the case of equilibrium liquid theory. In that case, the starting
point for systematic expansions is chosen to be the ideal gas, and thus the diagrammatic formulation is different from that
presented above. However, the idea is still to perform a double Legendre transform to obtain a functional of both 
mean-field and propagator. This leads to the standard hyper-netted chain approximation, which we will use in the following chapters
of this manuscript, and provides the basis for the replica formulation of Chapter \ref{chap:replicas}.

We consider spherical particles of diameter $\s$, the center of which is located at position $x_i$ in a \mbox{3-dimensional}
space. The particles interact via a pair potential $v$, have a chemical potential $\m$ and are placed in an external field 
$\Psi$. Finally the system is in contact with a thermal bath at temperature $T$, and can exchange particles with a particle 
reservoir, so that the probability $P_N$ of observing $N$ particles located at $x_1,\cdots,x_N$ is given by the following 
Boltzmann weight:
\begin{align}
P_N[x_1,\cdots,x_N] = e^{- \b \sum_{i<j} v(x_i-x_j) + \sum_{j=1}^N \left( \b \m - \b \P(x_i) \right)} , \label{boltzmann}
\end{align}
where $\b$ is the inverse temperature $1/ k_B T$, $k_B$ is the Boltzmann constant.
We will be interested in computing the microscopic density defined in Eq.(\ref{def_micro_density_liq}) and its cumulants.
For simplicity in the following, three-dimensional space coordinates will be
replaced by numerical subscripts, we gather the chemical potential and
external field in a single field $\n$, and we define a dimensionless potential $w$:
\begin{align}
& \n_1 = \b \m - \b \P_1 , \\
& w_1 = -\b v_1 .
\end{align}
Note that we will most of the time stick to the notation $w$ in the rest of the manuscript. \\

The probability $P_N$ can be rewritten in terms of these quantities as:
\begin{align}
P_N[\hr,\n] = e^{\frac 12 w_{12} \left( \hr_1 \hr_2 - \hr_1 \d_{12} \right) + \n_1 \hr_1} ,
\end{align}
so that the equivalent of the action $S$ in the previous section is the potential term 
$\frac 12 w_{12} \left( \hr_1 \hr_2 - \hr_1 \d_{12} \right)$, the field $\f$ is replaced by $\hr$ and the field $J$ is replaced 
by $\n$.

Finally the partition function is obtained by summing over all possible configurations, i.e. summing over the number of 
particles $N$ and integrating over all variables $x_i$:
\beq \begin{split}
& Z[\n,w] = \Tr  e^{ \frac 12 w_{12} \left( \hr_1 \hr_2 - \hr_1 \d_{12} \right) 
+ \n_1 \hr_1} , \\
& W[\n,w] = \ln Z[\n,w] \ .
\end{split} \eeq
where $\Tr$ must now be understood as $\sum_{N=0}^\io \frac 1 {N!} \int dx_1 \cdots dx_N$.

Although it may seem redundant, it will prove useful in the following to consider that the Boltzmann weight is a functional
of two fields, $\hr$ and $\hr^{(2)}$, with $\hr^{(2)}$ defined as:
\beq
\hr^{(2)} = \hr_1 \hr_2 - \hr_1 \d_{12} . \label{def_ro2}
\eeq
This gives:
\beq
Z[\n,w] = \Tr  e^{ \frac 12 w_{12} \hr^{(2)}_{12} + \n_1 \hr_1} .
\eeq
Since $\n$ is the field conjugated to $\hr$, $W[\n]$ is the generating functional of the cumulants of the density, defined in the same
way as in Eqs.(\ref{cumulants_1}--\ref{cumulants_3}):
\begin{align}
\r_1 \equiv \frac{\d W[\n,w]}{\d \n_1} ,
\end{align}
and the propagator and third order cumulant are defined as in the previous section.
More interestingly, $w$ is coupled to the function $\hr^{(2)}$:
\begin{align}
& \frac{\d W[\n,w]}{\d w_{12}} = \frac 12 \r^{(2)}_{12} \equiv  \frac 12 \la \hr^{(2)}_{12} \ra . \label{link_w_ro2}
\end{align}
The correlation function $\r^{(2)}$ is called the two-point density.
From this correlation function we simply recover the radial distribution \mbox{function $g$} and the pair correlation function $h$:
\begin{align}
& g_{12} = \frac {\r^{(2)}_{12}}{\r_1 \r_2} , \label{def_g} \\
& h_{12} = g_{12} - 1 . \label{def_h}
\end{align}
The advantage of the pair correlation function $h$ over $g$ is that it is a clustering function: it decays to zero at large 
separation between $1$ and $2$. Since $\hr^{(2)} = \hr_1 \hr_2 - \hr_1 \d_{12}$, the three functions are related to the propagator $G$ of the theory:
\begin{align}
& G_{12} = \la \hr_1 \hr_2 \ra - \r_1 \r_2 , \\
& \r^{(2)}_{12} = G_{12} + \r_1 \r_2 - \r_1 \d_{12} , \\
& g_{12} = \frac{G_{12}}{\r_1 \r_2} - \frac 1{\r_1} \d_{12} + 1 , \\
& h_{12} = \frac{G_{12}}{\r_1 \r_2} - \frac 1{\r_1} \d_{12} .
\end{align}
Finally, for homogeneous liquids, the structure factor of Eq.(\ref{S_k_intro}) is related to the Fourier transform of $h$ by:
\beq
S_k = 1 + \r h_k = \frac 1 \r G_k \label{def_Sk} .
\eeq
Note that the structure factor is sometimes defined as $1+\r g_k$, the difference between the two definitions being a 
delta function at $k=0$. This term represents the unscattered light in a light diffusion experiment, and is most of the times
eliminated by looking at the transmitted light under a small angle.

In the same way than in the field-theoretic case, we also define the first Legendre transform of the free-energy
(note the difference in the choice of sign):
\beq 
\left\{ \begin{array}{ll}
& \displaystyle \G_1[\r,w] = \ln Z[\n^*[\r],w] - \int_1 \n^*_1[\r] \r_1 \ , \\
& \\
& \displaystyle \n^*[\r] \text{ such that } \left. \frac{\d \ln Z[\n,w]}{\d \n_1} \right|_{\n^*} = \r_1 \ ,
\end{array} \right.
\label{LT1_liquid}
\eeq
As well as the second Legendre transform:
\beq
\left\{ \begin{array}{ll}
& \displaystyle \G_2[\r,\r^{(2)}] = \G_1[\r,w^*[\r,\r^{(2)}]] - \frac 12 \int_{12} w^*_{12}[\r,\r^{(2)}] \r^{(2)}_{12} \ , \\
& \\
& \displaystyle w^*[\r,\r^{(2)}] \text{ such that } \left. \frac{\d \ln Z[\n,w]}{\d w_{12}} \right|_{w^*} = \frac 12 \r^{(2)}_{12} \ .
\end{array} \right.
\label{LT2_liquid}
\eeq
Although it is fully equivalent, we will see that in the case of liquid theory, it is simpler to perform the two Legendre transformations 
successively.

\subsection{The case of the ideal gas}

In the case $v=0$, the partition function simplifies into:
\begin{align}
Z[\n,w=0] & = \sum_{N=0}^\io \frac 1 {N!} \int_{1,\cdots,N} e^{\sum_{i=1}^N \n_i} 
= \sum_{N=0}^\io \frac 1 {N!} \left(\int_1 e^{\n_1} \right)^N \nonumber \\
& = \exp \left( \int_1 e^{\n_1} \right) . \label{z_ideal_gas}
\end{align}
Note that we have thus $Z[\n=0,w=0]=e^V$, which is a diverging normalization constant in the thermodynamic limit
Which leads to:
\begin{align}
W[\n,w=0] = \int_1 e^{\n_1} .
\end{align}
We can thus calculate the density at fixed chemical potential and external field:
\beq
\r_1[\n,w=0] = e^{\n_1} ,
\eeq
and inverting this relation we find that the value of $\n$ that fixes $\r_1$ as density is:
\beq
\n^*_1[\r] = \ln \r_1.
\eeq
This allows to perform the Legendre transform of $W[\n,w=0]$ with respect to $\n$ to get:
\begin{align}
\G_1[\r,w=0] & = W[\n^*[\r]] - \int_1 \r_1 \n^*_1[\r] =  \int_1 e^{\ln \r_1} - \int_1 \r_1 \ln \r_1 \nonumber \\
& = \int_1 \r_1 \left[ 1 - \ln \r_1 \right] ,
\end{align}
which is the expected ideal gas term in the free-energy.
We can also calculate the different two-point functions defined above:
\begin{align}
& G_{12}[\n,w=0] = e^{\n_1} \d_{12} , \\
& \r^{(2)}_{12}[\n,w=0] = e^{\n_1} e^{\n_2} , \\
& g_{12}[\n,w=0] = 1 , \\
& h_{12}[\n,w=0] = 0 ,
\end{align}
and the derivatives of the functional $\G$ to get:
\begin{align}
& \G^{(1)}_1[\r,w=0] = -\ln \r_1 , \\
& \G^{(n)}_{1 \cdots n}[\r,w=0] = -\frac{(-1)^{n}}{(n-1) \r_1^{n-1}} \prod_{i=2}^n \d_{1i} .
\end{align}
We also see that the correlation functions defined by successive functional differentiation with respect to $\r$ are non-zero 
at every order, which leads to define the non-ideal gas part of these correlation functions, that are called the direct correlation
functions in liquid theory:
\begin{align}
c^{(n)}_{1 \cdots n}[\r,w] \equiv \frac{\d \left[ \G[\r,w] - \G[\r,w=0] \right]}{\d \r_1 \cdots \d \r_n} 
= \G^{(n)}_{1 \cdots n}[\r,w] + \frac{(-1)^{n}}{(n-1) \r_1^{n-1}} \prod_{i=2}^n \d_{1i} \label{def_direct-corr-func}
\end{align}
In the same way, we can compute the higher-order correlation functions of the density by successive differentiation, and we obtain:
\beq
W^{(n)}_{1 \cdots n}[\n,w=0] = e^{\n_1} \prod_{i=2}^n \d_{1i}
\eeq
This shows that the ideal gas, as simple as it may appear, is a highly non-gaussian theory: it has non-vanishing cumulants at all order.
An expansion around the ideal gas is thus bound to have a very different structure than an expansion around a Gaussian theory, as
we did in the previous section, and we can not blindly use this approach. However, the procedure of diagrammatic reduction by successive Legendre 
transformation is still useful in order to obtain a microscopic theory in terms of the order parameter of the transition that we are interested in:
in our case it is the correlations of the density.

\subsection{First Legendre transform: the virial expansion}

Coming back to the initial expression of the partition function, we see that it is naturally expressed as an infinite sum 
of integrals. A natural choice for the diagrams if we want to expand around the ideal gas is to choose the activity $z=e^\n$ for the nodes
of the diagrams, since a Legendre transform with respect to $\n$ will replace these nodes by simply $\r$, our field of 
interest. For the links of the diagrams, it would seem logical to choose $w$, but it is obviously unadapted to the 
canonical hard-sphere case, since in that case $w$ is singular. For hard-spheres, despite the singularity of $w$, the partition 
function is well defined since the potential only appears through factors $e^{w}$, which is a step function. Thus we will choose 
$e^{w}$ as link functions in our diagrammatic expansion. However, it decays to $1$ 
at large separations and thus may lead to divergent integrals. Finally, the good choice is to introduce the Mayer function 
$f_{12} = e^{w_{12}} - 1$, which has all the desired properties: it decays to zero
at large separation between $1$ and $2$, is strictly equal to zero in the ideal gas limit, and is finite in the hard-sphere
case. We will see in the following that it is also a good choice to perform the second Legendre transform described in 
Section \ref{sec_generic_fieldtheory}. Noticing now that any term in the expression of $Z$ is of the form:
\begin{align}
\frac 1{N!} \int_{1,\cdots,N} \left( \prod_{i=1}^N e^{\n_i} \right) \prod_{i < j} \left(1 + f_{ij} \right) ,
\end{align}
we see that they can be expressed as a sum of diagrams \cite{MayerMayer}, with each integration point having a factor 
$z$ attached to it,
and the points can either be separated (which is understood as a product as in the previous section) or joined by a 
$f$ link. Here two points cannot be joined by more than one line. The partition function can thus be pictured as:
\begin{align}
Z = & 1 +  \begin{minipage}[1,1]{1cm} \includegraphics[width=.31cm]{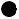} \end{minipage}
\hspace{-.7cm}+ \begin{minipage}[1,1]{1cm} \includegraphics[width=1cm]{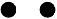} \end{minipage}
+ \begin{minipage}[1,1]{1cm} \includegraphics[width=1cm]{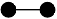} \end{minipage} 
+ \begin{minipage}[1,1]{1cm} \includegraphics[width=1cm]{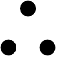} \end{minipage} 
+ \begin{minipage}[1,1]{1cm} \includegraphics[width=1cm]{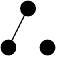} \end{minipage} 
+ \begin{minipage}[1,1]{1cm} \includegraphics[width=1cm]{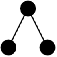} \end{minipage} 
+ \begin{minipage}[1,1]{1cm} \includegraphics[width=1cm]{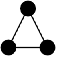} \end{minipage} 
+ \begin{minipage}[1,1]{1cm} \includegraphics[width=1cm]{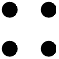} \end{minipage}
+ \begin{minipage}[1,1]{1cm} \includegraphics[width=1cm]{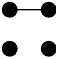} \end{minipage} 
+ \begin{minipage}[1,1]{1cm} \includegraphics[width=1cm]{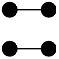} \end{minipage} \nonumber \\
& + \begin{minipage}[1,1]{1cm} \includegraphics[width=1cm]{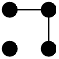} \end{minipage} 
+ \begin{minipage}[1,1]{1cm} \includegraphics[width=1cm]{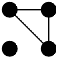} \end{minipage} 
+ \begin{minipage}[1,1]{1cm} \includegraphics[width=1cm]{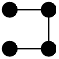} \end{minipage} 
+ \begin{minipage}[1,1]{1cm} \includegraphics[width=1cm]{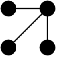} \end{minipage} 
+ \begin{minipage}[1,1]{1cm} \includegraphics[width=1cm]{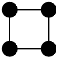} \end{minipage} 
+ \begin{minipage}[1,1]{1cm} \includegraphics[width=1cm]{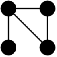} \end{minipage} 
+ \begin{minipage}[1,1]{1cm} \includegraphics[width=1cm]{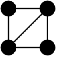} \end{minipage} 
+ \begin{minipage}[1,1]{1cm} \includegraphics[width=1cm]{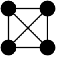} \end{minipage} 
+ \ldots
\label{Z_mayer}
\end{align}
Here we have shown all diagrams up to the order $N=4$ in the number of particles in the system.
The value of a given diagram, if it was labelled, would be that of the integral it represents:
\beq
\begin{minipage}[1,1]{1cm} \includegraphics[width=1cm]{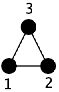} \end{minipage} 
~ = \int_{123} z_1 f_{12} z_2 f_{23} z_3 f_{31} \ .
\eeq 
An unlabeled diagram is the sum of all possible corresponding labelled diagrams.
For a diagram with $N$ black dots, there are $N!$ such relabelings, and thus
unlabeling the diagrams takes care of the $N!$ factor present in the definition of the 
partition function.
We see that the ideal gas limit is obtained by setting all lines to zero, thus there only remains the diagrams that are 
composed of
disconnected points, the sum of which gives the exponential form of $Z$ given in Eq.(\ref{z_ideal_gas}). 
If we were to expand
around the ideal gas now, at each order in, for example, $w$, we would similarly have to re-sum an infinity of diagrams to 
obtain the correct value of $Z$ to this order, which is not satisfactory. However, we will see that performing the reductions 
of diagrams described in Section \ref{sec_generic_fieldtheory} naturally corrects this problem.

We can now follow the prescription of Section \ref{sec_generic_fieldtheory} and begin by taking the logarithm of the 
partition function. This suppresses the disconnected diagrams (a diagrammatic proof can be found in \cite{MH61}), i.e. the
ones that are products of others, which gives
\beq
\ln Z = \begin{minipage}[1,1]{1cm} \includegraphics[width=.31cm]{liq1} \end{minipage}
\hspace{-.7cm} + ~ \begin{minipage}[1,1]{1cm} \includegraphics[width=1cm]{liq3} \end{minipage} ~ 
+ \begin{minipage}[1,1]{1cm} \includegraphics[width=1cm]{liq6} \end{minipage} 
+ \begin{minipage}[1,1]{1cm} \includegraphics[width=1cm]{liq7} \end{minipage} 
+ \begin{minipage}[1,1]{1cm} \includegraphics[width=1cm]{liq8} \end{minipage} 
+ \begin{minipage}[1,1]{1cm} \includegraphics[width=1cm]{liq9} \end{minipage} 
+ \begin{minipage}[1,1]{1cm} \includegraphics[width=1cm]{liq10} \end{minipage} 
+ \begin{minipage}[1,1]{1cm} \includegraphics[width=1cm]{liq11} \end{minipage} 
+ \begin{minipage}[1,1]{1cm} \includegraphics[width=1cm]{liq12} \end{minipage} 
+ \begin{minipage}[1,1]{1cm} \includegraphics[width=1cm]{liq13} \end{minipage} 
+ \ldots
\label{logZ_mayer}
\eeq
The ideal gas partition function is reduced to the single dot at the beginning of the expansion. 
For the moment, $\ln Z$ is a functional of both $\n$ and $w$, though only indirectly via $z=e^{\n}$
and $f=e^w-1$.

\subsubsection{First Legendre transform via an expansion in powers of $f$}

To perform the first Legendre transform defined in Eq.(\ref{LT1_liquid}), we will use the $f$ lines
as an organizing device for the diagrammatic series we will be using.
We will show the result to first orders, then generalize the result to all orders in $f$.
At third order in $f$ we have:
\beq
\ln Z = \begin{minipage}[1,1]{1cm} \includegraphics[width=.31cm]{liq1} \end{minipage}
\hspace{-.7cm} + ~ \begin{minipage}[1,1]{1cm} \includegraphics[width=1cm]{liq3} \end{minipage} ~ 
+ \begin{minipage}[1,1]{1cm} \includegraphics[width=1cm]{liq6} \end{minipage} 
+ \begin{minipage}[1,1]{1cm} \includegraphics[width=1cm]{liq7} \end{minipage} 
+ \begin{minipage}[1,1]{1cm} \includegraphics[width=1cm]{liq8} \end{minipage} 
+ \begin{minipage}[1,1]{1cm} \includegraphics[width=1cm]{liq9} \end{minipage} 
+ \OO(f^4) \ .
\eeq
We can evaluate the density by differentiating with respect to $\n$, or in a simpler way 
with respect to $z$:
\beq
\r_1 = \frac{\d \ln Z}{\d \n_1} = \int_3\frac{\d z_3}{\d \n_1} \frac{\d \ln Z}{\d z_3} = z_1 \frac{\d \ln Z}{\d z_1} \ .
\eeq
Diagrammatically differentiating with respect to $z_1$ amounts to replace 
one black dot by a white dot labeled by $1$ (which is just an unintegrated constant weight equal to $1$).
We thus find:
\beq
\frac{\r_1}{z_1} = 1 + ~ \begin{minipage}[1,1]{1cm} \includegraphics[width=1cm]{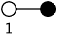} \end{minipage} ~ 
+ \begin{minipage}[1,1]{1cm} \includegraphics[width=1cm]{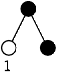} \end{minipage} 
+ \begin{minipage}[1,1]{1cm} \includegraphics[width=1cm]{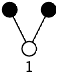} \end{minipage} 
+ \begin{minipage}[1,1]{1cm} \includegraphics[width=1cm]{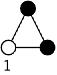} \end{minipage} 
+ \begin{minipage}[1,1]{1cm} \includegraphics[width=1cm]{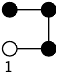} \end{minipage} 
+ \begin{minipage}[1,1]{1cm} \includegraphics[width=1cm]{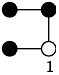} \end{minipage} 
+ \begin{minipage}[1,1]{1cm} \includegraphics[width=1cm]{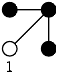} \end{minipage} 
+ \begin{minipage}[1,1]{1cm} \includegraphics[width=1cm]{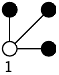} \end{minipage} 
+ \OO(f^4) \ .
\label{density_asafunction_of_z}
\eeq
Our strategy to compute the Legendre transform will be to find a diagrammatic expression of 
$\n^*[\r]$, which we know is by definition (minus) the first derivative of the $\G_1$ functional, then integrate
this expression with respect to density to obtain the result.

In order to obtain an expression for $\n$, we have to take the logarithm of $z$.
Taking the logarithm of Eq.(\ref{density_asafunction_of_z}) will suppress the diagrams in which the white circle 
is an articulation circle, since such diagrams are product of simpler subdiagrams. We will thus obtain:
\beq
\ln \left( \frac{\r_1}{z_1} \right) =  \begin{minipage}[1,1]{1cm} \includegraphics[width=1cm]{liq3_ro} \end{minipage} ~ 
+ \begin{minipage}[1,1]{1cm} \includegraphics[width=1cm]{liq6_ro} \end{minipage} 
+ \begin{minipage}[1,1]{1cm} \includegraphics[width=1cm]{liq7_ro} \end{minipage} 
+ \begin{minipage}[1,1]{1cm} \includegraphics[width=1cm]{liq8_ro} \end{minipage} 
+ \begin{minipage}[1,1]{1cm} \includegraphics[width=1cm]{liq9_ro} \end{minipage} 
+ \OO(f^4) \ .
\eeq
We put together the two results:
\beq
\left\{ \begin{array}{ll}
& \displaystyle \n_1 = \ln \r_1[\n] - \begin{minipage}[1,1]{1cm} \includegraphics[width=1cm]{liq3_ro} \end{minipage} ~ 
- \begin{minipage}[1,1]{1cm} \includegraphics[width=1cm]{liq6_ro} \end{minipage} 
- \begin{minipage}[1,1]{1cm} \includegraphics[width=1cm]{liq7_ro} \end{minipage} 
- \begin{minipage}[1,1]{1cm} \includegraphics[width=1cm]{liq8_ro} \end{minipage} 
- \begin{minipage}[1,1]{1cm} \includegraphics[width=1cm]{liq9_ro} \end{minipage} 
+ \OO(f^4) \ , \\
& \\
& \displaystyle z_1 = \r_1 - ~ \begin{minipage}[1,1]{1cm} \includegraphics[width=1cm]{liq3_ro} \end{minipage} ~ 
- \begin{minipage}[1,1]{1cm} \includegraphics[width=1cm]{liq6_ro} \end{minipage} 
- \begin{minipage}[1,1]{1cm} \includegraphics[width=1cm]{liq6_2_ro} \end{minipage} 
- \begin{minipage}[1,1]{1cm} \includegraphics[width=1cm]{liq7_ro} \end{minipage} 
- \begin{minipage}[1,1]{1cm} \includegraphics[width=1cm]{liq8_ro} \end{minipage} 
- \begin{minipage}[1,1]{1cm} \includegraphics[width=1cm]{liq8_2_ro} \end{minipage} 
- \begin{minipage}[1,1]{1cm} \includegraphics[width=1cm]{liq9_ro} \end{minipage} 
- \begin{minipage}[1,1]{1cm} \includegraphics[width=1cm]{liq9_2_ro} \end{minipage} 
+ \OO(f^4) \ .
\end{array} \right.
\label{nuandz_asafunctionof_z}
\eeq
Note that in the second line, white circles are weighted by $z$, while in the first one they are not weighted at all.
We can now solve iteratively the second line in order to obtain the expression of $z$ as a function only of density, then
reinsert this expression in the expression of $\n$ to obtain:
\beq
\n_1 = \ln \r_1[\n] - \begin{minipage}[1,1]{1cm} \includegraphics[width=1cm]{liq3_ro} \end{minipage} ~ 
- \begin{minipage}[1,1]{1cm} \includegraphics[width=1cm]{liq7_ro} \end{minipage} 
+ \OO(f^4) \ ,
\eeq
where the black dots are now $\r[\n]$ functions, lines are still $f$ functions, and white dots are constant (equal to 1) functions.
We can now evaluate this equation at $\n^*[\r]$, which will have the effect to replace the black $\r[\n]$ dots by $\r$ dots, which 
gives:
\beq
\n_1^*[\r] = - \frac{\d \G_1[\r]}{\d \r_1} = \ln \r_1 
- \begin{minipage}[1,1]{1cm} \includegraphics[width=1cm]{liq3_ro} \end{minipage}
- \begin{minipage}[1,1]{1cm} \includegraphics[width=1cm]{liq7_ro} \end{minipage} 
+ \OO(f^4) \ .
\eeq
Finally integrating this with respect to density leads to the final expression of $\G_1$ (the integration constant 
is strictly zero as can be checked by evaluating in the ideal-gaz):
\beq
\G_1[\r] = \int_1 \r_1 \left[ 1 - \ln \r_1 \right] + 
\begin{minipage}[1,1]{1cm} \includegraphics[width=1cm]{liq3} \end{minipage} 
+ \begin{minipage}[1,1]{1cm} \includegraphics[width=1cm]{liq7} \end{minipage} 
+ \OO(f^4) \ .
\eeq

This procedure can be continued at all orders in $f$, and one can show \cite{MayerMayer,MH61} that 
the full Legendre transform $\G_1$ is represented by the sum of all unlabeled, connected diagrams 
with $\r$ nodes and $f$ lines, that are free of articulation circles, i.e. circles that cut the diagrams in two 
separate parts when removed. 
This can easily be checked order by order, or by iteration, but the complete proof is more tedious.
Thus we obtain the result:
\begin{align}
\G_1[\r] = \int_1 \r_1 \left[ 1 - \ln \r_1 \right] 
+ \begin{minipage}[1,1]{1cm} \includegraphics[width=1cm]{liq3} \end{minipage} 
+ \begin{minipage}[1,1]{1cm} \includegraphics[width=1cm]{liq7} \end{minipage} 
+ \begin{minipage}[1,1]{1cm} \includegraphics[width=1cm]{liq10} \end{minipage} 
+ \begin{minipage}[1,1]{1cm} \includegraphics[width=1cm]{liq12} \end{minipage} 
+ \begin{minipage}[1,1]{1cm} \includegraphics[width=1cm]{liq13} \end{minipage} 
+ \ldots \label{expansion_1_liq}
\end{align}
This is the so-called virial expansion of liquid theory. The diagrams carry $\r$ nodes and $f$ lines.
They are all the connected 1-irreducible diagrams, i.e. that do not disconnect upon removal of
a node.

In the same way than the previous section, if we want to obtain approximations 
for two-point functions, we would have to perform another Legendre transform, this time with respect to $w$.

\subsection{Second Legendre transform: Morita \& Hiroike functional}

We now want to perform another Legendre transformation, this time with respect to $w$.
Before turning to this, we will define a new functional, the excess free-energy, defined as
the non-ideal gaz part of $\G_1$:
\beq
\G_{\rm{ex}}[\r,w] = \G_1[\r,w] - \int_1 \r_1 \left[ 1 - \ln \r_1 \right] \ .
\eeq
Inserting the diagrammatic expression of $\G_1$ we get:
\beq
\G_{\rm{ex}}[\r,w] = \begin{minipage}[1,1]{1cm} \includegraphics[width=1cm]{liq3} \end{minipage} 
+ \begin{minipage}[1,1]{1cm} \includegraphics[width=1cm]{liq7} \end{minipage} 
+ \begin{minipage}[1,1]{1cm} \includegraphics[width=1cm]{liq10} \end{minipage} 
+ \begin{minipage}[1,1]{1cm} \includegraphics[width=1cm]{liq12} \end{minipage} 
+ \begin{minipage}[1,1]{1cm} \includegraphics[width=1cm]{liq13} \end{minipage} + \cdots \ .
\eeq
The Legendre transform of $\G_{\rm{ex}}$ with respect to $w$ is the same as that of $\G_1$ since they differ only
by a term independant of $f$.

We can no longer use the $f$ lines as an organizing device for the diagrammatic expansions, since we want to
Legendre transform with respect to $w$, but we can now use the density as an organizing device.
We will thus compute everything at order $4$ in density, and then present the result generalized at all orders in density. 

We note first again that the dependance of $\G_{\rm{ex}}$ on $w$ is only through $f=e^w-1$.
We already know with Eq.(\ref{link_w_ro2}) that $w$ is coupled to the two-point density 
$\r^{(2)}$, so if we perform a differentiation of $\G_{\rm{ex}}$ with respect to $f$ we will obtain:
\begin{align}
& \frac{\d \G_{\rm{ex}}}{\d f_{12}} = \int_{34} \frac{\d w_{34}}{\d f_{12}} \frac{\d \G_1}{\d w_{34}} = \int_{34} \frac{\d \ln(1+f_{34})}{\d f_{12}} 
\left. \frac{\d \ln Z[\n]}{\d w_{34}} \right|_{\n^*} \nonumber \\
\Leftrightarrow ~ & (1+f_{12}) \frac{\d \G_{\rm{ex}}}{\d f_{12}} = \frac 12 \r^{(2)}_{12} \ . 
\label{def_ro2_gamma1-of-f}
\end{align}
Differentiating a diagram of Eq.(\ref{expansion_1_liq}) with respect to $f_{12}$ amounts to erasing a line and labeling 
$1$ and $2$ the two points that were joined by this line. In every diagram, the two densities attached to the nodes $1$ 
and $2$ will factor out and give an overall factor $\r_1 \r_2$. In the following an open point will be a point without a
density weight whose coordinates are not integrated over. The definition of the two point density gives then \cite{MM41a}:
\begin{align}
g_{12}[w] = \frac{\r^{(2)}_{12}[w]}{\r_1 \r_2} = e^{w_{12}} \left( 1 
+ \begin{minipage}[1,1]{1cm} \includegraphics[width=1cm]{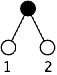}  \end{minipage}
+ \begin{minipage}[1,1]{1cm} \includegraphics[width=1cm]{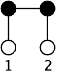}  \end{minipage}
+ \begin{minipage}[1,1]{1cm} \includegraphics[width=1cm]{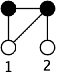}  \end{minipage}
+ \begin{minipage}[1,1]{1cm} \includegraphics[width=1cm]{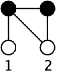}  \end{minipage}
+ \begin{minipage}[1,1]{1cm} \includegraphics[width=1cm]{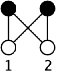}  \end{minipage}
+ \begin{minipage}[1,1]{1cm} \includegraphics[width=1cm]{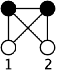}  \end{minipage} \right) + \OO(\r^3) \ ,
\label{g_asafunctionof_f}
\end{align}
which means that the function $h$ behaves like $e^{w}-1 = f$ to lowest order in density. This is also an {\it a-posteriori}
justification for the choice of $f$ as links in the diagrammatic expansion: Legendre transforming with respect to $w$ will
have the effect of replacing $f$ links by $h$ functions \cite{Mee60}, which have the good behavior for us: it cancels in the 
ideal gas limit and decays to zero at large separations in all cases, giving rise to well-behaved integrals.

We can now obtain a diagrammatic expression of $w$ by taking the logarithm of Eq.(\ref{g_asafunctionof_f}).
As can be seen order by order, and as expected, taking the logarithm will cancel diagrams that are the product of subdiagrams, and 
we obtain:
\beq 
w_{12} = \ln ( 1 + h_{12}[w] ) 
- \begin{minipage}[1,1]{1cm} \includegraphics[width=1cm]{liq_g_1}  \end{minipage}
- \begin{minipage}[1,1]{1cm} \includegraphics[width=1cm]{liq_g_2}  \end{minipage}
- \begin{minipage}[1,1]{1cm} \includegraphics[width=1cm]{liq_g_3}  \end{minipage}
- \begin{minipage}[1,1]{1cm} \includegraphics[width=1cm]{liq_g_4}  \end{minipage}
- \begin{minipage}[1,1]{1cm} \includegraphics[width=1cm]{liq_g_6}  \end{minipage}
+ \OO(\r^3) \  .
\label{w_asafunctionof_f}
\eeq
In order to replace all $f$ lines in this expression by $h$ lines, which will allow for the Legendre transformation,
we also need the corresponding diagrammatic expansion of $f$ as a function of $h$. This is obtained by iteration of
Eq.(\ref{g_asafunctionof_f}), rewritten in a slightly different way:
\beq
f_{12} = h_{12}[w] - (1+f_{12}) \left( 
\begin{minipage}[1,1]{1cm} \includegraphics[width=1cm]{liq_g_1}  \end{minipage}
+ \begin{minipage}[1,1]{1cm} \includegraphics[width=1cm]{liq_g_2}  \end{minipage}
+ \begin{minipage}[1,1]{1cm} \includegraphics[width=1cm]{liq_g_3}  \end{minipage}
+ \begin{minipage}[1,1]{1cm} \includegraphics[width=1cm]{liq_g_4}  \end{minipage}
+ \begin{minipage}[1,1]{1cm} \includegraphics[width=1cm]{liq_g_5}  \end{minipage}
+ \begin{minipage}[1,1]{1cm} \includegraphics[width=1cm]{liq_g_6}  \end{minipage} \right) + \OO(\r^3)
\eeq
We can solve this iteratively in order to obtain $f$ as a function of $h$ only, then reinject this in the expression of $w$
to obtain $w$ as a function of $h$. Since $f$ lines are always accompanied by at least one $\r$ weight, we only have
to compute the first order expansion of $f$ in order to obtain the second order expansion of $w$. We simply find:
\beq
f_{12} = h_{12}[w] - \begin{minipage}[1,1]{1cm} \includegraphics[width=1cm]{liq_g_1}  \end{minipage}
- \begin{minipage}[1,1]{1cm} \includegraphics[width=1cm]{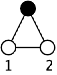}  \end{minipage} + \OO(\r^2) \ , 
\eeq
where the lines on the diagrams are now $h[w]$ lines.
Reinserting in the expression of $w$, we find:
\beq
w_{12} = \ln ( 1 + h_{12}[w] ) 
- \begin{minipage}[1,1]{1cm} \includegraphics[width=1cm]{liq_g_1}  \end{minipage}
+ \begin{minipage}[1,1]{1cm} \includegraphics[width=1cm]{liq_g_2}  \end{minipage}
- \begin{minipage}[1,1]{1cm} \includegraphics[width=1cm]{liq_g_6} \end{minipage} + \OO(\r^3) \ .
\eeq
The lines on the diagrams are of course again $h[w]$ lines.
We can now evaluate this safely at $w^*[\r^{(2)}]$, which will simply amount to replace the $h[w]$ nodes by $h$ nodes:
\beq
w_{12}^*[\r^{(2)}] = \ln ( 1 + h_{12} ) 
- \begin{minipage}[1,1]{1cm} \includegraphics[width=1cm]{liq_g_1}  \end{minipage}
+ \begin{minipage}[1,1]{1cm} \includegraphics[width=1cm]{liq_g_2}  \end{minipage}
- \begin{minipage}[1,1]{1cm} \includegraphics[width=1cm]{liq_g_6} \end{minipage} + \OO(\r^3) \ .
\label{wstar_asafunctionof_h}
\eeq
The lines are now $h$ lines.
To come back to the expression of the free-energy, we notice that:
\beq
- \frac 12 w_{12}^*[\r^{(2)}] = \frac{\d \G_2[\r,\r^{(2)}]}{\d \r^{(2)}_{12}} = \int_{34} \frac{\d h_{34}}{\d \r^{(2)}_{12}} \frac{\d \G_2[\r,\r^{(2)}]}{\d h_{34}} 
= \frac 1{\r_1 \r_2} \frac{\d \G_2[\r,\r^{(2)}]}{\d h_{12}} \ .
\eeq
We can thus multiply Eq.(\ref{wstar_asafunctionof_h}) by $\r_1 \r_2/2$ and integrate with respect to $h_{12}$ to get $\G_2$:
\beq \begin{split}
\G_2[\r,\r^{(2)}] = & \int_1 \r_1 \left[ 1 - \ln \r_1 \right] + \frac 12 \int_{12} \r_1 \r_2 \left[ \frac{}{} h_{12} - (1+h_{12}) \ln (1+h_{12}) \frac{}{} \right] \\
& + \begin{minipage}[1,1]{1cm} \includegraphics[width=1cm]{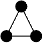}  \end{minipage}
- \begin{minipage}[1,1]{1cm} \includegraphics[width=1cm]{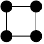}  \end{minipage}
 + \begin{minipage}[1,1]{1cm} \includegraphics[width=1cm]{liq13} \end{minipage} + \OO(\r^2) \ .
\end{split} \eeq
We have fixed the normalization constant by the ideal-gaz case, where we must have $\G_2[\r,h=0]$ that coincides with the ideal gaz result.
Here we see that the Legendre transformation does not give rise to usual 2PI diagrams, meaning diagrams that do not separate in several 
parts when differentiated twice with respect to their lines: there seems to remain ``ring diagrams" in addition to 2PI diagrams.
This is again a consequence of the fact that we did not perform an expansion around a Gaussian theory, but around a Poissonian one.
We can nonetheless get a generalization of this result to all orders in density.

\subsubsection{Generalization to all orders: Morita \& Hiroike functional}

Before doing the second Legendre transformation, we can obtain the expression of the second derivative of $\G_1$ with respect to $\r$, which
is $c_{12}$ by definition. It reads:
\beq
c_{12}[w] = f_{12} + 
\begin{minipage}[1,1]{1cm} \includegraphics[width=1cm]{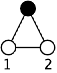}  \end{minipage}
+ \begin{minipage}[1,1]{1cm} \includegraphics[width=1cm]{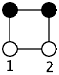}  \end{minipage}
+ \begin{minipage}[1,1]{1cm} \includegraphics[width=1cm]{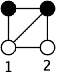}  \end{minipage}
+ \begin{minipage}[1,1]{1cm} \includegraphics[width=1cm]{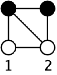}  \end{minipage}
+ \begin{minipage}[1,1]{1cm} \includegraphics[width=1cm]{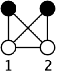}  \end{minipage}
+ \begin{minipage}[1,1]{1cm} \includegraphics[width=1cm]{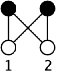}  \end{minipage}
+ \begin{minipage}[1,1]{1cm} \includegraphics[width=1cm]{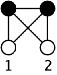}  \end{minipage}
+ \begin{minipage}[1,1]{1cm} \includegraphics[width=1cm]{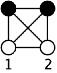}  \end{minipage} + \OO(\r^3)
\eeq
We can go back to the expression of $h$ as a function of $f$ to compare the two:
\beq \begin{split}
h_{12}[w] = f_{12} & + \begin{minipage}[1,1]{1cm} \includegraphics[width=1cm]{liq_g_1}  \end{minipage}
+ \begin{minipage}[1,1]{1cm} \includegraphics[width=1cm]{liq_g_2}  \end{minipage}
+ \begin{minipage}[1,1]{1cm} \includegraphics[width=1cm]{liq_g_3}  \end{minipage}
+ \begin{minipage}[1,1]{1cm} \includegraphics[width=1cm]{liq_g_4}  \end{minipage}
+ \begin{minipage}[1,1]{1cm} \includegraphics[width=1cm]{liq_g_5}  \end{minipage}
+ \begin{minipage}[1,1]{1cm} \includegraphics[width=1cm]{liq_g_6}  \end{minipage} \\ 
& + \begin{minipage}[1,1]{1cm} \includegraphics[width=1cm]{liq_c_1}  \end{minipage}
+ \begin{minipage}[1,1]{1cm} \includegraphics[width=1cm]{liq_c_2}  \end{minipage}
+ \begin{minipage}[1,1]{1cm} \includegraphics[width=1cm]{liq_c_3}  \end{minipage}
+ \begin{minipage}[1,1]{1cm} \includegraphics[width=1cm]{liq_c_4}  \end{minipage}
+ \begin{minipage}[1,1]{1cm} \includegraphics[width=1cm]{liq_c_5}  \end{minipage}
+ \begin{minipage}[1,1]{1cm} \includegraphics[width=1cm]{liq_c_6}  \end{minipage} + \OO(\r^3) \ .
\end{split}
\eeq
We see that in the expression of $w$ in Eq.(\ref{w_asafunctionof_f}) we can recognize exactly, at this order in density,
the diagrammatic expansion of $h-c$, with the exception of one diagram:
\beq
w_{12} = \ln( 1 + h_{12}[w] ) - h_{12}[w] + c_{12}[w] 
- \begin{minipage}[1,1]{1cm} \includegraphics[width=1cm]{liq_c_5_3} \end{minipage} + \OO(\r^3) \ .
\eeq
Since $f$ is $h + \OO(\r)$, we can at lowest order replace again the $f$ lines in the diagram above by $h$ lines, and we obtain
upon evaluating at $w^*[\r^{(2)}]$ the expression:
\beq
w_{12}^*[\r^{(2)}] = \ln( 1 + h_{12} ) - h_{12} + c_{12}[w^*[\r^{(2)}]] 
- \begin{minipage}[1,1]{1cm} \includegraphics[width=1cm]{liq_c_5_3} \end{minipage} + \OO(\r^3) \ .
\eeq
The lines in the diagram above are now $h$ lines.
Morita \& Hiroike \cite{MH61} have shown that to all orders in density, this property is still true, i.e. that we have:
\beq
w_{12}^*[\r^{(2)}] = \ln( 1 + h_{12} ) - h_{12} + c_{12} + \{ \text{ a class of diagrams } \} \ .
\eeq
The diagrams contributing to $w_{12}^*$ are diagrams that have two white circles labelled 1 and 2, $\r$ nodes, $h$ lines, 
and that do not disconnect upon removal of a line and the pair of circles that it connects.

What is the expression of $c$ in function of $h$ ? 
we know that the propagator and the two-point vertex function $\G^{(2)}$ are 
inverse of each other by properties of the Legendre transform. Eq.(\ref{W2_G2_formal}) reads, in the notations of 
liquid state theory:
\begin{align}
& \int_3 G_{13} \G^{(2)}_{32} = \d_{12} , \nonumber \\ 
\Leftrightarrow \quad & c_{12} = h_{12} - \int_3 h_{13} \r_3 c_{32} \label{OZ2} .
\end{align}
This equation can be solved iteratively in powers of $h$ to get:
\begin{align}
c_{12} & = h_{12} + \sum_{n'=1}^\io (-1)^{n'} ~ \int_{1',\ldots,n'} h_{11'} \r_{1'} h_{1'2'} \r_{2'} \cdots \r_{n'} h_{n'2} ,
\label{OZ2_expanded}
\end{align}
Or diagrammatically:
\beq
c_{12} = h_{12} 
- \begin{minipage}[1,1]{1.7cm} \includegraphics[width=1.7cm]{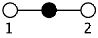}  \end{minipage}
+ \begin{minipage}[1,1]{2.4cm} \includegraphics[width=2.4cm]{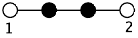}  \end{minipage}
- \begin{minipage}[1,1]{3cm} \includegraphics[width=3cm]{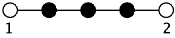}  \end{minipage}
+ ~ \cdots \ ,
\eeq
where the lines of the diagrams are $h$ lines. Thus $c$ is a sum of chain diagrams.
Eq.(\ref{OZ2}) is called the Ornstein-Zernike equation \cite{OZ14,RS53} and was introduced independently of its
interpretation in terms of Legendre transformations. Note that this equation is valid for any choice of potential $w$,
including $w^*[\r^{(2)}]$, and thus is valid also after the second Legendre transformation.

Now if we integrate $h-c$ with respect to $h$, we see that we get an alternate sum of ring diagrams, the first two terms
of which we already obtained when we calculated the Legendre transform at order $\r^4$. The final expression of the
full Legendre transform is thus:
\begin{align}
\G_2[\r,\r^{(2)}] = & \int_1 \r_1 \left[ 1 - \ln \r_1 \right] 
+ \frac 12 \int_{12} \r_1 \r_2 \left[ \frac{}{} h_{12} - \left( 1+h_{12} \right) \ln \left( 1 + h_{12} \right) \frac{}{} \right] \nonumber \\
& + \begin{minipage}[1,1]{1cm} \includegraphics[width=1cm]{liq_ring_1}  \end{minipage}
- \begin{minipage}[1,1]{1cm} \includegraphics[width=1cm]{liq_ring_2}  \end{minipage} 
+ \begin{minipage}[1,1]{1cm} \includegraphics[width=1.5cm]{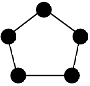}  \end{minipage}\quad \quad +\ldots \\
& - \{ \textrm{2PI diagrams} \} \label{morita}
\end{align}
Although this result was present in Morita's work, a clearer presentation in terms of Legendre transforms is found in
\cite{Do62,Ba64}. The diagrams above have $\r$ on the nodes and $h$ links, are connected, and more importantly, are
2-irreducible: they do not disconnect upon two differentiations with respect to $\r^{(2)}$.

Note that even though the natural parameter entering the diagrammatic expressions is $h$, we emphasize that
$\G_2$ is a functional of $\r$ and $\r^{(2)}$: these are the two independent variables. For example $h$ depends on $\r$ even
when $\r^{(2)}$ is kept constant since we have $\r^{(2)}_{12}/(\r_1 \r_2) - 1 = h_{12}$.
The diagrams above have $\r$ on the nodes and $h$ links, and they do not separate in several parts if one or two $h$ links 
are removed.

\subsubsection{Variational principle and inverse Legendre transform}

Thus to get the variational principle that will give a self-consistent equation on $h$, we perform the functional differentiation
with respect to $h_{12}$ and use Eq.(\ref{stat_principle_2_generic}) to come back again to:
\begin{align}
w_{12} = \ln (1 + h_{12} ) - h_{12} + c_{12} + \frac{2}{\r_1\r_2} \frac{\d }{\d h_{12}} \{ \textrm{2PI diagrams} \} . \label{morita_var-principle}
\end{align}
Since the 2PI diagrams are functionals of $\r$ and $h$, to perform approximations, one way is to select a class of 
2PI diagrams and discard the others, then perform the differentiation, to finally obtain a self-consistent equation bearing 
on $h$ alone. Taking the exponential of this equation, we obtain:
\begin{align}
h_{12} = e^{w_{12}} e^{h_{12} - c_{12} - \frac{2}{\r_1\r_2}\frac{\d }{\d h_{12}} \{ \textrm{2PI diagrams} \}} - 1 \ ,
\label{percusmethodexactequation}
\end{align}
Under this form, this equation is applicable to the hard sphere case since the singular hard-sphere potential
only appears through $e^w$, which is a Heaviside function. We see that this kind of approximations 
will naturally enforce the condition that the pair correlation function is zero for distance smaller than the
particle diameter: two hard-spheres cannot overlap !
The functional derivative of the 2PI diagrams appears under the terminology of ``bridge function".

Eq.(\ref{percusmethodexactequation}) is an exact equation which was derived originally by the Percus method \cite{Pe62}. 
The Percus method uses the following property of the partition function of the liquid: consider a fluid of $N$ particles 
subject to the influence of a $(N+1)$th particle at a fixed position. The effect of this additional particle can be treated in 
perturbation. But a crucial property for homogeneous systems is that when one computes the one-particle distribution 
function of the perturbed liquid, it is in fact equal to the two-particle distribution function of the unperturbed liquid. Thus 
performing a perturbation expansion in powers of the density will give integral equations in terms of the pair correlation 
function of the unperturbed liquid. By this mean, one can show that the bridge function, usually noted $b$, is equal to the 
sum of all third and higher order direct correlation functions:
\begin{align}
b_{12} = - \frac{2}{\r_1 \r_2} \frac{\d }{\d h_{12}} \{ \textrm{2PI diagrams} \} 
= \sum_{n=2}^\io \frac{1}{n!} \int_{1',\ldots,n'} c^{(n+1)}_{11' \ldots n'} \r_{1'} h_{1'2} \cdots \r_{n'} h_{n'2} .
\label{bridge}
\end{align}

\subsection{Hyper-Netted-Chain approximation}

The Hyper-Netted-Chain (HNC) approximation, which was derived independently by Morita \& Hiroike \cite{MH59} and Van Leeuwen, 
Groeneveld and De Boer \cite{LGB59}, simply consists in discarding all the 2PI diagrams in Eq.(\ref{morita}). 
By looking at 
Eq.(\ref{bridge}), we see that it is equivalent to setting all the $c^{(n)}$'s with $n \ge 3$ to zero. Since a large use of this approximation will be made in the rest of the manuscript, a brief overview of the results 
within HNC approximation is presented here.

Note again that the approximation we perform keeps the higher-order correlations of the fluid equal to their ideal-gas values, since it is only 
the $c^{(n)}$ that are approximated to zero. This is a natural consequence of the fact that Mayer diagrams perform an expansion around the 
ideal-gas ground state, and not around a Gaussian theory.

Discarding the 2PI diagrams in Eq.(\ref{morita_var-principle}), we obtain a self consistent equation on $h$, which involves
the direct correlation function $c$, which is related to $h$ by Eq.(\ref{OZ2}). In the homogeneous case, two-point
functions depend only on $r=|1-2|$, the modulus of the difference between the two points, and the HNC approximation 
thus reads in terms of the 1-dimensional functions $h$ and $c$:
\begin{align}
\left\{ \begin{array}{ll}
c(r) & = \displaystyle e^{w(r)}e^{h(r)-c(r)} - 1 - [h(r)-c(r)] \\
h(k) & \displaystyle = \frac{\r c(k)}{1 - \r c(k)} 
\end{array} \right.
\end{align}
These equations are more easily solved in terms of the function $k c(k)$ since, for isotropic functions, the 
three-dimensional Fourier transform reads:
\begin{align}
k c(k) = 4 \p \int_0^\io dr ~ r c(r) \sin(kr) , \\
r c(r) = \frac 1{2 \p^2} \int_0^\io dk ~ k c(k) \sin(kr) .
\end{align}
We will need to solve these equations in the next three chapters: either in the dynamical context to solve the equations 
obtained for the non-ergodicity factor, which needs as an input the liquid structure, or in the context of replica theory where 
a replicated version of HNC is used, or when computing the thermodynamics of the jammed states of harmonic spheres.
The scheme that we used for solving the HNC equations is an adaptation of a Picard iterative scheme:
\begin{itemize}
\item Start from a good guess for $k c(k)$
\item Deduce the corresponding $k h(k)$ and thus $k[h(k) - c(k)]$
\item Inverse Fourier transform to obtain $r[h(r)-c(r)]$
\item Plug into the HNC equation to obtain a new estimate of $r c(r)$
\item Fourier transform and mix the new solution with the old one to get the new estimate for $k c(k)$
\item Iterate the procedure until the absolute value of the maximum of the difference between old and new solution is 
smaller than some prescribed accuracy
\end{itemize}
This kind of iterative scheme can be very sensitive to variations of $c(k)$. Indeed if an abrupt change in $c(k)$ 
occurs, the denominator in the Ornstein-Zernike equation Eq.(\ref{OZ2}) can reach zero, and make the calculation diverge. 
This is why at each step, 
the new solution is incorporated gradually within the old one, to avoid unphysical values of the functions. Also, the starting 
guess in the calculation is important, since the calculation can become unstable if we stand far away from the physical 
solution. Thus, we systematically started our calculations at high temperatures and/or low density, where the virial 
expansion is valid, and where the function $g$ is equal approximately to $e^{w}$. Then when convergence is achieved at 
this high temperature/low density, we gradually increased the density or lowered the temperature to attain the $\r,T$ point 
of interest.

\begin{figure}[ht]
\includegraphics[width=14cm]{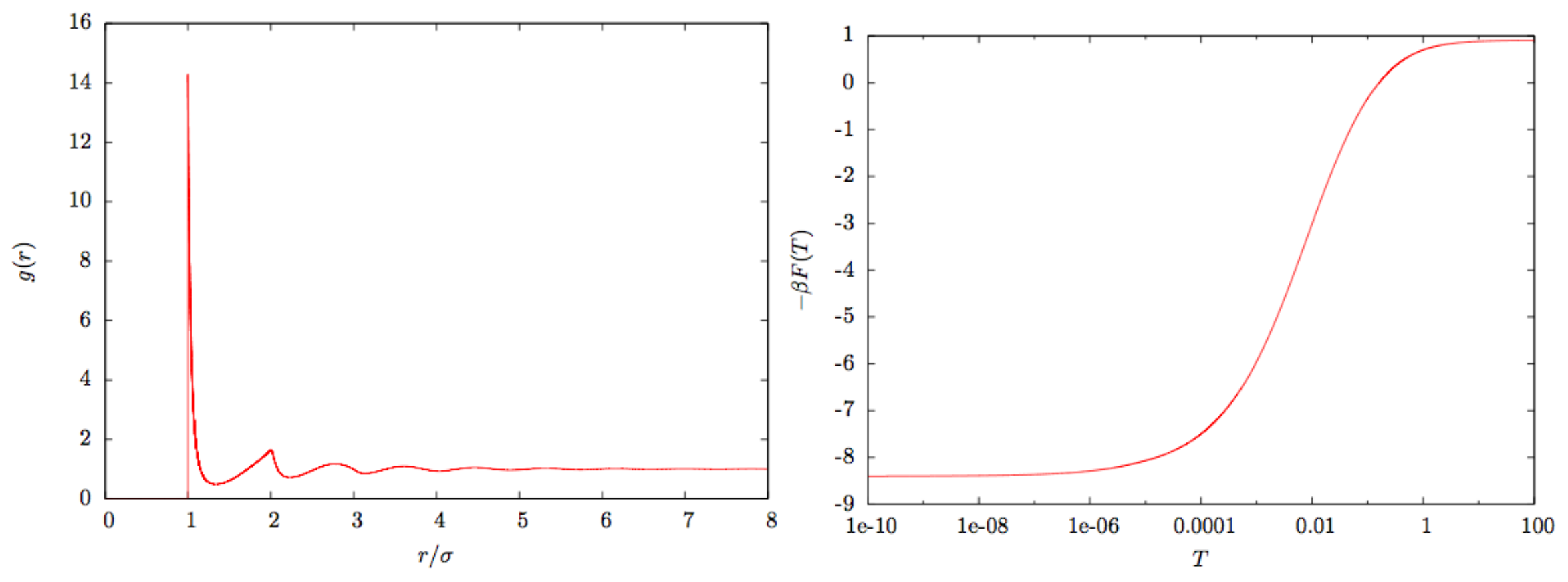}
\caption[Radial distribution function and free-energy of a fluid of harmonic spheres obtained with the HNC approximation]
{Left panel: pair correlation function $g$ as a function of $r/\s$ in the HNC approximation at density $\r \s^3 = 1.1$
and temperature $k_B T / \varepsilon = 3.58 \times 10^{-4}$. Right panel: free energy $-\b F$ as a function of 
temperature, at fixed density $\r \s^3 = 1.1$.}
\label{fig:HNC_1}
\end{figure}

We show in the left frame of Figure \ref{fig:HNC_1} the result for the pair correlation function of harmonic spheres at high 
density $\r = 1.2$ and very low a-dimensional temperature $k_B T / \epsilon = 3.58 \times 10^{-4}$.
This pair correlation function satisfies all the physical requirements: it decays to $1$ at large $r$, is almost equal to zero
for $r < \s$, since at very low temperature, harmonic spheres behave like hard spheres, and thus no overlaps are 
authorized for $r<1$, except on a very narrow regime close to $1$. It presents a very pronounced first peak at $r=\s$, and 
secondary peaks at $r=2\s, 3\s, \ldots$, reflecting the shells of neighbors that develop in very dense liquids. 

Inserting this result into the approximate functional $\G_2$ obtained by discarding the 2PI diagrams, we obtain the 
approximate equation of state of the liquid. We can notice that, in the isotropic and homogeneous case, a Fourier 
transform of one ring diagram reads, for example:
\begin{align}
\begin{minipage}[1,1]{1cm} \includegraphics[width=1cm]{liq_ring_1}  \end{minipage} 
= \frac 1{3!} \int_{123} \r_1 h_{12} \r_2 h_{23} \r_3 h_{31} = \frac V2 \frac 13 \r^3 \int_{k} h(k)^3 ,
\end{align}
so that the sum of all diagrams is a logarithm function in Fourier space, which leads to the following analytic expression:
\begin{align}
\G_2^{\rm{HNC}}[\r,h] = & V \r \left[ 1 - \ln \r \right] - \frac V2 \r^2 \int_r \left( \frac{}{} [1+h(r)] \ln[1+h(r)] - h(r) \right) \nonumber \\
& + \frac V2 \int_k \left( \ln[1+\r h(k)] - \r h(k) + \frac 12 \r^2 h(k)^2 \right) .
\end{align}
When this functional is evaluated at $h^*$, the solution of the variational equation, and the inverse Legendre 
transformation is performed, we obtain the $\G_1$ functional, which is related to the free-energy of the liquid by:
\begin{align}
\b F = - \ln Z = - \G_1[\r] & = - \G_2[\r,h^*] - \frac 12 w_{12} \r_1 \r_2 (1+h^*_{12}) , \\
\Rightarrow \b F^{\rm{HNC}} = & V \r \left[ \ln \r - 1 \right] + \frac V2 \r^2 \int_r \left( \frac{}{}[1+h(r)] (\ln[1+h(r)] - w(r)) - h(r) \right) 
\nonumber \\
& - \frac V2 \int_k \left( \ln[1+\r h(k)] - \r h(k) + \frac 12 \r^2 h(k)^2 \right) .
\end{align}
We show in the right frame of Fig.(\ref{fig:HNC_1}) the resulting free-energy as a function of temperature, at fixed density
$\r \s^3 = 1.1$. When $T \to 0^+$, it converges to the entropy of hard spheres, while increasing towards an ideal gas value
for large temperatures.

\chapter{Dynamics: a field theory approach}
\label{chap:dynamics1}

Independently from an active community that attempts to improve Mode-Coupling Theory, it is believed that a convenient
way to formulate the dynamics of colloids is by resorting to field theory, i.e. finding a diagrammatic approach inspired
from quantum field theory. This step forward would allow us to use all the powerful diagrammatic tools developed in this 
context, such as those described in Chapter \ref{chap:formalism}.

The search for a good field theory for studying the dynamics near the glass transition has a long history by now, and was 
pioneered by Das and Mazenko for coarse grained models \cite{DM86}. However application to realistic models such
as colloidal glasses has been delayed until a Langevin equation for the time-dependent density was derived by Dean 
\cite{De96}. This provided a starting point to formulate field theories for colloidal glasses, but they were soon shown by 
Miyazaki and Reichman \cite{MR05} to be inconsistent with the time-reversibility of the dynamics. A first attempt by Andreanov, 
Biroli and Lef\`evre to go beyond that problem gave unsatisfactory results \cite{ABL06}, due to the difficulty to
correctly deal with the entropic $\r \ln \r$ term inherent to the description of the ideal gas. Finally Kawasaki and Kim \cite{KK07b}
managed to settle this last issue, but at the cost of a very heavy formalism, leaving little hope to extend their approach
to higher orders.

In this chapter we briefly review the field-theoretic methods described above, and introduce an 
original way to formulate a field theory, that allows us to overcome some of their limitations, and more importantly 
provide a compact formalism.

\section{Field theory for supercooled liquids}

A system of $N$ harmonic spheres is placed in a solvent at temperature $T$, that plays the role of a thermal bath.
Each particle of our system is therefore a Brownian particle, that moves according to Brownian dynamics, more easily
written in terms of Langevin equations. The space position of particle $i$ at time $t$, will be denoted by $x_i(t)$, and 
we suppose that each trajectory satisfies the following over-damped Langevin equation:
\begin{align}
\left\{ \begin{array}{ll}
& \displaystyle  \frac{d x_i(t)}{dt}  = F_i(t) + \zeta_i(t) , \\
\vspace{-.2cm} & \\
& \displaystyle F_i(t) = - \nabla_{x_i} \sum_{j \ne i} v(x_i(t) - x_j(t)) , \\
& \displaystyle \la \zeta_i(t) \zeta_j(t) \ra = 2T \d_{ij} \d(t-t) .
\end{array} \right. \label{langevin}
\end{align}
The field $\zeta$ is a three-dimensional Gaussian white noise of variance $2T$, as indicated in Eq.(\ref{langevin}). This 
particular value of the variance ensures that the system converges towards equilibrium at long times, as given by the 
Boltzmann weight of Eq.(\ref{boltzmann}), i.e. $\exp \left( - \b \sum_{i<j} v(x_i-x_j) \right)$. 

This type of Brownian dynamics are much simpler to analyze than the usual Hamiltonian dynamics since no velocities
are involved, and the presence of a fluctuating noise allows us for more direct statistical descriptions. However, apart from differences
between the short time evolutions of Hamiltonian and Brownian dynamics, the glassy features at long times are largely 
independent of the choice of dynamics \cite{SF04}.

\subsection{Formulation in terms of the density and field-theoretic formulation}

We are not interested in being able to follow the trajectory of each individual particle of the system, but rather in being able to 
accurately predict collective quantities, such as correlation functions of the fluctuations of density, as described in the 
introduction. In analogy with the equilibrium theory of liquids, we can define the time-dependent microscopic density of the 
system as:
\begin{align}
\hr(x,t) = \sum_{i=1}^N \d(x-x_i(t)) ,
\label{def_micro_dens_dynamic}
\end{align}
and seek for an evolution equation for $\hr$ rather than for all the $x_i$'s. Dean \cite{De96} showed that the $N$ 
Langevin equations
written in Eq.(\ref{langevin}) exactly give the following equation of evolution for $\hr$:
\begin{align}
\left\{ \begin{array}{ll}
& \displaystyle \partial_t \hr(x,t) = \nabla_{x} \cdot \left( \hr(x,t) \int_y \hr(y,t) \nabla_x v(x-y) \right) + T \nabla_x^2 \hr(x,t) 
+ \nabla_x \cdot \left( \sqrt{\hr(x,t)} \eta(x,t) \right) , \\
& \\
& \displaystyle \la \eta(x,t) \eta(y,t) \ra = 2T \d(x-y) \d(t-t) .
\end{array} \right. \label{dean}
\end{align}
Of course these dynamics conserves the number of particles, since they can be written as:
\begin{align}
\left\{ \begin{array}{ll}
& \displaystyle \partial_t \hr(x,t) + \nabla_x j_L(x,t) = 0 , \\
& \\
& \displaystyle  j_L(x,t) = - \hr(x,t) \int_y \hr(y,t) \nabla_x v(x-y) - T \nabla_x \hr(x,t) - \sqrt{\hr(x,t)} \eta(x,t) . 
\end{array} \right. 
\end{align}
The $-T \nabla_x \hr(x,t)$ term expresses Fick's law of diffusion, while each particle is additionally driven by the force field 
$- \int_y \nabla_x v(x-y) \hr(y,t)$ created by the others particles. The current $j_L$ can alternatively be put under the form:
\begin{align}
\left\{ \begin{array}{ll}
& \displaystyle  j_L(x,t) = - \hr(x,t) \nabla_x \frac{\d \FF[\hr]}{\d \hr(x,t)} , \\
& \\
& \displaystyle  \FF[\hr] = T \int_x \hr(x,t) \left[ \ln \left( \frac{\hr(x,t)}{\r_0} \right) - 1 \right] + \frac 12 \int_{x,y} \hr(x,t) v(x-y) \hr(y,t) .
\end{array} \right. 
\end{align} 

By insisting on working with a collective field, we turned the equations on
$x_i$ with additive noise into an equation 
for $\hr$ with a multiplicative noise. This equation must thus be understood
in the It\={o} sense \cite{vanKampen}: at each new time step, the new value of
the multiplicative noise is computed with the values of the density at the
previous time step.  Note that this expression superficially seems to coincide
with a phenomenological equation derived by Kawasaki \cite{Ka94}, where the pair
potential has to be replaced by the two-body direct correlation function of
the liquid. It was later shown by Kawasaki and Kim \cite{KK08} that the use of
Eq.(\ref{dean}) provides a correct treatment of the static quantities in a
dynamical calculation, whereas using of a phenomenological equation does not.

Another difficulty is in the evaluation of the density profile at the beginning of the dynamics, that requires an input
at time $t=0$. To clarify this, the experiment that we have in mind is the following: at time $t=-\io$, we prepare a system of 
harmonic spheres in a bath at temperature $T$, then let them evolve freely towards equilibrium. At $t=0$, equilibrium is 
achieved, and we start monitoring the dynamics, which are modeled by our equation (\ref{dean}). In this interpretation, the 
initial value of $\hr(x,t)$ must be obtained from an equilibrium distribution. This distribution is known, at least formally, from 
the discussion in the previous section. Indeed, exactly at $t=0$, there will be fluctuations of the space dependent field 
$\hr(x,t=0)$ around its equilibrium value $\r_0$, which is constant in time and space because of translational and 
time-translational invariance. Defining the density fluctuation field $\d \r$ as:
\begin{align}
\d \r(x,t) = \hr(x,t) - \r_0 ,
\end{align}
we thus have to know the probability distribution of $\d \r(x,0)$. It is given by the $\G_1$ functional of the liquid state 
theory described in the previous subsection. Indeed, expanding $\G_1$ around a flat profile of density, we get:
\begin{align}
\G_1[\d \r] = \G_1[\r_0] & + \sum_{N=1}^\io \int_{x_1,\ldots,x_N} \frac{\d \G_1[\r]}{\d \r(x_1) \cdots \d \r(x_N)} 
\d \r(x_1) \cdots \d \r(x_N) \nonumber \\
= \G_1[\r_0] & + \int_x (\r_0 + \d \r(x)) \left[ \ln \left( 1 + \frac{\d \r(x)}{\r_0} \right) - 1 \right] \nonumber \\
& - \sum_{N=1}^\io \int_{x_1,\ldots,x_N} 
c^{(N)}(x_1,\ldots,x_N) \d \r(x_1) \cdots \d \r(x_N). \label{DFT_liquid}
\end{align}
Now performing the inverse Legendre transformation to come back to the free-energy of the liquid simply cancels the 
$c^{(1)}$ term, in order to obtain the following probability of observing a density profile at 
equilibrium:
\begin{align}
P[\d \r(x,t=0)] \propto \exp \left( \begin{array}{ll} 
& \displaystyle - \int_x (\r_0 +\d \r(x)) \left[ \ln \left(1+\frac{\d \r(x)}{\r_0}\right) - 1 \right] \\
& \displaystyle + \sum_{N=2}^\io \int_{x_1,\ldots,x_N} \hspace{-.7cm}
c^{(N)}(x_1,\ldots,x_N) \d \r(x_1) \cdots \d \r(x_N) 
\end{array} \right) . \label{distribution_rho_t=0}
\end{align}
By evaluating this functional in the HNC approximation, where all $c^{(N)}$'s are set to zero for $N \ge 3$, we get the 
Ramakrishnan-Yussouf \cite{RY79} expression for the probability of observing a density profile.

We show now for future use a standard procedure to turn the dynamical equation Eq.(\ref{dean}) into a diagrammatic
computational scheme similar to the one described in the first section of this chapter. 
In order to obtain a particular density profile between time $0$ and time $t_f$, we first have to draw an initial density profile 
out of the distribution in Eq.(\ref{distribution_rho_t=0}), then we have at each time step to draw a value of the noise out 
of its distribution $P[\eta]$, which is Gaussian, and finally we must find the correct value of the density profile that satisfies
the Langevin equation given the value of the noise that was chosen. Grouping all these contributions together, we can 
write down the probability of observing any density profile between time $0$ and $t_f$ (probability which is equal to one by 
definition !) as:
\begin{align}
1 = Z & = \int \DD \r_0 P[\r_0] \int \DD \eta P[\eta] \int \DD \r \prod_{t=0^+}^{t_f} \prod_{x}  
\delta \left( \text{Langevin equation at time t and point } x \right) \\
& = \int \DD \r \DD \olr \DD \eta \DD \r_0 P[\r_0] P[\eta] 
\exp \left( - \int_{t=0}^{t_f} \int_{x} \olr(x,t) \times \text{Langevin equation at time } t \right) \\
& = \int \DD \r_0 P[\r_0] \int \DD \r \DD \olr ~ \exp \left( - \int_{t,x} \left[ \begin{array}{ll}
& \displaystyle \olr(x,t) \partial_t \r(x,t) - T ~  \olr (x,t) \nabla_x^2 \r (x,t)   \\
& \displaystyle - \olr(x,t) \nabla_x . \left( \r(x,t) \int_y \r(y,t) \nabla_x v(x-y)\right) 
\end{array} \right]\right) \nonumber \\
& \quad \int \DD \eta ~ \exp \left( - \frac{1}{4 T} \int_{t,x} \left[ \eta^2(x,t) 
- 4 T ~ \olr(x,t) \nabla_x \cdot \left( \sqrt{\r (x,t)} \eta(x,t)  \right) \right] \right). 
\end{align}
In order to go from the first to the second line, we used the integral representation of the delta function, as well as a change of 
variables in the delta function, which has determinant 1 when the equation is understood in the It\={o} sense.
After an integration by parts, the integral over the noise is Gaussian, and we obtain :
\begin{align}
\left\{ \begin{array}{ll}
& \displaystyle Z = \int \DD \r_0 e^{- \b \FF_{\rm{equ}}[\r_0]} \int \DD \r \DD \olr 
e^{-S[\r,\olr]} \\
& \displaystyle S[\r,\olr] = \int_{t=0}^{t_f} \int_x \left[ \olr(x,t) \partial_t \r(x,t) 
- \olr(x,t) \nabla_x \cdot \left( \r(x,t) \nabla_x \frac{\d \b \FF[\r]}{\d \r(x,t)} \right) 
- T ~ \r(x,t) \left(\nabla_x \olr(x,t) \right)^2 \right] , \\
& \displaystyle \b \FF_{\rm{equ}}[\r] = \int_x \r(x) \left[ \ln \frac{\r(x)}{\r} - 1 \right] 
+ \sum_{N=2}^\io \frac 1{N!} \int_{x_1,\ldots,x_N} \hspace{-.7cm} c^{(N)}(x_1,\ldots,x_N) \r(x_1) \cdots \r(x_N) , \\
& \displaystyle \b \FF[\r] = \int_x \r(x,t) \left[ \ln \frac{\r(x,t)}{\r} - 1 \right] + \frac 12 \int_{x,y} \r(x,t) \b v(x-y) \r(y,t) . 
\end{array} \right.
\label{MSR_action}
\end{align}
This formalism was first formulated by Martin, Siggia, and Rose \cite{MSR73}, and put to work by De Dominicis 
\cite{Do76a} and Janssen \cite{Ja79,Ja92}. Now we find ourselves in the 
position of formulating a statistical field theory in the form described in the first chapter.

\subsection{Diagrammatic structure of the theory}

Following the line described in the first section, we look at the saddle point of the functional integral, defined as:
\begin{align}
\left\{ \begin{array}{ll}
0 = & \displaystyle (-\partial_t - \nabla_x^2) \olr(x,t) - \nabla_x \left( \r(x,t) \frac{\d \b \FF[\r]}{\d \r(x,t)} \right) 
- T \left( \nabla_x \olr \right)^2  \\
0 = & \displaystyle (\partial_t - \nabla_x^2) \r(x,t) - \nabla_x \frac{\d \FF[\r]}{\d \r(x,t)} \cdot \nabla_x \olr(x,t) 
- \int_y \olr(y,t) \partial_y \r(y,t) \cdot \partial_y \frac{\d^2 \FF[\r]}{\d \r(x,t) \d \r(y,t)} .
\end{array} \right.
\end{align}
Time translation invariance, isotropy and translation invariance enforce, if they are not spontaneously broken, that 
averages of one point quantities are constant, therefore leaving us with only $\r(x,t) = \r_0$ and $\olr(x,t) = 0$.
Now we gather the two fluctuations of the fields around their mean values into a vector $\f$ defined by:
\begin{align}
\f(x,t) = \left( \begin{array}{ll}
\d \olr(x,t) \\
\d \r(x,t) 
\end{array} \right)
\end{align}
and obtain a field theory described by:
\begin{align}
& S[\f] = \frac 12 \int_{k,t} \f(k,t) G_0^{-1}(k,t) \f(-k,t) + S_{\rm{ng}}[\f] , \\
& G_0^{-1}(k,t) = \begin{pmatrix}
- 2 T \r_0 k^2 & \partial_t + k^2 (1 + \b \r_0 v(k)) \\
-\partial_t + k^2(1 + \b \r_0 v(k)) & 0
\end{pmatrix} , \\
& S_{\rm{ng}}[\f] = \int_{k_1,k_2,k_3} \d_{k_1+k_2+k_3} \left( \begin{array}{ll}
& \displaystyle T k_2 \cdot k_3 \d \r(k_1,t) \d \olr(k_2,t) \d \olr(k_3,t) \\
& \\
+ & \displaystyle \frac 12 \left[ k_1 \cdot k_3 \b v(k_1) + k_2 \cdot k_3 \b v(k_2) \right] \d \r(k_1,t) \r(k_2,t) \d \olr(k_3,t)
\end{array} \right)
\label{S_MSR}
\end{align}

\subsection{Structure of the perturbative expansion}

In a pertubative expansion as described in the first section of this chapter, the main quantity to compute is the bare 
propagator $G^{(0)}$, which is the inverse of the quadratic term in the action. In our case, we find:
\begin{align}
G^{(0)}(k,t) = \begin{pmatrix}
0 & \theta(-t) \\
\theta(t) & T \r_0 k^2 
\end{pmatrix} e^{-k^2(1- \r_0 w(k)) |t|}
\end{align}
This form of propagator is crucial to obtain a causal structure, as was pointed out by De Dominicis. Indeed, if we look 
now at the diagrammatic structure of our theory, we see that we will have two kind of vertices to build diagrams, pictured 
as follows:
\begin{align}
& \d \olr(k,t_1) \equiv \begin{minipage}[1,1]{2cm} \includegraphics[width=1cm]{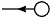} 
\end{minipage} \\
& \d \r(k,t_1) \equiv \begin{minipage}[1,1]{2cm} \includegraphics[width=1cm]{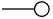}  
\end{minipage} \\
& \int_{k_1,k_2,k_3} \d_{k_1+k_2+k_3} T k_2 \cdot k_3 \d \r(k_1,t) \d \olr(k_2,t) \d \olr(k_3,t) \equiv
\begin{minipage}[1,1]{2cm} \includegraphics[width=1.5cm]{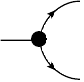} \end{minipage} \\
& - \int_{k_1,k_2,k_3} \d_{k_1+k_2+k_3} \frac 12 \left[ k_1 \cdot k_3 w(k_1) + k_2 \cdot k_3 w(k_2) \right] 
\r(k_1,t) \r(k_2,t) \d \olr(k_3,t) \equiv 
\begin{minipage}[1,1]{2cm} \includegraphics[width=1.5cm]{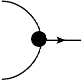}  \end{minipage} \\
& G_{11}(k,t) \equiv 0 \equiv  
\begin{minipage}[1,1]{2cm} \includegraphics[width=2cm]{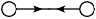}  \end{minipage} \\
& G_{12}(k,t) \equiv \theta(t_1-t_2) e^{-k^2(1- \r_0 w(k)) |t_2-t_1|} \equiv 
\begin{minipage}[1,1]{2cm} \includegraphics[width=2cm]{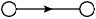}  \end{minipage} \\
& G_{22}(k,t) \equiv T \r_0 k^2 e^{-k^2(1- \r_0 w(k)) |t_2-t_1|} \equiv 
\begin{minipage}[1,1]{2cm} \includegraphics[width=2cm]{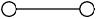}  \end{minipage}
\end{align}
Given the value of the bare propagator, we see that joining one $\d \r$ line with a $\d \olr$ line gives a causal step function. 
Thus the arrows of the lines indicate the arrow of time, and if a diagram
contains two arrows that run in opposite directions, or make a loop, the diagram is equal to zero. 

Inevitably, diagrams of the free-energy are then all equal to zero, 
because by definition they are closed, and we recover the fact that $Z=1$ in this dynamical theory.

We also see that the causal structure of the propagators is conserved when we add diagrams to the Gaussian estimate
$G^{(0)}$: indeed a diagram that contributes to renormalize $G_{11}$ has conflicting arrow of times and is therefore equal to 
zero, hence the property $G_{11} = 0$ holds non perturbatively. And a diagram that contributes to $G_{12}(k,t_2-t_1)$
 forces $t_2$ to be greater than $t_1$, and thus $G_{12}$ is always proportional to a Heaviside function.

\subsection{Symmetries and approximations}

Now that we have a field-theoretic formulation, we can come back to the prescription of Chapter \ref{chap:formalism}, 
and build up self consistent equations for the correlation functions. 
We use the variational principle in Eq.(\ref{stat_principle_2_sigma}) to obtain the two-point vertex function $\Si$ as
the derivative of the sum of all 2PI diagrams of the theory.
Since our formulation contains only cubic interactions, the first terms in this sum of diagrams will be of the form:
\begin{align}
\begin{minipage}[1,1]{2cm} \includegraphics[width=2cm]{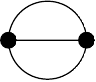} \end{minipage} ~ 
+ \quad \begin{minipage}[1,1]{2cm} \includegraphics[width=2cm]{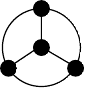} 
\end{minipage} + \ldots
\label{diags_2PI_dyn_generic}
\end{align}
where the double lines represent the full propagators. We show here all the diagrams corresponding to the first 
term of Eq.(\ref{diags_2PI_dyn_generic}):
\begin{align}
\begin{minipage}[1,1]{2cm} \includegraphics[width=2cm]{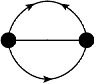} \end{minipage} , \quad
\begin{minipage}[1,1]{2cm} \includegraphics[width=2cm]{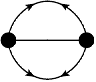} \end{minipage} , \quad
\begin{minipage}[1,1]{2cm} \includegraphics[width=2cm]{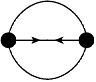} \end{minipage} , \quad
\begin{minipage}[1,1]{2cm} \includegraphics[width=2cm]{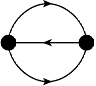} \end{minipage} , \quad
\begin{minipage}[1,1]{2cm} \includegraphics[width=2cm]{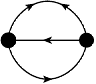} \end{minipage} .
\label{diags_2PI_dyn}
\end{align}
Note that since we have performed the second Legendre transform, the propagators on the lines of these diagrams 
are arbitrary functions; they are not necessarily causal. To obtain a self consistent equation on the propagators, we have 
to differentiate with respect to $G$, which leads to an expression for the $\Si$ functional described in Chapter 
\ref{chap:formalism} via Eq.(\ref{stat_principle_2_sigma}). In our case, we have two fields, so that the propagator and 
two-point vertex functions $\Si$ are $2 \times 2$ matrices. Differentiating the diagrams in Eq.(\ref{diags_2PI_dyn}) with 
respect to $G_{11}$ for example leads to:
\begin{align}
\Si_{11}(1,2) = 
\begin{minipage}[1,1]{2cm} \includegraphics[width=2cm]{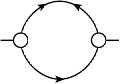} \end{minipage} +
\begin{minipage}[1,1]{2cm} \includegraphics[width=2cm]{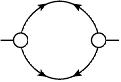} \end{minipage} +
\begin{minipage}[1,1]{2cm} \includegraphics[width=2cm]{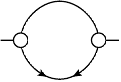} \end{minipage} +
\begin{minipage}[1,1]{2cm} \includegraphics[width=2cm]{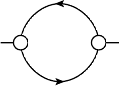} \end{minipage} + \cdots
\end{align}
Note that it is customary to represent the vertex functions with external lines to recall which vertex function is considered, 
but these lines must be omitted when the diagram is computed. These are called amputated diagrams in the field-
theoretic language.

When evaluated at the physical propagators, which have been proved to be causal, we see that all these contributions
will disappear. We can see that this property is conserved whatever the number of diagrams we keep in our expansion is, 
so that $\Si_{11} = 0$ non-perturbatively as a consequence of causality. In the same way we will find that $\Si_{12}$ is 
non-zero but is proportional to a step function of the difference between its two time arguments, and $\Si_{22}$ 
is generically non zero. These causality properties for $G$ and $\Si$ are non-perturbative features, and have to be 
conserved by a truncation of the $\G_2$ functional in order to obtain a consistent approximation for the propagators.

Another important physical constraint imposed on dynamical theories is the time-reversal invariance: i.e. the invariance
of the dynamical equations with respect to a change $t \to -t$. It has been shown by Janssen\cite{Ja92} to correspond, in 
the formalism that we present here, to a non-linear symmetry relating the density field and the response field:
\begin{align}
\left\{ \begin{array}{ll}
& t \to -t \\
& \\
& \d \r(x,t) \to \d \r^R(x,t) = \d \r(x,-t) , \\
& \\
& \displaystyle \d \olr(x,t) \to \d \olr^R(x,t) = - \d \olr(x,-t) + \frac{\d \b \FF[\r]}{\d \r(x,-t)} 
\end{array} \right.
\label{TR_ro_robar}
\end{align}
Because of the purely entropic ideal gas term in $\FF[\r]$, this symmetry is non-linear in the sense that it relates the fields
to their higher 
order cumulants. While a truncation of the 2PI functional $\G_2$ can easily be shown to automatically conserve linear 
symmetries of the action (if evaluated at propagators that verify themselves these symmetries), this is not the case 
automatically for non-linear symmetries. Thus an arbitrary truncation will surely violate micro reversibility \cite{MR05},
which is a crucial 
symmetry in our case: indeed we want to obtain an approximation that is able to predict a transition from an ergodic phase
to a non-ergodic one. If we use an approximation that explicitly breaks time-reversal symmetry, a possible transition may 
be an artifact of this approximation. To be sure that we find (or do not find !) \emph{spontaneous} symmetry breaking, 
we have to make sure time-reversal is properly enforced, as well as causality.

\subsection{Linearizing the time-reversal}

There are currently two different ways to linearize the time-reversal symmetry. Within the MSR formalism, the original
idea was introduced by Andreanov Biroli and Lef\`evre in \cite{ABL06}, but was properly implemented later by
Kawasaki and Kim in \cite{KK08}. One introduces a Lagrange multiplier $\theta$ in the path integral, which will 
constrain all the non-linear contributions coming from $\displaystyle \frac{\d \FF[\r]}{\d \r}$ to be equal to $\theta$:
\begin{align}
& Z  = \int \DD \r \DD \olr ~ \d \left( \theta - \frac{\d \b \FF[\r]}{\d \r} + \frac{\d \r}{\r_0} \right) e^{- S[\r,\olr]} , \\
& = \int \DD \r \DD \olr \DD \theta \DD \olth ~ e^{- S_{\rm{KK}}[\r,\olr,\theta,\olth]} , \\
& S_{\rm{KK}}[\f] = \frac 12 \int_{k,t} \f(k,t) G_0^{-1}(k,t) \f(-k,t) + S_{\rm{ng}}[\f] , \\
& G_0^{-1}(k,t) = \begin{pmatrix}
0 & -i \partial_t + k^2(1 - \r_0 w(k)) & 0 & 0 \\
i \partial_t + k^2(1 - \r_0 w(k)) & 2 \r_0 k^2 & \r_0 k^2 & 0 \\
0 & \r_0 k^2 & 0 & - 1 \\
0 & 0 & -1 & 0
\end{pmatrix} , \\
& S_{\rm{ng}}[\f] = - \int_{x,t} \left[ \begin{array}{ll}
& \displaystyle \d \olr(x,t) \nabla_x \cdot \left( \d \r(x,t) \nabla_x \left[ \theta(x,t) + \frac{\d \r(x,t)}{\r_0} - (w \otimes \d \r)(x,t) \right] \right) \\
& \displaystyle - \d \r(x,t) \left( \nabla_x \d \olr(x,t) \right)^2 + \olth(x,t) \left( \sum_{n=2}^\io \frac{(-1)^n}{(n-1)} 
\frac{\d \r(x,t)^n}{\r_0^{n-1}} \right)
\end{array}\right] ,
\end{align}
where $\otimes$ denotes space convolution.
The time reversal is now a linear transformation of the fields:
\begin{align}
\left\{ \begin{array}{ll}
t & \to \quad -t \\
\r(x,t) & \to \quad \r^R(x,t) = \r(x,-t) \\
\olr(x,t) & \to \quad \displaystyle \r^R(x,t) = - \olr(x,-t) + \frac 1 {\r_0} \d \r(x,-t) - \left( w \otimes \d \r \right)(x,-t) 
+ \theta(x,-t) \\
\theta(x,t) & \to \quad \theta^R(x,t) = \theta(x,-t) \\
\olth(x,t) & \to \quad \olth^R(x,t) = \olth(x,-t) - i \partial_t \r(x,-t) 
\end{array} \right.
\end{align}
and will therefore automatically be conserved by a truncation of the 2PI functional $\G_2$. However the  
cost has been to introduce an infinite number of interaction vertices in the action, that are all related to the non-interacting 
system. Thus, to each order of the calculation, one has to calculate an infinite number of diagrams in order to be able 
to fully describe the correct non-interacting system.

Note that these symmetry considerations on the time-reversal do not apply if one works with the It\^o discretization, but
only in the Stratonovich discretization \cite{vanKampen}. Luckily, for the precise Langevin equation under scrutiny, 
the two discretizations are identical !

By cleverly taking into account the ideal gas contributions non-perturbatively, Kawasaki and Kim 
\cite{KK08} were able to 
derive back the MCT equation. The same analysis were later performed by Nishino and
Hayakawa \cite{NH08} for the fluctuating non-linear hydrodynamics.
However, the cost of taking correctly into account the ideal gas is a very 
heavy formalism,
leaving little hope to extend the calculation to next orders properly. It would be better to have a theory that directly starts
from non-interacting particles and treats the interactions as a perturbation. In the following sections  of this chapter we present 
a first step towards such a theory, that is provided by a mapping of the particles onto effective bosons.

\section{Quantum mechanical formulation}

\subsection{Master equation and Doi-Peliti formalism}

Coming back to a discretized version of the dynamics of our system, we now consider that our $N$ particles
are dispatched on an infinite, 3D, square lattice with mesh size $a$. The possible positions
of the spheres are denoted $x_i = a \left( p e_1 + q e_2 + r e_3 \right)$, where $\{p,q,r\}$ are positive or negative numbers
and $e_1,e_2,e_3$ are unitary vectors which form an orthonormal basis. 
The ensemble of vectors belonging to the lattice is called $\LL$ in the following.

The equivalent of the microscopic density that we are interested in is now the set of occupation numbers of the lattice sites. 
We call $n_x$ the occupation number of the site located at $x$. A microscopic configuration of the system $\CC$ is then the 
set of all numbers of occupations :
\begin{equation}
\CC = \{ n_x,x \in \LL \}
\end{equation}
We call $E_p(\CC)$ the total potential energy of a given configuration $\CC$, defined by : 
\begin{equation}
E_p(\CC) = \frac{1}{2} \sum_{y \ne z} n_y v\left( y-z \right) n_z ,
\label{potential_energy_discrete}
\end{equation}
where $v$ is still the pair potential of our system, that can be for example the pair potential of harmonic spheres
given in Eq.(\ref{potential_harmS}).
We choose the following dynamic rules : within a time interval $dt$, a particle can jump from 
a site $x$ to another site $y$ (changing the microscopic configuration from $\CC$ to $\CC '$) 
with probability $\exp\left( -\frac{\b}{2} [ E_p(\CC ')-E_p(\CC) ] \right) n_x dt$. We consider the 
evolution with time of the probability of observing a particular configuration $\CC$ at time $t$: 
$P(\CC,t)$. We call $\CC '$ the configuration that differs 
 from $\CC$ only by a jump of one particle to a closest neighbor :
\begin{align}
& \CC = \{ \ldots , n_x , \ldots , n_{x+e} , \ldots \} \nonumber ,\\
& \CC ' = \{ \ldots , n_x - 1 , \ldots , n_{x+e} + 1 , \ldots \},\nonumber\\
& \text{with } e = \pm ~ a ~ e_i , i \in \{1,2,3\}
\end{align}
Note that this type of dynamics is compatible with the hard-sphere case, where $e^{-\b v}$ is equal to $0$ or $1$ 
depending whether the configuration has overlaps or not. The only restriction is that the initial configuration must be an 
authorized one. In that case, the rate to go from the current configuration to a forbidden one is always $0$, and the 
hard-sphere constraint is always satisfied.

These dynamics obviously satisfy detailed balance, and thus lead to a unique stationary distribution, which is the 
Boltzmann one:
\begin{align}
P(\CC,t) \underset{t \to \io}{\sim} P_{\rm{eq}}(\CC) = e^{-\b E_p(\CC)}
\end{align}
The continuum limit (letting $a$ go to $0$ while properly rescaling the physical quantities of interest) of these dynamics 
can be shown to be equivalent to the Langevin approach presented above \cite{ABBL06}. However, we are interested in 
finding an alternative formulation, in which the non-interacting particles are exactly treated and time-reversal invariance is linear.

The microscopic dynamics described here corresponds to an evolution equation for $P(\CC,t)$ that reads:
\begin{align}
\left\{ \begin{array}{ll}
& \partial_t P(\CC,t) = W P(\CC,t) \\
& \\
& \displaystyle W = \sum_{x \in \LL} \sum_{e} \left[ (n_{x+e}+1) e^{- \frac \b 2 ( E_p(\CC)-E_p(\CC'))} P(\CC',t) 
- n_x e^{\frac \b 2 ( E_p(\CC') - E_p(\CC))} P(\CC,t) \right] ,
\end{array} \right.
\label{master_eq}
\end{align}

We follow now the work of Doi \cite{Do76b}, popularized in the context of reaction-diffusion processes by Peliti \cite{Pe85}.
It consists in introducing a Fock space corresponding to the states $\CC$ and making use of the ladder operators of
quantum mechanics to obtain the evolution of our system in a second quantized form, i.e. in terms of occupation numbers,
which are the discrete versions of the microscopic densities defined in Eq.(\ref{def_micro_dens_dynamic}).
We introduce the lowering and raising operators $\ha_x$ and $\hac_x$, which satisfy bosonic commutation relations:
\begin{align}
& [ \ha_x , \ha_y ] = 0 \nonumber \\
& [ \hac_x , \hac_y ] = 0 \nonumber \\
& [ \ha_x , \hac_y ] = \delta_{x,y}
\end{align}
Defining $\hn_x = \hac_x \ha_x$, and choosing convenient normalizations we then construct 
the Fock space of states from the vacuum state $\ket0$ with no particles in it:
\begin{equation}
\ha_x ~ \ket0 = 0 \quad \forall x .
\end{equation}
Starting from the vacuum state $\ket0$, the creation operators $\hac_x$ generate all the microscopic states $\CC$, that 
have their corresponding kets $\ketC$:
\begin{align}
\ketC \equiv | \ldots , n_x , \ldots \rangle = \prod_{x} (\ha_x^+)^{n_x} ~ \ket0 .
\end{align}
Application of the operators $\ha$,$\hac$ and $\hn$ on these states, and on their corresponding bra's are given by the 
following relations (note that the normalizations are different from usual conventions for the quantum harmonic 
oscillator):
\begin{align}
& \hn_x ~ | \ldots , n_x , \ldots \rangle = n_x ~ | \ldots , n_x , \ldots \rangle \nonumber \\
& \hac_x ~ | \ldots , n_x , \ldots \rangle = | \ldots , n_x + 1 , \ldots \rangle \nonumber \\
& \ha_x ~ | \ldots , n_x , \ldots \rangle = n_x ~ | \ldots , n_x - 1 , \ldots \rangle \nonumber \\
& \langle \ldots , n_x , \ldots | \ha_x = \langle \ldots , n_x+1 , \ldots | (n_x + 1) \nonumber \\
& \langle \ldots , n_x , \ldots | \hac_x = \langle \ldots , n_x-1 , \ldots | \nonumber \\
& \bra0 \hac_x = 0 \quad \forall x .
\end{align}
In this way, the states are normalized so that:
\begin{align}
\la \CC | \CC' \ra = \d_{\CC,\CC'} = \prod_x \d_{n_x,n'_x}
\end{align}
We can now re-write the master equation Eq.(\ref{master_eq}) in the Fock space, by defining a wave function $\ketPsi$ 
that contains all the information on the statistical state of the system:
\begin{equation}
\ketPsi_t = \sum_{\CC} P(\CC,t) \ketC
\end{equation}
We can then rewrite the master equation as:
\begin{align}
& \partial_t \ketPsi_t = \hat{W} \ketPsi_t , \\
& \hW = \sum_{x \in \LL} \sum_{e}  e^{\frac{\b V(e)}2}
\left[ \hac_x \ha_{x+e} e^{\frac{\b}{2} \D \hat{E}_p} - \hac_x \ha_x e^{-\frac{\b}{2} \D \hat{E}_p} \right] , \\
& \D \hat{E_p} = \sum_{y \ne x} \left( \hn_{y+e} - \hn_y \right) v \left( x - y \right) ,
\end{align}
which is functionally solved in time as:
\begin{align}
\ketPsi_t = e^{\hat{W} t} \ketPsi_{t=0} .
\end{align}
The average of an observable $A(\CC)$ that depends on occupation numbers, such as $E_p$ in 
Eq.(\ref{potential_energy_discrete}) is defined as:
\begin{align}
\la A \ra_t = \sum_{\CC} A(\CC) P(\CC,t) .
\end{align}
To compute this kind of object, we introduce the projection state:
\begin{align}
& \braP = \sum_{\CC} \braC = \bra0 \prod_{x} e^{a_x} ,
\end{align}
which has the following properties:
\begin{align}
& \braP \hac_x = \braP , \\
& \braP (\hac_x-1) = 0 , \\
& \braP \Psi \rangle_t = \sum_{\CC,\CC'} P(\CC',t) \braC \CC' \rangle = \sum_{\CC} P(\CC,t) = 1 \quad \forall t , \\
& \braP \CC \rangle = 1 \quad \forall \CC,
\end{align}
and thus compute the average of an observable, as:
\begin{align}
\la A \ra_t = \braP \hat{A} \ketPsi_t,
\end{align}
where $\hat{A}$ is the canonically quantized operator obtained by replacing all $n_x$ in $A$ by $\hn_x$.

\subsection{Passage to a field theory}

Given a set of states $| \phi \rangle$ that form a representation of the identity,
\begin{align}
\rm{Id} = \sum_\phi | \phi \rangle \langle \phi | ,
\end{align}
we can construct a path integral similar to the Feynman path integral by splitting the time interval $[0,t]$ into $N$ slices 
of size $\D t = t/N$ and write:
\begin{align}
\la A \ra_t & = \braP \hat{A} ~ e^{\hat{H} \D t} \ldots e^{\hat{H} \D t} ~ \ketPsi_{t=0} \nonumber \\
& = \sum_{\phi_0,\ldots,\phi_N} \langle P | \hat{A} | \phi_N \rangle \langle \phi_N | e^{\hat{H} \D t} | \phi_{N-1} \rangle \ldots \langle \phi_1 | 
e^{\hat{H} \D t} | \phi_0 \rangle \langle \phi_0  \ketPsi_{t=0} \nonumber \\
& = \sum_{\phi_0,\ldots,\phi_N} \langle P | \hat{A} | \phi_N \rangle \prod_{i=1}^N \langle \phi_i | \phi_{i-1} \rangle \langle \phi_0  \ketPsi_{t=0} ~ 
e^{\D t \sum_{i=1}^N \frac{\langle \phi_i | \hat{H} | \phi_{i-1} \rangle}{\langle \phi_i | \phi_{i-1} \rangle}} \label{decomposition}
\end{align}

We specify now the complete set of states that we use to construct the path integral. We construct coherent states:
\begin{align}
& | a \rangle = | \ldots , a_x , \ldots \rangle = e^{-\frac 1 2 \sum_x |a_x|^2} e^{\sum_x a_x \hac_x} | 0 \rangle , \\
& \langle a | = \langle \ldots , a_x , \ldots | = e^{-\frac 1 2 \sum_x |a_x|^2} \langle 0 | e^{ \sum_x \overline{a}_x \ha_x} .
\end{align}
The action of $\ha$ and $\hac$ on these states are:
\begin{align}
\ha_x | a \rangle & = e^{-\frac 1 2 \sum_x |a_x|^2} \prod_{y \ne x} e^{a_y \hac_y} \ha_x e^{a_x \hac_x} | 0 \rangle 
\nonumber \\
& = e^{-\frac 1 2 \sum_x |a_x|^2} \prod_{x} e^{a_x \hac_x} (a_x + \ha_x) | 0 \rangle \nonumber \\
& = a_x | a \rangle . \\
\langle a | \hac_x & = e^{-\frac 1 2 \sum_x |a_x|^2} \langle 0 | (\overline{a}_x + \hac_x) \prod_x e^{\overline{a}_x \ha_x} 
\nonumber \\
& = \langle a | \overline{a}_x .
\end{align}
The normalisation is given by:
\begin{align}
\langle a | a \rangle & = 1 , \\
\langle a' | a \rangle & = \exp \left( - \frac 1 2 \sum_x \left[ |a'_x|^2 + |a_x|^2 \right] \right) \exp \left(  \sum_x \overline{a}'_x 
a_x \right)  , \\
& = \exp \left( \frac 1 2 \sum_x \left[ | a'_x |^2 - | a_x |^2 \right] \right) \exp \left( \sum_x \overline{a}'_x \left[ a_x - a'_x \right] 
\right) .
\end{align}
Finally these states form a complete set of the space, as we wanted. To show it, first we remark that (defining $a=u+iv$):
\begin{align}
\int \frac {dudv}{\pi} {\overline{a}}^n a^m e^{-|a|^2} & = \frac 1 \pi \int_{r=0}^\io \int_{\theta=0}^{2\p} dr d\theta ~ r e^{-r^2} r^{n+m} 
e^{i(n-m)\theta} \nonumber \\
& = 2 \d_{n,m} \int_0^\io r^{2n+1} e^{-r^2} dr = n! \d_{n,m} . \label{identity}
\end{align}
It is easy to verify, by resorting to Eq. (\ref{identity}), that we have the unity representation:
\begin{align}
\rm{Id} = \int \prod_x \left( \frac {d \text{Re} z_x d \text{Im} z_x }{\p} \right) ~ | a \rangle \langle a | .
\end{align}
An important remark is that:
\begin{align}
\langle P | = e^{\frac M 2} \langle a =1 | ,
\end{align}
where $M$ is the number of lattice sites.

Coming back to Eq. (\ref{decomposition}), we define $t_i = i \Delta t$ and we use those coherent states to get:
\begin{align}
\la A \ra_t & = \int \prod_{i,x} \frac{d \Re a_{x,t_i}{d \Im a_{x,t_i}}}{\pi} \left[ \langle P | \hat{A} | a_t \rangle \langle a_0  \ketPsi_{t=0} 
\prod_{i=1}^N \langle a_{t_i} | a_{t_{i-1}} \rangle  \right] e^{\D t \sum_{i=1}^N \frac{\langle a_{t_i} | \hat{W} | a_{t_{i-1}} \rangle}
{\langle a_{t_i} | a_{t_{i-1}} \rangle}}
\end{align}
We have:
\begin{align}
\prod_{i=1}^N \langle a_{t_i} | a_{t_{i-1}} \rangle & = \prod_{i=1}^N e^{\frac 1 2 \sum_x \left( |a_{x,t_i}|^2 - |a_{x,t_{i-1}}|^2 
\right)} e^{\sum_{i=1}^N \sum_x \overline{a}_{x,t_i} \left( a_{x,t_i} - a_{x,t_{i-1}} \right)} \nonumber \\
& = \exp \left( \frac 12 \sum_x \left[ |a_{x,t}|^2 - |a_{x,0}|^2 \right] \right) \exp \left( \sum_{i=1}^N \sum_x \overline{a}_{x,t_i}
\left( a_{x,t_i} - a_{x,t_{i-1}} \right) \right)
\label{time_discretized}
\end{align}
There is still the term $\langle P | \hat{A} | a_t \rangle$ to calculate. The properties of our coherent states and of
the projection state are such that, if in $\hat{A}$, all $\hac$ operators are placed on the left and all $\ha$ operators on the 
right, this average will be easy to calculate. The operation of putting $\hac$ on the left and $\ha$ on the right is called
normal ordering \cite{BHPSD07}. Defining as $\tilde{A}$ the normal ordered form of $\hat{A}$, we will have:
\begin{align}
\langle P | \hat{A} | a_t \rangle & = e^{\frac N 2} \langle 1 | \hat{A} | a_t \rangle \nonumber \\
& = e^{\frac N 2} \tilde{A}(1,a_t) e^{\frac 1 2 \sum_x \left( 1 - |a_{x,t}|^2 \right)} e^{\sum_x \left( a_{x,t} - 1 \right)} 
\nonumber \\
& = \tilde{A}(1,a_t) \exp \left( \sum_x \left[ a_{x,t} - \frac 12 | a_{x,t} |^2 \right] \right) ,
\end{align}
where $\tilde{A}(1,a_t)$ is a short-hand notation for the function of the $\overline{a}$'s and the $a$'s obtained by replacing 
all $\hac$ operators by their corresponding eigenvalues $\overline{a}$ and all $\ha$ operators by their eigenvalues $a$,
and finally setting all $\overline{a}$ to unity.
Finally, after taking the continuum limit, we arrive at:
\begin{align}
& \la A \ra_T  = \int \DD \overline{a} \DD a ~ \tilde{A}(1,a(x,T)) \langle a(x,0) \ketPsi_{t=0} ~ e^{-S[\overline{a},a]} , \\
& S[\overline{a},a] = \int_x \left( |a(x,0)|^2 - a(x,T) \right) + \int_0^T \int_x \left( \overline{a}(x,t) \partial_t a(x,t) 
- \tilde{W}(\overline{a}(x,t),a(x,t)) 
\right) .
\end{align}

We thus have found a field theory representation of our stochastic process. In the same way as above, we have had to 
normal order the operator $\hW$ to be able to easily calculate the matrix elements $\langle a_{t_i} | \hat{W} | a_{t_{i-1}} 
\rangle$, and we denoted the resulting function of $\overline{a}$ and $a$ by $\tilde{W}(\overline{a}(x,t),a(x,t))$.
Note that the $\overline{a}$'s always have to be calculated at an earlier time than the $a$'s, a distinction which has 
disappeared when taking the continuum limit, but has to be remembered when calculating functions at equal times
in this formalism. This is the equivalent of the It\^o's interpretation of the Langevin equation.

\subsection{Cole-Hopf transformation}

Since the representation in terms of a master 
equation is fully equivalent to the Langevin equation in Eq.(\ref{dean}), all this procedure is for now only a complicated way
to reformulate an existing theory. A Cole-Hopf change of variables (first introduced to deal with the Burgers equation
in fluid mechanics \cite{Co51,Ho50}) defined by:
\begin{align}
\left\{ \begin{array}{ll} 
\overline{a}(x,t) & \displaystyle \to ~  \exp \left( \olr(x,t) \right) \\
a(x,t) & \displaystyle \to ~ \r(x,t) \exp \left( -\olr(x,t) \right)
\end{array} \right.
\end{align}
will lead exactly to the action Eq.(\ref{S_MSR}). The terms involving the pair potential that come from the evolution 
operator $\hW$ will be obvious since they will depend only on $\overline{a} a$ which becomes $\r$ in this change of 
variables. Let us look at the other bulk terms. We see that they transform exactly into the ideal gas part of the action in 
Eq.(\ref{S_MSR}):
\begin{align}
\int_0^T \int_x \overline{a}(x,t) \left( \partial_t - \nabla_x^2 \right) a(x,t) \to \int_0^T \int_x \left[ \olr(x,t) \left( \partial_t - \nabla_x^2 \right) \r(x,t) 
- \r(x,t) \left( \nabla_x \olr(x,t) \right)^2 \right]
\end{align}
up to boundary terms.
This change of formulation, although mathematically equivalent to the Langevin formulation is very convenient since
we see that the ideal gas part of the problem is now represented by a quadratic action, which is trivially solvable, 
whereas in the Langevin representation, an analysis in terms of causality of the diagrams must be performed.
What about the time reversal ?

The transformation Eq.(\ref{TR_ro_robar}) when considered for the case of non-interacting particles simply becomes:
\begin{equation}
\left\{ \begin{array}{ll}
a(x,-t) & \to ~ \overline{a}(x,t) \\
\overline{a}(x,-t) & \to ~ a(x,t) 
\end{array} \right. ,
\end{equation}
which is a linear transformation.
However the interacting part of the free-energy involved in the transformation makes the symmetry look more 
complicated, giving:
\begin{align}
\left\{ \begin{array}{ll}
a(x,-t) & \to ~ \overline{a}(x,t) \\
\overline{a}(x,-t) & \displaystyle \to ~ a(x,t) \exp \left( \int_y \b v(x-y) \overline{a}(y,t) a(y,t) \right)
\end{array} \right.
\end{align}
We see that in this new formulation, all the difficulty of conserving the time-reversal is now contained in the interacting part
of the problem. In a sense this second quantized form has solved one of our problems, the ideal gas contribution, but
not yet the other one: the non-linear expression of the time-reversal.

\subsection{An alternative linearization of the time-reversal}

In order to render time-reversal a linear invariance, we can make use of the well-known similarity transformation 
\cite{vanKampen},
that corresponds in our formalism to a rotation in the Fock space:
\begin{align}
\ketPsi \to \hat{P}_{\rm{eq}}^{1/2} \ketPsi ,
\end{align}
where $\hat{P}_{\rm{eq}}$ is the canonical quantized form of $P_{\rm{eq}}(\CC)$, i.e.:
\begin{equation}
\hat{P}_{\rm{eq}} = \exp \left( \frac 1 2 \sum_{x \ne y} \hn_x w(x-y) \hn_y \right) .
\end{equation}
This has the effect of modifying the evolution operator as follows:
\begin{align}
& \hW \to \hW_{\rm{sym}} \equiv \hat{P}_{\rm{eq}}^{-1/2} \hW \hat{P}_{\rm{eq}}^{1/2} , \\
& \hW_{\rm{sym}} =  \sum_{x \in \LL} \sum_{e} \left[ \hac_x \ha_{x+e} - \hac_x \ha_x ~ e^{-\frac{\b V(e)}{2}}  
e^{-\frac{\b}{2} \D \hat{E}_p} \right] .
\label{W_sym_operator}
\end{align}
Taking the continuum limit, which we define as:
\begin{align}
\left\{ \begin{array}{ll}
\ha_x & \to \quad  \ha(x) a^{3/2} \\
\hac_x & \to \quad \hac(x) a^{3/2}  \\
\hn_x & \to \quad \hn(x) a^3 \\
\displaystyle \sum_{y \ne x} a^3  & \to \quad  \int_y 
\end{array} \right.
\label{continous_limit}
\end{align}
and expanding the exponentials to second order in the lattice spacing leads to the following evolution equation:
\begin{align}
\left\{ \begin{array}{ll}
& \displaystyle \partial_t \ketPsi = \hWs \ketPsi \\
& \\
& \displaystyle \hWs = e^2 \int_x \left( \hac(x,t) \partial^2_x \ha(x,t) - U_{\rm{eff}} [\hn] (x,t) \right) \\
& \\
& \displaystyle U_{\rm{eff}} [\hn] (x,t) = \frac{1}{4} \hn(x,t) \int_{y,z} \partial_y \hn(y,t) \cdot \partial_z \hn(z,t) w \left( x - y \right) 
w \left( x - z \right) - \frac 12 \hn(x,t) \int_y \partial^2_y \hn(y) w \left( x - y \right)
\end{array} \right. .
\label{final_doi_action}
\end{align}
The appearance of the effective potential $U_{\rm{eff}}$ is at this stage very satisfying. Indeed this quantity can be seen as
the local rate of escaping from a state, and therefore seems to be a good dynamical quantity in 
terms of which formulating our dynamical theory.
Following the procedure described earlier, Eq.(\ref{W_sym_operator}) leads to a field theory in the continuum in terms of 
the two fields $\overline{a}$ and $a$ that read:
\begin{align}
& Z = \int \DD \overline{a} \DD a ~ e^{-S[\overline{a} , a]} , \\
& S[\overline{a},a] = \int_{x,t} \left[ \overline{a}(x,t) \left( \partial_t - \nabla_x^2 \right) a(x,t) 
+ U_{\rm{eff}}[\overline{a}(x,t) a(x,t)] \right] .
\end{align}
A drawback of this formulation is that the density $\ola a$ is not the physical density, unless we place ourselves in the 
bulk of time: it is easily seen from Eq.(\ref{time_discretized}) that the only modification due to the rotation that we 
performed is in the initial and final terms in the time discretization. The averages in the functional integral will thus 
correspond to the physical averages only if we send the initial time in the far past and the final observation time in the far 
future.

What has time-reversal become in this formalism ? Now, thanks to the symmetrization of the evolution operator, the 
dynamics can fully be interpreted as a quantum mechanical evolution (albeit in imaginary time), and time reversal is simply
guaranteed by the symmetry of the action upon exchange of $\overline{a}$ and $a$, i.e. we are back to the much simpler
symmetry:
\begin{align}
\left\{ \begin{array}{ll}
\overline{a}(x,-t) & \to \quad a(x,t) \\
a(x,-t) & \to \quad \overline{a}(x,t) 
\end{array} \right. ,
\label{time_reversal_symmetrized}
\end{align}
which is finally a linear symmetry. Thus the second of our problems is settled, and we can now safely perform a 
diagrammatic analysis of our theory. The strength of the interactions will be taken as an expansion parameter, and since
all the non-quadratic parts of the action depend only on $\overline{a} a$ through the effective potential $U_{\rm{eff}}$, we
are guaranteed never to break the time-reversal symmetry.

A last problem that has to be dealt with is the fact that we want to compute a density-density correlation function, 
which in our formalism is a four-point function. We dealt with that by constraining $\ola a$ to be equal to $\r$ by a 
Lagrange multiplier $\lam(x,t)$:
\begin{align}
Z & = \int \DD \ola \DD a ~ e^{- \int_{x,t} \overline{a}(x,t) \left( \partial_t - \nabla_x^2 \right) a(x,t) } \int \DD \r ~ 
\d \left( \r - \ola a \right) e^{- \int_{x,t} U_{\rm{eff}}[\rho(x,t)] } \\
& = \int \DD \ola \DD a \DD \r \DD \lam ~ e^{-S[\ola,a,\r,\lam]} , \\
S[\ola,a,\r,\lam] & = \int_{x,t} \left( \frac{}{} \overline{a}(x,t) \left( \partial_t - \nabla_x^2 \right) a(x,t) + \lam(x,t) \left[ \r(x,t) 
- \ola(x,t) a(x,t) \right] + U_{\rm{eff}}[\r(x,t)] \right)
\label{action_particles_aabar}
\end{align}

Introducing by hand the density has had the effect of adding a new cubic vertex in the action. Finally we are left with
a four-field theory that contains two cubic interaction vertices, which is a much simpler situation than in the case of the
Langevin representation, since in that case there is an infinity of interacting vertices in the action, that must be 
manipulated with care in order to take properly into account the non-interacting particles. 
The advantages of our formulation are two-fold:
\begin{itemize}
\item The ideal gas case is the natural starting point of the perturbative expansion, thus will be correct at all order of 
approximation, and the strength of the interactions will serve as an organizing device for the diagrammatic expansion
\item The time-reversal is a ``rapidity" symmetry between $\ola$ and $a$, and the introduction of the density $\r = \ola a$
ensures that these two fields will always play a symmetric role when truncations will be performed
\end{itemize}

\section{Fredrickson-Andersen model}

The Fredrickson-Andersen (FA) model \cite{FA84,FA85} is a model of independent spins on a lattice that have two states, 
up and down, or zero and one, whose dynamics is constrained by a dynamical rule saying that a spin can flip only if one of 
its neighbors is in the up state. This model represents a coarse grained version
of structural glasses. If one considers that the glass transition is a purely dynamical phenomenon, it must have a finite
static correlation length. One then separates the system into cells of size of the order of this length, and consider these 
cells as independent at equilibrium. However, dynamically, one supposes that for a rearrangement to 
occur in one given cell (for a spin to flip), at least one of the neighboring cells must be in a mobile state (spin up).
It is the idea of dynamical facilitation \cite{RS03}. 

These models are called a kinetically constrained model, and although they usually have trivial thermodynamics, 
as a consequence of the kinetic constraints, their dynamics can reproduce various aspects of the glassy dynamics.

We study the FA model only as a toy model, in which the statics are trivial, and the dynamics present glassy features. We 
will see that in terms of field theory it will correspond to a simplification of the interaction terms when compared to the 
complicated case of interacting particles in Eq.(\ref{action_particles_aabar}). We will thus use this model as a benchmark 
for our method. The following is yet unpublished material.

\subsection{The model: dynamics without the statics}

We start from the standard action in the Doi-Peliti formalism for the FA model \cite{WBG04}, and symmetrize it using the 
similarity transformation described in the previous section. We obtain as a starting point the following field theory:
\begin{align}
 S[\ola,a] = \int_{x,t} ~ \left[ \ola \left( \partial_t - \Delta \right) a + \z ~ \ola a + \ola^2 a^2 
 - \sqrt{\z} ~ \ola a (\ola + a ) \right] . \label{action_abara}
 \end{align}
 Introducing the density, we get:
 \begin{align}
 S[\ola,a,\lam,\r] = \int_{x,t} ~ \left[ \ola \left( \partial_t - \Delta \right) a + \z ~ \ola a 
 + \r^2 - \sqrt{\z} ~ \r (\ola + a) + \lam(\ola a-\r) \right] . \label{action}
 \end{align}
 In the following, we will group the four fields in a vector $\varphi_i(1)$, defined by:
\begin{align}
\varphi(1) \equiv \left( \begin{array}{l}
a(x,t) \\
\ola(x,t) \\
\lam(x,t) \\
\r(x,t) 
\end{array} \right)
\end{align}
The propagators are defined as $G_{ij}(1,2) \equiv \la \varphi_i(1) \varphi_j(2) \ra - \la \varphi_i(1) \ra \la \varphi_j(2) \ra$.

\subsubsection{Consequences of time-reversal}

As described in the previous section, the time-reversal symmetry of the action reads:
\begin{align}
\left\{ \begin{array}{l} a(x,t) \to \ola(x,-t) \\ 
\ola(x,t) \to a(x,-t) \\ 
\lam(x,t) \to \lam(x,-t) \\
\r(x,t) \to \r(x,-t)
\end{array} \right.
\end{align}
This symmetry has consequences on the propagators, expressed in the Fourier space for space
directions and real space for the time direction: 
\begin{align}
\left\{ \begin{array}{l}
G_{11}(k,\t) = G_{22}(k,\t) \\
G_{13}(k,\t) = G_{23}(k,-\t) \\
G_{14}(k,\t) = G_{24}(k,-\t) \\
G_{34}(k,\t) = G_{34}(k,-\t)
\end{array} \right. , \label{time_reversal_G}
\end{align}
We introduce a source term $J$ for the fields, and define as in Chapter \ref{chap:formalism} the functional $\G$, the 
Legendre transform of the logarithm of the partition function with respect to $J$. Defining as $\ph$ the means of the 
fields, the functional derivatives of $\G$ with respect to $\ph$ are defined as:
\begin{align}
& \G^{(1)}_i(1) \equiv \frac{\d \G[\ph]}{\d \ph_i(1)} , \\
& \G^{(2)}_{ij}(1,2) \equiv \frac{\d^2 \G[\ph]}{\d \ph_i(1) \d \ph_j(2)} .
\end{align}
The time reversal has the same consequences on the two-point functions $\G^{(2)}$ than on $G$:
\begin{align}
\left\{ \begin{array}{l}
\G^{(2)}_{11}(k,\t) = \G^{(2)}_{22}(k,\t) \\
\G^{(2)}_{13}(k,\t) = \G^{(2)}_{23}(k,-\t) \\
\G^{(2)}_{14}(k,\t) = \G^{(2)}_{24}(k,-\t) \\
\G^{(2)}_{34}(k,\t) = \G^{(2)}_{34}(k,-\t)
\end{array} \right. , \label{time_reversal_Gamma}
\end{align}

\subsubsection{Averages of the field and mean-field approximation}

The derivative of the action is:
\begin{align}
\frac{\d S[\varphi]}{\d \varphi(1)} = \begin{pmatrix}
\left[- \partial_t - \D_x + \z + \lam(1) \right] \ola(1) - \sqrt{\z} \r(1) \\
\left[- \partial_t - \D_x + \z + \lam(1) \right] a(1) - \sqrt{\z} \r(1) \\
\ola(1) a(1) - \r(1) \\
2 \r(1) - \sqrt{\z} \left[ \ola(1) + a(1) \right] - \lam(1)
\end{pmatrix} \label{deltaS_deltaPhi}
\end{align}
The saddle-point is defined by the value of $\varphi$ where this derivative is zero. 
We assume translational invariance, so that means of fields are independent of space and time, 
to obtain:
\begin{align}
\left\{ \begin{array}{l}
a = \ola = \sqrt{\z} \\
\lam = 0 \\
\r = \z
\end{array} \right.
\end{align}
Shifting the fields by their mean field values, we find that the propagator of the theory is:
\begin{align}
G_0(k,\o) = \frac 1 {\o^2 + \O^2} \begin{pmatrix}
0 & - i \o + \O & \sqrt{\z} (-i\o+\O) & \sqrt{\z} (-i \o + \O) \\
i \o + \O & 0 & \sqrt{\z} (i \o + \O) & \sqrt{\z} (i \o + \O) \\
\sqrt{\z} (i \o + \O) & \sqrt{\z} (-i \o + \O) & -2 ( \o^2 + \O^2 ) + 2 \z \O & -(\o^2 + \O^2) +2 \z \O \\
\sqrt{\z} (i \o + \O) & \sqrt{\z} (-i \o + \O) & -(\o^2 + \O^2) +2 \z \O & 2 \z \O \\
\end{pmatrix} , 
\end{align}
which reads, in the real time domain:
\begin{align}
G_0(k,\t) = \begin{pmatrix}
0 & \theta(\t) & \sqrt{\z} \theta(\t) & \sqrt{\z}\theta(\t) \\
\theta(-\t) & 0 & \sqrt{\z}\theta(-\t) & \sqrt{\z}\theta(-\t) \\
\sqrt{\z}\theta(-\t) & \sqrt{\z}\theta(\t) & -2 e^{\O|\t|} + \z & -e^{\O|\t|} + \z \\
\sqrt{\z}\theta(-\t) & \sqrt{\z}\theta(\t) & -e^{\O|\t|} + \z  & \z
\end{pmatrix} e^{-\O |\t|}
\label{G_0_FA}
\end{align}
We see that, concerning the fields $a$, $\ola$ and $\lam$, we have a causal structure of the bare
propagator. In the very same way than the analysis performed in the case of the MSR formalism, an analysis of the 
diagrams that may renormalize $G_{13}$ at all order of perturbation shows that all corrections to $G_{13}(k,\t)$ remain 
proportional to $\theta(\t)$. By symmetry, it is the same for $G_{23}$, and the analysis can be carried for $G_{12}$ also, 
leading to the same result.

Since, for the Doi-Peliti construction we used, the action Eq. (\ref{action}) must be understood in
It\^o's sense, we must conclude that $G_{12}$, $G_{13}$ and $G_{23}$ must be zero when 
evaluated
at equal times.

We can compute evolution equations for the average fields by exploiting the fact that, when the source $J$ is set to zero, 
we have:
\begin{align}
\left\langle \frac{\d S[\varphi]}{\d \varphi(1)} \right\rangle = 0 ,
\end{align}
which gives four evolution equations:
\begin{align}
\left\{ \begin{array}{l} 
\sqrt{\z} \r = ( \z + \lam ) \ola + G_{23}(1,1) \\
\sqrt{\z} \r = ( \z + \lam ) a + G_{13}(1,1) \\
\r = \ola a + G_{12}(1,1) \\
\lam = 2 \r - \sqrt{\z} (\ola + a )
\end{array} \right.
 \end{align}
 The rapidity symmetry imposes that $\ola = a$ and $G_{13}(1,1) = G_{23}(1,1) = G_{12}(1,1) = 0$ by 
 causality so we are left with:
 \begin{align}
 \left\{ \begin{array}{l} 
 a = \ola = \sqrt{\z} \\
 \lam = 0 \\
 \r = \z
 \end{array} \right. ,
 \label{means_of_fields_FA}
 \end{align}
 i.e. the mean-field averages are exact at all order of perturbation.

\subsubsection{Interaction vertices}

Defining $\f = (a-\sqrt{\z},\ola-\sqrt{\z},\lam,\r-\z)$, the field theory reads:
\begin{align}
& Z = \int \DD \f ~ e^{-\frac 12 \int_{1,2} \f(1) G_0^{-1}(1,2) \f(2) - S_{\rm{ng}}[\f]} , \\
& G_0^{-1}(k,\t) \equiv \begin{pmatrix}
0 & \left[ - \partial_{\t} + k^2 +  \z \right] & \sqrt{\z}  & - \sqrt{\z}  \\
\left[ \partial_{\t} + k^2 +  \z \right]  & 0 & \sqrt{\z}  & - \sqrt{\z} \\
\sqrt{\z} & \sqrt{\z} & 0 & -1 \\
-\sqrt{\z} & -\sqrt{\z} & -1 & 2 
\end{pmatrix} , \\
& S_{\rm{ng}}[\f] = \int_{x,t} \f_3(x,t) \f_1(x,t) \f_2(x,t) .  
\end{align}
We see now that this model has the very same structure than the interacting particles, but without the interaction term 
coming from the pair potential, which involves the density field $\f_4$. Here the absence of this term will greatly simplify
the formalism. However, this suppresses the possibility of organizing the diagrammatic expansion in powers of the pair 
potential.

\subsection{Building self-consistent equations for correlation functions}

Following the prescription of Chapter \ref{chap:formalism}, we will write self-consistent equations on the two-point 
functions by performing the double Legendre transformation. 
We define the vertex functions $\Si$:
\begin{align}
& \Si \equiv \begin{pmatrix}
\Si_{11} & \Si_{12} & \Si_{13} & \Si_{14} \\
\Si_{21} & \Si_{22} & \Si_{23} & \Si_{24} \\
\Si_{31} & \Si_{32} & \Si_{33} & \Si_{34} \\
\Si_{41} & \Si_{42} & \Si_{43} & \Si_{44} 
\end{pmatrix}
\end{align}
By definition $G_0^{-1} - \Si$ is the inverse of the full (unknown) propagator $G$ of the system. A self consistent equation
for $G$ is obtained by resorting to the stationary condition of the Legendre transform Eq.(\ref{stat_principle_2_sigma}), 
which gives an expression of $\Si$ as a function of $G$, and then inserting this expression in the Dyson equation:
\begin{align}
( G_0^{-1} - \Si ) G = 1 
\label{dyson_FA}
\end{align}
$\Si$ is bound to respect the symmetries of the action, so that it simplifies its matrix composition:
\begin{equation}
\Si \equiv \begin{pmatrix}
\Si_{11} & \Si_{12} & \Si_{13} & \Si_{14} \\
\Si_{12}^+ & \Si_{11} & \Si_{13}^+ & \Si_{14}^+ \\
\Si_{13}^+ & \Si_{13} & \Si_{33} & \Si_{34} \\
\Si_{14}^+ & \Si_{14} & \Si_{34} & \Si_{44} 
\end{pmatrix} ,
\label{def_Sigma}
\end{equation}
where the superscript $\cdot^+$ denotes time-reversed quantities.

\subsubsection{Mori-Zwanzig equation}

Before resorting to a diagrammatic expansion, we face the fact that we want to obtain equations that depend solely
on the density-density correlation function, whereas we have a four-field field theory. Without making any approximation,
we can obtain an equation similar to the Mori-Zwanzig equation Eq.(\ref{mori-zwanzig_MCT}) by simply rearranging the
elements of the Dyson equation (\ref{dyson_FA}) that contain $G_{44}$.

Using the fact that the interaction vertices are independent of the density field, we have that $\Si_{i4}$ and $\Si_{4i}$ are
equal to zero whatever the value of $i$ is, which allows us to extract from Eq.(\ref{dyson_FA}) 
an exact relation between the propagators:
\begin{align}
G_{34}(1,2) = -\d(1,2) + 2 G_{44}(1,2) - \sqrt{\z} \left[ G_{14}(1,2) + G_{14}^+(1,2) \right] . 
\label{expr_G34}
\end{align}

We get another set of three equations that contain $G_{44}$ plus corrections coming from the vertex functions:
\begin{align}
& \left[ -\partial_\t + \O(k) \right] G_{14}^+(k,\t) + \sqrt{\z}\left( G_{34}(k,\t) - G_{44}(k,\t) \right) 
- \left[ \quad \right]_{14} = 0 , \label{eq1} \\
& \left[ \partial_\t + \O(k) \right] G_{14}(k,\t) + \sqrt{\z}\left( G_{34}(k,\t) - G_{44}(k,\t) \right) 
- \left[ \quad \right]_{24} = 0 , \label{eq2} \\
& \sqrt{\z} \left[ G_{14}(k,\t) + G_{14}^+(k,\t) \right] - G_{44}(1,2) 
- \left[ \quad \right]_{34} = 0 . \label{eq3}  
\end{align}
with the notation $\left[ \quad \right]_{ij} = \sum_k \int_3 \Si_{ik}(1,3) G_{kj}(3,2)$ and $\O(k) = k^2 + \z$. Combinations of
Eqs. (\ref{expr_G34},\ref{eq1}--\ref{eq3}) gives a equation that is closed on $G_{44}$ if all vertex functions are neglected:
\begin{align}
2 \z \O(k) \d(\t) = & \left[-\partial_\t^2 + \O(k)^2 \right] G_{44}(k,\t) \nonumber \\
&  - \left[ 2 \z \O(k) + \partial_\t^2 - \O(k)^2 \right] \left[\quad\right]_{34} 
- \sqrt{\z} \left[ \partial_\t + \O(k) \right] \left[ \quad \right]_{14} 
- \sqrt{\z} \left[ - \partial_\t + \O(k) \right] \left[ \quad \right]_{24}
\label{mori_zwanzig_FA}
\end{align}
We recover the free-diffusion behavior of $G_{44}$ if $\Si = 0$, and the 
effect of the non-linearities of the theory are encoded in the matrix elements $\left[ \quad \right]_{ij}$, that involve the 
vertex functions and all the propagators. We arrive at the same situation than in the Mode-Coupling context, where the 
exact memory kernel depends on all correlation functions of the system, and not only on the density-density one.

\subsection{Mode-Coupling like equation}

The 2PI diagrams that we must consider are built up from a cubic vertex, and thus have the same structure than those 
shown in Eq.(\ref{diags_2PI_dyn})
\begin{align}
& \Si_{11} = \begin{minipage}[1,1]{2cm} \includegraphics[width=2cm]{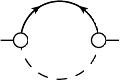} 
\end{minipage} + \begin{minipage}[1,1]{2cm} \includegraphics[width=2cm]{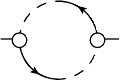} 
\end{minipage} , \\
& \Si_{12} = \begin{minipage}[1,1]{2cm} \includegraphics[width=2cm]{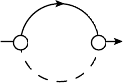} 
\end{minipage} + \begin{minipage}[1,1]{2cm} \includegraphics[width=2cm]{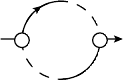} 
\end{minipage} , \\
& \Si_{13} = \begin{minipage}[1,1]{2cm} \includegraphics[width=2cm]{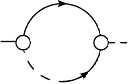} 
\end{minipage} + \begin{minipage}[1,1]{2cm} \includegraphics[width=2cm]{FA_Sigma_13_1} 
\end{minipage} , \\
& \Si_{14} = 0 , \\
& \Si_{33} = \begin{minipage}[1,1]{2cm} \includegraphics[width=2cm]{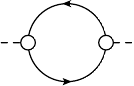} 
\end{minipage} + \begin{minipage}[1,1]{2cm} \includegraphics[width=2cm]{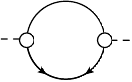} 
\end{minipage} , \\
& \Si_{34} = \Si_{44} = 0 .
\end{align}
The diagrams above must be understood with the following convention:
\begin{align}
& \ola(k,t_1) - \sqrt{\z} \equiv 
\begin{minipage}[1,1]{2cm} \includegraphics[width=1cm]{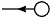} \end{minipage} \\
& a(k,t_1) - \sqrt{\z} \equiv 
\begin{minipage}[1,1]{2cm} \includegraphics[width=1cm]{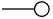}  \end{minipage} \\
& \l(k,t_1) \equiv \begin{minipage}[1,1]{2cm} \includegraphics[width=1cm]{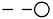}  
\end{minipage} \\
& \int_{k_1,k_2,k_3} \d_{k_1+k_2+k_3} \lambda(k_1,t) ( \ola(k,t_1) - \sqrt{\z} ) ( a(k,t_1) - \sqrt{\z} ) \equiv
\begin{minipage}[1,1]{2cm} \includegraphics[width=1.5cm]{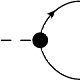} \end{minipage} \\
& G_{11}(k,t) \equiv  
\begin{minipage}[1,1]{2cm} \includegraphics[width=2cm]{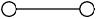}  \end{minipage} \\
& G_{12}(k,t) \equiv 
\begin{minipage}[1,1]{2cm} \includegraphics[width=2cm]{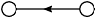}  \end{minipage} \\
& G_{13}(k,t) \equiv 
\begin{minipage}[1,1]{2cm} \includegraphics[width=2cm]{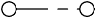}  \end{minipage} \\
& G_{22}(k,t) \equiv 
\begin{minipage}[1,1]{2cm} \includegraphics[width=2cm]{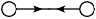}  \end{minipage} \\
& G_{23}(k,t) \equiv 
\begin{minipage}[1,1]{2cm} \includegraphics[width=2cm]{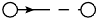}  \end{minipage} \\
& G_{33}(k,t) \equiv 
\begin{minipage}[1,1]{2cm} \includegraphics[width=2cm]{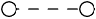}  \end{minipage} 
\end{align}

The natural thing to do in a field-theoretic context would be then to compute the first correction to mean-field in terms 
of all the propagators and look at the result for $G_{44}$. However, in order to understand the structure of the 
Mode-Coupling theory, we are interested in obtaining an equation solely on $G_{44}$, that leads us to perform another
approximation allowing us to replace all the $G_{ij}$s with $(i,j) \ne (4,4)$ in terms of $G_{44}$. This step is rather 
arbitrary in the context on the FA model, but will take its sense in the context of particle systems.

At the level of mean-field, all propagators are proportional to $G_{44}$, as can be seen in Eq.(\ref{G_0_FA}), and we will
use these relations in order to eliminate them in favor of $G_{44}$:
\begin{align}
& G_{11}(k,\t) = G_{22}(k,\t) = 0 , \\
& G_{13}(k,\t) = \frac 1{2 \sqrt{\z} \O(k)} \left[-\partial_\t + \O(k)\right] G_{44}(k,\t) , \\
& G_{23}(k,\t) = \frac 1{2 \sqrt{\z} \O(k)} \left[\partial_\t + \O(k)\right] G_{44}(k,\t) , \\
& G_{12}(k,\t) = \frac 1 {2 \z \O(k)} \left[- \partial_\t + \O(k) \right] G_{44}(k,\t) , \\
& G_{33}(k,\t) = -2 \d(\t) + G_{44}(k,\t)  , \\
& G_{14}(k,\t) = \frac 1{2 \sqrt{\z} \O(k)} \left[-\partial_\t + \O(k)\right] G_{44}(k,\t)
\end{align}
Furthermore, we observe that, when neglecting all vertex functions, we have the following relations:
\begin{align}
& G_{44}(k,\t) = \z e^{-\O(k) | \t |} , \\
\Rightarrow & \left\{ \begin{array}{l}
G_{44}(k,\t) = - \frac 1 {\O(k)} s(\t) \partial_\t G_{44}(k,\t) \\
\left[ \partial_\t + \O(k) \right] G_{44}(k,\t) = 2 \theta(-\t) \O(k) G_{44}(k,\t) \\
\left[ -\partial_\t + \O(k) \right] G_{44}(k,\t) = 2 \theta(\t) \O(k) G_{44}(k,\t)
\end{array} \right.
\end{align}
Which induce:
\begin{align}
\left\{ \begin{array}{l}
G_{11}(k,\t) = G_{22}(k,\t) = 0 , \\
G_{13}(k,\t) = \frac 1 {\sqrt{\z}} \theta(\t) G_{44}(k,\t) , \\
G_{23}(k,\t) = \frac 1{\sqrt{\z}} \theta(-\t) G_{44}(k,\t) , \\
G_{12}(k,\t) = \frac 1 {\z} \theta(\t) G_{44}(k,\t) , \\
G_{33}(k,\t) = -2 \d(\t) + G_{44}(k,\t) , \\
G_{14}(k,\t) = - \frac 1 {\sqrt{\z} \O(k)} \theta(\t)  G_{44}(k,\t)
\end{array} \right.
\end{align}
Replacing these relations in the diagrammatic expressions of the $\Si_{ij}$s, we obtain:
\begin{align}
\left\{ \begin{array}{ll}
\Si_{12}(k,\t) & \displaystyle = \frac 2 \z \theta(-\t) \int_q G_{44}(k-q,\t) G_{44}(q,\t) \\
& \\
\Si_{ij}(k,\t) & = 0 \quad \text{if } (i,j) \ne (1,2)
\end{array} \right.
\end{align}
We thus see that:
\begin{align}
\left[ \quad \right]_{14} & = \int_{t'} \Si_{12}(k,t') G_{14}^+(k,t-t') \nonumber \\
& = \frac 2 {\z^{3/2}} \int_{t'} \theta(-t') \theta(t'-t) \int_q G_{44}(k-q,t') G_{44}(q,t') G_{44}(k,t-t') \nonumber \\
& = 0  \\
\left[ \quad \right]_{24} & = \int_{t'} \Si_{12}^+(k,t') G_{14}(k,t-t') \\
& = \frac 2 {\z^{3/2}} \int_{t'} \theta(t') \theta(t-t') \int_q G_{44}(k-q,t') G_{44}(q,t') G_{44}(k,t-t') \nonumber \\
& = \frac 2 {\z^{3/2}} \int_0^t dt' ~ \int_q G_{44}(k-q,t') G_{44}(q,t') G_{44}(k,t-t') \\
\left[ \quad \right]_{ij} & = 0 \quad \text{otherwise}
\end{align}
Inserting these relations in Eq. (\ref{mori_zwanzig_FA}), we get:
\begin{align}
2 \z \O(k) \d(\t) = \left[-\partial_\t^2 + \O(k)^2 \right] G_{44}(k,\t) + \frac 4 {\z} \int_0^t dt' ~ \int_q G_{44}(k-q,t-t') 
G_{44}(q,t-t') \partial_{t'} G_{44}(k,t') ,
\end{align}
which has the structure of a Mode-Coupling equation, albeit with a trivial kernel, which is due to the trivial statics of this
particular model.

Suppose that the two-point correlation function can have a non-zero value at $t \to \io$, which would signal a loss of
ergodicity. Then we can obtain an equation on the non-ergodicity factor by Laplace transform. Defining:
\begin{align}
G_{44}(k,t) = f(k) G_{44}(k,t=0) + g(k,t) , 
\end{align}
with $g$ a function that decreases to $0$ for $t \to \io$, for all $k$. We get in that case:
\begin{align}
\frac{f(k)}{1-f(k)} = \frac {4}{\z \O(k)^2} \int_q G_{44}(q,t=0) G_{44}(k-q,t=0) f(q) f(k-q).
\end{align}
At $t=0$, the correlation function takes its equilibrium value, which is:
\begin{align}
G_{44}(k,t=0)=\z .
\end{align}
We obtain an equation for $f^*=f(k=0)$:
\begin{align}
\frac {f^*}{1-f^*} = \frac 4 {\z} {f^*}^2 , 
\end{align}
which has a trivial solution $f^*=0$, but also, for $\z \le 1$, two solutions:
\begin{align}
f^* = \frac {1 \pm \sqrt{1 - \z}}2
\end{align}

This analysis shows that by a well defined set of approximations, a Mode-Coupling like equation can be found. This 
procedure is directly applicable to the case of interacting particles, as we show below, but provides a very compact 
formalism when compared to the only existing field-theory approach, that of Kawasaki and Kim \cite{KK07b}.

It was shown in \cite{WBG05,Sz04} that the slow dynamics of the FA model are governed by a zero-temperature fixed 
point, whereas the initial predictions of Fredrickson and Andersen \cite{FA85} was a divergence at finite temperature of 
the relaxation time of the system. 

The transition predicted by our theory is thus spurious. Indeed, this is a consequence of the absence of small
parameter in the theory, that made our approximations quite unjustified. However, this encourages us to apply now this 
method to the case of particle systems, where the strength of the interaction can be used as an organizing device for 
the perturbation expansion.

\section{Application to harmonic spheres}

Now that we presented the approximation method in a simplified model, we present here its application in the case of
the dynamics of harmonic spheres, or any equivalent pair potential that has a repulsive finite-range interaction. The steps
of the derivation are essentially the same, but complicated by the appearance of a second interaction vertex that depends
on the pair potential, and with a modification of the bare action. A conceptual difference with the derivation above is that 
the pair potential naturally serves as an expansion parameter around the ideal-gas contribution.

\subsection{Diagrammatic structure and time-reversal}

The time-reversal transformation is unchanged when compared to the FA case, so that Eqs.(\ref{time_reversal_G}) and 
(\ref{time_reversal_Gamma}) hold. The mean values of the fields are now constrained, since the density is a conserved
quantity, so that we have at all times:
\begin{align}
& \la a(x,t) \ra = \la \ola(x,t) \ra = \sqrt{\r_0} , \\
& \la \r(x,t) \ra = \r_0 , \\
& \la \lam(x,t) \ra = \la \olr(x,t) \ra = 0 .
\end{align}
Expanding the action Eq.(\ref{action_particles_aabar}) around these saddle point solutions, we get the expression of the 
inverse bare propagator for our system of particles:
\begin{align}
& G_0^{-1}(k,\t) = \begin{pmatrix}
0 & \partial_\t + k^2 & - \sqrt{\r_0} & 0 \\
-\partial_\t + k^2 & 0 & -\sqrt{\r_0} & 0 \\
- \sqrt{\r_0} & - \sqrt{\r_0} & 0 & 1 \\
0 & 0 & 1 & u(k)
\end{pmatrix},
\end{align}
with
\begin{align}
u(k) = \frac{k^2}{2 \r_0} \left[ \left( 1 + \b \r_0 v(k) \right)^2 - 1 \right] .
\end{align}
There is now an additional vertex in the theory, that contains the pair potential. Diagrammatically, the fields are defined as:
\begin{align}
& \ola(k,t_1) - \sqrt{\r_0} \equiv 
\begin{minipage}[1,1]{2cm} \includegraphics[width=1cm]{FA_abar} \end{minipage} \\
& a(k,t_1) - \sqrt{\r_0} \equiv 
\begin{minipage}[1,1]{2cm} \includegraphics[width=1cm]{FA_a}  \end{minipage} \\
& \lam(k,t_1) \equiv \begin{minipage}[1,1]{2cm} \includegraphics[width=1cm]{FA_lambda}  
\end{minipage} \\
& \r(k,t) - \r_0 \equiv \begin{minipage}[1,1]{2cm} \includegraphics[width=1cm]{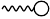}  ,
\end{minipage}
\end{align}
the interaction vertices as:
\begin{align}
& \int_{k_1,k_2,k_3} \d_{k_1+k_2+k_3} \lambda(k_1,t) ( \ola(k,t_1) - \sqrt{\z} ) ( a(k,t_1) - \sqrt{\z} ) \equiv
\begin{minipage}[1,1]{2cm} \includegraphics[width=1.5cm]{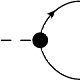} \end{minipage} \\
& \int_{k_1,k_2,k_3} \d_{k_1+k_2+k_3} (\r(k_1,t)-\r_0) ( \r(k_2,t_1) - \sqrt{\r_0} ) ( a(k_3,t_1) - \sqrt{\r_0} ) \equiv
\begin{minipage}[1,1]{2cm} \includegraphics[width=1.5cm]{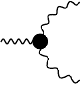} \end{minipage}
\end{align}
and the propagators as:
\begin{align}
& G_{11}(k,t) \equiv  
\begin{minipage}[1,1]{2cm} \includegraphics[width=2cm]{FA_G11}  \end{minipage} \\
& G_{12}(k,t) \equiv 
\begin{minipage}[1,1]{2cm} \includegraphics[width=2cm]{FA_G12}  \end{minipage} \\
& G_{13}(k,t) \equiv 
\begin{minipage}[1,1]{2cm} \includegraphics[width=2cm]{FA_G13}  \end{minipage} \\
& G_{14}(k,t) \equiv 
\begin{minipage}[1,1]{2cm} \includegraphics[width=2cm]{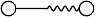}  \end{minipage} \\
& G_{22}(k,t) \equiv 
\begin{minipage}[1,1]{2cm} \includegraphics[width=2cm]{FA_G22}  \end{minipage} \\
& G_{23}(k,t) \equiv 
\begin{minipage}[1,1]{2cm} \includegraphics[width=2cm]{FA_G23}  \end{minipage} \\
& G_{24}(k,t) \equiv 
\begin{minipage}[1,1]{2cm} \includegraphics[width=2cm]{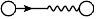}  \end{minipage} \\
& G_{33}(k,t) \equiv 
\begin{minipage}[1,1]{2cm} \includegraphics[width=2cm]{FA_G33}  \end{minipage} \\
& G_{34}(k,t) \equiv 
\begin{minipage}[1,1]{2cm} \includegraphics[width=2cm]{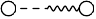}  \end{minipage} \\
& G_{44}(k,t) \equiv 
\begin{minipage}[1,1]{2cm} \includegraphics[width=2cm]{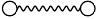}  \end{minipage} .
\end{align}

\subsection{Mori-Zwanzig equation}

Now we can write the equivalent of Eq.(\ref{mori_zwanzig_FA}), i.e. the Mori-Zwanzig equation for the density density
correlation function by the very same procedure to obtain:
\begin{align}
\d(\t) & = \frac1{2 \r_0 k^2} \left( -\partial_\t^2 + \O(k)^2 \right) G_{44}(k,\t) - [ \phantom{ab} ]_{44} 
- \frac{-\partial_\t^2+k^4}{2 \r_0 k^2} [ \phantom{ab} ]_{34} - \frac{\partial_\t+k^2}{2 \sqrt{\r_0} k^2} [ \phantom{ab} ]_{24} 
- \frac{-\partial_\t+k^2}{2 \sqrt{\r_0} k^2} [ \phantom{ab} ]_{14} , \label{mori_obvious_sym}
\end{align}
where $\O(k)$ is now $k^2(1+\b \r_0 v(k))$ and $[ \phantom{ab} ]_{ij}$ is still the $i,j$ matrix element of $\Sigma G$.
By replacing these matrix elements by their actual values we obtain the expression:
\begin{align} 
\d(\t) & = \frac{1}{2 \r_0 k^2} \left( -\partial_\t^2 + \O(k)^2 \right) G_{44}(k,\t) \nonumber \\
& \phantom{ = } - \int_{t} \left[ \Sigma_{14}^+(k,t) + \frac{- \partial_\t^2+k^4}{2 \r_0 k^2} \Sigma_{13}^+(k,t) 
+ \frac{-\partial_\t+k^2}{2 \sqrt{\r_0} k^2} \Sigma_{11}(k,t) 
+ \frac{+\partial_\t+k^2}{2 \sqrt{\r_0} k^2} \Sigma_{12}^+(k,t) \right] G_{14}(k,\t-t) \nonumber \\
& \phantom{ = } - \int_{t} \left[ \Sigma_{14}(k,t) + \frac{- \partial_\t^2+k^4}{2 \r_0 k^2} \Sigma_{13}(k,t) 
+ \frac{-\partial_\t+k^2}{2 \sqrt{\r_0} k^2} \Sigma_{12}(k,t) 
+ \frac{+\partial_\t+k^2}{2 \sqrt{\r_0} k^2} \Sigma_{11}(k,t) \right] G_{14}^+(k,\t-t) \nonumber \\
& \phantom{ = } - \int_{t} \left[ \Sigma_{34}(k,t) + \frac{- \partial_\t^2+k^4}{2 \r_0 k^2} \Sigma_{33}(k,t) 
+ \frac{-\partial_\t+k^2}{2 \sqrt{\r_0} k^2} \Sigma_{13}(k,t) 
+ \frac{+\partial_\t+k^2}{2 \sqrt{\r_0} k^2} \Sigma_{13}^+(k,t) \right] G_{34}(k,\t-t) \nonumber \\
& \phantom{ = } - \int_{t} \left[ \Sigma_{44}(k,t) + \frac{- \partial_\t^2+k^4}{2 \r_0 k^2} \Sigma_{34}(k,t) 
+ \frac{-\partial_\t+k^2}{2 \sqrt{\r_0} k^2} \Sigma_{14}(k,t) 
+  \frac{+\partial_\t+k^2}{2 \sqrt{\r_0} k^2} \Sigma_{14}^+(k,t) \right] G_{44}(k,\t-t) . 
\label{mori-zwanzig_particles}
\end{align}
As expected, the memory kernel depends on all correlation functions, and a substitution has to be performed in order
to express them as a function of $G_{44}$. In Kawasaki and Kim, this substitution is performed in a loop-expansion
scheme, substituting in the one-loop result, the 0-loop relations between propagators. This also what we did in the case
of the FA model. However, here, we have the strength of the interaction as an organizing device, and we can keep 
only the lowest orders in $\e(k) = \b \r_0 v(k)$.

\subsection{Calculation to lowest order in the potential}

Let us first have a look at the lowest order diagrams. There are more of them than in the case of the FA model, but we
will see that only one will contribute at lowest order.
\begin{equation}
\Si_{11} = \begin{minipage}[1,1]{2cm} \includegraphics[width=2cm]{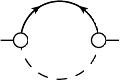} 
\end{minipage} + \begin{minipage}[1,1]{2cm} \includegraphics[width=2cm]{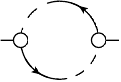} 
\end{minipage} ,
\end{equation}
\begin{equation}
\Si_{12} = \begin{minipage}[1,1]{2cm} \includegraphics[width=2cm]{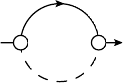} 
\end{minipage} + \begin{minipage}[1,1]{2cm} \includegraphics[width=2cm]{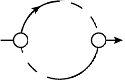} 
\end{minipage} , 
\end{equation}
\begin{equation}
\Si_{13} = \begin{minipage}[1,1]{2cm} \includegraphics[width=2cm]{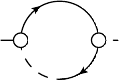} 
\end{minipage} + \begin{minipage}[1,1]{2cm} \includegraphics[width=2cm]{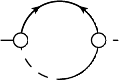} 
\end{minipage} , 
\end{equation}
\begin{equation}
\Si_{14} = \begin{minipage}[1,1]{2cm} \includegraphics[width=2cm]{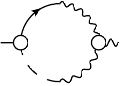} 
\end{minipage} , 
\end{equation}
\begin{equation}
\Si_{33} = \begin{minipage}[1,1]{2cm} \includegraphics[width=2cm]{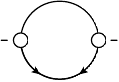} 
\end{minipage} + \begin{minipage}[1,1]{2cm} \includegraphics[width=2cm]{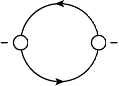} 
\end{minipage}, 
\end{equation}
\begin{equation}
\Si_{34} = \begin{minipage}[1,1]{2cm} \includegraphics[width=2cm]{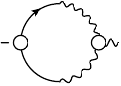} 
\end{minipage} , 
\end{equation}
\begin{equation}
\Si_{44} = \begin{minipage}[1,1]{2cm} \includegraphics[width=2cm]{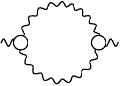} 
\end{minipage} .
\end{equation}
All propagators verify a Mori-Zwanzig equation such as Eq.(\ref{mori-zwanzig_particles}) that relates them to $G_{44}$
at lowest order, with loop corrections that depend on $\Si$, in the same way as in the case of the FA model. 
The expression of the inverse propagator is rather complicated, but we will only need it at the lowest order in $\e$.
Setting $\e=0$ gives:
\begin{align}
G_0(k,\t) = \frac 1 {\r_0} \begin{pmatrix}
0 & \theta(-\t)  & 0 & \sqrt{\r_0} \theta(-\t) \\
\theta(\t) & 0 & 0 & \sqrt{\r_0} \theta(\t) \\
0 & 0 & 0 & 0 \\
\sqrt{\r_0} \theta(\t) & \sqrt{\r_0} \theta(-\t) & 0 & 1 
\end{pmatrix} G_{44}(k,\t)
\end{align}
Using the lowest order diagrams combined with the lowest order expressions of the propagators in function of $G_{44}$,
we get that all vertex corrections can be expressed as integrals of memory kernels multiplied by density-density 
correlation functions:
\begin{align}
\Si_{ij}(k,\t) = \int_q \MM_{ij}(k,q,\t) G_{44}(q,\t) G_{44}(k-q,\t) ,
\end{align}
with the memory kernels given by:
\begin{align}
& \MM_{11}(k,q,\t) = - \frac{\e(q) \e(k-q)}{4 \r_0^3} \left[ q^4 \e(q) + (k-q)^4 \e(k-q) \right] , \\
& \MM_{12}(k,q,\t) = \frac{\theta(\t)}{2 \r_0^3} \left[ q^2 \e(q) + (k-q)^2 \e(k-q) \right]^2 , \\
& \MM_{13}(k,q,\t) = \frac{\theta(\t)}{4 \r_0^{5/2}} \left[ \e(k-q) - \e(q) \right] \left[ q^2 \e(q) - (k-q)^2 \e(k-q) \right], \\
& \MM_{14}(k,q,\t) = \frac{\theta(\t)}{4 \r_0^{3/2}} \left[ k \cdot q \e(k) \e(q) + k \cdot (k-q) \e(k) \e(k-q) 
- q \cdot(k-q) \e(q) \e(k-q) \right] \\
& \phantom{\MM_{14}(k,q,\t) = \frac{\theta(\t)}{4 \r_0^{3/2}}} \left[ q^2 \e(q) + (k-q)^2 \e(k-q) \right] .
\end{align}
\begin{align}
& \MM_{33}(k,q,\t) = \frac1{8 \r_0^2} \left[ \e(q) + \e(k-q) \right]^2 , \\
& \MM_{34}(k,q,\t) = \frac1{4 \r_0} \left[ k \cdot q \e(k) \e(q) + k \cdot (k-q) \e(k) \e(k-q) 
- q \cdot(k-q) \e(q) \e(k-q) \right] \left[ \e(q) + \e(k-q) \right]  , \\
& \MM_{44}(k,q,\t) = \frac12  \left[ k \cdot q \e(k) \e(q) + k \cdot (k-q) \e(k) \e(k-q) - q \cdot(k-q) \e(q) \e(k-q) \right] ^2 , \\
\end{align}
We see that all vertex functions are proportional at least to $\e^2$, even when evaluating the 
propagators at zero order
in the potential. Therefore the relations between propagators that we used are correct to 
$\OO(\e)$.

Finally, we see that we can drop, at order $\OO(\e^2)$, all contributions except $\Si_{12}$,$\Si_{13}$ and $\Si_{33}$. 
Upon substituting their expressions in Eq.(\ref{mori-zwanzig_particles}), we also see that we can neglect the contribution
from $\Si_{33}$ since it is convoluted with $G_{34}$, which is $\OO(\e)$.
Only the two first lines of the loop corrections in Eq.(\ref{mori-zwanzig_particles}) survive
at this order, and we can also replace $G_{14}$ and $G_{14}^+$ by expressions at the lowest 
order in the potential
in function of $G_{44}$:
\begin{align}
& (\partial_\t + k^2) G_{14}(k,\t) = 2 \sqrt{\r_0} \theta(-\t) \partial_\t G_{44}(k,\t) , \\
& (-\partial_\t + k^2) G_{14}^+(k,\t) = - 2 \sqrt{\r_0} \theta(\t) \partial_\t G_{44}(k,\t) .
\end{align}
If $\t>0$ the Heaviside functions restrict the time interval to $[0,\t]$ for the term containing $\Si_{12} G_{14}^+$
and cancels the term containing $\Si_{12}^+ G_{14}$, and we finally obtain:
\begin{align}
0 = & \left( -\partial_\t^2 + \O(k)^2 \right) G_{44}(k,\t) \nonumber \\
& + \frac{k^2}{2 \O(k) \r_0^3} \int_0^{\t} dt \int_q \left[ q^2 \e(q) + (k-q)^2 \e(k-q) \right]^2 
G_{44}(q,\t-t) G_{44}(k-q,\t-t) \partial_{t} G_{44}(k,t) 
\label{MCT_final_particules}
\end{align}
We note that $G_{44}$ is equal to $\r S(k,t)$, where $S(k,t)$ is the dynamical structure factor.
By Laplace transforming Eq.(\ref{MCT_final_particules}) and focusing on the long-time limit, we get an equation for the 
non-ergodicity parameter:
\begin{align}
\frac{f(k)}{1-f(k)} = \frac{\r_0 k^2}{2 \O(k)^3} \int_q \left[ q^2 \b v(q) + (k-q)^2 \b v(k-q) \right]^2 S(q) S(k-q) f(q) f(k-q) .
\end{align}
\begin{figure}[t]
\centering
\includegraphics[width=8cm]{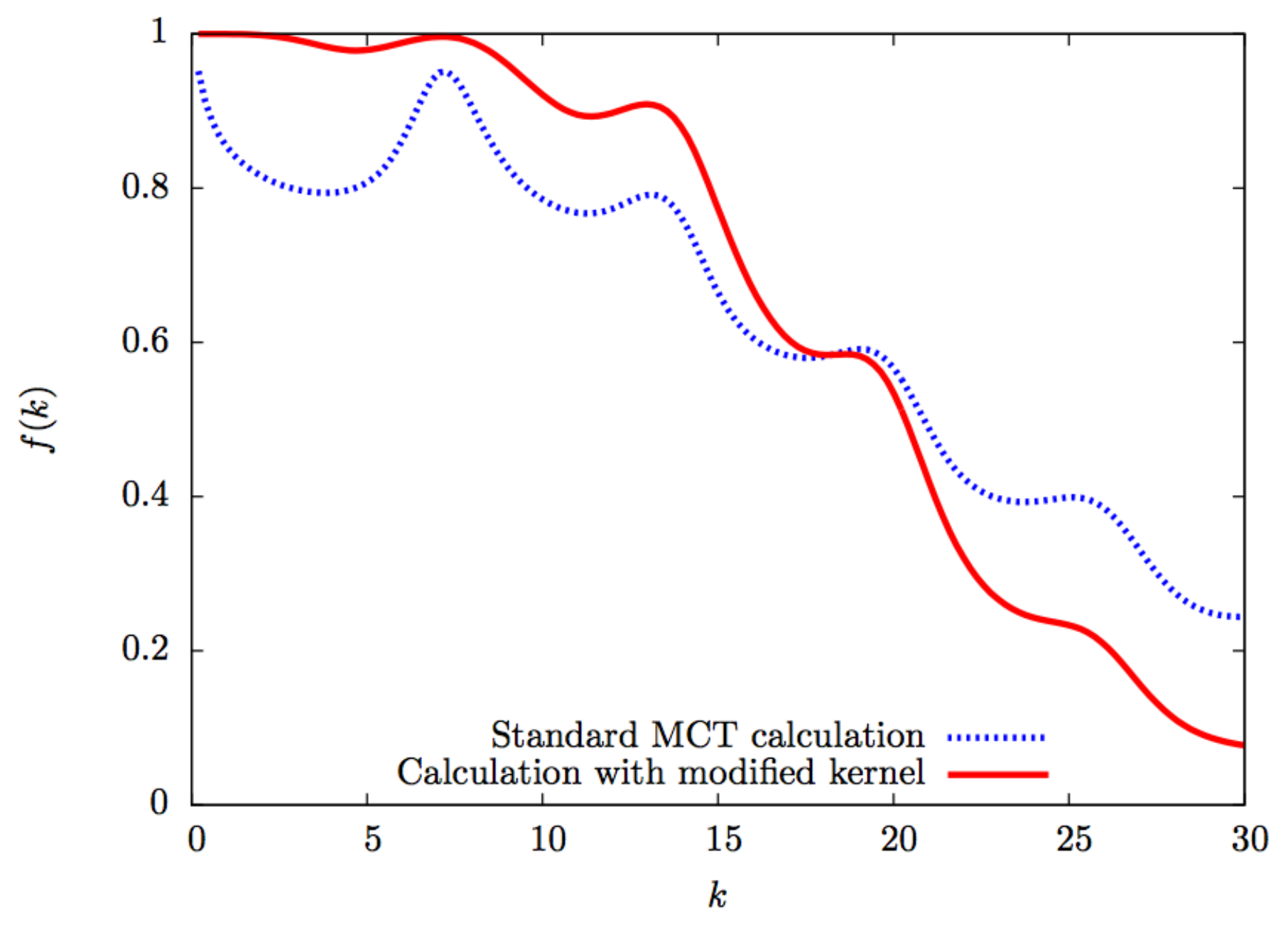}
\caption{Non-ergodicity factor of harmonic spheres obtained by numerically solving Eq.(\ref{f_q_final_me}) (solid line) and 
the original mode-coupling result Eq.(\ref{true_MCT}) (dashed line), for density $\r_0=1.01$ and reduced temperature 
$\b \e = 10^{-4}$.}
\label{fig:f_k_me-MCT}
\end{figure}
This equation has a flavor of the Mode-Coupling equation (\ref{true_MCT}), but because we insisted on staying at the 
lowest order in the potential, we have that the direct correlation function is approximated by $c(k) = - \b v(k)$, which is
equivalent to setting $S(k) = 1/(1+\b \r_0 v(k)) \Leftrightarrow \O(k) = k^2 / S(k)$. This is the RPA approximation of liquid theory \cite{hansen}. Restoring 
the renormalized values of the static quantities we end up with:
\begin{align}
\frac{f(k)}{1-f(k)} = \frac{\r S(k)}{2 k^4} \int_q \left[ q^2 c(q) + (k-q)^2 c(k-q) \right]^2 S(q) S(k-q) f(q) f(k-q) .
\end{align}
Put under this form, the only difference with the true MCT equation is that in the kernel we have $q^2$ instead of 
$k \cdot q$ and $(k-q)^2$ instead of $k \cdot (k-q)$. We can simplify the integral by switching to bi-polar coordinates, i.e.
defining $u=|q|,v=|k-q|$ and making the change of variables $q \to (u,v,\theta)$, where $\theta$ is the angle between 
$k$ and $q$. The Jacobian of this transformation is $uv / k$, and we thus obtain:
\begin{align}
\frac{f(k)}{1-f(k)} = \frac{\r S(k)}{2 k^5} \int_0^\io du \int_{|k-u|}^{k+u} dv ~ uv \left[ u^2 c(u) + v^2 c(v) \right]^2 
S(u) f(u) S(v) f(v)
\label{f_q_final_me}
\end{align}
Numerically solving this equation for harmonic spheres gives the result shown in Fig. \ref{fig:f_k_me-MCT}.
We see that $f$ goes to one when $k$ is small, which seems to constrain the form of the result in the vicinity of
$k = 0$. Indeed, it is a direct mathematical consequence of the form of the kernel: because of the replacement
$k \cdot q \to q^2$, the small $k$ behavior of the right hand side of Eq.(\ref{f_q_final_me}) is singular, which forces the
left hand side to be singular too, which means that $f(k=0)$ must be equal to $1$.

\subsection{A new symmetry}

This small $k$ divergence is related to the appearance in the formalism of a new symmetry. Upon the transformation:
\begin{align}
\left\{ \begin{array}{ll}
 \ola(x,t) & \to \ola(x,t) e^{i \theta} \\
 a(x,t) & \to a(x,t) e^{i \theta}
\end{array} \right. ,
\end{align}
the original action is invariant. This gauge symmetry naturally enforces that $\la \ola \ra = \la a \ra = 0$, a condition
that we break in the beginning of our calculation. Even if this symmetry is not as crucial as time-reversal, and we could
obtain reasonable results while breaking it, it is obvious that in order to have a consistent formalism, we need to find a way
to properly include it in the formalism. The search for a 2PI action that would be operational in the symmetry-broken 
phase of Bose-Einstein condensate is of current interest in condensed matter, see in particular \cite{Ki06,Ki09}, and we 
hope that such extension of the existing tools will be developed in the next years.

\subsection{Discussion}

The approach presented here has three main advantages: 
\begin{itemize}
\item it takes into account the ideal gas part of the dynamics without
having to resort to complicated non-perturbative manipulations as in \cite{KK07b}
\item it possesses a small parameter when the pair potential of the liquid under scrutiny has an energy scale
\item and it gives an operational result with few calculations when staying at the lowest order in the potential
\end{itemize}
It shows in a transparent way what kind of approximations are made in order to obtain a mode-coupling like equation. In particular
it emphasizes the role of projection onto the density-density correlation functions, also present in the formulation
of \cite{KK07b}: a more natural thing to do would be to keep all correlators and solve their self-consistent equations all at 
once, without performing substitutions. In the case of an expansion at the lowest order in the potential this procedure is well
defined, however in most cases in the study of glasses, such expansion would not be possible, for example for 
hard-spheres an expansion in powers of the potential would clearly break down. Also the glass transition happens at quite
low temperature, so that the parameter $\e$ of the theory, which is $\b v(k)$, is rarely small ! A possible exception could be
provided by the Gaussian core model when studied at very high densities, where MCT has been found to be quite 
accurate \cite{IM11b,IM11a,IM11c}.

In the case of the FA model, we have found that when solving the whole set of Dyson equations without focusing on
the density-density sector, the ergodic phase with $f(k)=0$ is always stable at finite temperature. Whether the situation is
the same for the particles is much harder to analyse. 

Another intriguing aspect is that it has been proved, when studying a schematic version of the Mode-Coupling equations,
where space indices are dropped, that upon including successively higher order corrections, the transition is gradually 
shifted to larger and larger densities, and eventually disappears when the whole series is re-summed \cite{MMR06}, while
the MCT fits still provide a good representation of the exact result for a limited time-interval. This is exactly the current 
experimental and numerical situation, and extending such results to the full dynamics of the system would be particularly
satisfying.

There are two main drawbacks to the present theory: 
\begin{itemize}
\item first the use of the similarity transformation forces one to consider a calculation
in the bulk of time, therefore loosing track of the initial conditions, which are ordinarily used to recover the full static 
correlations of the system \cite{ABL06,KK07b}.
\item secondly, a U(1) symmetry has been added in the problem, which is broken by hand by our approximation scheme, 
and is speculated to deteriorate the results at large wave-length.
\end{itemize}
The first drawback is easily dealt with thanks to the accumulation of knowledge in the field theory formulation of glassy
dynamics, since we know on independent grounds how the static correlations get renormalized, but the second issue is 
more serious and prevents an attempt to go to next order in the calculation.

\chapter{Statics: the replica method}
\label{chap:replicas}

After focusing on the treatment of the glass transition by dynamical approaches, this chapter is devoted to the presentation
of results obtained by using the replica method in the vicinity of the glass transition. Starting from assumptions
on the structure of the system in the glass phase inspired by the RFOT scenario, and working in a static framework, 
one is able to derive self-consistent equations for the non-ergodicity factor \cite{PZ10}, a quantity which is originally defined 
in terms of dynamical quantities. 

The link between such equations in dynamical and static theories is still unclear, in particular because the assumptions
made by RFOT on the metastable states of the system are not explicitly {\it a priori} related to a dynamical calculation, 
except in some mean-field disordered models \cite{KW87a,CK93,BM98}. A first attempt to relate the two approaches
by Szamel \cite{Sz10}, by characterizing metastable states by a vanishing current condition, is promising, but suffers from 
the same short-comings than MCT itself, in the sense that the truncation procedure is again rather arbitrary.

We adopt here a complementary approach, by staying in the hypotheses of RFOT, and setting up an expansion of the 
free-energy in powers of the static order parameter defined within replica-theory. The replicated version of the Hyper-Netted-Chain
approximation of liquid theory discussed at the end of Chapter \ref{chap:formalism} provides us with the correct starting 
point for this expansion, and we calculated the lowest order (in powers of the static order parameter) correction to this approximation.
This lowest order correction corresponds to a quadratic term in the self-consistent equation for the non-ergodicity parameter, 
and thus we find the {\it exact} quadratic part of this equation within replica theory: this should be equivalent to the two-mode approximation
made within dynamical formulations.

Surprinsingly, we show that a part of the MCT kernel is recovered, while the other and most important can not be recovered within statics, 
thus disentangling its static and dynamic parts. We found that the qualitative picture of dynamical glass transition is stable
against inclusion of the lowest order correction to the Hyper-Netted-Chain result, but that these corrections are quantitatively relevant
at and away from the transition. Our expansion provides a starting point in order to take correctly these corrections into account, and 
possibly offer a unified picture with the results presented in Chapter~\ref{chap:jamming}.

\section{Focusing on the long time limit}

Motivated by the observations from dynamical calculations such as Mode-Coupling theory, or the calculation presented
in Chapter \ref{chap:dynamics1}, we will now make the assumption that there exists a well-defined transition from an 
ergodic 
liquid to a non-ergodic one, at a density $\r_d(T)$. Above this density, adopting the point of view of the RFOT, the system 
is supposed to be stuck in one of many metastable states, and is not able to visit its whole phase-space anymore. The 
partition function is thus postulated to be separated in many pure states, parts of the phase space that are disconnected
from one another. This picture is of course inspired from mean-field models, where it is exactly realized \cite{MPV87}.
In finite dimension, the time needed in order to jump from one state to the other will be finite. A phenomenological theory of 
nucleation \cite{KTW89,DSW09,BB04a,BB09} has been devised in order to correct 
the mean-field vision that we describe here. However, until now, no theoretic calculation on realistic models such as 
harmonic spheres or hard spheres have been able to put quantitative background on these ideas, and we will stay at the 
level of these mean-field concepts.

\subsection{Ergodicity breaking and complexity}

By definition, ergodicity breaking, if it takes place, separates the phase space into ``pure" states, i.e. states that can not
be connected by any dynamical process in the thermodynamic limit, even in the $t \to \io$ limit. Thus, we index these 
states by $\alpha$ that runs 
from $1$ to $\NN$, and we calculate the average of an observable $A$ as follows:
\begin{align}
\la A \ra & = \frac 1 Z \sum_{\CC} e^{-\b H(\CC)} A(\CC) = \sum_{\a = 1}^\NN \frac{\sum_{\CC \in \a} e^{-\b H(\CC)}}{Z} 
\frac 1 {\sum_{\CC \in \a} e^{-\b H(\CC)}} \sum_{\CC \in \a} e^{-\b H(\CC)} A(\CC) ,
\end{align}
where $\CC$ is a microscopic state of the system (here the set of positions of the particles). Defining
\begin{align}
& Z_\a = \sum_{\CC \in \a} e^{-\b H(\CC)}, \\ 
& w_\a = \frac 1 Z \sum_{\CC \in \a} e^{-\b H(\CC)} ,
\end{align}
We thus define the average inside a state $\a$:
\begin{equation}
\la A \ra_\a \equiv \frac 1 {Z_\a} \sum_{\CC \in \a} A(\CC) e^{-\b H(\CC)} ,
\end{equation}
and an average over all states:
\begin{align}
\overline{A_\a} \equiv \sum_{\a = 1}^\NN w_\a A_\a .
\end{align}
Finally any average will be denoted by $\overline{\la ~ \cdot ~ \ra_\alpha}$. 

Defining $f_\alpha$, the effective free-energy of state $\alpha$, by:
\begin{align}
f_\alpha = - \frac{k_B T}N \ln Z_\a ,
\end{align}
where $N$ is the number of particles in the system, and inserting this definition in the partition function we get:
\begin{align}
Z = \sum_{\alpha=1}^\NN e^{-N \beta f_{\alpha}} .
\end{align}
We define now the complexity as the extensive component of the number of metastable states. For a given free-energy 
level $f$, supposing that there are 
$\NN(f)$ metastable states that have this precise free-energy, we introduce $\Si(f)$ as:
\begin{align}
\Si(f) = \frac{1}{N} \ln \NN(f) .
\end{align}
By introducing a delta function in the partition function, we obtain:
\begin{align}
Z & = \sum_{\alpha=1}^\NN \sum_f ~ \d(f-f_\alpha) e^{-N \b f}  \\
& = \sum_f  e^{-N \b f} \underset{= \NN(f)}{\underbrace{\sum_{\alpha=1}^\NN \d(f-f_\alpha)}} \\
& = \sum_f  e^{-N \b \left[ f - T \Si(f) \right]} .
\end{align}
In the thermodynamic limit, the free-energy $F$ of the system will be dominated by the free-energy that minimizes the 
argument of the exponential:
\begin{align}
& F \equiv \lim_{N \to \io} - \frac{k_B T}{N} \ln Z \\
\Rightarrow  & \left\{  \begin{array}{ll}
& F = f^* - T \Si(f^*) \\
& \\
& \displaystyle f^* \text{ defined as } \left. \frac{\partial \Si(f)}{\partial f} \right|_{f^*} = \frac 1 T
\end{array} \right. \label{complexity_1}
\end{align}
If the number of metastable states is sub-exponential, we will have $\Si = 0$, and the partition function will be dominated
by the true free-energy minimum. But if we have $\Si \ne 0$, the free-energy of the glass in Eq.(\ref{complexity_1}) will be
smaller than the free energy of this isolated free-energy minimum, and thus the glass phase will be preferred 
thermodynamically.

We have now reduced the problem of detecting the glass transition to checking whether there exists or not a finite 
complexity. It is worth noting that, if there exists a transition from zero to non-zero complexity, this won't be associated
to a true thermodynamic phase transition: indeed the entropic term $-T \Si(f^*)$ will act, thermodynamically, as a driving 
force that allows the system to visit all metastable states, thus restoring ergodicity. 

However, dynamically, and in the thermodynamic limit, the time needed to perform these jump goes to infinity at the 
transition
$\r_d(T)$, and the system is effectively stuck forever in this particular state.

Of course Eq.(\ref{complexity_1}) is useless since we still do not know the free-energy of the states $f^*$ and the 
complexity $\Si(f^*)$, and we need a tool to go further. An adaptation of the replica method used in the study of mean-field 
models provides such a tool, as was realized by Monasson \cite{Mo95}.

\subsection{Order parameter for the glass transition}

In analogy with the ferromagnet-paramagnet transition, detecting the transition in that case with 
thermodynamical methods is easy, since we know exactly the structure of the two equivalent ordered states: all spins up
or all spins down. Thus, we add a small external field which is coupled to the magnetization to explicitly break the symmetry
between the two states, calculate the free-energy with this small symmetry breaking field, then let the strength of the field
go to zero. If the system is in the disordered phase, letting go of the external field averages the magnetization to zero. However, if the 
system is in the ordered (ferromagnetic) state, the total magnetization will be equal to a non-zero value even in the 
limit of zero external field. This provides a scalar order parameter for the ferromagnet-paramagnet transition: the 
magnetization.

In the case of the glass transition, we do not know the possible final states after the transition: they are infinitely numerous,
and are frozen liquid configurations. Thus it is much harder to pin the system towards one of these states. However, one 
possible way is to introduce exact copies of the system, say for example $m$ copies. The copies are indexed by 
$a=1,\ldots,m$. If the copies are uncorrelated, then the partition function of the replicated liquid is just $Z^m$. Now 
consider introducing a small attractive coupling between the copies. More precisely, particle $i$ in copy $a$ will interact 
with particle $j$ in copy $b$ (if $a \ne b$) via an attractive pair potential $\tilde{w}(x_i^a - x_j^b)$, that has a small 
amplitude $\e$. 

Exactly as in liquid theory, we are interested in global quantities such as the density, and we can generalize the definitions
of the density and correlation functions to the case of the replicated liquid, defining:
\begin{align}
\hr_a(x) & = \sum_{i=1}^N \d(x-x_i^a) , \label{def_hroa} \\
\r_a(x) & = \la \hr_a(x) \ra. \label{def_roa}
\end{align}
Of course by translational invariance and because all copies are identical, in the vanishing coupling limit, $\r_a(x)=\r$. 
We have to look at a two-point quantity, as in the dynamical framework. We now search for a quantity playing the role of
an order parameter for the glass transition, and we expect it to be a two-point quantity. We naturally turn to the 
generalization for copies of the pair correlation function defined in Eq.(\ref{def_h}) for copies:
\begin{align}
h_{ab}(x,y) & \equiv \frac{\la \hr_a(x) \hr_b(y) \ra}{\r^2} - 1 - \d_{ab} \frac 1{\r} \d(x-y) \label{def_hab} ,
\end{align}
and we show below that this is indeed the correct order parameter.
Note that, for $a \ne b$, the coinciding point term is absent, because particle $i$ of copy $a$ interacts with all particles of
copy $b$, including particle $i$, whereas within copy $a$, particle $i$ does not interact with itself. 
At finite $N$, ergodicity is guaranteed and the small attraction will ensure that the two copies will be 
in the same state. Now perform the large system size limit $N \to \io$: 
\begin{itemize}
\item if $\r < \r_d$, ergodicity is maintained, and letting $\e$ to zero will lead the two copies to de-correlate, i.e. 
$h_{ab} = 0$;
\item if $\r > \r_d$, the two copies will be trapped into a metastable state $\a$. Finally letting $\e$ go to zero, the two 
copies will remain correlated and $h_{ab} \ne 0$.
\end{itemize}
We see that the $h_{ab}$ function, provided it is calculated with vanishingly small attractive coupling, is a 
good order parameter for the glass transition.

The average in Eq.(\ref{def_hab}) is, in our static interpretation, first an average inside a state $\alpha$, then an 
average over all states. The presence of the attractive coupling between the copies will force them into the same pure 
state $\a$, leading to:
\begin{align}
h_{ab}(x,y) = \frac{\overline{\la \hr_a(x) \hr_b(y) \ra_{\a}}}{\r^2} - 1 \quad \text{ for } a \ne b.
\end{align}
Letting the interaction go to zero will allow them to de-correlate inside the state, leading to:
\begin{align}
\lim_{\e \to 0} h_{ab}(x,y) =  \frac{\overline{\la \hr(x) \ra_\a \la \hr(y) \ra_\a}}{\r^2} - 1
\end{align}
Note that the average density inside a state is the same for all copies, but needs not be constant, since translational 
invariance is restored only after summation over all the states.

Independently, the time-dependent density-density correlation, which serves as a dynamic order parameter for the glass
transition is defined in real-space by:
\begin{align}
S(x,y,t) & = \la (\hr(x,t) - \r) (\hr(y,0) - \r) \ra .
\end{align}
The RFOT interpretation is that, at a volume fraction above the glass transition volume fraction, and in the long time limit, 
the system is stuck in a pure state $\a$, and we have:
\begin{align}
S(x,y,t) \underset{t \to \io}{\to} \overline{\la \hr(x,t) \hr(y,0) \ra_{\a}} - \r^2
\end{align}
But the system is at least able to de-correlate inside the state, giving:
\begin{align}
S(x,y,t) \underset{t \to \io}{\to} \overline{ \la \hr(x) \ra_{\a} \la \hr(y) \ra_{\a} } - \r^2 = \r^2 h_{ab}(x,y)  \label{link_Fkt_htilde}.
\end{align}
Now we get in Fourier space:
\begin{align}
S(k,t) \underset{t \to \io}{\to} \r^2 h_{ab}(k) .
\end{align}
Finally we see that $h_{ab}$ is directly proportional to the non-ergodicity parameter \cite{PZ10} defined in 
Eq.(\ref{def_f_k_intro}) of Chapter \ref{chap:intro}:
\begin{align}
f(k) = \frac{\r h_{ab}(k)}{S(k)} \text{ with } a \ne b \label{link_f_hab}
\end{align}
We see that the static order parameter that we defined with the introduction of copies is the same physical observable
as the one defined by dynamical means. It is thus interesting to investigate the degree of relevance of the set of 
approximations made to obtain this static description (separation of the partition function into pure states, form of the 
spectrum of free-energies, number of metastable states, ...), in order to sort out what is purely static from what is purely
dynamic in the glass transition.

\subsection{Computing the complexity with replicas}

We have seen that introducing copies attracted to each other allows one to define an order parameter, but it also gives a 
way to compute the complexity and free energy of the system. 
Consider $m$ identical copies of the original system.
If copies are attracted to each other, they fall in the same state. Then in the limit of vanishing coupling, they de-correlate 
inside this state. Moreover, the free-energy landscape felt by all copies is the same as that of the original
one when the coupling is sent to zero, such that for the partition function $Z_m$ of the replicated liquid can now be written, 
in the same way as in Eq.(\ref{complexity_1}) as:
\begin{align}
& Z_m = \sum_f  e^{- N \b \left( mf - T \Si(f) \right)} , \\
\Rightarrow & \left\{ \begin{array}{ll} 
& \displaystyle F_m = f^*(m) - T \Si(f^*(m)) , \\
& \\
& \displaystyle \left. \frac{\partial \Si}{\partial f} \right|_{f^*(m)} = \frac m T
\end{array} \right. \label{complexity_m}
\end{align}
Performing now an analytic continuation for arbitrary values of $m$, the free energy and complexity can be 
calculated as \cite{Mo95}:
\begin{align}
\left\{ \begin{array}{ll}
& f^*(m) = \displaystyle \left. \frac{\partial F_m}{\partial m} \right|_{m=1} , \\
& \\
& \Si(f^*(m)) = \displaystyle \left. \frac{m^2}T \frac{\partial F_m/m}{\partial m} \right|_{m=1} .
\end{array} \right.
\end{align}
The program is now clear: 
\begin{itemize}
\item compute the free-energy of an $m$-time replicated liquid in the presence of an 
attractive coupling
\item take the large system size limit
\item let the coupling go to zero
\item make the free-energy analytic in $m$
\item take a derivative with respect to $m$ to get $\Si(f^*(m))$
\item take the limit $m \to 1$
\end{itemize}
If the resulting complexity is non zero, the glass transition has been detected and we can compute the corresponding 
structure factor by computing the correlation function between two copies.

\section{Replicated liquid theory}

To realize our program, we only need tools issued from liquid theory, as described in chapter 
\ref{chap:formalism}, since a replicated liquid can be seen as a mixture of $m$ different species of particles.
All results from chapter \ref{chap:formalism} have straightforward generalizations to multi-component systems.
The partition function of the replicated system is:
\beq
Z_m = \Tr e^{\frac{1}2 \underset{i,j,a,b}{\sum'} w_{ab}(x_i^a - x_j^b) + \underset{i,a}{\sum} \n_a(x_i^a)} \ ,
\eeq
where $w_{ab}$ is equal to $w$ for $a=b$, and $w_{ab}$ is a small attractive coupling when $a \ne b$. 
The prime on the summation sign $\sum'$ means that when $a=b$, the summation must exclude the case $i=j$, 
and the trace operation is now defined 
\mbox{as: $\Tr \bullet = \sum_{N=0}^\io \frac 1{N!^m} \int \prod_{a=1}^m \prod_{i=1}^N dx_i^a ~ \bullet$.} For the sake of 
simplicity, $\bfn$ in the following will denote the set of $m$ chemical potentials $\{\n_a\}_{a=1 \dots m}$, and $\bf{w}$ will 
denote the set of $m^2$ pair potentials $\{ w_{ab} \}_{a,b=1 \ldots m}$. Equivalently the family of $m$ average densities 
$\{ \r_a \}_{a=1 \ldots m}$ defined in 
Eq.~(\ref{def_roa}) will be denoted by $\bfr$ and the family of $m^2$ correlation functions $\{ h_{ab} \}_{a,b=1 \ldots m}$ will be denoted by 
$\bf{h}$. 
The replicated two-point density and pair potential are defined as \cite{MP96}:
\begin{align}
& \hr^{(2)}_{ab}(x,y) = \hr_a(x) \hr_b(y) - \d_{ab} \d(x-y) \hr_a(x) \ , \\
& w_{ab}(x,y) = w(x,y) \d_{ab} + \tilde{w}(x,y) (1-\d_{ab}) .
\end{align}
We see at that point that a replica index always appears with a space index, so that, depending on the context, we may 
group them in one subscript for simplicity.
As in the non-replicated case, the chemical potentials $\bfn$ is coupled to the replicated one-point density defined 
in Eqs.~(\ref{def_roa}--\ref{def_hroa}):
\beq
\frac{\d \ln Z_m[\bfn,{\bf w}]}{\d \n_a(x)} = \r_a(x) \ ,
\eeq
and the pair potentials $\bf{w}$ are coupled to the replicated two-point density:
\beq
\frac{\d \ln Z_m[\bfn,{\bf w}]}{\d w_{ab}(x,y)} = \frac 12 \r^{(2)}_{ab}(x,y) \ .
\eeq
The two-point density is trivially related to the pair correlation function defined in Eq.~(\ref{def_hab}):
\beq
h_{ab}(x,y) = \frac{\r^{(2)}_{ab}(x,y)}{\r_a(x) \r_b(y)} - 1 \ .
\eeq
The static order parameter of the glass transition is thus the two-point density.
We saw that in order to obtain accurate and consistent approximations on the free-energy (that is needed to access the
complexity) and pair correlation function (that we need to compute the non-ergodicity factor), it is better to start from
the double Legendre transform of the partition function with respect to the chemical potential and the pair potential
(note the different choice of sign when compared to Chap.~\ref{chap:formalism}) :
\beq
\left\{ \begin{array}{ll}
& \displaystyle \G_m[\bfr,{\bf h}] = \sum_a \int_x \r_a(x) \n^*_a(x) + \frac 1 2 \sum_{a,b}  \int_{x,y} \r^{(2)}_{ab}(x,y) w_{ab}^*(x,y) 
- \ln Z_m[\bfn^*,{\bf w}^*] \\
& \\
& \displaystyle \left. \frac{\delta \ln Z_m[\bfn,{\bf w}]}{\delta \n_{a}(x)} \right|_{\bfn^*,{\bf w}^*} = \r_{a}(x)
\quad \text{ and } \quad 
\left. \frac{\delta \ln Z_m[\bfn,{\bf w}]}{\delta w_{ab}(x,y)} \right|_{\bfn^*,{\bf w}^*} = \frac 1 2 \r^{(2)}_{ab}(x,y) \ .
\end{array} \right.
\eeq
As in the non-replicated case, this double Legendre transform was found by Morita \cite{Mo59a} to be given by:
\beq \begin{split}
& \G_m[\bfr,\bfh] = \G_{\rm{IG}}[\bfr,\bfh] + \G_{\rm{ring}}[\bfr,\bfh] + \G_{\rm{2PI}}[\bfr,\bfh] \ , \\
& \G_{\rm{IG}}[\bfr,\bfh] = \sum_a \int_x \r_a(x) \left[ \frac{}{}\! \ln \r_a(x) - 1 \right] \\
& \phantom{\G_{\rm{IG}}[\bfr,\bfh] =} + \frac 12 \sum_{a,b} \int_{x,y} \r_a(x) \r_b(y) \left( \frac{}{}\! [1+h_{ab}(x,y)] \ln[1+h_{ab}(x,y)] - h_{ab}(x,y) \right) \ , \\
& \G_{\rm{ring}}[\bfr,\bfh] = \frac 12 \sum_{n \ge 3} \frac{(-1)^n}{n} \Tr \r_{a_1}(x_1) h_{a_1 a_2}(x_1,x_2) \cdots \r_{a_n}(x_n) h_{a_n a_1}(x_n,x_1) \ , \\ 
\end{split}
\label{morita_rep}
\eeq
and $\G_{\rm{2PI}}$ is the sum of all 2-irreducible diagrams composed of black nodes $\r_a(x)$ and links $h_{ab}(x,y)$. They are 
such that when two links are removed from a diagram, it doesn't disconnect in two separate parts. 
As in the one-component case, $\G_{\rm{ring}}$ in Eq.(\ref{morita_rep}) is the sum of all ring diagrams, with appropriate weighting.
Note that we use ${\bf h}$ as a variable, but the natural pair of variables is $\r$ and $\r^{(2)}$, and differentiations with respect to $\r_a$
must be done at $\r^{(2)}_{ab}$ fixed instead of $h_{ab}$ fixed: this can be important since $h_{ab}$ is a function of both $\r^{(2)}$ and $\r$!

We define for convenience the propagator and self-energy of the replicated system:
\beq \begin{split}
& G_{ab}(x,y) = \la \r_a(x) \r_b(y) \ra - \la \r_a(x) \ra \la \r_b(y) \ra = \r_a(x) \d_{ab} \d(x,y) + \r_a(x) \r_b(y) h_{ab}(x,y) \ , \\
& \G^{(2)}_{ab}(x,y) = \frac{\d^2 \G_m[\bfr,{\bf h}]}{\d \r_a(x) \d \r_b(y)} = \frac1{\r_a(x)} \d_{ab} \d(x,y) - c_{ab}(x,y) \ ,
\end{split} \eeq
where $c_{ab}$ is the generalization to mixtures of the direct correlation function. 
We have the replicated version of the OZ relation in Eq.~(\ref{OZ2}), i.e. that: 
\beq
\sum_c \int_z \G^{(2)}_{ac}(x,z) G_{cb}(z,y) = \d_{ab} \d(x,y) \ , 
\eeq 
or equivalently:
\beq
h_{ab}(x,y) = c_{ab}(x,y) - \sum_c \int_z h_{ac}(x,z) \r_c(z) c_{cb}(z,y) \ .
\label{OZ2rep}
\eeq

\subsection{Replica symmetric ansatz and calculation of the complexity}

The replicated Morita-Hiroike functional in Eq.~(\ref{morita_rep}) must be extremalized with respect to the pair correlation $h_{ab}$, 
which gives a self consistent equation
on $h_{ab}$, making use of the replicated version of Eq.(\ref{morita_var-principle}) \cite{hansen}:
\begin{align}
w_{ab}(x,y) = \ln (1 + h_{ab}(x,y) ) - h_{ab}(x,y) + c_{ab}(x,y) + \frac 2{\r_a(x) \r_b(y)} \frac{\d }{\d h_{ab}(x,y)} \{ \textrm{2PI diagrams} \} , 
\label{morita_var-principle_rep}
\end{align} 
We already see that the form of pair potential that we have chosen implies an asymmetry between $h_{ab}$ with $a \ne b$
and $h_{aa}$. Otherwise, we will suppose that all replicas play the same role, and we define for the following:
\begin{align}
\left\{ \begin{array}{ll}
& h(x,y) \equiv h_{aa}(x,y) , \\
& \\
& \tilde{h}(x,y) \equiv h_{ab}(x,y) \quad \text{ with } a \ne b .
\end{array} \right. \label{1RSB}
\end{align}
This is equivalent to a 1RSB structure in the context of mean-field disordered spin glasses \cite{Mo95}, although we merely assumed
replica symmetry. 
We called the $aa$ correlation function $h$, since we anticipate that in the ergodic phase, $\tilde{h}=0$ and thus $h_{aa}$ is the liquid 
correlation function. For the density fields $\r_a(x)$, the chemical potentials are not supposed to be asymmetric, and we 
suppose that as a result $\r_a(x) = \r \quad \forall ~ a$.

With the expression of the double Legendre transform and the 1RSB ansatz, performing our program is quite easy starting from this set of equations:
\begin{itemize} 
\item The thermodynamic limit has been taken, since Mayer diagrams, apart from a trivial
volume dependence, will depend only on $\r = N/V$, so that $N$ and $V$ can safely be sent to infinity ;
\item Letting the couplings go to zero is trivially done by setting $w_{ab}$ to zero for $a \ne b$ in
Eq.(\ref{morita_var-principle_rep}) before evaluation of $h_{ab}$ ;
\item The equivalence between replicas assumed in Eq.(\ref{1RSB}) will render the free energy analytic in $m$, 
since the summations over replica indices will make the appearance of $m$ explicit, and the free-energy will then only 
depend on $m$, $h$ and $\tilde{h}$. For example we can perform the summations over the replica indices in 
Eq.(\ref{OZ2rep}) to get:
\begin{align}
\left\{ \begin{array}{ll}
c(x,y) & = \displaystyle h(x,y) -  \r \int_z h(x,z) c(z,y) - (m-1) \r \int_z \th(x,z) \tc(z,y) \\
& \\
\tc(x,y) & = \displaystyle \th(x,y) - \int_z \th(x,z) c(z,y) - \r \int_z h(x,z) \tc(z,y) - (m-2) \r \int_z \th(x,z) \tc(z,y)
\end{array} \right.
\end{align}
\item We can take the limit $m=1$ that we are interested in:
\begin{align}
\left\{ \begin{array}{ll}
c(x,y) & =  \displaystyle h(x,y) -  \r \int_z h(x,z) c(z,y) \\
& \\
\tc(x,y) & = \displaystyle \th(x,y) - \int_z \th(x,z) c(z,y) - \r \int_z h(x,z) \tc(z,y) + \r \int_z \th(x,z) \tc(z,y)
\end{array} \right. 
\label{OZ2_rep_1RSB}
\end{align}
\item The replicated free-energy is calculated in the end by inserting $h$ and $\tilde{h}$ that result from 
Eq.(\ref{morita_var-principle_rep}) and Eq.(\ref{OZ2_rep_1RSB}) back into Eq.(\ref{morita_rep}).
\end{itemize}

The first natural thing approximation to do is to neglect altogether the 2PI diagrams in the full expression of the free-energy.
This was done by M\'ezard and Parisi \cite{MP96}, and we review their results here, in the light of our analogy between the 
static order parameter $\tilde{h}$ and the dynamical one $f(k)$ provided by Eq.(\ref{link_f_hab}).

\subsection{Replicated Hyper-Netted-Chain approximation}

Now if we neglect the 2PI diagrams, Eq.(\ref{morita_var-principle_rep}) gives two self consistent equations on $h$ 
and $\th$:
\begin{align}
\left\{ \begin{array}{ll}
\ln (1 + h(x,y)) & = w(x,y) + h(x,y) - c(x,y)  \\
& \\
\ln (1 + \th(x,y)) & = \th(x,y) - \tc(x,y)
\end{array} \right. \label{HNC_rep_liq}
\end{align}
These equations must be supplemented with the replicated Orstein-Zernike equations evaluated at 
$m=1$ Eq.(\ref{OZ2_rep_1RSB}), that read in Fourier space:
\begin{align}
\left\{ \begin{array}{ll}
& c(k) = h(k) - \r h(k) c(k) , \\
& \\
& \tc(k) = \th(k) - \r h(k) \tc(k) - \r \th(k) c(k) + \r \th(k) \tc(k) .
\end{array} \right. \label{HNC_rep}
\end{align}
We see that the functions $h$ and $c$ do not depend on $\th$ and $\tc$, as expected since they must coincide with the 
corresponding liquid quantities when $\th = \tc = 0$. They are given by the HNC equations, that have been already 
discussed in chapter \ref{chap:formalism}.

The functions $\th$ and $\tc$ satisfy a similar set of equations, that also involve $h$ and $c$:
\begin{align}
\left\{ \begin{array}{ll}
& \tc(x,y) = \th(x,y) - \ln (1 + \th(x,y)) \\
& \\
& \tc(k) = \th(k) - \r h(k) \tc(k) - \r \th(k) c(k) + \r \th(k) \tc(k) .
\end{array} \right.
\label{HNCrep_offdiag}
\end{align}
We can rewrite the second of these equations as:
\begin{align}
\tc(k) = \frac{(1 - \r c(k)) \th(k)}{1+\r h(k) - \r \th(k)} ,
\end{align}
and use the definition of the structure factor Eq.(\ref{def_Sk}), the Ornstein-Zernike equation Eq.(\ref{OZ2}) and the link
between $h_{ab}$ for $a \ne b$ with the non-ergodicity factor Eq.(\ref{link_f_hab}) to obtain:
\begin{align}
\tc(k) = \frac 1{\r S(k)} \frac{f(k)}{1-f(k)} . \label{link_f_tc}
\end{align}
Since $\tc$ and $\th$ are both related to $f$, we see that Eq.(\ref{HNC_rep}) is a self-consistent equation for the 
non-ergodicity parameter. 

One may wonder if there is any resemblance between such result and the Mode-Coupling one. 
To make an explicit bridge between the two, we expand the logarithm term in powers of $f$, to get, to lowest order:
\begin{align}
\frac{f(k)}{1-f(k)} = \frac {S(k)}{2 \r} \int_q S(q) S(k-q) f(q) f(k-q) , \label{eq_f_HNC_expanded}
\end{align}
which has the very same structure as the long-time limit of the Mode-Coupling equation Eq.(\ref{true_MCT}), except 
that the kernel is missing here. For the sake of comparison, we note that we can rewrite the MCT equation as:
\beq \begin{split} 
& \frac{f(k)}{1-f(k)} = \frac{\r S(k)}2 \int_q \MM(k,q) S(q) S(k-q) f(q) f(k-q) \ , \\
& \MM(k,q) = \left[ \frac{k \cdot q}{k^2} \G^{(2)}_{liq}(q) + \frac{k \cdot(k-q)}{k^2} \G^{(2)}_{liq}(k-q) + \r \G^{(3)}_{liq}(q,k-q) \right]^2 \ .
\end{split} \eeq
Note here the presence of the three-body direct correlation function of the liquid, even though it is usually neglected, since it has been shown 
to be negligible with respect to the other, two-body term~\cite{BGL89}, except for special cases~\cite{AGA11}.
We see that Mode-Coupling only keeps $\OO(f^2)$ terms, whereas even a simple 
approximation like HNC retains an infinite number of those. However the HNC kernel to second order is trivial, and 
reminiscent of the result obtained by \cite{ABL06} by a dynamical field-theory calculation. The result of such equation
is that $f(k) = 1$ for all densities, a result that must be rejected when compared to the trivial solution $f(k)=0$. 
In contrast, if one keeps the full $\ln$ term, we obtain the following self-consistent
equation:
\begin{align}
\frac{f(k)}{1-f(k)} = & \r S(k) \FF \TT \left\{ \frac{(S \otimes f)(r)}{\r} - \ln \left( 1 + \frac{(S \otimes f)(r)}{\r} \right) \right\} ,
\label{f_k_repHNC}
\end{align}
where $\otimes$ means convolution in real space, and $\FF \TT\{ g(r) \}$ is the Fourier transform of $g(r)$. 

We solved this equation, combined with the HNC approximation for the
liquid part, for the hard-sphere pair potential:
\begin{align}
w(r) = \left\{ \begin{array}{ll}
- \io & \quad \text{ if } r < \s \\
0 & \quad \text{ otherwise }
\end{array} \right. ,
\label{potential_HS}
\end{align}
which is the $T \to 0$ limit of the harmonic spheres potential Eq.(\ref{potential_harmS}).

Similarly to the liquid HNC equations, the numerical scheme we used was the following:
\begin{itemize}
\item Start from a guess on $\tc(k)$;
\item Deduce the corresponding $f(k)$ with Eq.(\ref{link_f_tc});
\item Inverse Fourier transform to obtain $f(r)$;
\item Plug into Eq.(\ref{f_k_repHNC}) to get a new estimate of $f(k)$;
\item Deduce the corresponding $\tc(k)$;
\item Iterate this procedure until $\tc(k)$ does not change anymore between two steps.
\end{itemize}
\begin{figure}[htb]
\includegraphics[width=14cm]{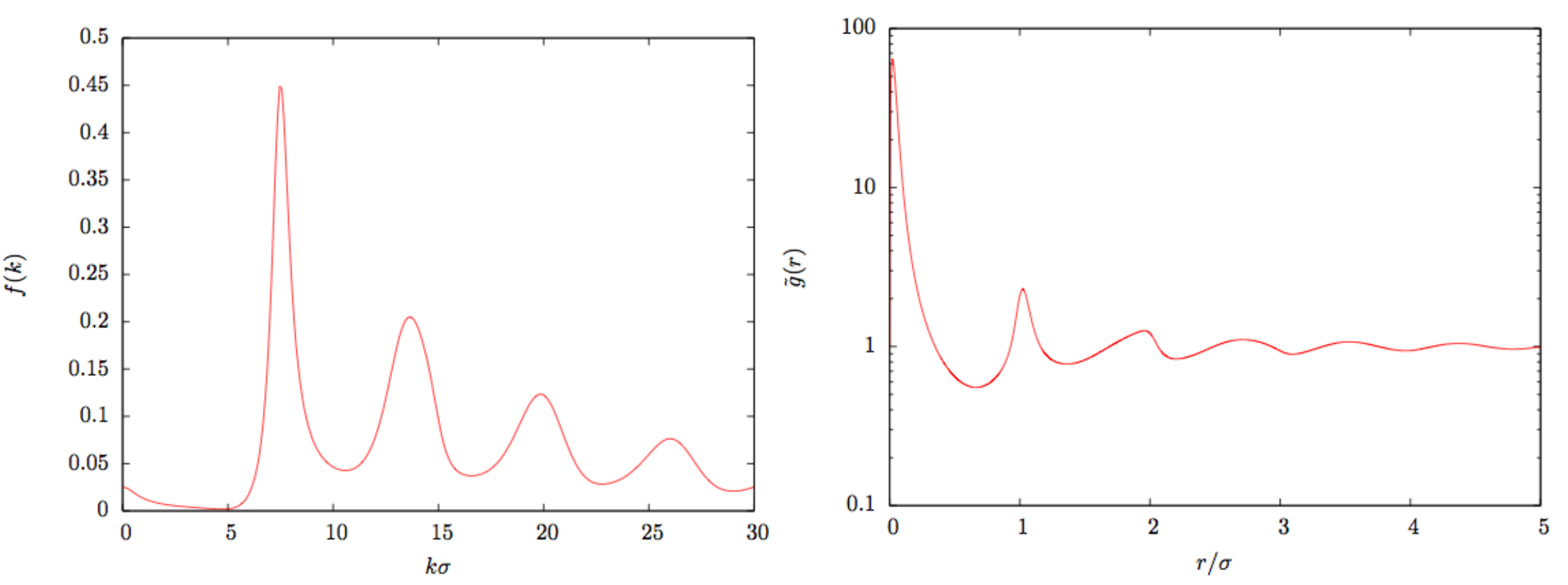}
\caption{Left: Non-ergodicity factor computed from Eq.(\ref{f_k_repHNC}) for hard-spheres in 3 dimensions, computed at the transition. 
Right: Corresponding $\tilde{g}$ function in real space.}
\label{fig:f_of_k_repHNC}
\end{figure}

Upon solving the RHNC equation, we, {\it without assuming it}, the existence of a dynamical transition~\cite{MP96}: 
for hard spheres, for densities lower than $\r_d \approx 1.14$,
the solution to Eq.~(\ref{f_k_repHNC}) is always $f(k) = 0$, i.e. a liquid phase, whereas 
$f(k)$ discontinuously jumps to a non zero value for $\r \ge \r_d$, indicating the glass transition. 
This results is quite good, because the dynamical transition for three-dimensional hard spheres is estimated to
be around $\r_d \approx 1.08$. 
Note that Mode-Coupling Theory instead strongly underestimates the transition~\cite{Go99}.

As described in Chapter \ref{chap:formalism}, the HNC equation was solved by gradually increasing the density and 
inserting the previous value for $c(k)$ in the next step. Here we can not do that since for all densities smaller than $\r_d$,
$\tc(k)$ sticks to $0$. At each density we thus started with a Gaussian guess for $\tc(k)$ until a first non-zero value of the 
function is spontaneously generated. To stabilize the value of the critical density, we decrease the density and iterate again
the scheme by using this obtained $f(k)$ as a starting point, until the solution spontaneously disappears. The obtained
non-ergodicity factor is shown in Fig.~\ref{fig:f_of_k_repHNC}. 

The qualitative behavior is generally good: the non-ergodicity factor has a strong peak around $k = 2 \pi / \s$ and 
oscillates in phase with $S(k)$ while gradually, decreasing to zero at large $k$.
However, the obtained structure factor is too small to be realistic: $f(k)$ is quite far from the numerical results 
(see~\cite{MUP91,Go99}) which are instead well captured by the Mode-Coupling theory.

We will demonstrate that the order parameter $h_{ab}$ with $a \ne b$ can be used as an organizing 
device for the theory in order to gradually incorporate higher-order correlations of the liquid into the replica result. 
Of course, we see in Fig~\ref{fig:f_of_k_repHNC} that the order parameter is not a small quantity, and thus an expansion 
in powers of $h_{ab}$ is not a priori justified. Note however that RHNC already re-sums an infinite number of diagrams 
containing arbitrary numbers of $h_{ab}$ links, which maybe explains its ability to predict a transition towards a non-small 
value of the order parameter. Our purpose here is to build from RHNC and incorporate more diagrams, and we will see 
that even keeping the lowest order correction modifies sensibly the results obtained. Furthermore this lowest order correction
allows for the calculation of the two-mode term in the equation for the non-ergodicity parameter.

\subsection{Improvement over the liquid quantities}

\begin{figure}[t]
\centering
\includegraphics[width=8cm]{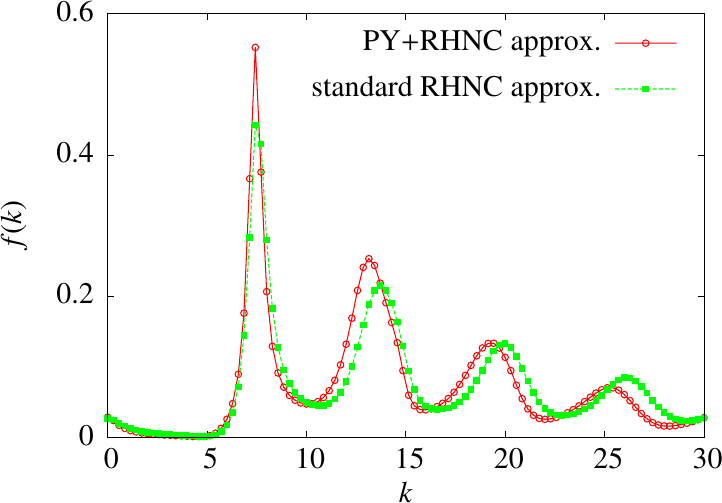}
\caption{Non-ergodicity factor as a function of the wave vector at packing fraction $0.6$, for the standard replicated HNC calculation, and from the 
combination of PY approximation for the diagonal part and RHNC for the off-diagonal correlation.}
\label{fig:compRHNC_PYRHNC_fk}
\end{figure}

A first natural question to ask is the role of the liquid HNC approximation that was plugged into the resolution of 
the self-consistent equation on the non-ergodicity parameter. We have seen that in the $m=1$ limit, the liquid quantities decouple
from the inter-replica correlations.
This allows us to use any liquid theory approximation to evaluate liquid quantities that appear in our equations. 
Formally, this is justified by writing the Gibbs free-energy $\Gamma_m$ in Eq.~(\ref{morita_rep}) as:
\beq
\G_m[\bfr,{\bf h}] = \G_{m}^{HNC}[\bfr,{\bf h}] + \G^{2PI}_{liq}[\bfr,{\bf h}] + \G^{2PI}_{glass}[\bfr,{\bf h}] \ , \label{glass_liquid_2PI}
\eeq
where $\G^{2PI}_{liq}$ is the sum of all 2PI diagrams that do not contain any $h_{ab}, a \ne b$ links. We can now use the variational principles, 
which will give the full
liquid correlation function for the $a = b$ components. For a given approximation of $\G^{2PI}_{glass}$, we will obtain a self consistent
equation for the $a \ne b$ components.
Neglecting altogether $\G^{2PI}_{glass}$, we recover the set of equations Eqs.(\ref{HNCrep_offdiag}), in which $c$ and $h$ are now the full 
liquid correlation functions. 
Of course, inside the glass phase, the liquid quantities cannot be obtained numerically or experimentally, 
so that such a full re-summation is useless. Instead, we need to have
an approximation that can be extrapolated from the liquid phase.
For example we can choose the PY approximation for the liquid quantities.
Choosing to work at a packing fraction (defined as $\varphi = \p \r / 6$) $\varphi = 0.6$, which is expected to be above the dynamical transition,
we solved the RHNC equation for the off-diagonal part, Eq.~(\ref{HNCrep_offdiag}) 
with the PY direct correlation function as an 
input for the diagonal part, and found that this brings about a little improvement over standard RHNC results. 
It is known that the PY approximation gives a less important underestimation of the peak of $S(k)$, 
which is the main ingredient that leads to the glass transition.
In standard RHNC as in the PY version of it, we find a transition from a liquid state at $\varphi < \varphi_d$ to a glass state at $\varphi \ge \varphi_d$, where 
the self-consistent equation on ${\tilde h}$ admits a non-zero solution. The value of the critical density is shifted downwards from 
$\varphi_c \approx 0.599$ to $0.591$ by the use of PY approximation, which is an improvement, even if modest.
We show the resulting non-ergodicity factor in the two sets of approximations in Fig. \ref{fig:compRHNC_PYRHNC_fk}. 
Even though use of the PY approximation gives a slightly larger non ergodicity factor, the result is still very small when compared to simulation and 
experimental data, where $f$ is much closer to $1$ at small wave vectors.
We conclude that the source of the problem in the static approach is not the diagonal part of the liquid. 
We should therefore seek for a way of improving
the equation for the off-diagonal correlation, Eq.~(\ref{HNCrep_offdiag}).

\subsection{Systematic expansion in powers of the order-parameter}

In order to make further progress, we must incorporate some ``glassy" 2PI diagrams (that contain $h_{ab}, a \ne b$ links)
into the free-energy. In the following, we will denote such links by $\tilde{h}$ links for simplicity, but the reasonings 
we perform are valid also in more general ans\"atze than the replica symmetric one.
We expand $\G_{glass}^{2PI}$ in powers of $\th$:
\beq
\G^{2PI}_{glass}[\bfr,{\bf h}] = \sum_{n=1}^\io \Tr' \frac 1{n!} 
\left. \frac{\d^n \G^{2PI}_{glass}[\bfr,{\bf h}]}{\d h_{a_1 b_1}(x_1,y_1) \cdots \d h_{a_n b_n}(x_n,y_n)} \right|_{\th=0} 
h_{a_1 b_1}(x_1,y_1) \cdots h_{a_n b_n}(x_n,y_n) \ , 
\label{taylor_2PIglass}
\eeq
where the prime on the trace means that  we must choose $a_i \ne b_i$ $\forall i$, 
and we have started the expansion at first order since the full functional must vanish in the liquid phase, where 
$\G^{2PI}_{glass} = 0$ by construction.

Since the diagrams are 2PI, it is easy to see that they must contain at least three of these $\th$ links. 
Indeed, a $\th$ link joins two nodes that have different replica indices, say $a$ and $b$. All the nodes connected to the 
$a$ node by a path of $h$ links must also have replica index $a$, and the same applies to the $b$ node. 
Thus all $\th$ links are nodal links: they separate the diagram into two parts, each of which has a different replica index. If a 
2PI diagram would contain one or two $\th$ links, then differentiating once or twice with respect to $\th$ would cut the 
diagram into two parts, which is in contradiction with the fact that the diagram is 2PI. Thus we proved that all 2PI diagrams contain 
either zero, or three, or more $\th$ links (this was already found in~\cite[Appendix A3]{BB09}, and our order parameter expansion
is thus the ``weak glass" expansion described in this reference). 

Thus the lowest order correction to the free-energy in an expansion in powers of $\th$ is a cubic one.
When the free-energy is differentiated with respect to $\th$ in order to obtain the equation for the non-ergodicity parameter, 
this cubic correction will give contributions at order $\OO(\th^2)$ to the missing kernel of Eq.(\ref{f_k_repHNC}). 
To make contact with the long-time 
limit of the MCT equation Eq.(\ref{true_MCT}), we thus have to re-sum all $\OO(\th^3)$ 2PI diagrams. 

\begin{figure}[ht]
\begin{center}
\includegraphics[width=12cm]{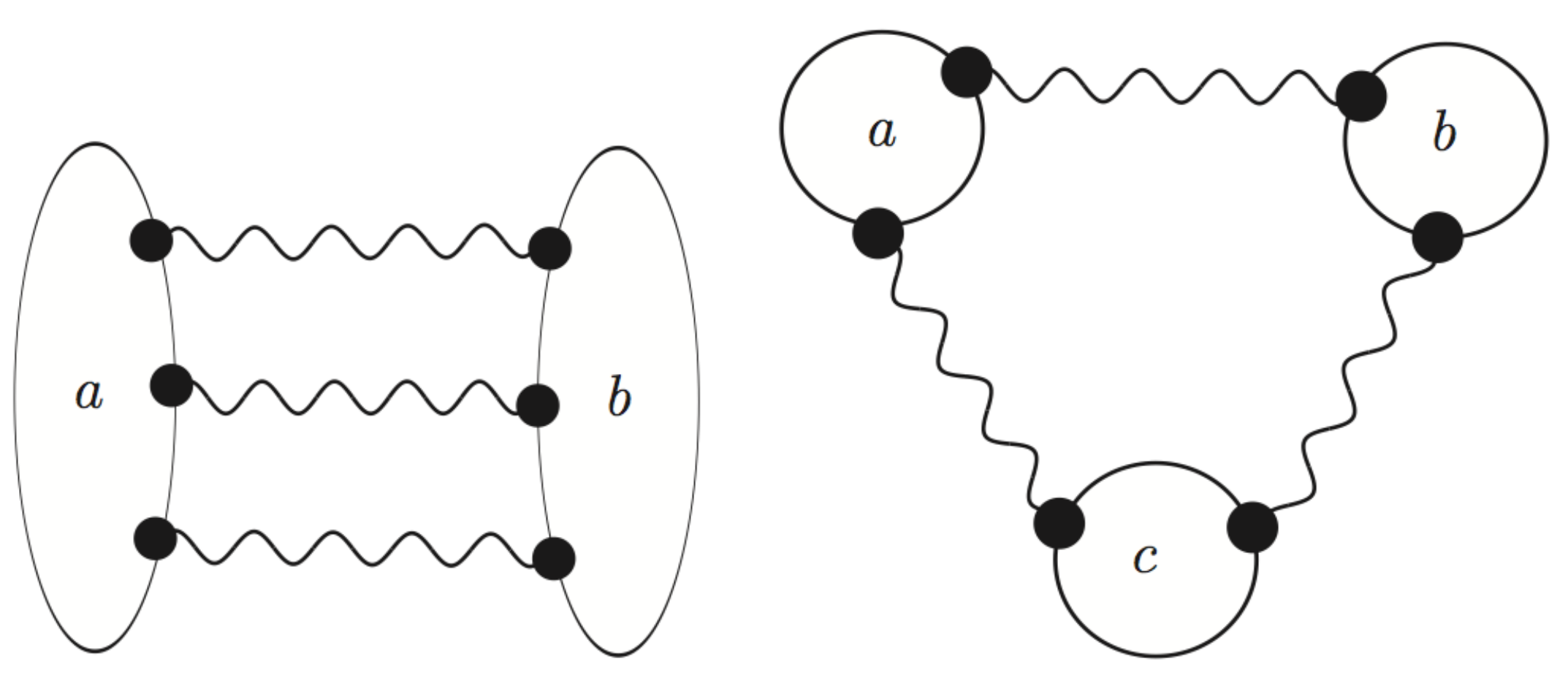}
\end{center}
\caption{Diagrams that contribute to the free-energy at order $\th^3$. A wiggly line joining two replica indices $a$ and $b$ 
is a $h_{ab}$ function, a black dot attached to a zone with replica index $a$ is an integration point weighted by a density 
factor $\r_a$.}
\label{diagrams}
\end{figure}

Let us look at the diagrammatic structure
of the $\OO(\th^3)$ diagrams of the free-energy. A diagram that contains three $\th$ links can have at most six parts 
composed of $h$ links and $\r$ nodes that all have the same replica index. But since the diagram is a free-energy 
contribution, it must be 1PI, as explained in Chapter \ref{chap:formalism}, i.e. it cannot be 
cut into two parts by cutting a line. The only 1PI possibilities to order $\th^3$ are the diagrams depicted in Fig.
\ref{diagrams}.

Note that the ring diagram in Fig. \ref{diagrams} is already contained in the re-summation performed by the HNC 
approximation since it is 1PI but not 2PI. Thus we only have to calculate the diagram on the left to get the contribution 
that corrects the HNC result in Eq.(\ref{f_k_repHNC}). It is important to understand that there are two distinct 
re-summations
involved in this calculation: first a re-summation of all the diagrams that are not 2PI (the RHNC approximation), 
then a further re-summation of the diagrams that contain exactly three $\th$ links (the lowest order correction in $\th$ to RHNC). 
In practice, we will calculate the sum of all $\OO(f^3)$ diagrams. This sum contains diagrams that are not 2PI, and thus will
pick a contribution coming from the RHNC approximation, and diagrams that are 2PI, that will give the desired lowest order
contribution.

\section{Expansion at third order}

Formally, we have to calculate the third derivative of the free-energy with respect to $h_{ab}$ with $a \ne b$, and 
evaluate this derivative at zero $\th$, to discard all the diagrams that are of higher order in $\th$ than $\th^3$. 
In the following we will need to 
distinguish the derivatives with respect to the density and with respect to the correlation 
functions, as well as derivatives with respect to chemical potentials and with respect to pair potentials. We define:
\beq \begin{split}
\G^{(1,0)}_a(x_1) & = \frac{\delta \G_m[\bfr,{\bf h}]}{\delta \r_a(x_1)} \ , \\
\G^{(0,1)}_{ab}(x_1,y_1) & = \frac{\delta \G_m[\bfr,{\bf h}]}{\delta h_{ab}(x_1,y_1)} \ , \\
\G^{(1,1)}_{a,cd}(x_1;x_2,y_2) & = \frac{\delta^2 \G_m[\bfr,{\bf h}]}{\delta \r_a(x_1) \delta h_{cd}(x_2,y_2)} \ ,
\end{split} \eeq
and so on. 
We will also need to define derivatives of $\ln Z_m$ with respect to the chemical potentials and pair potentials:
\beq \begin{split}
W^{(1,0)}_a(x_1) & = \left. \frac{\delta \ln Z_m[\bfn,{\bf w}]}{\delta \n_a(x_1)} \right|_{\bfn^*,{\bf v}^*} \ ,  \\
W^{(0,1)}_{ab}(x_1,y_1) & = \left. \frac{\delta \ln Z_m[\bfn,{\bf w}]}{\delta w_{ab}(x_1,y_1)} \right|_{\bfn^*,{\bf v}^*} \ ,  \\
W^{(1,1)}_{a,cd}(x_1;x_2,y_2) & = \left. \frac{\delta^2 \ln Z_m[\bfn,{\bf w}]}{\delta \n_a(x_1) \delta w_{cd}(x_2,y_2)} 
\right|_{\bfn^*,{\bf v}^*} \ ,
\end{split} 
\label{derivatives_lnZm_3}
\eeq
and so on. Because of our analysis, we see that the $n=1$ and $n=2$ terms in  Eq.~(\ref{taylor_2PIglass}) are necessary 
zero. We want to calculate the first non-zero term in this expansion, the third-order term. In practice, we found it
easier to calculate the third derivative of the total free energy, and substract from it the third order term of the 
RHNC free-energy, i.e. we will calculate, with the above notations:
\beq
\G^{(0,3)}_{ab,cd,ef} (x_1,y_1;x_2,y_2;x_3,y_3) = \frac{\d^3 \left[ \G_m^{HNC}[\bfr,{\bf h}] 
+ \G^{2PI}_{glass}[\bfr,{\bf h}] \right]}{\d h_{ab}(x_1,y_1) \d h_{cd}(x_2,y_2) \d h_{ef}(x_3,y_3)} \ ,
\eeq
with $a \ne b$, $c \ne d$ and $e \ne f$.
The third order term of the RHNC free-energy will be simply calculated from Eq.~(\ref{morita_rep}).
The third derivative we seek can be related
to cumulants of the density and the second derivative $\G^{(0,2)}$ by exploiting the properties of the double Legendre 
transform.

Considering, for simplicity, a discretized version of our theory, we have that $\bfr$ is an $m \times M$ matrix, where $M$ is 
the number of points of the underlying lattice, and $\boldsymbol \r^{(2)} \equiv \{ \r^{(2)}_{ab} \}_{a,b}$ 
is an $m \times m \times M \times M$ object, and the same 
applies to $\bfn$ and $\bfw$. We can write the two pairs $\bfr,{\boldsymbol \r^{(2)}}$ and $\bfn,\bfw$ in the form of two vectors:
\begin{align}
& \Psi \equiv \left( \bfr , {\boldsymbol \r^{(2)}}/2 \right) , \\
& \Phi \equiv \left( \bfn , \bfw \right) ,
\end{align}
with the convention that if $i > mM$ an index $i$ must be understood as a group of two space coordinates $(x_i,y_i)$ and 
two replica indices $(a_i,b_i)$, but if $i \le mN$, it must be understood as one space coordinate and one replica index.
The double Legendre transform $\G_m$ can then be written as:
\begin{align}
\left\{ \begin{array}{ll}
& \G_m[\Psi] = \Tr \Phi^*_1 \Psi_1 - \ln Z_m[\Phi^*] , \\
& \\
& \displaystyle \text{with } \Phi^* \text{ such that } ~ \left. \frac{\d \ln Z_m[\Phi]}{\d \Phi_1} \right|_{\Phi^*} = \Psi_1 .
\end{array} \right.
\end{align}
With these notations, we can make direct use of Eq.(\ref{W3_G3_formal}) which gives us:
\begin{align}
& \frac{\d^3 \G_m[\Psi]}{\d \Psi_1 \d \Psi_2 \d \Psi_3} = - \Tr \left( \frac{\d^2 \G_m[\Psi]}{\d \Psi_1 \d \Psi_{1'}} 
\frac{\d^2 \G_m[\Psi]}{\d \Psi_2 \d \Psi_{2'}} \frac{\d^2 \G_m[\Psi]}{\d \Psi_3 \d \Psi_{3'}} 
\left. \frac{\d^3 \ln Z_m[\Phi]}{\d \Phi_{1'} \d \Phi_{2'} \d \Phi_{3'}} \right|_{\Phi^*} \right).
\end{align}

We are interested in the continuum space limit of this expression, evaluated in the liquid, and with $1,2$ and $3$ that 
are all greater than $mM$. Then we get, after
replacing derivatives with respect to $\r^{(2)}$ with derivatives with respect to $h$:
\beq \begin{split}
& \G^{(0,3)}_{ab,cd,ef}(x_1,y_1;x_2,y_2;x_3,y_3) = \\
& - \G^{(1,1)}_{ab,a'}(x_1,y_1;x_1') \G^{(1,1)}_{cd,c'}(x_2,y_2;x_2') \G^{(1,1)}_{ef,e'}(x_3,y_3;x_3')
W^{(3,0)}_{a',c',e'}(x_1';x_2';x_3')  \\
& - \frac {2}{\r^2} ~ \G^{(0,2)}_{ab,a'b'}(x_1,y_1;x_2',x_3') \G^{(1,1)}_{cd,c'}(x_2,y_2;x_2') \G^{(1,1)}_{ef,e'}(x_3,y_3;x_3')
W^{(2,1)}_{a'b',c',e'}(x_1',y_1';x_2';x_3')  \\
& - \text{ two permutations } \{(a,b);(x_1,y_1)\} \leftrightarrow \{(c,d);(x_2,y_2)\} \leftrightarrow \{(e,f);(x_3,y_3)\}  
\label{derivee3_gamma}  \\
& - \frac {4}{\r^{4}} ~ \G^{(1,1)}_{ab,a'}(x_1,y_1;x_1') \G^{(0,2)}_{cd,c'd'}(x_2,y_2;x_2',y_2') 
\G^{(0,2)}_{ef,e'f'}(x_3,y_3;x_3',y_3') W^{(1,2)}_{a',c'd',e'f'}(x_1';x_2',y_2';x_3',y_3')  \\			
& - \text{ two permutations } \{(a,b);(x_1,y_1)\} \leftrightarrow \{(c,d);(x_2,y_2)\} \leftrightarrow \{(e,f);(x_3,y_3)\} 
  \\
& - \frac {8}{\r^{6}} ~ \G^{(0,2)}_{ab,a'\!b'}\!(x_1,y_1;x_1',y_1') \G^{(0,2)}_{cd,c'\!d'}\!(x_2,y_2;x_2',y_2') 
\G^{(0,2)}_{ef,e'\!f'}(x_3,y_3;x_3',y_3') W^{(0,3)}_{a'b'\!,c'd'\!,e'\!f'}\!(x_1',y_1';x_2',y_2';x_3',y_3'). 
\end{split}\eeq
Implicit summation and integration over repeated indices and variables, is assumed for compactness.

We note that this expression is correct independantly of the value of $m$, the replica ansatz chosen, and the value of the fields 
$\bfr$ and $\bfh$ chosen. However three simplifications will occur: we are performing a Taylor expansion in powers of $h_{ab}$, 
with $a \ne b$, so that the indices $a,b,c,d,e$ and $f$ in Eq.~(\ref{derivee3_gamma}) must be chosen so that $a \ne b$, $c \ne d$ 
and $e \ne f$. Secondly we must evaluate the derivatives at zero off-diagonal correlation ($\th \to 0$ in the RS ansatz). Finally
we are interested only in the dynamical transition point, which is described by the $m \to 1$ limit. These three features will greatly 
simplify the calculation.

There are two types of objects that we need to compute in order to use this relation: cumulants of the microscopic densities
that are generated by the differentiation of $\ln Z_m$ with respect to $\n_a$ and $w_{ab}$, and second derivatives of
$\G_m$ with respect to $\r_a$ and $h_{ab}$. In the end, we want to evaluate these objects in the liquid phase where $\th$
is equal to zero. But we know that the free energy can be written as the HNC free energy plus
2PI contributions, that are $\OO(\th^3)$, i.e. contain more than three $\th$ links. Thus when taking one or 
two derivatives of the 2PI diagrams with respect to $\th$, they still will contain at least one $\th$ link, and will all cancel out 
when evaluated at zero $\th$. This proves that Eq.~(\ref{derivee3_gamma}) when evaluated in the liquid phase can be
computed by replacing the $\G_m$ functionals in the r.h.s. by $\G_m^{\rm{HNC}}$.

Thus the needed derivatives of $\G_m$ can be computed starting from Eq.~(\ref{morita}) by dropping the 2PI diagrams.
The only difficult term is the sum of ring diagrams. We have seen that the derivative of the sum of chain diagrams 
with respect to a link is $h-c$. Thus we have that:
\beq
\frac{\d \G_m^{HNC}[\bfr,\bfh]}{\d h_{ab}(x_1,y_1)} = \frac 12 \r_a(x_1) \r_b(y_1) \left( \frac{}{} \ln (1 + h_{ab}(x_1,y_1)) 
 - h_{ab}(x_1,y_1) + c_{ab}(x_1,y_1) \frac{}{} \right) 
\label{dGammaHNC_dh} \ .
\eeq
In order to perform a second derivative with respect to $h$, one must resort to the expression of $c$ as a function of $h$ 
and $\r$ in Eq.~(\ref{OZ2rep}). We can expand it perturbatively to find the equivalent of Eq.~(\ref{OZ2_expanded}) 
to multicomponent systems:
\beq
c_{ab}(x,y) = h_{ab}(x,y) + \sum_{n=1}^\io (-1)^{n} h_{a a_1}(x,x_1) \r_{a_1}(x_1) \cdots \r_{a_n}(x_n) h_{a_n b}(x_n,y) \ ,
 \label{OZ2rep_expanded}
\eeq
We easily find upon differentiation:
\beq \begin{split}
\frac{\d c_{ab}(x_1,y_1)}{\d h_{cd}(x_2,y_2)} & = \r_c(x_2) \r_d(y_2) \left( \frac1{\r_a(x_1)} \d_{ac} \d(x_1,x_2) - c_{ac}(x_1,x_2) \right)
\left( \frac1{\r_b(y_1)} \d_{bd} \d(y_1,y_2) - c_{bd}(y_1,y_2) \right) \ , \\
& = \r_c(x_2) \r_d(y_2) \G^{(2)}_{ac}(x_1,x_2) \G^{(2)}_{bd}(y_1,y_2) \ .
\end{split} \eeq
In principle, this expression should be symmetrized with respect to a change of indices, but here everything will be traced in the end 
of the calculation, thus we can keep working with non-symmetrized quantities.
We obtain as a consequence:
\beq \begin{split}
\frac{\d^2 \G_m^{HNC}[\bfr,\bfh]}{\d h_{ab}(x_1,y_1) \d h_{cd}(x_2,y_2)} = & ~ 
\frac 12 \r_a(x_1) \r_b(y_1) \left( \frac{1}{1 + h_{ab}(x_1,y_1)} -1\right) \d_{ab,cd}(x_1,y_1;x_2,y_2) \\
& ~ + \frac 12 \r_a(x_1) \r_b(y_1) \r_c(x_2) \r_d(y_2) \G^{(2)}_{ac}(x_1,x_2)\G^{(2)}_{bd}(y_1,y_2) \ .
\end{split} \eeq
And finally taking the limit $\th \to 0$ with $a \ne b$ (a limit we denote by $\left. \bullet \right|_{liq}$ in the following) we get:
\beq \begin{split}
\G^{(0,2)}_{ab,cd} (x_1,y_1;x_2,y_2) \left. \right|_{liq} & = \frac 12 \r^4  \d_{ac} \d_{bd} \G^{(2)}_{liq}(x_1,x_2) \G^{(2)}_{liq}(y_1,y_2) \ , \\
& \equiv \frac 12 \r^4 \d_{ac}\d_{bd} \G_{HNC}^{(0,2)}(x_1,x_2;y_1,y_2) \ .
\label{Gamma_02}
\end{split} \eeq
The other needed derivative is $\G^{(1,1)}$, which is calculated from Eq.~(\ref{dGammaHNC_dh}) by noting that differentiating 
$c$ with respect to the density, at $h$ fixed, simply gives a product of $c$ functions:
\beq
\frac{\d c_{ab}(x_1,y_1)}{\d \r_c(x_2)} = - c_{ac}(x_1,x_2) c_{bc}(y_1,x_2) \ .
\eeq
Now recall that derivatives with respect to density must be done at $\r^{(2)}$ fixed instead, so we use the chain rule:
\beq 
\left. \frac{\d c_{ab}(x_1,y_1)}{\d \r_c(x_2)} \right|_{\r^{(2)} \text{ cte}} = \left. \frac{\d c_{ab}(x_1,y_1)}{\d \r_c(x_2)} \right|_{h \text{ cte}}
+ \sum_{e,f} \int_{u,v} \left. \frac{\d h_{ef}(u,v)}{\d \r_c(x_2)} \right|_{\r^{(2)} \text{ cte}} \left. 
\frac{\d c_{ab}(x_1,y_1)}{\d h_{ef}(u,v)} \right|_{\r \text{ cte}} \ ,
\eeq
to obtain the final result:
\beq
\left. \frac{\d^2 \G_m^{HNC}[\bfr,\bfh]}{\d h_{ab}(x_1,y_1) \d \r_c(x_2)} \right|_{liq} = 
-\frac 1{2 S(k=0)} \left[ \frac{}{} \d_{ac} \G^{(2)}_{liq}(y_1,x_2) + \d_{bc} \G^{(2)}_{liq}(x_1,x_2) \right] \ ,
\label{Gamma_11}
\eeq
where $S(k)$ is the structure factor of the liquid.

If we drop the space indexes, we can now perform the trace over replica indexes in
Eq.~(\ref{derivee3_gamma}), since all derivatives are delta functions with respect to replica indices. We obtain:
\beq \begin{split}
 \G^{(0,3)}_{ab,cd,ef} = & - \r^6 ~ \G^{(0,2)}_{HNC} \otimes \G^{(0,2)}_{HNC} \otimes \G^{(0,2)}_{HNC} \otimes W^{(0,3)}_{ab,cd,ef}  \\
& + \frac{\r^4}{2 S(0)} \G^{(2)}_{liq} \otimes \G^{(0,2)}_{HNC} \otimes \G^{(0,2)}_{HNC} \otimes W^{(1,2)}_{ab,cd,e} \\
& + \left\{ 2 \text{ perms.} \right\} \\
& - \frac{\r^2}{2 S(0)} \G^{(2)}_{liq} \otimes \G^{(2)}_{liq} \otimes \G^{(0,2)}_{HNC} \otimes W^{(2,1)}_{ab,c,e} \\
& - \left\{ 2 \text{ perms.} \right\} \\
& + \frac 1{8 S(0)} \G^{(2)}_{liq} \otimes \G^{(2)}_{liq} \otimes \G^{(2)}_{liq} \otimes W^{(3,0)}_{a,c,e} \ .
\label{Gamma_03_traced_replicas}
\end{split} \eeq
where $\otimes$ means space convolution with respect to two spatial indexes.

Finally, we only have left the task to compute the derivatives of the logarithm of the partition function with respect to pair 
potentials or chemical potentials. These terms are cumulants of microscopic one- or two-point densities, 
that are easily computed within the RS framework that we use here, as we explain in the following.

\subsection{Replica symmetric structure of the theory}

We have denoted by $\la \bullet \ra$ the equilibrium average for the replicated system.
Once again, in the limit $w_{ab} \to 0$ for $a \ne b$, all replicas fall in the same state 
but are otherwise uncorrelated inside the state. Finally, we want to evaluate
all our averages in the liquid phase.
This leads to the following rule to compute the average $\la \bullet \ra$: one should
\begin{itemize}
\item
factorize the averages $\la \bullet \ra$ when they involve different replicas, and
\item
remove the replica indexes.
\item
replace $\la \bullet \ra = \la \bullet \ra_{liq}$
\end{itemize}
For instance, for any spatial argument, and for $a \neq b$, we have that following the prescription above
\beq
\la \hat\r_a \hat\r_b \ra =  \la \hat\r_a \ra \la \hat\r_b \ra  =  \la \hr \ra_{liq} \la \hr \ra_{liq} = \r^2 \ .
\eeq
Similarly, assuming that different letters denote different values of the indexes:
\beq \begin{split}
& \la \hat\r_a \hat\r_a\hat\r_b \ra = \la \hat\r_a\hat\r_a\ra \la \hat\r_b \ra = \la \hr \hr \ra_{liq} \la \hr \ra_{liq} = \r (G_{liq} + \r^2) \ , \\
& \la \hr^{(2)}_{ab} \hr^{(2)}_{ac} \ra = \la \hr_a \hr_b ~ \hr_a \hr_c \ra = \la \hr_a \hr_a \ra \la \hr_b \ra \la \hr_c \ra 
= \la \hr \ra_{liq} \la \hr \ra_{liq} \la \hr \hr \ra_{liq} = \r^2 \left( G_{liq} + \r^2 \right) \ .
\end{split}\eeq
We will thus obtain quantities that do not depend on replica indices anymore, 
allowing to sum over these indices, and finally take the $m \to 1$ limit. The free-energy will 
have an overall factor $m(m-1)$, and thus we will consider the free energy divided by $m(m-1)$.
Indeed, recalling that we will calculate a free-energy correction of the form:
\beq \begin{split}
\d \G_m[\bfr,{\bf h}] & \equiv \frac 1{3!} \sum_{a \ne b} \sum_{c \ne d} \sum_{e \ne f} \int_{x_1,y_1,\cdots,x_3,y_3} 
\G^{(0,3)}_{ab,cd,ef}(x_1,y_1;x_2,y_2;x_3,y_3) h_{ab}(x_1,y_1) 
h_{cd}(x_2,y_2) h_{ef}(x_3,y_3) \\
& = \frac {1}{3!} \int_{x_1, \ldots, y_3} \left( \sum_{a \ne b, c \ne d, e \ne f} 
\G^{(0,3)}_{ab,cd,ef}(x_1,y_1;x_2,y_2;x_3,y_3) \right) \th(x_1,y_1) \th(x_2,y_2) \th(x_3,y_3) \label{derivee3_gamma_traced}
\end{split} \eeq
Now it is in order to remember that we will want to evaluate everything at $m=1$ at the end of the calculation. 
Everything will be proportional to $m-1$, thus we will first remove this
factor before evaluating. Afterwards, all terms that contain an additional factor $m-1$ will disappear. 
Now when we look at Eq.(\ref{Gamma_03_traced_replicas}), we see that
there are terms in which replica indices do not appear explicitely, for example in the term containing 
$W^{(3,0)}_{a,c,e}$, $b,d$ and $f$ do not appear. They are only constrained
to be different from their conjugate indices $a,c$ and $e$ respectively, thus when summing over all values of these indices, 
we will obtain three factors $m-1$, and the term will cancel in
the $m \to 1$ limit. This observation allow us to discard all terms but those containing $W^{(0,3)}$ in the limit $m \to 1$.

\subsection{Final calculation}

Before turning to the explicit evaluation of $W^{(0,3)}$, it is useful to remark that, within 
the RS structure that we have, we can parametrize its replica dependance in a simple way.

Take a matrix that depends on two pairs of replica indexes $M_{ab,cd}$, with $a \ne b$ 
and $c \ne d$ (we do not explicit the space indexes). Examination of the different possibilities for $a,b,c$ and $d$ shows that 
we have only three genuinely different possibilities:
\beq
M_{ab,cd} = \left\{ \begin{array}{ll}
& M_{ab,ab} \ , \\
& M_{ab,ac} \ , \\
& M_{ab,cd} \ .
\end{array} \right.
\eeq
This is a consequence both of the RS ansatz and of the symmetry of the functions 
with respect to permutations of indexes (and their associated space indexes).
This can be summarized in:
\beq
M_{ab,cd} = M_1  \frac{\d_{ac}\d_{bd} + \d_{ad}\d_{bc}}2 + M_2 \frac{\d_{ac} + \d_{ad} + \d_{bc} + \d_{bd}}4 + M_3 \ , 
\label{4replica_matrix_compact}
\eeq
where $M_1,M_2$ and $M_3$ are related to the above terms by:
\beq
\left\{ \begin{array}{ll}
& M_1 = 2 \left[ M_{ab,ab} - 2 M_{ab,ac} + M_{ab,cd} \right] \ , \\
& M_2 = 4 \left[ M_{ab,ac} - M_{ab,cd} \right] \ , \\
& M_3 = M_{ab,cd} \ .
\end{array}\right.
\eeq
The quantity that we are interested in is a matrix that depends on three pairs of indexes. In this case there are 
8 topologically different possibilities~\cite{TDP02}:
\beq \begin{split}
W^{(0,3)}_{ab,cd,ef} = \left\{ \begin{array}{ll}
W_1 = W_{ab,bc,ca} \ , \\
W_2 = W_{ab,ab,ab} \ , \\
W_3 = W_{ab,ab,ac} \ , \\
W_4 = W_{ab,ab,cd} \ , \\
W_5 = W_{ab,ac,bd} \ , \\
W_6 = W_{ab,ac,ad} \ , \\
W_7 = W_{ab,ac,de} \ , \\
W_8 = W_{ab,cd,ef} \ .
\end{array} \right. 
\end{split} \eeq
\vspace{2cm}
A relation like Eq.~(\ref{4replica_matrix_compact}) is again possible, but cumbersome,
and we do not write it explicitly because we do not need it.
By using the prescription for calculating averages of one and two-point densities 
described above, we can easily compute the $W_i$. We obtain:
\beq \begin{split}
& W_1 = \frac 1 8 \left[ \begin{array}{ll}
& G_{liq}(x_1,x_2)G_{liq}(y_1,x_3)G_{liq}(y_2,y_3) \\
& + \r^2 \left(G_{liq}(x_1,x_2)G_{liq}(y_1,x_3) + G_{liq}(x_1,x_2)G_{liq}(y_2,y_3) + G_{liq}(y_1,x_3)G_{liq}(y_2,y_3) \right) \end{array} \right] \ ,
\label{W_1}
\end{split}\eeq
\beq\begin{split}
& W_2 = \frac 1 8 \left[ \begin{array}{ll} 
& W^{(3)}_{liq}(x_1,x_2,x_3) W^{(3)}_{liq}(y_1,y_2,y_3) \\
& + \r W^{(3)}_{liq}(x_1,x_2,x_3) \left( \r^2 + G_{liq}(y_1,y_2) + G_{liq}(y_1,y_3) + G_{liq}(y_2,y_3) \right) \\
& + \r W^{(3)}_{liq}(y_1,y_2,y_3) \left( \r^2 + G_{liq}(x_1,x_2) + G_{liq}(x_1,x_3) + G_{liq}(x_2,x_3) \right) \\
& + \r^2 \left( G_{liq}(x_1,x_2)G_{liq}(y_1,y_3) + G_{liq}(x_1,x_2)G_{liq}(y_2,y_3) + G_{liq}(x_1,x_3)G_{liq}(y_1,y_2) \right) \\
& + \r^2 \left( G_{liq}(x_1,x_3)G_{liq}(y_2,y_3) + G_{liq}(x_2,x_3)G_{liq}(y_1,y_2) + G_{liq}(x_2,x_3)G_{liq}(y_1,y_3) \right) 
\end{array} \right] \ ,
\label{W_2}
\end{split} \eeq
\beq \begin{split}
& W_3 = \frac 18 \left[ \r^3 W^{(3)}_{liq}(x_1,x_2,x_3) + \r G_{liq}(y_1,y_2) \left( W^{(3)}(x_1,x_2,x_3) + \r G_{liq}(x_1,x_3) + \r G_{liq}(x_2,x_3) \right) \right] \ , 
\label{W_3}
\end{split} \eeq
\beq
W_5 = \frac 18 \r^3 W^{(3)}_{liq}(x_1,x_2,x_3) \ ,
\label{W_5}
\eeq
\beq
W_6 = \frac 18 \r^2 G_{liq}(x_1,x_2) G_{liq}(y_1,x_3) \ , 
\label{W_6}
\eeq
and we find $W_4 = W_7 = W_8 = 0$.
The factors $1/8$ come from the fact that a derivative of $\ln Z_m$ with respect to 
$w$ gives $\hr^{(2)}/2$ and not $\hr^{(2)}$. Of course, the choice of spatial indexes
is arbitrary, and one can make any permutation, as long as it respects the symmetry of the functions. We can now perform the trace 
over replica indexes in Eq.~(\ref{derivee3_gamma_traced}), which will give an expression analytic in $m$. We omit the space indices
in the following, but they are recovered by considering all permutations of the space indices written in Eqs.~(\ref{W_1}--\ref{W_6}).
The number of terms are obtained by considering the number of possible ways to choose a particular arrangement of replica 
indices among $m$ indices. For example to construct a term contributing to $W_2$, one must first pick a value of $a$ ($m$ 
possibilities), then a different value of $b$ ($m-1$ possibilities) which shows that the trace of $W_2$ let appear $m(m-1)$ identical
terms (with the exact same space indices structure). In addition to this multiplicity, because of the invariance by permutation of 
indices (together with their corresponding space indices), we have that:
\beq
W^{(0,3)}_{ab,ab,ab} = W^{(0,3)}_{ab,ba,ab} = W^{(0,3)}_{ab,ab,ba} = W^{(0,3)}_{ab,ba,ba} \ ,
\eeq
and this will give four terms that have the same replica structure, but permutations of space indices. The multiplicity of $W_2$ is
thus $4m(m-1)$. Performing the same counting on all cubic masses, we obtain:
\beq
\frac 1{m(m-1)} \sum_{a \ne b, c \ne d, e \ne f} \!\!\! W^{(0,3)}_{ab,cd,ef} = 
4 W_2 + (m-2) \left[ 8 W_1 + 24 W_3 \right] + (m-2)(m-3) \left[ 8 W_5 + 24 W_6 \right] + \OO(m-1) \ .
\eeq
After taking the $m \to 1$ limit, we obtain:
\beq\begin{split}
\lim_{m \to 1} \frac 1{m(m-1)} & \sum_{a \ne b}\sum_{c \ne d}\sum_{e \ne f} W^{(0,3)}_{ab,cd,ef} 
= 4 W_2 - 8 W_1 - 24 W_3 + 16 W_5 + 48 W_6  \\
& = \frac 12 W^{(3)}_{liq}(x_1,x_2,x_3) W^{(3)}_{liq}(y_1,y_2,y_3) - G_{liq}(x_1,x_2)G_{liq}(y_1,x_3)G_{liq}(y_2,y_3)  \ .
\end{split}\eeq
Finally we can perform the convolution with the derivatives of the HNC free energy in Eq.~(\ref{derivee3_gamma_traced}), and by making repeated use of the second- and third-order OZ equations we obtain:
\beq
\lim_{m \to 1} ~ \frac {\d \G_m[\bfr,{\bf h}]}{m(m-1)} 
= - \frac {\r^6}{6} \int_{x_1,\cdots,y_3} \hspace{-0.7cm} V(x_1,\cdots,y_3) \th(x_1,y_1) \th(x_2,y_2) \th(x_3,y_3) \ ,  
\eeq
\beq
V(x_1,\cdots,y_3) = \frac 12 \G^{(3)}_{liq}(x_1,x_2,x_3) \G^{(3)}_{liq}(y_1,y_2,y_3) 
-\G^{(2)}_{liq}(x_1,x_2) \G^{(2)}_{liq}(y_1,x_3) \G^{(2)}_{liq}(y_2,y_3) \ . 
\label{result_with_HNCterm} 
\eeq
This third order correction includes the RHNC term, which is recovered by setting $c^{(3)}_{liq} = 0$ \cite{Iy84}, but which we can 
also recover by directly differentiating Eq.~(\ref{Gamma_02}). In any case we find:
\beq
V(x_1,\cdots,y_3) = V^{HNC}(x_1,\cdots,y_3) + V^{2PI}(x_1,\cdots,y_3) \ ,
\eeq
where we defined:
\beq
V^{HNC}(x_1,\cdots,y_3) = - \G^{(2)}_{liq}(x_1,x_2) \G^{(2)}_{liq}(y_1,x_3) \G^{(2)}_{liq}(y_2,y_3) 
+ \frac 1{2 \r^4} \d(x_1,x_2) \d(x_1,x_3) \d(y_1,y_2) \d(y_1,y_3) \ , 
\eeq
which is the contribution coming from $\G_m^{HNC}$, and
\beq
V^{2PI}(x_1,\cdots,y_3) = \frac 12 c^{(3)}_{liq}(x_1,x_2,x_3) c^{(3)}_{liq}(y_1,y_2,y_3)
+ \frac 1{2\r^2} \d(x_1,x_2)\d(x_1,x_3) c^{(3)}_{liq}(y_1,y_2,y_3) \ ,
\eeq
which is the sought for contribution coming from $\G_m^{2PI}$.
We obtain finally:
\beq \begin{split}
\left. \frac {\G_m[\bfr,{\bf h}]}{m(m-1)} \right|_{m=1} = & ~ \left. \frac{\G_m^{HNC}[\bfr,{\bf h}]}{m(m-1)} \right|_{m=1} 
- \frac{\r^4}6 \int_{x_1,x_2,x_3,y} c^{(3)}_{liq}(x_1,x_2,x_3) \tilde{h}(x_1,y) \tilde{h}(x_2,y) \tilde{h}(x_3,y)  \\
& ~ - \frac{\r^6}{12} \int_{x_1, \cdots , y_3} c^{(3)}_{liq}(x_1,x_2,x_3) c^{(3)}_{liq}(y_1,y_2,y_3) \tilde{h}(x_1,y_1) \tilde{h}(x_2,y_2) \tilde{h}(x_3,y_3) \ .
\end{split} \eeq
This is the desired result: the next order term in the order-parameter expansion beyond
the RHNC approximation. From this approximation of the free-energy, we can make use 
of the variational principle to obtain a closed equation on $\tilde{h}$. For a translationally invariant system we get:
\beq \begin{split}
\tilde{c}(r) = & ~ \tilde{h}(r) - \ln [ 1 + \tilde{h}(r) ]  + \frac{\r^4}{2} \int_{r_1 \cdots r_4} \hspace{-0.5cm} c^{(3)}_{liq}(r,r_1,r_3) c^{(3)}_{liq}(0,r_2,r_4) 
\tilde{h}(r_1,r_2) \tilde{h}(r_3,r_4)  \\
& ~ + \frac{\r^2}2 \int_{r_1,r_2} \hspace{-0.3cm}c^{(3)}_{liq}(r,r_1,r_2) \tilde{h}(r_1) \tilde{h}(r_2) 
+ \frac{\r^2}2 \int_{r_1,r_2} \hspace{-0.3cm}c^{(3)}_{liq}(0,r_1,r_2) \tilde{h}(r-r_1) \tilde{h}(r-r_2) \ .  
\end{split} 
\label{final_rspace}
\eeq
which provides the first correction to Eq.~(\ref{HNCrep_offdiag}).
Using translational invariance as well as the invariance under permutation of the three variables of $c^{(3)}$, 
we get:
\beq\begin{split}
& c^{(3)}_{liq}(r_1,r_2,r_3) = c^{(3)}_{liq}(r_1-r_2;r_1-r_3) \equiv c^{(3)}_{liq}(r;s) \ , \\
& \text{with } r = r_1-r_2 \quad s=r_1-r_3 \ , \\
& c^{(3)}_{liq}(r;s) = c^{(3)}_{liq}(s,r) = c^{(3)}_{liq}(-r;s-r) \ . \label{properties_inv_trans}
\end{split}\eeq
Defining the double fourier transform of $c^{(3)}_{liq}$ as:
\beq
c^{(3)}_{liq}(k,q) = \int_{r,s} e^{-i k r} e^{-i q s} c^{(3)}_{liq}(r,s) \ , \label{fourier_2variables}
\eeq
we obtain two invariance principles:
\beq\label{c3inv}
c^{(3)}_{liq}(k,q) = c^{(3)}_{liq}(q,k) = c^{(3)}_{liq}(-k-q,q) \ .
\eeq
Performing a Fourier transformation on our equation we get
\beq\begin{split}
\tilde{c}(k) = & ~ \FF \left( \tilde{h} - \ln [ 1 + \tilde{h} ] \right)(k) + \frac{\r^4}2 \int_q c^{(3)}_{liq}(-k,q) c^{(3)}_{liq}(k,-q) \tilde{h}(q) 
\tilde{h}(k-q)  \\
& ~ + \frac{\r^2}{2} \int_{q}  c^{(3)}_{liq}(q,k-q) \tilde{h}(q) \tilde{h}(k-q) 
+ \frac{\r^2}{2} \int_{q} c^{(3)}_{liq}(-q,-k+q) \tilde{h}(q) \tilde{h}(k-q) \ , 
\end{split}\eeq
which using the invariances in Eq.~(\ref{c3inv}) is simplified to:
\beq
\tilde{c}(k) =  \FF \left( \tilde{h} - \ln [ 1 + \tilde{h} ] \right)(k) 
+ \frac{1}{2} \int_{q} \left(\left[ 1 + \r^2  c^{(3)}_{liq}(q,k-q) \right]^2 -1 \right) \tilde{h}(q) \tilde{h}(k-q) \ .
\eeq
Before turning to a numerical resolution of this equation, we can again make the naive 
expansion of the HNC term in powers of $\th$, and keep only the lowest order term, 
to obtain:
\beq
\frac{f(k)}{1-f(k)} = \frac{S(k)}{2 \r} \int_{q} \r^4 \G^{(3)}_{liq}(k,-q)^2 S(q) S(k-q) f(q) f(k-q) \ .
\eeq
This recovers exactly the three-body term in the MCT kernel Eq.~(\ref{kernel_MCT}).
This result is highly surprising because it is the outcome of a purely static calculation, 
while the MCT kernel arises from a dynamical calculation.

\subsection{Three-body correlations and numerical solving}

We thus have obtained a closed equation of the order-parameter, that necessitates as an input the two- and three-body direct correlation 
functions of the liquid. We already quoted that we decided here to work within the PY approximation for the two-point functions. 
It is known~\cite{hansen} that the PY approximation~\cite{PY58}, which amounts to treat the fluid in a mean-field approximation, but under the exact constraint that the pair 
correlation function should vanish for distances smaller than $1$~\cite{ES03}, is particularly efficient for hard spheres.
 Furthermore, we dispose of an analytic expression for the two-body direct correlation function 
in that approximation~\cite{We63,Th63}. However, the three-body direct correlation function still needs to be approximated. 
Computing the third-order direct correlation function is in itself a hard problem of liquid theory. The best approximation available was shown by 
numerical works~\cite{BK94} to be the 
HNC3 approximation developed by Attard~\cite{At90}. However, this approximation is very computationally demanding, and because our purpose here
is merely to demonstrate the 
importance of higher-order terms in the expansion in powers of the order parameter, we do not aim at quantitative efficiency, and wish to 
find a simpler approximation scheme.

\subsection{Denton \& Ashcroft approximation}

A good compromise between simplicity and efficiency for evaluating the third-order direct correlation function~\cite{BK93} 
is the Denton-Ashcroft approximation~\cite{DA89}. This approximation gives an analytic form that necessitates as an input only the second-order
direct correlation function, for which we can use the PY result. Within their approximation, $c^{(3)}_{liq}$ is given by:
\beq
c^{(3)}_{DA}(k,q) = \frac 1{c^{(2)}(0)} \left[ c^{(2)}_{liq}(k) \partial_\rho c^{(2)}_{liq}(q) + c^{(2)}_{liq}(q) \partial_{\rho} c^{(2)}_{liq}
(k) \right]  - \frac{\partial_\rho c^{(2)}_{liq}(0)}{\left( c^{(2)}_{liq}(0) \right)^2} c^{(2)}_{liq}(k) c^{(2)}_{liq}(q) \ .
\label{DA_nonsymmetrized}
\eeq
Within this approximation, angular dependance is neglected. We can recover it by symmetrizing the expression:
\beq
c^{(3)}_{DAS}(k,q) = \frac 13 \left[ c^{(3)}_{DA}(k,q) + c^{(3)}_{DA}(k,|k+q|) + c^{(3)}_{DA}(q,|k+q|) \right] \ .
\eeq

As stated before, we use as an input the Percus-Yevick direct correlation function, that reads (in units of the hard-sphere diameter):
\beq \begin{split}
c^{(2)}_{PY}(r) = & \left\{ \begin{array}{ll}
\displaystyle - a - 6 ~ \varphi ~ b ~ r - \frac 12 \varphi ~ a ~ r^3 & r \le 1 \ , \\
\displaystyle 0 &  r > 1 \ . \end{array} \right. 
\label{c_PY}
\end{split} \eeq
where $ \varphi$ is the packing fraction, $a=(1+2\varphi)^2/(1-\varphi)^4$ and \mbox{$b=-(1+\varphi/2)^2/(1-\varphi)^4$.}
The corresponding Fourier transforms, and derivatives with respect to density are simply computed analytically.
Using this approximation for $c^{(3)}_{liq}$, and writing the integrals in bipolar coordinates by using the isotropy of the liquid,
we obtain the following set of equations:
\beq
\left\{ \begin{array}{l l}
\tilde{c}(k) & \displaystyle = F(k) + H(k) \ , \\
& \\
F(r) & = \displaystyle \tilde{h}(r) - \ln \left( 1 + \tilde{h}(r) \right) \ , \\
& \\
H(k) & \displaystyle = \frac{\r S(k)}{8 \p^2 k}  \int_0^\infty du \int_{|k-u|}^{k+u} dv ~ u ~ v \tilde{h}(u) \tilde{h}(v) 
\left( \left[ 1+ \rho^2 {\hat c}^{(3)}_{DAS}(u,v;k) \right]^2 - 1 \right) \ , \\
\end{array} \right.  \label{RHNC3body} 
\eeq
where we have defined
\beq
{\hat c}^{(3)}_{DAS}(u,v;k) = \frac 13 \left[ c^{(3)}_{DA}(k,u) + c^{(3)}_{DA}(k,v) + c^{(3)}_{DA}(u,v) \right] \ .
\label{final_4}
\eeq
Equations (\ref{DA_nonsymmetrized}) -- (\ref{final_4}) now completely specify our approximation. We present in the following a preliminary numerical 
resolution in order to demonstrate the importance of the correction $H$.

\subsection{Numerical resolution methodology}

The usual calculation for HNC alone is more stable if we write the iteration procedure in terms of $\tilde{c}$ and $\tilde{\chi}=\tilde{h}-\tilde{c}$, 
which would give in our case:
\beq
\tilde{c}(r) = e^{\tilde{\chi}(r)+H(r)}-1 - \tilde{\chi}(r) \ ,
\label{HNC_3body}
\eeq
or the same equation with $H=0$ in the case of RHNC.
We first solve the RHNC equation Eq.~(\ref{RHNC3body}) with $H=0$, by using PY approximation for the two-point functions. 
Once a stable solution of the RHNC equations has been found
we introduce the three-body correction $H$, and solve by the very same Picard iterative scheme that is used for solving HNC and RHNC.
Explicitly, the resolution will give:
\begin{itemize}
\item Start from the value of $k \tilde{c}(k)$, obtained with the previous iteration, or from the old HNC value if first iteration
\item Use Eq.~(\ref{OZ2rep}) to deduce the value of $k \tilde{h}(k)$ and $k \tilde{\chi}(k)$
\item Use $k \tilde{h}(k)$ to evaluate $k H(k)$
\item Inverse Fourier transform $k \tilde{\chi}(k)$ and $k H(k)$ to obtain $r \tilde{\chi}(r)$ and $r H(r)$
\item Use it to evaluate the new $r \tilde{c}(r)$ with Eq.~(\ref{HNC_3body})
\item Fourier transform $r \tilde{c}(r)$ to obtain $k \tilde{c}(k)$
\item Mix it with the old $k \tilde{c}(k)$ to avoid rapid changes 
\item Repeat these steps until $k \tilde{c}(k)$ has converged
\end{itemize}
We used a grid of $2^{10}$ equally spaced points on a box of size $11$, and a mixing parameter $0.01$. We note that evaluation of the 
correcting term $H$ in Eq.~(\ref{RHNC3body}) has a computational cost of order of the square number of points on the grid, significantly slowing down 
the resolution of the equation, since to avoid instabilities, $\tilde{c}$ is made to evolve very slowly by the mixing procedure.

\begin{figure}[t]
\centering
\includegraphics[width=7.5cm]{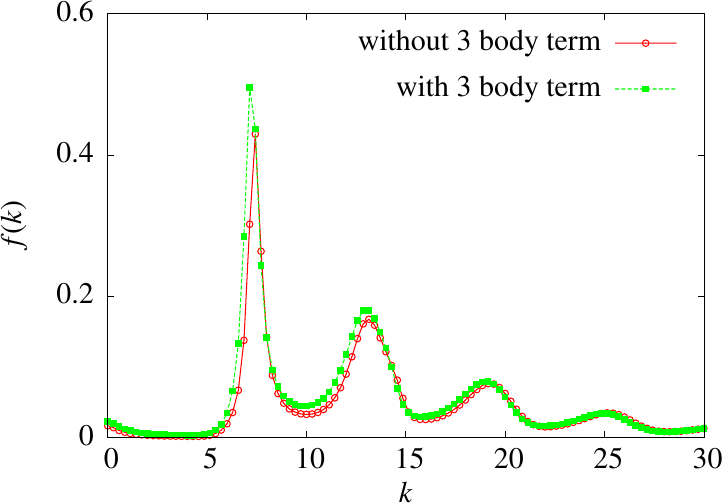}
\includegraphics[width=7.5cm]{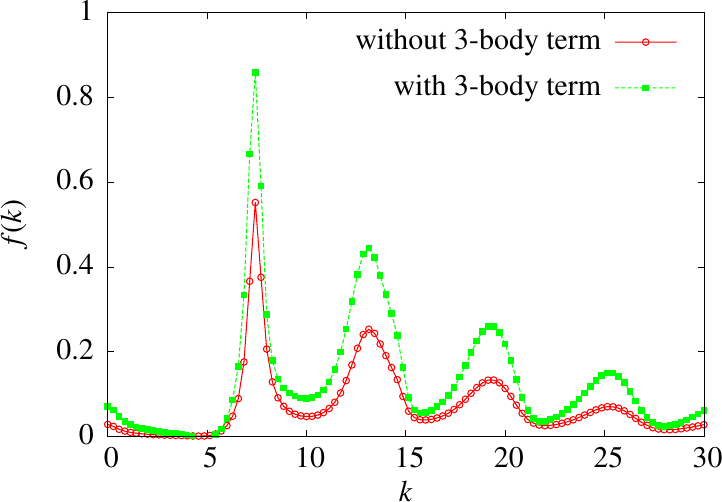}
\caption{Left: Non-ergodicity factor as a function of the wave vector very close to the critical point $\varphi=\varphi_d$, 
for the PY + HNC calculation, and with inclusion of the three body term. Right: Non-ergodicity factor as a function of the wave 
vector at fixed packing fraction $0.6$, for the PY + RHNC calculation, and with inclusion of the three body term.}
\label{fig:compPYRHNC_3bodyRHNC}
\end{figure}

\subsection{Results and discussion}

We solved our improved equation~(\ref{RHNC3body}) for several packing fractions 
starting from $\varphi = 0.6$, and decreasing it until the non-trivial solution for
$\th$ disappeared. We found that the inclusion of the three-body terms, even when 
using a crude approximation such as the symmetrized Denton \& Ashcroft 
approximation, leads to a strong shift of the dynamical transition point from
$\varphi_d = 0.591897$ to $\varphi_d = 0.555860$, which is closer to the 
predicted MCT transition but farther from the numerically estimated transition.
It is hard to guess what would be the evolution of $\varphi_d$ when additional corrections are added.

The resulting non-ergodicity factors at the transition are 
depicted in the left panel of 
Fig.~\ref{fig:compPYRHNC_3bodyRHNC}. 
The inclusion of the three body term has 
significantly enhanced the first peak of $f(k)$ upon inclusion of our correction.
These preliminary results must be treated with caution, because it was 
found~\cite{FJPUZ13} that the numerical solving of the RHNC equation is sensitive to the discretization used, and 
very large grids with a very large number of points must be used in order to obtain
stable results. This situation is expected to be the same with the presence of the 3 body 
term, but the $\OO(N^2)$ scaling of the numerical resolution in our case prevented us
from performing a stability analysis. The qualitative picture is nevertheless not expected 
to be modified by these considerations.

In the right panel of Fig.~\ref{fig:compPYRHNC_3bodyRHNC} we show the results inside the glass phase, at a fixed packing fraction
$\varphi = 0.6$, with and without inclusion of the three-body term.
In this case the effect of including the 3-body term is much bigger and goes in the right direction of increasing $f(k)$ at small
$k$ (although still not enough at very small $k$ below the first peak).

\subsection{Higher orders}

Our calculation provides the {\it exact} free-energy and correlation function at order $\tilde{h}^2$. This feature allows us to put into
correspondence the MCT kernel, which is also $\OO(f^2)$, with the replica result. We have seen that the MCT kernel in 
Eq.~(\ref{kernel_MCT}) contains exactly the three-body contribution we obtain with replica theory, which is already a quite surprising
result, given the differences that exist between the two approaches. However, it shows that the two-body terms contained in 
the MCT kernel are thus impossible to obtain within a static framework. The peculiar wave-vector dependance of these terms arise 
from the calculation of forces, inherent in dynamical theories, but absent in static ones. One could however wonder whether these
two-body terms could arise in a dynamical calculation because of the factorization of a dynamical four-point vertex function, that would 
be forced to be expressed as a $\OO(f^2)$ function. Indeed, the main approximation involved in Mode-Coupling theories is the 
factorization of a four-point function, and since MCT breaks down at high dimensions, it is possible that in the process, ``glassy" 
correlations are factorized along with ``liquid" ones, forcefully introducing these new $\OO(f^2)$ terms. This is nevertheless highly
speculative, and no satisfying dynamical theory exists yet, that would be able to investigate these considerations (note however two
formulations in \cite{Sz07} and \cite{Ma11} that may have this potential). 

\begin{figure}[ht]
\begin{center}
\includegraphics[width=6cm]{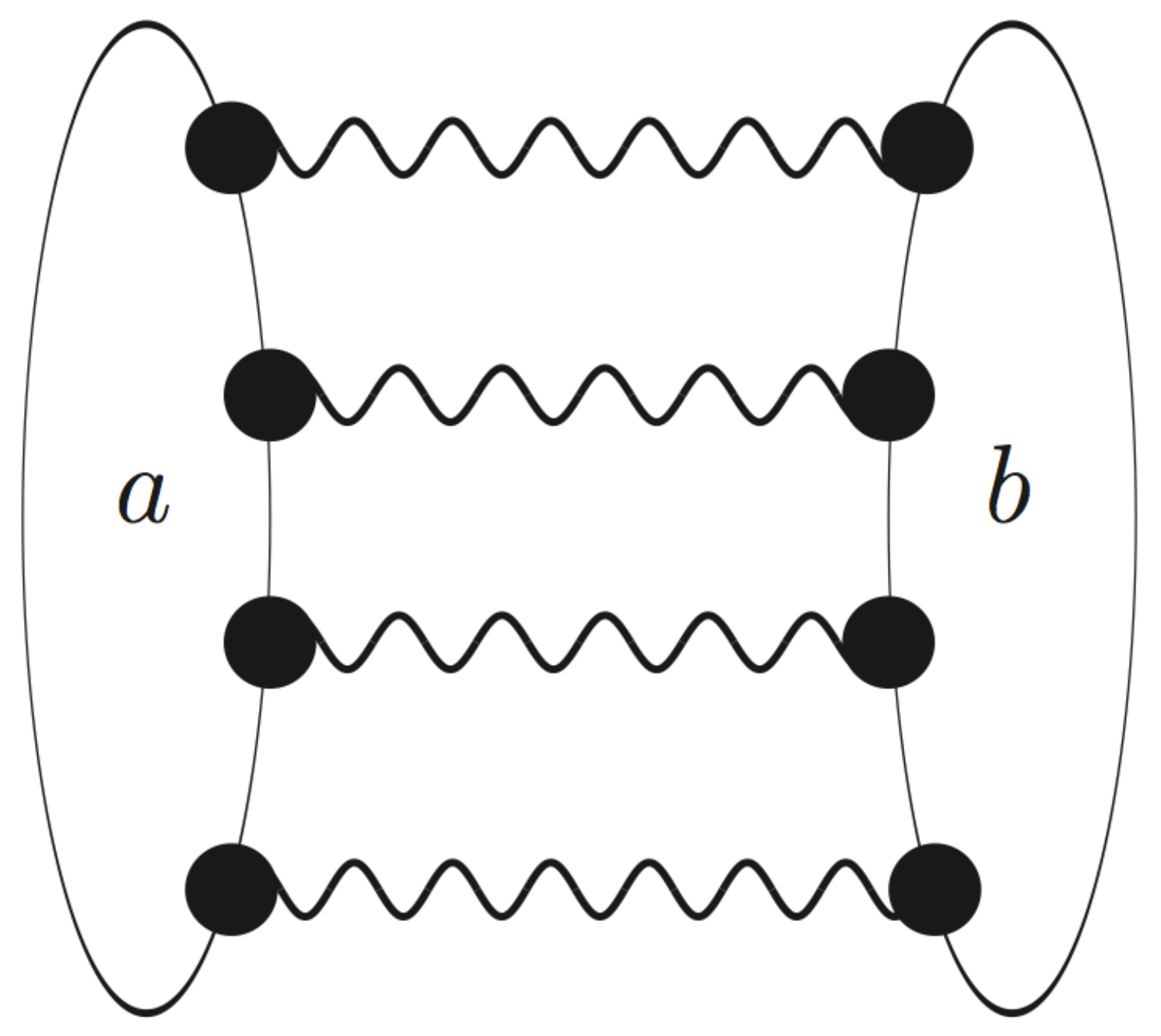}
\end{center}
\caption{Diagrams that contribute to the free-energy at order $\tilde{h}^4$. A wiggly line joining two replica indices $a$ and $b$ 
is a $h_{ab}$, with $a \ne b$ function, a black dot attached to a zone with replica index $a$ is an integration point weighted by a 
density factor $\r_a$.}
\label{fig:order_4}
\end{figure}

An interesting feature of our calculation is that, once the theory is set up, we can already uncover the next terms with a 
diagrammatical visualization of the expansion. To go further, we can now wonder what is the term $\OO(\tilde{h}^4)$ in the 
free-energy (that would correspond to a $\OO(f^3)$ term for the correlation function). The requirement that we must have 2PI 
diagrams is quite strong, and it is easy to convince one-self that the only possible diagram that we can construct is shown in 
Figure~\ref{fig:order_4}. The four-point functions that are connected by $\tilde{h}$ lines are made of 1PI diagrams, and we will
thus obtain a new contribution to the free-energy that contains the 4-point vertex functions of the liquid:
\beq
\text{const} \times \int_{r_1,\ldots,r_8} \G^{(4)}_{liq}(r_1,\ldots,r_7) \G^{(4)}_{liq}(r_2,\ldots,r_8) \times \tilde{h}(r_1,r_2) \cdots \tilde{h}(r_7,r_8) \ .
\eeq
Interestingly, we don't even have to work out the precise diagrammatics behind this procedure, since RHNC gives a contribution
in every diagram, and at all orders in $\tilde{h}$, so that we can use RHNC to fix the prefactor of each diagram we compute.

Computing the $n$-th order direct correlation function is still a difficult problem, but with approximations such as the Denton-Ashcroft
approximation, that are issued from density functional theory, we can deduce from the approximation for $c^{(3)}$ the corresponding
approximation for $c^{(n)}$ by successive differentiation with respect to density. The last difficulty is that an $\OO(f^n)$ kernel will
require the numerical evaluation of an $n$-dimensional integral, that have a computational cost of order $\OO(N^n)$.

\begin{figure}[ht]
\centering
\includegraphics[width=7cm]{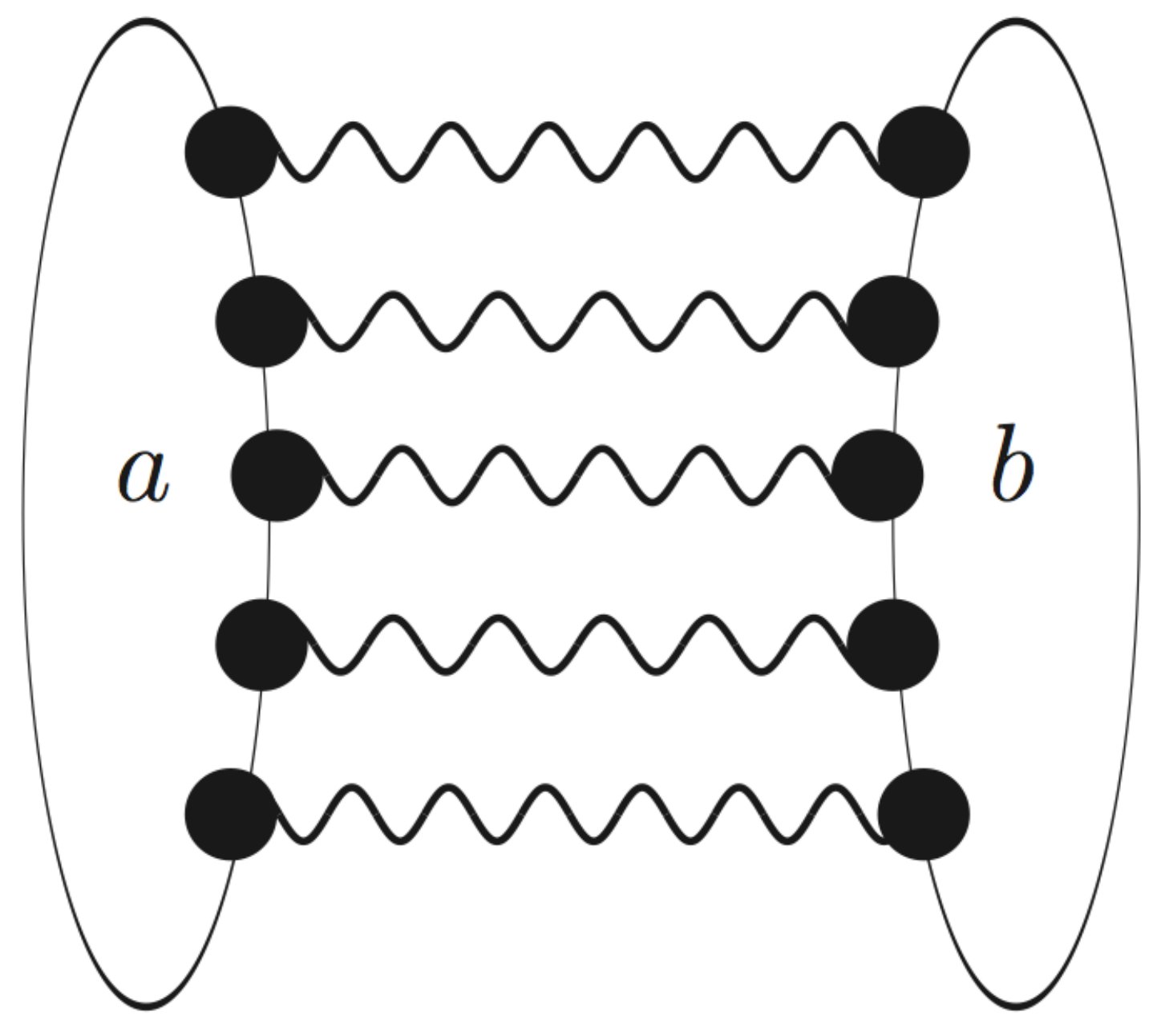}
\includegraphics[width=5.5cm]{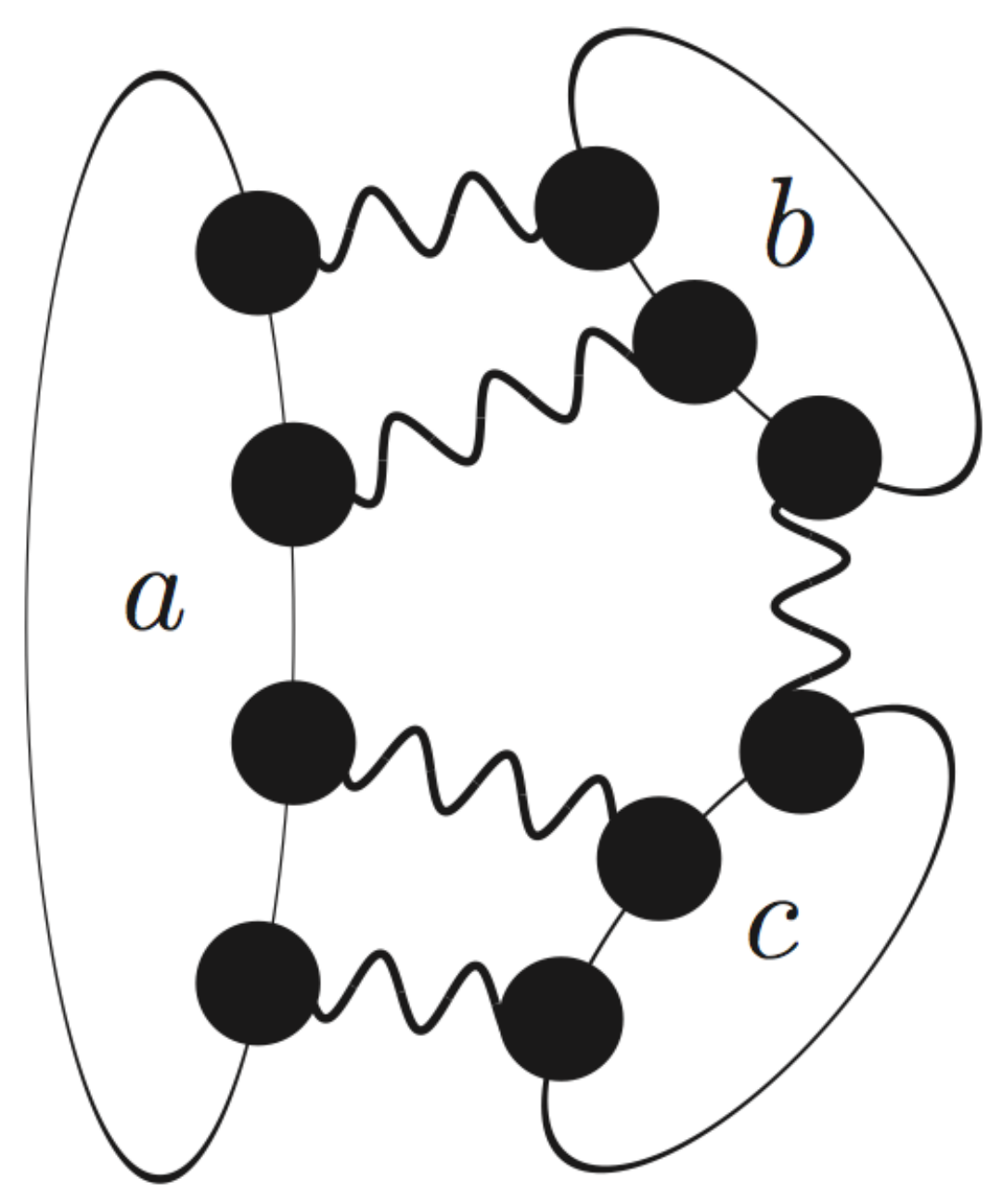}
\caption{Diagrams that contribute to the free-energy at order $\tilde{h}^5$. A wiggly line joining two replica indices $a$ and $b$ 
is a $h_{ab}$, with $a \ne b$ function, a black dot attached to a zone with replica index $a$ is an integration point weighted by a 
density factor $\r_a$.}
\label{fig:order_5}
\end{figure}

In order to visualize the increasing difficulty of going to the next orders, we show in Figure~\ref{fig:order_5} the possible diagrams at 
order $\tilde{h}^5$. We see that new, intricate terms arise, and that in general, the $\OO(\tilde{h}^n)$ contribution will contain all 
vertex functions of the liquid of orders ranging from $3$ to $n$.

\section{Conclusion and discussion}

Our analysis shows that the 2PI corrections to the RHNC free energy are quantitatively
relevant at the dynamical transition, and must be properly taken into account in order to
obtain an accurate static description of the transition. It seems then that only by doing this 
properly we will be able to make a clear connection between dynamical and static theories of the 
dynamical transition of glasses. 
In this respect our results are the following:
\begin{itemize}
\item At the level of the RHNC approximation, the equation for $f(k)$ can be developed
in powers of $f(k)$. The result is Eq.~(\ref{eq_f_HNC_expanded}), which corresponds to the Mode-Coupling
equation with a kernel equal to 1. The latter is also the result of a zero-th order field theory calculation as
reported in~\cite{ABL06}.
\item The first correction to the RHNC approximation provides an additional contribution
to the kernel which happens to correspond exactly to the three-body term of the Mode-Coupling kernel.
\item The next corrections will give terms proportional to $f^3$ in the right hand side of Eq.~(\ref{eq_f_HNC_expanded}),
hence no additional contributions to the Mode-Coupling kernel can be generated by these terms.
We are forced to conclude that there is no way of generating the terms of the Mode-Coupling kernel
proportional to $c_{liq}(k)$ by means of a static computation.
\item We have shown the way to compute higher-order terms, although we expect the numerical resolution of the corresponding
approximations to be hard.
\end{itemize}
It would be therefore very important to perform a similar calculation (namely, a systematic expansion in powers of $f(k)$) 
also on the dynamical side. This would allow for a systematic comparison of the results. 
One would then obtain a proper theory for the ergodicity breaking that 
occurs at $\varphi_d$, free of the ambiguities of MCT, and systematically improvable. Work 
has been done in this direction in the last years~\cite{ABL06,KK07b,ABL09,JW11}, but 
the situation is still unsatisfactory. 

\begin{figure}[ht]
\centering
\includegraphics[width=8cm]{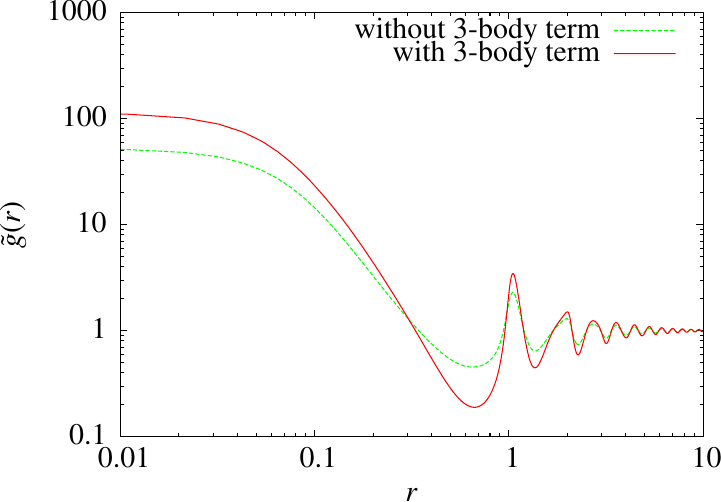}
\caption{Inter-replica pair correlation function $\tilde{g}(r)$ at packing fraction $\varphi=0.6$, with and without three-body correction.}
\label{fig:gtilde_r}
\end{figure}

On the static side, the first task to be performed is to reconcile the small cage expansion
and the 2PI approach that we use in this work in order to have a unified static theory.
The way to do this is indicated by the shape of the pair correlation function between 
two different replicas at the transition, depicted in Fig. \ref{fig:gtilde_r}.
We see that $\tilde{g}$ develops a strong peak at $r=0$, that dominates the rest of its
features. This peak simply reflects the fact that, in the glass phase, two replicas tend
to be very close from each other. In the small cage expansion, this idea is reflected in
the introduction of a cage size parameter $A$, which serves as a expansion parameter.
In our case a direct introduction of the cage size $A$ would be more difficult, because it 
would amount to parametrize the pair correlation function by the cage size, which is difficult
to do without specifying a given shape of cages. In the high dimensional limit, the 
cage can safely be approximated by a Gaussian as far as the free energy is concerned~\cite{KPZ12}, 
but this is not expected to hold in our low dimensional case~\cite{CIPZ12}.

It is interesting to note that we have observed, as in Fig.~\ref{fig:gtilde_r} that the inclusion of the three-body term systematically
increases the spatial separation between the first, glassy, peak of $\tilde{g}(r)$ and the subsequent, liquid peaks.
A clear separation between these contributions allows for an unambiguous definition of the molecules
introduced in order to perform small-cage expansions \cite{MP99b,PZ10}, and is clearly a sign that reconciling
RHNC with small-cage expansions may not be out of reach.

The large contribution coming from the peak at $r=0$ in $\tilde{g}$ shows 
that the diagrams that contribute the most to the free-energy are the most connected 
ones~\cite{PZ10}. It should be possible to put these diagrams in correspondence with the diagrams
re-summed in the small cage expansions of~\cite{MP99b} and~\cite{PZ10} in order to 
make progress. Our work has set up the tools necessary to perform such resummations 
and we believe this is the natural line of work to follow in the future.

\chapter{Jamming transition}
\label{chap:jamming}

In this chapter we present the results obtained when applying the ideas of replica theory as presented in Chapter 
\ref{chap:replicas} to the vicinity of the jamming point, with the goal to assess whether the jamming transition is 
the zero-temperature mechanism that controls the glass transition at finite temperature. 
We study harmonic spheres at very large volume fractions, around $0.64$, and at very low or zero temperature, in order to study
the jamming phase diagram of Fig. \ref{fig:liu_diagram}. 
It is found that for these very dense states, at very low temperature, 
replica theory can be cast into a computationally efficient approximation scheme, allowing us to argue, within a given set of 
approximations, that the jamming transition is a phenomenon conceptually different from the glass transition, in the 
line of the calculations on mean-field models of \cite{KK07a} and \cite{MKK09} and hard spheres \cite{PZ10}, and 
numerical experiments on harmonic spheres \cite{BW09a,BW09b}. 

We predict, without assuming it, the existence of a jamming transition, solely from microscopic calculations, and 
successfully reproduce a large set of observed characteristics of the Jamming transition.

\section{Mean-field hypothesis}

\begin{figure}[htb]
\centering
\includegraphics[width=8cm]{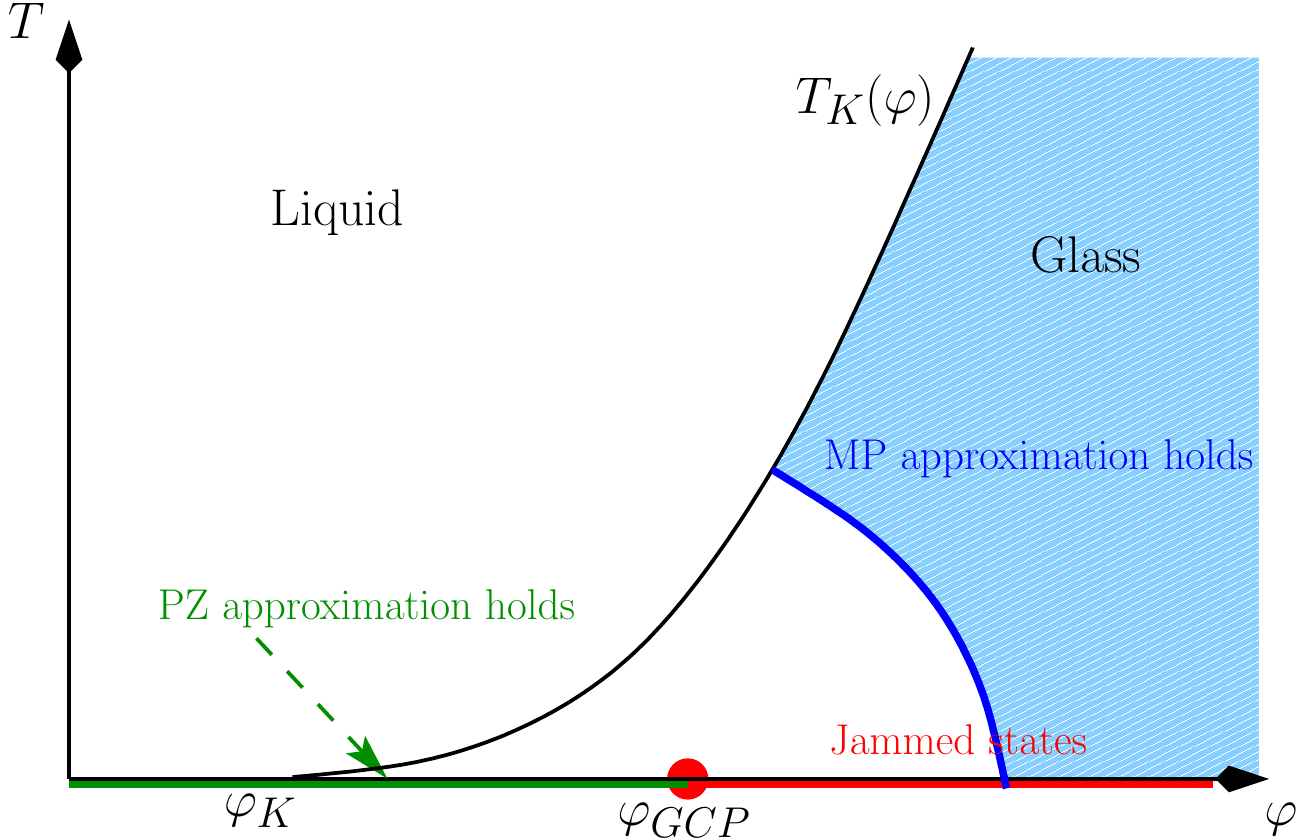}
\caption{ A schematic phase diagram for the glassy states of the 
model defined by Eq.~(\ref{potential_harmS}) showing both the 
glass transition line at $T_K(\varphi)$ and the jamming point 
at $T=0$ and $\varphi = \varphi_{GCP}$.
The Parisi-Zamponi (PZ) approach 
only holds at $T=0$ and $\varphi < \varphi_{GCP}$, while the 
M\'ezard-Parisi (MP) approach holds for large enough densities 
and temperatures. We developed a method to fill the gap between
the PZ and MP analytical schemes to explore the vicinity of 
the jamming transition.}
\label{fig:PDsch}
\end{figure}

Based on previous works cited above, we have in mind the schematic phase diagram reported in Fig.~\ref{fig:PDsch}.
In this picture, that we will derive in the following, the liquid undergoes an ideal glass transition at the Kauzmann
temperature $T_{K}(\varphi)$ (note that in the following, we will mostly use the volume fraction $\varphi = \p \r \s^3 /6$ 
instead of the density, because it is more suited to the study of hard-spheres and jamming). 

The line $T_K(\varphi)$ vanishes at a volume fraction $\varphi_K$ which also corresponds to the ideal glass
transition for hard spheres \cite{PZ10}. The point $\varphi_K$ {\it is not the jamming transition}: the ground state energy and 
pressure remain zero across $\varphi_K$. Above $\varphi_K$, the system enters, at zero temperature, 
a {\it hard sphere glassy state}. In this state, particles vibrate near well-defined (but random) positions, 
and the system is not jammed yet~\cite{PZ10}. Jamming occurs when the glass reaches its close packing density, which 
we shall call ``glass close packing'' (GCP), following \cite{PZ10}. We {\it identify the jamming transition with 
$\varphi_{GCP}$}. Note that at this point this is only an assumption, although a reasonable one, and quantitative 
arguments will be derived in the following.

The main theoretical difficulty in the study of the jamming transition at $\varphi_{GCP}$ is that it occurs deep inside 
the glass phase. Therefore, we first need to develop an accurate theoretical description of the glass phase. 
As discussed in detail in \cite{BJZ10}, previous theories of the glass phase fail (see Fig.~\ref{fig:PDsch}). 
The M\'ezard-Parisi approach (denoted by MP in the following) \cite{MP96,MP99a} holds only at high enough temperature or density,
where a simple harmonic approximation for vibrations in the glass holds. The
Parisi-Zamponi approach (denoted by PZ) \cite{PZ10} holds only in the 
hard sphere region, $T=0$ and $\varphi < \varphi_{GCP}$.
The central theoretical achievement that we we describe here is an approximation scheme that can naturally 
interpolate between those of MP and PZ and is correct in the entire vicinity of the jamming transition, thus
allowing us to fully explore the phase diagram of Fig.~\ref{fig:PDsch},
and in particular the region of low temperature, $T \ll T_K(\varphi)$  and $\varphi \sim \varphi_{GCP}$. 
We theoretically obtain from a first principle calculation a large number of the observed behaviors of harmonic 
spheres in numerical experiments \cite{OLLN02,OSLN03,DTS05a,SLN06,ZXCYAAHLNY09}, and predict new results for the correlation functions of this system around the jamming transition. 

Our approach is very different from, but complementary to, recent theoretical works on jamming.
The aim of our work is to show, directly from the Hamiltonian, that the jamming transition exists, and determine from first 
principles its location and properties. Other approaches assume the existence
of jammed states and try to obtain geometrical informations on them~\cite{CCSB09,SWM08}, or develop a scaling picture of the jamming transition by showing that the transition is characterized by a diverging correlation length~\cite{WSNW05}. 

Before proceeding further, it is worth noting that even though the mean-field picture on which we will base our 
calculations is expected to be tempered, in finite dimensional models by nucleation processes \cite{LW07,Ca09}, 
we expect that the local and static properties that we intend to compute (such as the number of contacts, energy of a 
packing, etc.) will not be affected by these processes in the very low temperature and high density
regime near the jamming transition.

For this reason, in this work we follow M\'ezard and Parisi \cite{MP96,MP99a} and directly apply the mean-field concepts 
described above to three dimensional glass-formers.

\subsection{Thermodynamics of the glass phase}

In Chapter \ref{chap:replicas}, we used to replica trick to detect the emergence of a non-zero complexity, signaling the 
appearance of an exponential number of metastable states, responsible for the slowing down of the dynamics. However, 
the idea of replicating the system can be pushed further, in order to compute the thermodynamics of the glass phase, as 
was shown by Monasson \cite{Mo95}. 

Returning to Eq.(\ref{complexity_1}), the ideal glass transition is reached at the Kauzmann temperature $T_K(\varphi)$, 
defined by the point where the saddle point of Eq.(\ref{complexity_1}) for which the complexity vanishes:
\begin{align}
\left\{ \begin{array}{ll}
& \displaystyle \left. \frac{\partial \Si(f)}{\partial f} \right|_{f^*(T_K,\varphi)} = \frac{1}{T_K(\varphi)} , \\
& \\
& \displaystyle \Si(f^*(T_K(\varphi),\varphi)) = \Si(f_{min}) = 0 .
\end{array} \right.
\label{Sivanishes}
\end{align}
In the following we will work directly with the entropy of the system, i.e. the logarithm of its partition function.
It can be read off from the free-energy by mutiplying by $-\b$. We will call this object the ``free-entropy":
\beq
\SS = - \b F = \lim_{N \to \io} \frac 1N \ln Z
\eeq
The free-entropy has a simple expression for 
$T < T_K(\varphi)$, where the complexity sticks to zero:
\begin{align}
\SS_{\rm{glass}}(T,\varphi) = - \b f_{\rm{min}}(T,\varphi) .
\label{ss_glass_1}
\end{align}
We have shown in Chapter \ref{chap:replicas} that introducing $m$ replicas coupled together allows for an explicit 
computation of the complexity. After introducing $m$ replicas, Eq.(\ref{complexity_1}) was modified to obtain 
Eq.(\ref{complexity_m}), which contained an $m$ dependance that allows for the computation of $\Si$. We rewrite here 
this equation in terms of the quantity $\SS$:
\begin{align}
\left\{ \begin{array}{ll}
& \displaystyle \SS(m;T,\varphi) = \Si(f^*(m;T,\varphi)) - m \b f^*(m;T,\varphi) ,  \\
& \\
& \displaystyle \frac{\partial \Si}{\partial f}(f^*(m;T,\varphi)) = \frac{m}{T} , \\
& \\
& \displaystyle \Si(f^*(m;T,\varphi)) = - m^2 \frac{\partial}{\partial m} \left( \frac{\SS(m;T,\varphi)}{m} \right) .
\end{array} \right.
 \label{def_ssglass_replicated} 
\end{align}
Eq.(\ref{ss_glass_1}) shows that for $T < T_K$ the free energy of the glass is equal to $f_{min}(T)$, the point where 
$\Si(f)=0$. Note also that this value of $f$ depends only on the free-energy landscape, which is the same for all replicas, and 
it is thus independent of $m$. Inside the glass phase, if there are several copies, the free-energy will simply be equal to
$m f_{min}(T)$.

The parameter $m$ allows, without changing the temperature or density, to modify the value of $f^*$ the saddle point in
Eq.(\ref{def_ssglass_replicated}), i.e. to displace the Kauzmann temperature artificially. If we choose $m^*$ it so that 
$f^*(m^*;T,\varphi) = f_{min}(T,\varphi)$, the complexity will be zero when evaluated at this free-energy, and the expression 
of the free-entropy will simply become:
\begin{align}
\SS(m^*;T,\varphi) = - m^* \b f_{min}(T,\varphi) ,
\end{align}
and, recalling Eq.(\ref{ss_glass_1}), the free-entropy of the glass will we obtained by dividing by $m^*$:
\begin{align}
\SS_{glass}(T,\varphi) = \frac{\SS(m^*;T,\varphi)}{m^*} \quad \text{if } T < T_K .
\end{align}
Still, $m^*$ is an unknown, but we remark that it is made to constrain that $f^* = f_{min}$. Since we expect the complexity to 
be a monotonic function of $f$, this is equivalent to say that $m^*$ is made up so that $\Si(f^*) = \Si(f_{min}) = 0$. Thus
$m^*$ is calculated as the value of $m$ that cancels the complexity. But the complexity can be calculated as the derivative
of $\SS/m$ as shown in Eq.(\ref{def_ssglass_replicated}). Thus $m^*$ is so that:
\begin{align}
\left. \frac{\partial}{\partial m} \left( \frac{\SS(m;T,\varphi)}{m} \right) \right|_{m^*} = 0 ,
\end{align}
and thus is the value of $m$ that minimizes $\SS/m$, which finally gives:
\begin{align}
\SS_{glass}(T,\varphi) = \min_{m} \frac{\SS(m;T,\varphi)}{m}.  
\label{def_SS_glass}
\end{align}

Now we see that, in order to compute the thermodynamics inside the ideal glass phase, one must be able to calculate the 
free entropy of a replicated liquid, and extremize it with respect to the number of copies.

\subsection{The small cage expansion}

Computing the free-energy and complexity of the glass amounts to computing the free entropy of an $m$-time replicated 
liquid. One can use a replicated version of standard liquid theory approximations such as the replicated 
Hyper-Netted-Chain (HNC) \cite{MP96,CFP99} presented in Chapter \ref{chap:replicas}, but we have seen that these 
approximations break down deep in the glass phase \cite{PZ10}. 
One can also follow M\'ezard and Parisi (MP)~\cite{MP99a,MP99b} and start 
at high density, then perform cage expansions: at high densities the copies stay close to the originals, forming molecules of 
size $A$. One can then expand the replicated free entropy with respect to this parameter $A$. 
To do this, one has to switch from a liquid of copies from a liquid of molecules, as described below.

\subsubsection{M\'ezard-Parisi approach}

We will denote by $x_i^a$ the $d$-dimensional position of particle $i$ of replica $a$, and collect 
all replica indices in $m \times d$ dimensional vectors, defining:
\beq
\bx \equiv \{ x^1 , \ldots , x^m \} \ .
\eeq
We define the partition function of the replicated liquid:
\beq \begin{split}
Z_m[z] = \sum_{N=0}^\io \frac 1{N!} \int d\bx_1 \cdots d\bx_N ~ 
\prod_{i<j} \prod_{a=1}^m ~ e^{w(x_i^a-x_j^a)} \prod_{i=1}^N z(\bx_i) \ ,
\end{split} \eeq
where we have defined:
\beq
d \bx \equiv d^d x^1 ~ \cdots ~ d^d x^m \ .
\eeq
Note that a $(N!)^{m-1}$ factor has been absorbed in all the possible relabelings of the molecules.
$z$ a generalized activity for the replicated liquid. If we take $z$ independant of the replica indices,
we obtain a set of $m$ uncoupled copies, that all have the same chemical potential.
We can make $z$ depend on the replica positions to constrain the replicas to be close.
For example the small cage expansion of M\'ezard and Parisi \cite{MP99a} is obtained by 
choosing:
\beq
z^{\rm{MP}}(\bx) =  z ~ \exp \left( - \frac{\e}{m} \sum_{a < b} \left[ x^a - x^b \right]^2 \right) \ .
\eeq
This corresponds to setting a quadratic attraction of intensity $\e$ between the copies.
We define the usual Mayer function plus its molecular counterpart:
\beq \begin{split}
& f(x,y) \equiv e^{w(x,y)}-1 \ , \\
& \bff(\bx,\by) \equiv e^{w(x^1,y^1)} \cdots e^{w(x^m,y^m)} -1 = \left( \prod_{a=1}^m e^{w(x^a,y^a)} \right) - 1 \ .
\end{split} \eeq
and in the hard sphere case $f$ is a Heaviside function. For harmonic spheres, 
$f$ looks like a smoothed Heaviside function.
We rewrite the partition function as:
\beq
Z_m[z,\bff] = \sum_{N=0}^\io \frac 1{N!} \int d\bx_1 \cdots d\bx_N ~ 
\prod_{i<j} \left( \frac{}{}\! 1 + \bff(\bx_i,\bx_j) \right) \prod_{i=1}^N z(\bx_i) \ ,
\eeq
The MP small cage expansion chooses the above particular form of activity, then Legendre 
transform the free-energy with respect to the intensity of the coupling to obtain a functional
of the parameter coupled to the coupling intensity, the cage size.
Indeed
\beq
\frac{\partial \ln Z_m[z^{\rm{MP}}]}{\partial \e} = \la - \frac 1m \sum_{i=1}^N \sum_{a<b} \left( x_i^a - x_i^b \right)^2 \ra \ .
\eeq
By replica symmetry, and homogeneity, we have that:
\beq
\sum_{i=1}^N \sum_{a<b} \left( x_i^a - x_j^b \right)^2 = \frac {m(m-1)}{2} \la (x_i^a - x_i^b)^2 \ra
\eeq
which is directly related to the cage size, as defined in a replica scheme:
\beq
A \equiv \frac 1{2d} \la (x_i^a - x_i^b)^2 \ra \ ,
\eeq
which is, in a dynamical point of view, the height of the long-time plateau in the mean-square 
displacement.Finally we see that:
\beq
\frac {\partial \ln Z_m[z^{\rm{MP}}]}{\partial \e} = -(m-1) d A \ ,
\eeq
which shows that $\e$ and $A$ are indeed a pair of coupled variables through a Legendre 
transformation.

The small cage expansion amounts to Legendre transforming the free-energy, to obtain a 
functional of the cage size, then expanding in powers of $A$ this functional. In practice this is done
via a large $\e$ expansion before Legendre transforming.

This scheme is not suited to the pair potentials that lead to a jamming transition, since they 
must contain a singularity, preventing one to follow the original strategy of M\'ezard and Parisi.
In practice this leads to a singular expansion in powers of $A$~\cite{PZ10}.

\subsubsection{Molecular formulation}

To correctly treat the replicated hard sphere system, Parisi and Zamponi (PZ)~\cite{PZ10} developed an effective potential 
approximation, which amounts to performing an expansion in powers of $\sqrt{A}$ instead of $A$. We explain in 
the following the extension of the effective potential approximation to finite temperature harmonic spheres. By construction, 
we will see that we are bound to recover both high density MP results and zero temperature PZ results
in a unified treatment. 

The strategy adopted by Paris and Zamponi was to perform a 
Legendre transform with respect to the full activity $z(\bx)$. As described in Chapter \ref{chap:formalism}, one 
then obtains the free energy as a functional of the single molecule density and the interaction potential.

We can see that $z$ is coupled to the density of molecules by a direct computation:
\beq \begin{split}
\frac{\d \ln Z_m}{\d z(\bx)} & = \frac 1{Z_m} \sum_{N=0}^\io \frac 1{N!} 
\int \prod_{i=1}^N d\bx_i \prod_{i<j} \left( \frac{}{}\! 1+ \bff(\bx_i,\bx_j) \right) \frac{\d}{\d z(\bx)} \left( \prod_{i=1}^N z(\bx_i) \right) \ , \\ 
& = \sum_{j=1}^N \frac 1{Z_m} \sum_{N=0}^\io \frac 1{N!} 
\int \prod_{i=1}^N d\bx_i \prod_{i<j} \left( \frac{}{}\! 1+ \bff(\bx_i,\bx_j) \right) \d(\bx-\bx_i)  \prod_{\underset{i \ne j}{i=1}}^N z(\bx_i) \ , \\
& = \sum_{j=1}^N \frac 1{Z_m} \sum_{N=0}^\io \frac 1{N!} 
\int \prod_{i=1}^N d\bx_i \prod_{i<j} \left( \frac{}{}\! 1+ \bff(\bx_i,\bx_j) \right) \d(\bx-\bx_i) \frac 1{z(\bx_j)} \prod_{i=1}^N z(\bx_i) \ , \\
& = \frac 1{z(\bx)} \sum_{j=1}^N \frac 1{Z_m} \sum_{N=0}^\io \frac 1{N!} 
\int \prod_{i=1}^N d\bx_i \prod_{i<j} \left( \frac{}{}\! 1+ \bff(\bx_i,\bx_j) \right) \d(\bx-\bx_i) \prod_{i=1}^N z(\bx_i) \ , \\
& =  \frac 1{z(\bx)} \la \sum_{i=1}^N \d(\bx-\bx_i) \ra \ .
\end{split}\eeq
Defining:
\beq
\r(\bx) \equiv \la \sum_{i=1}^N \d(\bx-\bx_i) \ra \ , 
\eeq
we obtain:
\beq
\frac{\d \ln Z_m}{\d \ln z(\bx)} = \r(\bx) \ .
\eeq
We emphasize again that this is not the same object than the replicated density define in the previous chapter.
The precise diagrammatic relation between the expansions performed in this chapter and thos performed in the previous
chapter are still to be established.

We perform now the Legendre transform of $\ln Z$ with respect to the molecular activity:
\beq
\left\{ \begin{array}{ll}
& \displaystyle \SS[\r] = \ln Z_m \left[\ln z^*[\r] \right] - \int d\bx ~ \r(\bx) \ln z^*[\r](\bx) \ , \\
& \\
& \displaystyle z^*[\r] \text{ such that } \left. \frac{\d \ln Z_m}{\d \ln z(\bx)} \right|_{z^*} = \r(\bx) \ .
\end{array} \right.
\eeq
Here we have omitted the thermodynamic limit, we will take it later to ease the notations.

\subsubsection{Ideal gaz case}

As usual in liquid theory one must start with the ideal gaz case.
In the ideal gaz, we have:
\beq \begin{split}
Z_m^{IG}[z] & = \sum_{N=0}^\io \frac 1{N!} \int d\bx_1 \cdots d\bx_N ~ \prod_{i=1}^N z(\bx_i) \\
& = \sum_{N=0}^\io \frac 1{N!} \left( \int d\bx ~  z(\bx) \right)^N \\
& = \exp \left( \int d \bx ~ z(\bx) \right) \ .
\end{split} \eeq
This allows the explicit Legendre transformation:
\beq
\frac{\d \ln Z_m^{IG}}{\d \ln z(\bx)} = e^{\ln z(\bx)} \ ,
\eeq
which shows that:
\beq
z^*[\r](\bx) = \r(\bx) \ ,
\eeq
and thus we get directly as expected:
\beq
\SS^{IG}[\r] = \int d\bx ~ \r(\bx) \left[ 1 - \ln \r(\bx) \right] \ .
\eeq
By using Mayer diagrammatics, we will thus obtain that our replicated entropy is expressed
as the sum of this ideal gaz contribution, plus diagrams with $\r(\bx)$ nodes and $\bff$ links.
\beq
\SS[\r] = \SS^{IG}[\r]
+ \begin{minipage}[1,1]{1cm} \includegraphics[width=1cm]{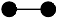} \end{minipage}
+ \begin{minipage}[1,1]{1cm} \includegraphics[width=1cm]{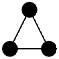} \end{minipage}
+ \begin{minipage}[1,1]{1cm} \includegraphics[width=1cm]{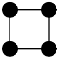} \end{minipage}
+ \cdots
\eeq

\subsubsection{Gaussian ansatz for the molecule density}

We can not extremize the free-energy in the full space of $\RRR^m$ functions, and we must
find a simplifying assumption. Since $\e$ and $A$ are coupled parameters, we will suppose that
$\r$ keeps the original Gaussian form of the MP choice of activity, but with $\e$ replaced by 
$1/A$:
\beq
\r^G(\bx) = \text{Cte} ~  \exp \left( - \frac 1{2mA} \sum_{a<b} (x^a - x^b)^2 \right) \ .
\eeq
This is a strong assumption: we suppose that the cage seen by a particle is well approximated by a Gaussian.
We rewrite this as:
\beq \begin{split}
\exp \left( - \frac 1{2mA} \sum_{a<b} (x^a - x^b)^2 \right) 
& = \exp \left( - \frac 1{2A} \sum_a (x^a)^2 \right) \exp \left( \frac 12 \left(\frac 1A \sum_a x^a\right) \frac Am \left(\frac 1A \sum_a x^a\right)
\right)
\end{split} \eeq
We can represent the second exponential by an integral under a Gaussian weight of variance 
$A/m$ by noting that:
\beq
\exp \left( \frac 12 \left(\frac 1A \sum_a x^a\right) \frac Am \left(\frac 1A \sum_a x^a\right) \right) = 
\frac 1{(2 \p A/m)^{d/2}} \int dX ~ \exp \left( - \frac 12 \frac{X^2}{A/m} + X \frac 1A \sum_a x^a\right) \ .
\eeq
We obtain:
\beq \begin{split}
\r^G(\bx) & = \text{Cte} ~ \frac {m^{d/2}}{(2 \p A)^{d/2}}  \int dX ~ \exp \left( - \frac{1}{2A} \sum_a 
\left( X - x^a \right)^2 \right) \ .
\end{split} \eeq
The molecule density must be normalized to $N$ the number of molecules, thus we can fix the 
prefactor by calculating the normalization:
\beq \begin{split}
N & = \int d\bx ~ \r^G(\bx) = \text{Cte} ~ \frac {m^{d/2}}{(2 \p A)^{d/2}}  \int dX \left[\left( 2 \p A \right)^{d/2}\right]^m \\
\end{split} \eeq
And thus:
\beq
\text{Cte} = \frac N V \frac{m^{-d/2}}{(2 \p A)^{(m-1)d/2}} \ , 
\eeq
and we get the final Gaussian ansatz for the molecule density~\cite{PZ10}:
\beq
\r^G(\bx) = \r \int dX ~ \prod_{a=1}^m \g_A(X-x^a) \ ,
\label{rhoGauss}
\eeq
where $\g_A$ is a normalized and centered Gaussian of variance $A$:
\beq
\g_A(x) = \frac 1{(2 \p A)^{d/2}} \exp \left( - \frac {x^2}{2A} \right) \ .
\eeq
These functions will tend to delta functions in the jamming limit, where the cage size is expected to 
go to zero, and we will make a repeated use of them.

\subsubsection{Ideal gaz contribution in the Gaussian ansatz}

We can now evaluate the functional $\SS$ at this particular value of the molecule density.
First we note that:
\beq
\ln \r^G(\bx) = \ln \r - \frac d2 \ln m - \frac {(m-1)d}2 \ln (2 \p A) - \frac 1{2mA} \sum_{a<b} 
\left( x^a - x^b \right)^2 \ .
\eeq
We can now compute:
\beq \begin{split}
\int d\bx ~ \r(\bx) \ln \r(\bx) = & \r \int_{X,\bx} \!\! \left(\prod_{a=1}^m \g_A(X-x^a)\right) \!\!
\left[ \ln \left( \frac{\r}{m^{d/2} (2 \p A)^{(m-1)d/2}} \right) - \frac 1{2mA} \sum_{a<b} 
\left( x^a - x^b \right)^2 \right] \\
= & \r V \left[ \ln \r - \frac d2 \ln m - \frac {(m-1)d}2 \ln (2 \p A) - \frac d2(m-1) \right] \ ,
\end{split} \eeq
and finally:
\beq \begin{split}
& \SS^{IG}[\r^G] = \r V \left[ \frac{}{}\! 1 - \ln \r \right] + \SS_{\rm{harm}}(m,A) \ , \\
& \SS_{\rm{harm}}(m,A) = \r V \left[ \frac{}{}\! \frac d2 \left( \frac{}{} 1-m + \ln m \right) + \frac {(m-1)d}2 \ln (2 \p A) \right] \ .
\label{Sharm}
\end{split} 
\eeq

\subsubsection{Reorganizing the Mayer expansion}

Now we want to express the free-energy of the replicated liquid as the free-energy of an effective liquid.
The stratey to do so is to average over the positions of $m-1$ replicas, and keep only one replica, say replica 1,
as reference. This procedure will give rise to effective interactions, two-body ones, but also many-body interactions.
We have seen that the free-entropy is given by the ideal gaz result plus the sum of all connected and irreducible Mayer diagrams, 
which carry $\r(\bx)$ nodes and $\bff(\bx,\by)$ links. 
How will the small cage expansion affect the diagrams ? At lowest order, the Gaussian ansatz for the density will give 
delta functions, and thus all replicas will be at the same integration point.
Thus we have that:
\beq
\bff(\bx,\by) \underset{A \to 0^+}\to e^{m w(x^1,y^1)} -1 = f_1^{(m)}(x^1,y^1) \ ,
\eeq
where we have defined:
\beq
f_1^{(m)}(x,y) \equiv e^{m w(x,y)} - 1 \ .
\eeq
This is the Mayer function of a liquid interacting via a potential $w$, but at temperature $T/m$.
Thus we expect the lowest order result to give an effective liquid interacting via the original potential, at temperature $T/m$,
as was already found in \cite{MP99a}. Note that here we have chosen a given replica as a reference, but we could also have chosen
the center of mass of the molecule as a reference. In our translationally invariant case, this has no consequence, but simplifies the
calculations. If translational invariance does not hold, an expansion around the center of mass is required.

What we can thus do is to separate the integrations on the $x_i^1$ from those 
on the others replicas, and to decompose the total Mayer function in two contributions:
\beq \begin{split}
\bff & = f_1^{(m)} + \left[ \frac{}{}\! \bff - f_1^{(m)} \right] \ .
\end{split}\eeq
Each node is integrated upon, with weight $\r^G$, and we separate this into:
\beq
\int d\bx \r^G(\bx) \left(\frac{}{}\bullet\frac{}{}\right) = \int \r ~ dx^1 \int \DD {\mathbf x} \left(\frac{}{}\bullet\frac{}{}\right) \ ,
\eeq
where we have defined the measure $\DD {\mathbf x}$ as:
\beq
\DD {\mathbf x} \equiv \frac{\r^G(\bx)}{\r} dx^2 \cdots dx^m \ .
\eeq
And we have trivially that $\int \DD {\mathbf x} = 1$.

Now consider one diagram of the free-entropy, and make the replacement of the links described above.
This amounts to take the same diagram, but with links that can be either $f_1^{(m)}$ or $\bff - f_1^{(m)}$:
\beq \begin{split} 
\SS[\r^G] = \SS^{IG}[\r^G] & ~
+ \begin{minipage}[1,1]{1cm} \includegraphics[width=1cm]{diag1_jamm} \end{minipage}
+ \begin{minipage}[1,1]{1cm} \includegraphics[width=1cm]{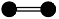} \end{minipage}
+ \begin{minipage}[1,1]{1cm} \includegraphics[width=1cm]{diag2_jamm} \end{minipage}
+ \begin{minipage}[1,1]{1cm} \includegraphics[width=1cm]{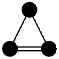} \end{minipage}
+ \begin{minipage}[1,1]{1cm} \includegraphics[width=1cm]{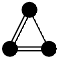} \end{minipage}
+ \begin{minipage}[1,1]{1cm} \includegraphics[width=1cm]{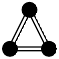} \end{minipage} \\
& ~ + \begin{minipage}[1,1]{1cm} \includegraphics[width=1cm]{diag3_jamm} \end{minipage}
+ \begin{minipage}[1,1]{1cm} \includegraphics[width=1cm]{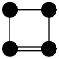} \end{minipage}
+ \begin{minipage}[1,1]{1cm} \includegraphics[width=1cm]{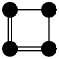} \end{minipage}
+ \begin{minipage}[1,1]{1cm} \includegraphics[width=1cm]{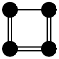} \end{minipage}
+ \begin{minipage}[1,1]{1cm} \includegraphics[width=1cm]{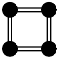} \end{minipage}
+ \ldots
\end{split}\eeq
The double lines are $\bff-f_1^{(m)}$ links, while the normal lines are $f_1^{(m)}$ links.

We can classify these diagrams by the number of $\bff-f_1^{(m)}$ links they contain, since they are a small quantity.
At lowest order, we get a sum of diagrams with only $f_1^{(m)}$ links. All integrations over the replicas other than
replica 1 give 1 because of our choice of measure. Adding to these diagrams the result from the ideal gaz, 
we recover as anticipated the free-entropy of a liquid interacting 
via a potential $mw$, plus the harmonic part described earlier:
\beq
\SS[\r^G,w] \underset{A \to 0^+}\to \SS_{liq}[\r,mw] + \SS_{\rm{harm}} \ .
\eeq

\subsubsection{Expansion at first order}

At order 1 in the numbers of $\bff-f_1^{(m)}$ links, we obtain the sum of all diagrams of the free-entropy
$\SS[\r,mw]$, where one has replaced one of the $f_1^{(m)}$ links by a $\bff-f_1^{(m)}$ link. 
Thus we get:
\beq \begin{split}
& \SS[\r^G,w] = \SS_{liq}[\r,mw] + \SS_{\rm{harm}} + \SS^{(2)} + \OO((\bff-f_1^{(m)})^2) \ , \\
& \SS^{(2)} = \begin{minipage}[1,1]{1cm} \includegraphics[width=1cm]{diag1_2} \end{minipage}
+ \begin{minipage}[1,1]{1cm} \includegraphics[width=1cm]{diag2_2} \end{minipage}
+ \begin{minipage}[1,1]{1cm} \includegraphics[width=1cm]{diag3_2} \end{minipage}
+ \ldots
\end{split} \eeq
We can write $\SS^{(2)}$ as an integral:
\beq \begin{split}
& \SS^{(2)} = \int d\bx d\by \r^G(\bx) \r^G(\by) \left[ \frac{}{}\! \bff(\bx,\by) - f_1^{(m)}(x^1,y^1) \right] F(x^1,y^1) \ , \\
& F(x,y) = \begin{minipage}[1,1]{1cm} \includegraphics[width=1cm]{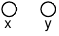} \end{minipage}
+ \begin{minipage}[1,1]{1cm} \includegraphics[width=1cm]{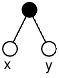} \end{minipage}
+ \begin{minipage}[1,1]{1cm} \includegraphics[width=1cm]{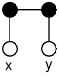} \end{minipage}
+ \ldots
\end{split} \eeq
The links of the diagrams are $f_1^{(m)}$, and the white labeled nodes don't carry any weight.
This sum of diagrams is obviously equal to the derivative of $\SS[\r,mw]$ with respect to $f_1^{(m)}$ 
divided by $\r^2$. But we know that:
\beq
\left( 1+ f_1^{(m)}(x,y) \right) \frac{\d \SS[\r,mw]}{\d f_1^{(m)}(x,y)} = \frac 12 \r^{(2)}(T/m;x,y) \ .
\eeq
Thus we find:
\beq
F(x,y) = \frac 1{\r^2} \frac 12 \r^{(2)}(T/m;x,y) e^{-m w(x,y)}
\eeq
And we obtain the final result:
\beq \begin{split}
& \SS[\r^G,w] = \SS_{liq}[\r,mw] + \SS_{\rm{harm}} + \frac {\r^2}2 \int_{x,y}  g(T/m;x,y) Q(x,y) \ , \\
& Q(x,y) = e^{(1-m)w(x,y)} \int dX dY \g_A(X-x) \g_A(Y-y) \int \left( \prod_{a=2}^m dx^a \g_A(X-x^a) \right) \left( \prod_{a=2}^m dy^a \g_A(Y-y^a) \right) \\
& \phantom{Q(x,y) = \int dX dY} \left[ \prod_{a=2}^m e^{w(x^a,y^a)} - e^{(m-1)w(x,y)} \right] \ .
\end{split} \eeq 
We can simplify a bit the expression of $Q$:
\beq \begin{split} 
& 1+Q(x,y) = e^{(1-m)w(x-y)} \int dX dY \g_A(X-x) \g_A(Y-y) q(X-Y)^{m-1} \ , \\
& q(x,y) = \int du dv ~ \g_A(x-u) \g_A(y-v) e^{w(u-v)} \ .
\end{split} \eeq
Using translational invariance we can rewrite all this as:
\beq\begin{split}
& \SS[\r^G,w] = \SS_{liq}[\r,mw] + \SS_{\rm{harm}}(m,A) + \frac {\r^2 V}2 \int_r g(T/m;r) Q(T,m,A;r) \ , \\
& Q(T,m,A;r) = e^{(1-m)w(r)} \int_{r'} \g_{2A}(r-r') q(A,T;r')^{m-1} ~-1\ , \\
& q(A,T;r) = \int_{r'} \g_A(r-r') e^{w(r')} \ ,
\label{S_of_Q}
\end{split}\eeq
where we have made explicit the dependances on temperature, $m$ and $A$. To come back to the initial free-entropy, one has to 
do the inverse Legendre transform, and thus find the optimal Gaussian density, which amounts to finding the optimal 
cage size $A^*$.

Due to the Gaussian {\it ansatz}, and the quadratic form of the harmonic potential Eq.(\ref{potential_harmS}), 
the function $q$ has a tractable expression \cite{BJZ11}, that depends on both parameters $A$ and $m$:
\beq \begin{split}
q(A,T;r) = & ~ \frac 12 \left( 2 + \erf \left[ \frac{r-1}{2\sqrt{A}} \right] - \erf \left[ \frac{r+1}{2\sqrt{A}} \right] \right) 
+ \frac{4A^{3/2}\b}{(1+4A\b) \sqrt{\p} r} \left( \frac{}{}\! e^{-\frac{(r-1)^2}{4A}} - e^{-\frac{(r+1)^2}{4A}} \right) \\
& ~ + e^{-\frac{(r-1)^2 \b}{1+4 A \b}} \frac{r+4A\b}{2r(1+4A\b)^{3/2}} \left( \erf \left[ \frac{r+4A\b}{2 \sqrt{A(1+4A\b)}} \right]
+ \erf \left[ \frac{1-r}{2 \sqrt{A(1+4A\b)}} \right] \right) \\
& ~ - e^{-\frac{(r+1)^2 \b}{1+4 A \b}} \frac{r-4A\b}{2r(1+4A\b)^{3/2}} \left( \erf \left[ \frac{r-4A\b}{2 \sqrt{A(1+4A\b)}} \right]
- \erf \left[ \frac{1+r}{2 \sqrt{A(1+4A\b)}}\right] \right)
\end{split} \eeq

We can see easily that the M\'ezard-Parisi approximation scheme is recovered if we suppose that the potential $w(r)$ is differentiable in $r=1$.
Indeed coming back to the expression of $q$ we have:
\beq
q(A,T;r) = \int d^3 r' ~ \g_{2A}(r') e^{w(r-r')} \underset{A \to 0}{\sim} e^{w(r)} 
\left[ \frac{}{}\!1 + A \D_r w(r) + A \left[ \nabla_r w(r) \right]^2 + \OO(A^2) \right] \ .
\eeq
Plugging this into the expression of the free-energy we get the first-order correction:
\beq
G(m,A;T) = - 3 A(m-1) \int_0^\io dr ~ r^2 \D_r w(r) g(T/m;r) \ ,
\eeq
which reproduces the MP result when introduced in the expression of the free-entropy.

\subsubsection{Formulation in terms of two-body effective potential}

Parisi and Zamponi~\cite{PZ10} have shown that the molecular liquid could be expressed as an effective liquid, albeit 
with modified interaction: the two-body potential $w$ is replaced by an effective potential $w_{\rm{eff}}$, and higher-order 
interaction potentials also arise. Here we neglect the higher-order potentials. 
They are expected to be negligible at high dimensions
but not necessarily in finite dimensions. However taking them into account makes the calculation much harder, already at next order.

\begin{figure}[htb]
\begin{center}
\includegraphics[width=7cm]{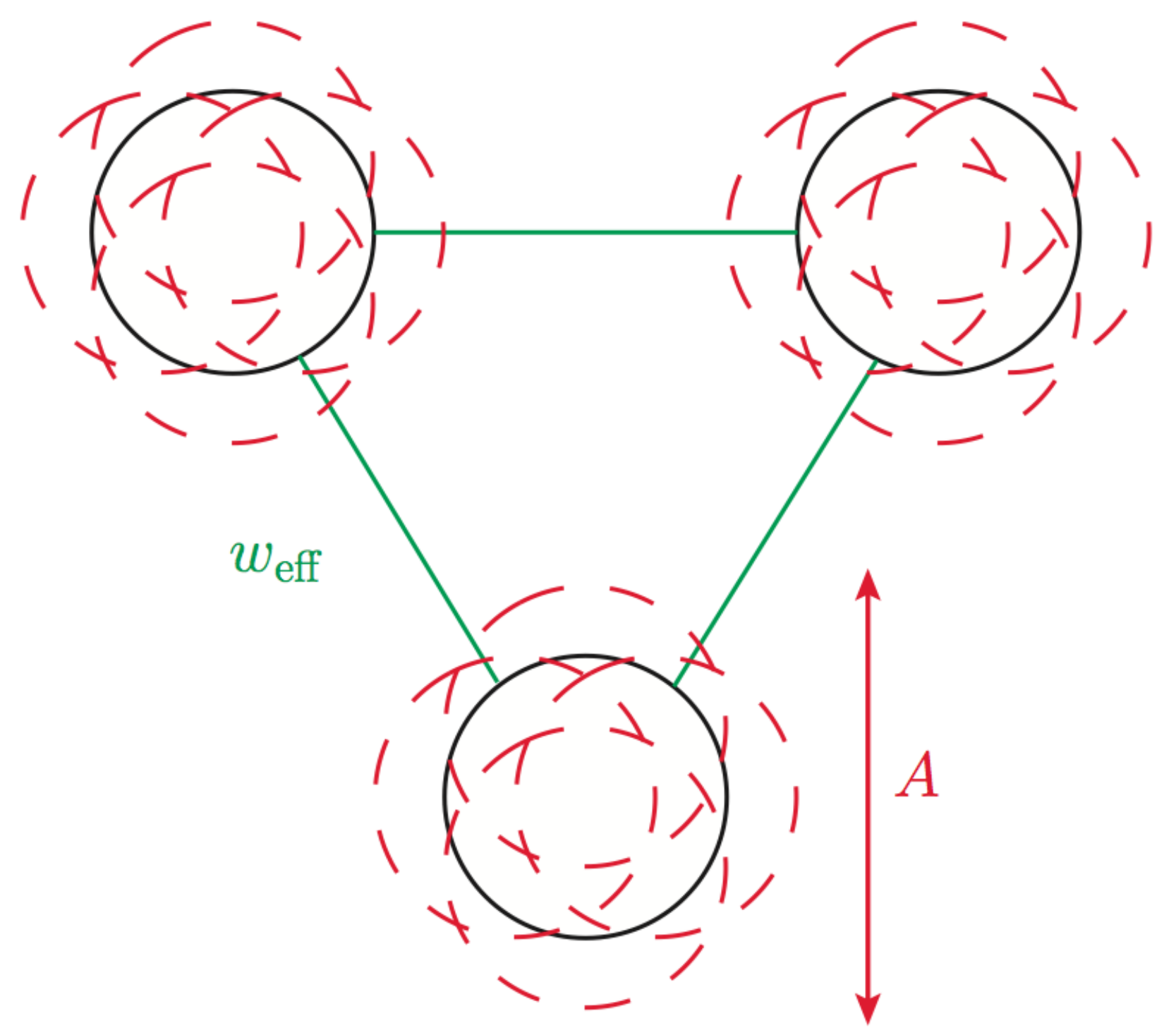}
\end{center}
\caption{A schematic representation of the effective potential 
approximation. Each particle in the
original liquid is replicated $m$ times (dashed spheres). Assuming
that the replicated particles form a molecule of average cage
size $A$, in the partition sum, we trace out the degrees of freedom
of ($m-1$) copies of the liquid to obtain an effective 
one-component liquid (black spheres) interacting with an effective
pair potential $w_{\rm{eff}}$ (green lines).}
\label{fig:sketchveff}
\end{figure}

The calculation is schematically represented in Fig.~\ref{fig:sketchveff}.
The two body effective potential $w_{\rm{eff}}$ is defined as:
\beq
e^{w_{\rm{eff}}(x,y)} = e^{w(x,y)} \la \prod_{a=2}^m e^{w(x^a,y^a)} \ra_{x,y} \ ,
\eeq
where the average is over the molecule probability density, at fixed $x^1=x$ and $y^1=y$:
\beq
\la \bullet \ra_{x,y} = \int d\bx d\by ~ (\bullet) ~ 
\d(x^1 - x) \d(y^1-y) \frac{\r^G(\bx)\r^G(\by)}{\r^2} \ .
\eeq
It is easy to check that in the $A \to 0$ limit, this effective potential reduces to 
$e^{mw}$, and that we get:
\beq
e^{w_{\rm{eff}}(x,y)} = e^{mw(x,y)} \left[ \frac{}{} 1  + Q(x,y) \right]  \ .
\eeq
In three dimensions, this can be rewritten, for future use, as:
\beq
e^{w_{\rm{eff}}(r)} = e^{w(r)} \frac 1{r \sqrt{4 \p A}} \int_{0}^\io du ~ u \left[ e^{-\frac{(r-u)^2}{4A}} - e^{-\frac{(r+u)^2}{4A}} \right] q(A,T;r)^{m-1} \ .
\label{final_expression_effpot}
\eeq

Equation~(\ref{S_of_Q}) allows for a full calculation of the replicated free entropy, and allows us to recover both the MP small 
cage expansion in powers of $A$ at high density, and the PZ expansion in powers of $\sqrt{A}$ at zero temperature, as we 
will show below. In the following, we do not discuss the details of the calculations, that are extensively described in 
\cite{BJZ11}. Again we emphasize that, due to the Gaussian approximation that we make for the density, and the quadratic
form of the potential, the detailed calculations only involve Gaussian integrals. Numerical minimizations must also be 
performed, for example to compute the optimal number of replicas $m^*$, or the optimal cage size.

\subsection{Low temperature liquid theory approximation}

Our purpose in interpolating between zero and finite temperature was to be able to probe the vicinity of the jamming point 
in the $(\varphi, T)$ phase diagram, thus we are interested mainly in the low temperature behavior of the glass. We can 
exploit this by making an approximation that will allow us to push the analytical calculation much further. Defining the cavity 
distribution function $y$ by $g(r)=e^{w(r)} y(r)$ \cite{hansen}, we make the approximation of taking 
the cavity function of the liquid as a constant and equal to its $T=0,r=1$ value, which we call $y^{HS}_{liq}(\varphi)$. As 
$T$ goes to zero, we see that the exponential factor converges towards a step function around $r=1$, so that in all 
integrals that are cut at $r=1$ by the pair potential, we can safely evaluate $y$ at its $r=1$ value. Expanding $y$ in powers 
of $T$, it is easy to convince oneself that the temperature dependence leads to subleading contributions in the integrals.
Thus we suppose:
\begin{align}
g(T/m;r) \sim e^{mw(r)} y^{HS}_{liq}(\varphi) \ ,  \label{y_approx_rep}
\end{align}
which simplifies further the expression of the free-entropy. We have made explicit the dependance of $y$ on the packing fraction.

The last quantities that we need to compute are the free entropy $\SS_{liq}$ and 
the pair correlation function $g_{liq}(r)$ 
of the liquid. 
Given the pair correlation function, which represents the probability of finding a particle at distance $r$ from a particle 
fixed at the origin, we can express the internal energy $U$ as:
\begin{align}
U(T,\varphi) = 12 \varphi \int_0^{\io}dr \, r^2 g(T,\varphi;r) v(r) \ .
\label{def_U}
\end{align}
Plugging the low temperature approximation Eq.~(\ref{y_approx_rep}) for the liquid into this, we obtain:
\begin{align}
U_{liq}(T,\varphi) & \underset{T \to 0}{\sim} 12 \varphi y^{HS}_{liq}(\varphi) \int_0^1 dr \,  r^2 (1-r)^2 e^{ - \b (1-r)^2} \ .
\label{y_approx}
\end{align}
Making use of the standard identity $\displaystyle U(T) = \frac{\partial (F/T)}{\partial (1/T)}$, we can derive the low 
temperature approximation for the free entropy of the liquid:
\begin{align}
& \SS_{liq}(T,\varphi) \underset{T \to 0}{\sim} S_{liq}^{HS}(\varphi) + 6 \varphi y^{HS}_{liq}(\varphi) \left[ \frac{\sqrt{\pi}}{2} 
\sqrt{T} \left( 2 + T \right) \text{erf} \left( \frac{1}{\sqrt{T}} \right) \right.  \left.+ T \left(e^{-1/T} - 2 \right) \right] ,
\label{dev_energy}
\end{align}
where $y^{HS}_{liq}(\varphi)$ and $\SS_{liq}^{HS}(\varphi)$ are short-hand notations for $y_{liq}(T=0,\varphi;r=1)$ and 
$\SS_{liq}(T=0,\varphi)$.

At this level of approximation, it is clear that the only input that is needed from liquid theory is the equation of state of 
the hard sphere liquid. From any given equation of state one can easily deduce
the hard sphere free entropy $\SS_{liq}^{HS}$ and cavity function $y^{HS}_{liq}$.
The most reasonable choice would be to use the phenomenological Carnahan-Starling (CS) equation of state, as 
in~\cite{PZ10}, that provides the best fit to the hard sphere pressure. 

We instead used the Hyper-Netted Chain approximation described in the end of Chapter \ref{chap:formalism}. 
Although HNC is known to be less accurate for the hard sphere system, using HNC allows us to also compare our results 
to the ones obtained from the MP approach valid at large density and finite temperatures. 
The HNC approximation overestimates
$y^{HS}_{liq}$ by $20\%$ in the relevant range of volume fraction $\varphi \sim 0.64$. This has the effect of reducing the 
glass transition density obtained from the theory, from the value $\varphi_K = 0.62$ obtained from CS~\cite{PZ10} to $
\varphi_K = 0.58$ obtained with HNC. 

Therefore the reader should keep in mind that the glass densities reported in the following are lower than the correct ones.
In any case, here we are more interested in the low-temperature scaling in the glass phase than to the actual value of the 
glass transition density. We also checked that the scaling results are insensitive to the precise choice of the equation
of state of the hard sphere liquid.

With all these approximations, the free-entropy simplifies further into
\beq \begin{split}
& \SS(m,A;T,\varphi) = \SS_{liq}(T/m,\varphi) + \SS_{\rm{harm}}(m,A) + 4 \varphi y^{HS}_{liq}(\varphi) G(m,A;T) \ , \\
& G(m,A;T,\varphi) = 3 \int_0^\io dr ~ r^2 \left[ q(A,T;r)^m - e^{m w(r)} \right] \ .	
\label{final_expression_entropy}
\end{split} \eeq
Optimization over $A$ gives a self-consistent equation for the optimal cage radius $A^*(m;T,\varphi)$ that reads:
\beq \begin{split}
& J(m,A^*(m;T,\varphi);T) = \frac{9}{4 \p \r y^{HS}_{liq}(\varphi)} \ , \\
& J(m,A;T) \equiv \frac A{1-m} \frac{\partial G(m,A;T)}{\partial A} = \frac{3mA}{1-m} \int_0^\io dr ~ r^2 q(A,T;r)^{m-1} \frac{\partial q(A,T;r)}{\partial A} \ .
\end{split} \eeq

\section{Thermodynamics of the glass} 
\label{glass}

\subsection{Complexity and phase diagram}

As discussed above, the glass transition is signaled by the point where the saddle point in
Eq.~(\ref{complexity_1}) reaches the minimum $f_{min}$ at which the complexity $\Si(f)$ vanishes.
In the replica formalism, the equilibrium complexity $\Si(f^*)$ in Eq.~(\ref{Sivanishes}) corresponds
to the complexity in Eq.~(\ref{complexity_m}) evaluated at $m=1$. We call this
quantity the ``equilibrium'' complexity of the liquid $\Si_{eq}(T,\varphi)$. The latter is easily computed
by expanding the equations around $m=1$:

\beq \begin{split}
& \Si_{eq}(T,\varphi) = \SS_{liq}(T,\varphi) - \frac 32 \ln \left( \frac{}{}\! 2 \p A^*(m=1;T,\varphi) \right) - 3 - 12 \varphi y^{HS}_{liq}(\varphi) ~ 
H \! \left( \frac{}{}\!\! m=1, A^*(m=1;T,\varphi);T \! \right), \\
& H(m,A;T) \equiv \frac 1m \frac{\partial G(m,A;T)}{\partial m} = \frac 1m \int_0^\io dr ~ r^2 q(A,T;r)^m \ln q(A,T;r) 
+ \frac 1m \int_0^\io dr ~ r^2 w(r) e^{m w(r)} \ .
\end{split} \eeq

\begin{figure}[htb]
\centering
\includegraphics[width=14cm]{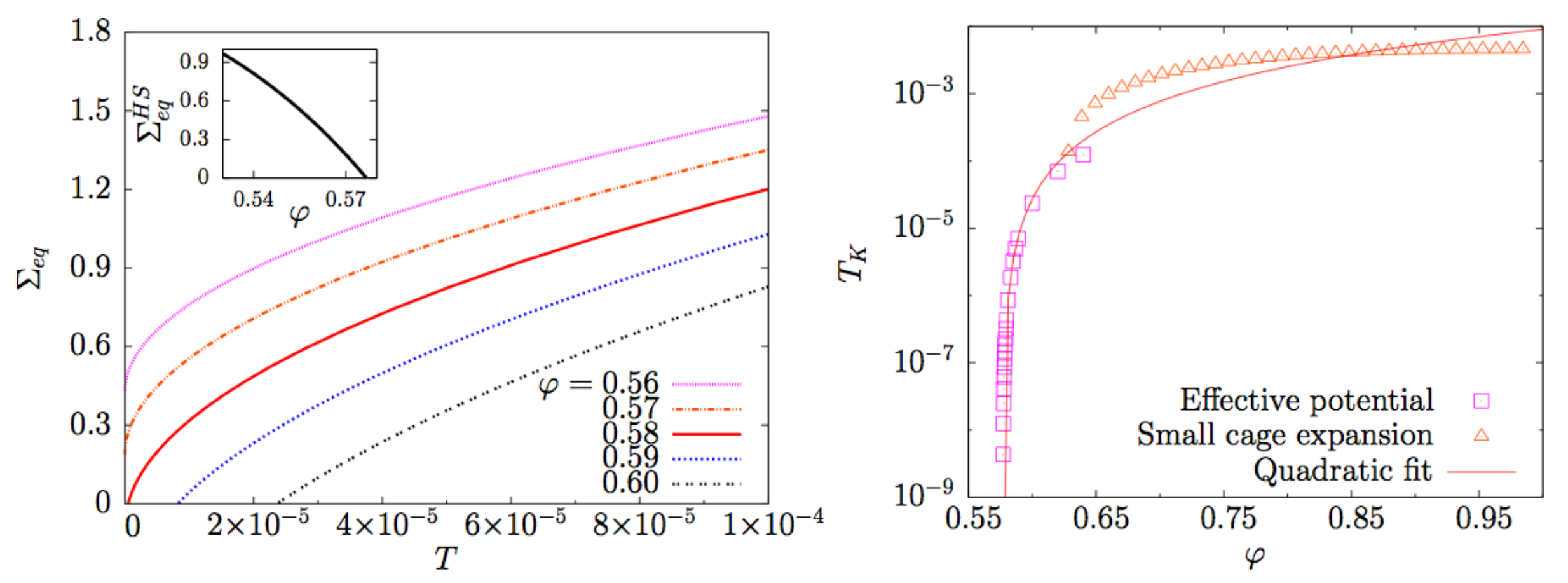}
\caption{Left: Equilibrium complexity against temperature for several volume fractions around $\varphi_K$, calculated 
within the effective potential approximation. Inset: complexity of hard spheres, obtained as the $T \to 0$ limit of the 
complexity $\Si_{eq}$, plotted against the volume fraction. Right: Kauzmann temperature against volume fraction, within 
the the effective potential approach developed in this work (open squares). The quadratic fit (red line) gives an estimated 
Kauzmann transition for hard spheres at $\varphi_K = 0.576898$. The small cage M\'ezard-Parisi approach breaks down 
at low temperatures and low densities \cite{MP99b}.}
\label{fig:sigma_tk}
\end{figure}

For $\varphi \ge \varphi_K$, the complexity vanishes at a temperature $T_K(\varphi)$, called the Kauzmann temperature, 
that increases with the density. These results are summarized in the left frame of Fig.~\ref{fig:sigma_tk}, where we show 
the complexity as a function of temperature for several volume fractions around $\varphi_K$. In the inset we show the zero 
temperature limit of the complexity $\Si_{eq}^{HS}$ as a function of the density, that vanishes at $\varphi_K$.

From the complexity we deduce the phase diagram 
shown in the right frame of Fig.~\ref{fig:sigma_tk}. 
In this figure we report $T_K(\varphi)$, as obtained in the framework of the effective potential approximation, and we 
compare it with the result obtained
from the M\'ezard-Parisi small cage expansion~\cite{BJZ10}.
In the effective potential case we find that the Kauzmann temperature goes to zero at $\varphi_K  \approx 0.5769$, 
quadratically with the temperature. On the other hand, the result from the small cage expansion is a finite $T_K(\varphi)$ 
which jumps to zero 
abruptly at a value of $\varphi$ which
is unrelated to hard sphere results. This is due to fact that 
the small cage expansion is valid only in the region indicated 
in Fig.~\ref{fig:PDsch}.
On the other hand, our effective potential computation becomes
inaccurate when the temperature is too high because of the 
approximation in Eq.~(\ref{y_approx_rep}). 
Still we obtain a reasonable matching of the two approximation schemes for intermediate densities around 
$\varphi = 0.64$.
It would be easy, in principle, to reconcile the two approximations at all volume fractions, including the crossover regime, 
by avoiding the low temperature approximation made in Eq.~(\ref{y_approx_rep}), but these computations would require a much heavier numerical treatment.

\subsection{Relaxation time and glass fragility}

The calculation of the liquid complexity, apart from signaling the glass transition, can be related to the relaxation time and 
fragility of the glass as described in the introduction, through Eq.(\ref{adam-gibbs}) and its mean-field counterpart 
Eq.(\ref{adam-gibbs_RFOT}).

\begin{figure}[hbt]
\centering
\includegraphics[width=8cm]{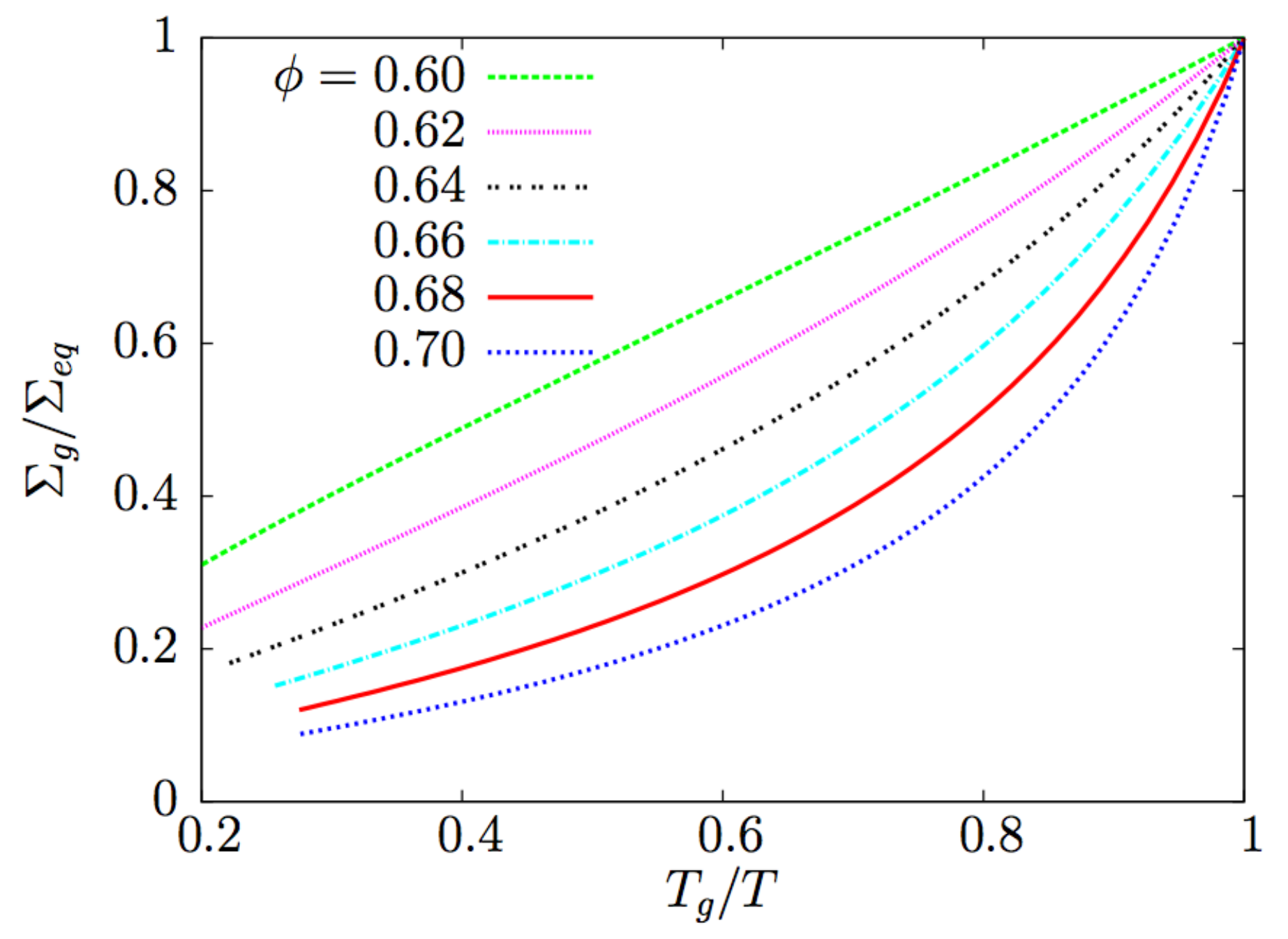}
\caption{A thermodynamic Angell plot of $\Si_g/\Si_{eq}$, with $\Si_g=1$ as a definition of the glass transition, plotted against $T_g/T$, for different volume fractions. The thermodynamic fragility is predicted to increase 
rapidly with volume fraction, in agreement with numerical 
simulations~\cite{BW09b}.}
\label{fig:angell}
\end{figure}

In Refs.~\cite{BW09b,BW09a}, Berthier and Witten showed that, for harmonic spheres, compressing the system leads to a 
very sensible increase of the fragility. Using our results for the complexity, we are able to qualitatively check whether 
replica theory can reproduce this feature. 
Indeed, thanks to Eq.~(\ref{adam-gibbs_RFOT}), the fragility can be extracted equivalently from the relaxation time or from 
the complexity.
Since the fragility is usually evaluated at the conventionally defined laboratory glass transition,
we arbitrarily define the glass transition temperature $T_g(\varphi)$ as the temperature at which the equilibrium complexity 
is equal to one, $\Si(T_g(\varphi),\varphi) = 1$, which is a typical value of the configurational entropy at the glass transition 
in most numerical simulations and experiments (its precise value is immaterial for our purposes).
Using these values of $T_g$ and $\Si_g=1$, we can construct an Angell plot similar to Fig. \ref{fig:angell} of the introduction, but for the complexity.
In Fig.~\ref{fig:angell} we show the inverse of the complexity, 
linked to the logarithm of the relaxation time by Eq.~(\ref{adam-gibbs}), plotted against $T_g/T$, for several densities. 
The fragility is the slope of the curves in $T_g/T=1$~\cite{MA01}.
One can clearly see that increasing the density drastically increases the fragility of the glass-former.

\subsection{Free-energy of the glass}

We now turn to the calculation of the free entropy of the glass, 
$\SS_{glass}$. We have seen that in order to compute $\SS_{glass}(T,\varphi)$, one needs to optimize the replicated free 
entropy $\SS(m,A;T,\varphi)/m$ with respect to $A$ first (in order to perform the inverse Legendre transform), then with 
respect to $m$, via Eq.~(\ref{def_SS_glass}). Calling $A^*(T,\varphi)$ and $m^*(T,\varphi)$ the optimal values of $A$ and 
$m$, we obtain:
\begin{align}
\SS_{glass}(T,\varphi) = \frac{\SS(m^*(T,\varphi),A^*(T,\varphi);T,\varphi)}{m^*(T,\varphi)}.
\label{eqSSglass_final}
\end{align}

\begin{figure}
\centering
\includegraphics[width=8cm]{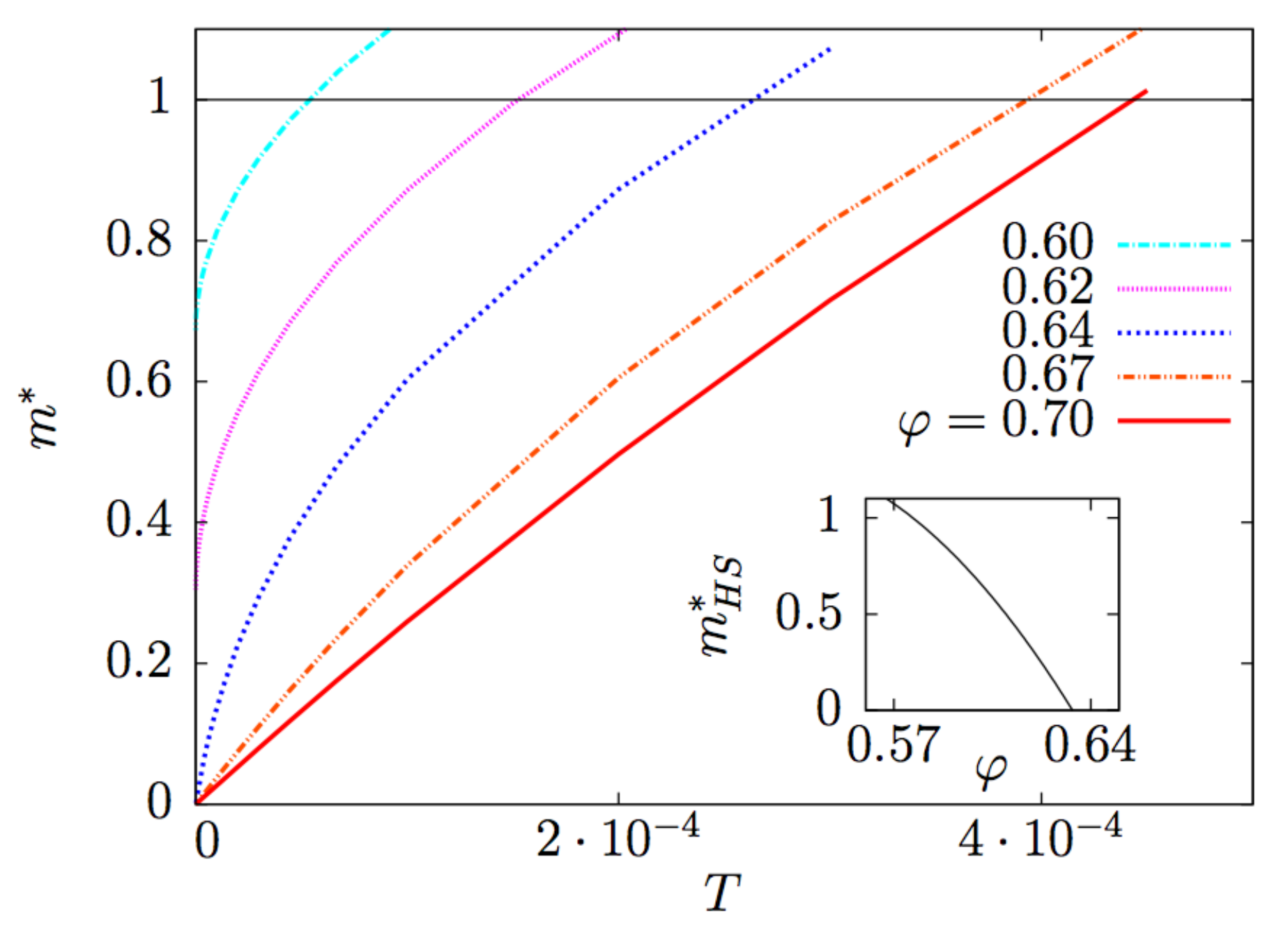}
\caption{Optimal number of replicas $m^*$ as a function of $T$, for several volume fractions. The inset shows the $T \to 0$ limit of $m^*$ as a function of the volume fraction.
}
\label{fig:mstar}
\end{figure}

Starting from Eq.~(\ref{eqSSglass_final}), we can deduce all quantities of interest, for instance the pressure 
and energy of the glass and its specific heat. 

We start from the general thermodynamic relations
$U = - d\SS/d\b$ and $p = \b P/\r = -\varphi d\SS/d\varphi$, where $p$ is the so-called ``reduced pressure'' or ``compressibility factor''. 
We have that the free-energy of the glass is $\SS(m^*,A^*;T,\varphi)/m^*$, where $m^*$ and $A^*$ are determined by 
optimization of $\SS(m,A;T,\varphi)/m$. Therefore, we do not need to take explicit derivatives with respect to $m$ and $A$, 
and we can directly compute the derivatives with respect to density and temperature to obtain explicit expressions for the 
energy and the pressure \cite{BJZ11}.

The pressure is thus obtained simply as:
\beq \begin{split}
p_{glass}(T,\varphi) & = - \frac{\varphi}{m^*} \frac{\partial \SS(m^*,A^*;T,\varphi)}{\partial \varphi} \\
& = \frac 1{m^*} p_{liq}(T/m^*,\varphi) 
- \frac{4 \varphi}{m^*} \left[ y^{HS}_{liq}(\varphi) + \varphi \frac{d y_{liq}^{HS}(\varphi)}{d \varphi} \right] G(m^*,A^*;T) \ .
\label{pressure_glass}
\end{split} \eeq
The energy is obtained as:
\beq
U_{glass}(T,\varphi) = - \frac 1{m^*} \frac{\partial \SS(m^*,A^*;T,\varphi)}{\partial \beta} = 
- 12 \varphi y_{liq}^{HS}(\varphi) \int_0^\io dr ~ r^2 q(A^*,T;r)^{m^*-1} \frac{\partial q(A^*,T;r)}{\partial \beta} \ .
\eeq
As expected, we can check that this can be rewritten as:
\beq
- \b U_{glass}(T,\varphi) = 12 \varphi y_{liq}^{HS}(\varphi) \int_0^\io dr ~ r^2 w(r) e^{w_{\rm{eff}}(r)} \ .
\eeq

An important quantity for the following is the optimal number of replicas $m^*$. In Fig.~\ref{fig:mstar}, we show its behavior 
as a function of the temperature, for different volume fractions. We see that the $T \to 0$ limit of $m^*(T,\varphi)$ converges 
when $\varphi$ is not too large to a finite value, which we call $m^*_{HS}(\varphi)$, and which is shown in the inset. 
The replica parameter of hard spheres vanishes at a density $\varphi_{GCP}$, the glass close packing. This point is the 
equivalent, in our mean-field picture, of the jamming point of harmonic spheres. The behavior of the cage radius $A^*$ is 
similar to that of $m^*$. 

Since our goal is to study the vicinity of the jamming point, we have to take into account that the optimal number of replicas,
as well as the optimal cage size, vanish at jamming.

\begin{figure}[hbt]
\centering
\includegraphics[width=8cm]{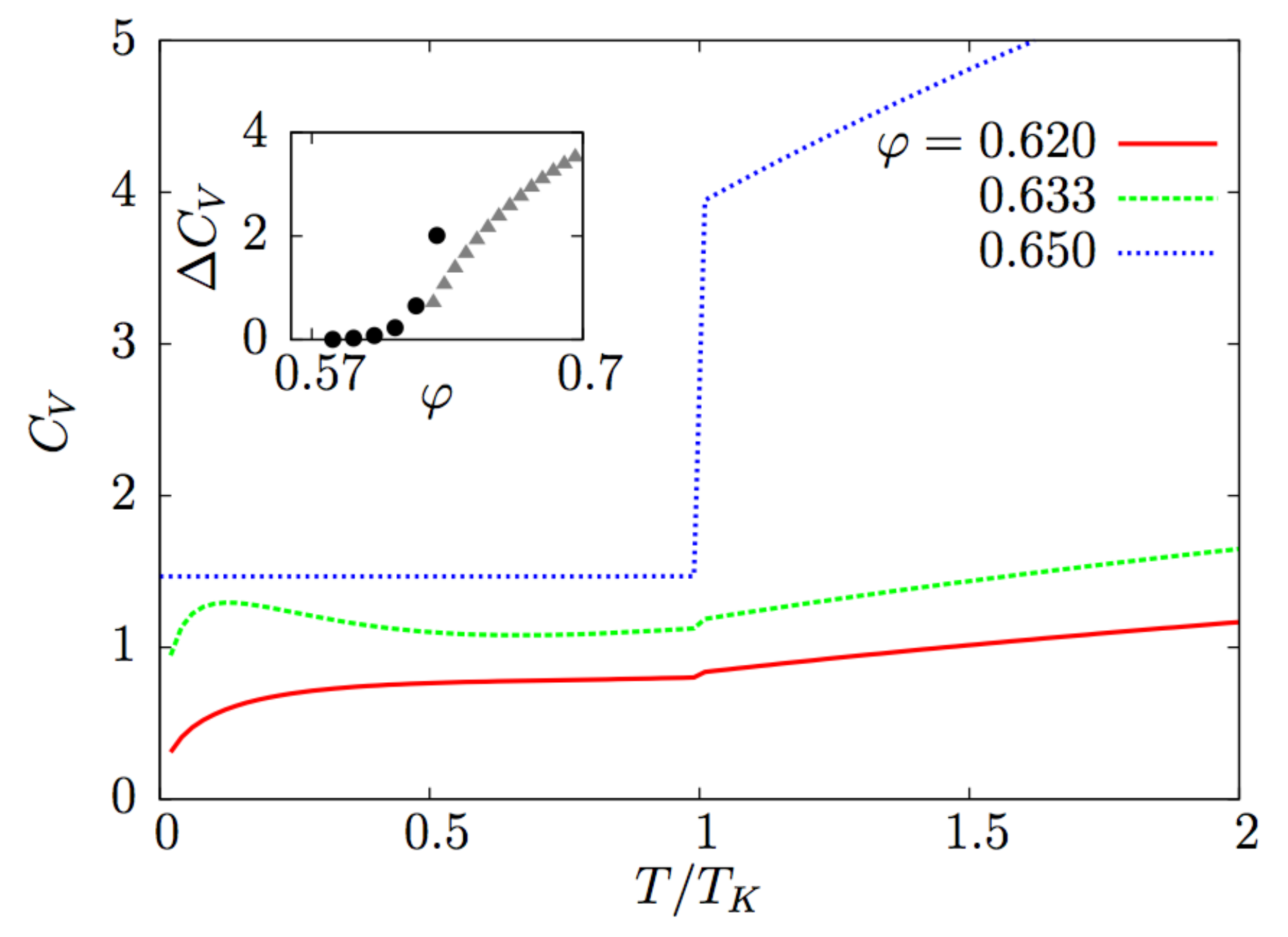}
\caption{Specific heat $C_V$ of the system plotted against $T/T_K$, for several volume fractions.
The curves for $\varphi = 0.620, 0.633$ are obtained with the effective potential method. The curve for 
$\varphi = 0.650$ is obtained with the M\'ezard-Parisi small cage expansion~\cite{MP99b},
since at this density $T_K(\varphi)$ is too high and the effective potential method is not reliable. 
{Inset:}~Specific heat jump at $T_K$; the black dots
are obtained with the effective potential approximation, the gray triangles are obtained with the
M\'ezard-Parisi small cage expansion.
}
\label{fig:cv}
\end{figure}

From the knowledge of $m^*$ and $A^*$, we can deduce the energy and specific heat of the glass. We show in 
Fig.~\ref{fig:cv} the temperature evolution of the 
specific heat for three densities, one below $\varphi_{GCP}$, one very close to it, and one above. Upon crossing the ideal 
glass transition, the specific heat undergoes a finite jump. We find that the amplitude of this jump 
increases continuously with $\varphi$ from $\varphi_K$. 
A qualitatively similar result was obtained in numerical 
simulations~\cite{BW09b}, 
where the jump of specific heat was studied at the 
numerical glass transition temperature. The behavior of the specific heat 
correlates well with the evolution of the thermodynamic fragility discussed 
in Fig.~\ref{fig:angell}.
Finally, we note that the $T \to 0$ limit of the specific heat jumps discontinuously from $0$ to $3/2$ at $\varphi_{GCP}$, 
which reveals that the ground state properties of the glass phase abruptly change at the jamming transition, as
we now study in more detail.

\section{Jamming point of harmonic spheres}
\label{jamming}

We now turn to the study of the region of the phase diagram deep in the glass
phase, inside the line $T_K(\varphi)$ and close to the jamming point $\varphi_{GCP}$, see Fig.~\ref{fig:PDsch}.
In this region, $m^*(T,\varphi)$ is very small, as we discussed in the last section (see Fig.~\ref{fig:mstar}). In principle, one 
could just compute $m^*(T,\varphi)$ and then take the limit $T\to 0$ and $\varphi \to \varphi_{GCP}$ (``jamming limit''). 
However, both numerically and analytically, it is much more convenient to exchange the optimization with respect to $m$ 
and $A$ with the jamming limit, take the latter first, and then optimize the free energy. This is because, in the jamming 
limit, many of the integrals that appear in the free energy simplify considerably. However, the scaling of $m$ and $A$ in 
the jamming limit is different depending on the order of the limits, which emphasizes the asymmetry that exists between 
both sides of the jamming point. In this section we discuss how to exchange the jamming limit with the optimization 
procedure.

\subsection{Zero temperature, below close packing}
\label{sec:IVA}

We consider first the limit $T\to 0$ for fixed $\varphi < \varphi_{GCP}$. In this case, we are bound to recover the results of 
\cite{PZ10} for hard spheres. We reproduce here these results, in order to fix the notations and for the sake of 
completeness. This is easily seen by considering the $T \to 0$ limit of the expression of the function $q$:
\beq
q(A,T;r) \underset{T \to 0}{\to} q(A;r) = \frac 12 \left( 2 + \erf \left[ \frac{r-1}{2\sqrt{A}} \right] - \erf \left[ \frac{r+1}{2\sqrt{A}} \right] \right) 
+ \sqrt{\frac{A}{\p}} \frac 1r \left( \frac{}{}\! e^{-\frac{(r-1)^2}{4A}} - e^{-\frac{(r+1)^2}{4A}} \right)  \ , 
\eeq
which recovers the result found in Parisi and Zamponi for the hard-sphere case.
We expect the optimal number of replicas $m^*(T,\varphi)$  and cage radius 
$A^*(T,\varphi)$ to tend to finite values $m^*_{HS}(\varphi)$ and $A^*_{HS}(\varphi)$ when $T \to 0$. Therefore, to 
recover hard spheres results, one has to take the limit $T \to 0$ at fixed $A$ and $m$, and optimize the resulting free 
entropy over $A$ and $m$. 

The corresponding expression for the replicated free entropy coincides with that of 
\cite{PZ10} and reads:
\beq \begin{split}
& \SS(m,A;T,\varphi) \underset{T \to 0}{\to} \SS(m,A;\varphi) = \SS_{\rm{harm}}(m,A) + \SS_{liq}^{HS}(\varphi) + 4 \varphi y_{liq}^{HS}(\varphi) G(m,A) \ , \\
& G(m,A) = 3 \int_0^\io dr ~ r^2 \left[ \frac{}{}\! q(A;r)^m - \theta(r-1) \right] \ .
\end{split} \eeq
We numerically solved the optimization equations; the result for $m^*_{HS}(\varphi)$ is plotted in the inset of 
Fig.~\ref{fig:mstar}. Approaching $\varphi_{GCP}$ by the left, $m^*_{HS}$ goes to zero linearly:
\beq\label{eq:mstHS}
m^*_{HS} \underset{\varphi \to \varphi_{GCP}}{\propto} (\varphi_{GCP} - \varphi),
\eeq
From the inset of Fig.~\ref{fig:mstar} we find the value of the glass close packing to be:
\begin{equation}
\varphi_{GCP} = 0.633353, 
\label{obtained_phiGCP}
\end{equation}
recovering the result of \cite{PZ10}. 

We also find that $A^*_{HS}$ also goes to zero with $m$ as $\a m$. Thus the close packing limit for hard spheres can be computed by 
taking first the limit $T \to 0$, and then $m \to 0$ with $A = \a m$. Optimization on $A$ will now become an optimization on 
$\a$. 
The $q$ function, when $A=\a m$ and $m \to 0$ has for limits:
\beq
q(m\a;r) \underset{m \to 0}{\to} \left\{ \begin{array}{ll} 0 & \quad r<1 \ , \\ 1 & \quad r \ge 1 \end{array} \right.
\eeq
Going to next order by resorting to the asymptotic expansion of the error function, and neglecting subleading terms, we obtain:
\beq
q(m \a;r) \underset{m \to 0}{\sim} \theta(r-1) + \sqrt{\frac{A}{\p}} e^{-\frac{(r-1)^2}{4A}} \left[ \frac 1r + \frac 1{1-r} 
R \left( \frac{|1-r|}{\sqrt{A}} \right) \right] \ .
\eeq
From this we can obtain the expansion of $G$ in powers of $m$:
\beq \begin{split}
& G(m,\a m) \underset{m \to 0}{\sim} m^{m/2} G_0^{HS}(\a) \left( 1 + R_1 m + R_2 m^2 + \cdots \right) + S_1 m^2 + S_2 m^{5/2} + \cdots \ , \\
& G_0^{HS}(\a) = 3 \int_0^1 dr ~ r^2 e^{-\frac{(r-1)^2}{4A}}  \ .
\end{split} \eeq
The first part comes from the $r<1$ part of the integral, and the other, subdominant one comes from the $r>1$ part of the integral.
Finally the free-entropy can be expressed only with analytic terms:
\beq \begin{split}
& \SS_0^{HS}(\a;\varphi) \equiv \lim_{m \to 0} \SS(m,A=\a m;\varphi) = - \frac 32 \left[ \frac{}{}\! \ln (2 \p \a) +1 \right]
+ \SS_{liq}^{HS}(\varphi) + 4 \varphi y^{HS}_{liq}(\varphi) G_0^{HS}(\a) \ , \\
& G_0^{HS}(\a) = 3 \left[ \frac{}{}\! \sqrt{\p \a} (1+2\a) ~ \erf \left( \frac{1}{2 \sqrt{\a}} \right) + 2 \a ~ e^{-\frac 1{4\a}} - 4 \a \right] \ .
\end{split}\eeq
The optimization over $\a$ gives a self consistent equation for the optimal value $\a^*_{HS}$:
\beq \begin{split}
& J_0^{HS}(\a^*_{HS}(\varphi)) = \frac{3}{8 \varphi y_{liq}^{HS}(\varphi)} \ , \\
& J_0^{HS}(\a) = \a \frac{d G_0^{HS}(\a)}{d \a} \ .
\end{split} \eeq
To find the complexity, we must go the the leading order in the small $m$ expansion 
(since we need to take a derivative with respect to $m$). This reads:
\beq
\SS(m,\a m;\varphi) = \SS_0^{HS}(\a;\varphi) 
+ \frac 12 \left[ \frac{}{}\! 3 + 4 \varphi y_{liq}^{HS}(\varphi) G_0^{HS}(\a) \right] m \ln m + \OO(m,(m \ln m)^2) \ .
\eeq
We compute the complexity as:
\beq \begin{split}
\Si(m,\a;\varphi) & = -m^2 \frac {d}{dm} \left[ \frac{}{}\! \SS(m,\a ~ m;\varphi)/m \right] \\
& = \SS_0^{HS}(\a;\varphi) - \frac m2 \left[ \frac{}{}\! 3 + 4 \varphi y_{liq}^{HS}(\varphi) G_0^{HS}(\a) \right] \ .
\label{complexity_HS_nearjamming}
\end{split}\eeq
Taking the $m \to 0$ limit, and evaluating this at the optimal value of $\a$, we get the complexity of hard spheres at $m=0$:
\beq
\Si_0^{HS}(\varphi) = \SS_0^{HS}(\a^*_{HS}(\varphi);\varphi) \ .
\eeq
This complexity vanishes linearly at $\varphi_{GCP}$, signaling the jamming transition.
The optimal value of $\a$ is found to be very small at the transition, which will prove to be convenient in the following:
\beq
\a^*_{HS}(\varphi_{GCP}) = 9.72187 \times 10^{-5} \ . 
\eeq
We can also obtain the leading term of the optimal number of replicas from Eq.~(\ref{complexity_HS_nearjamming})
by noting that $m^*$ is defined as $\Si(m^*) = 0$ leading to:
\beq
m^*_{HS}(\varphi) = \frac{2 \Si_0^{HS}(\varphi)}{3+4\varphi y_{liq}^{HS}(\varphi) G_0^{HS}(\a^*_{HS}(\varphi))} \ .
\eeq
The value of the cavity function of hard spheres at $GCP$ is found to be:
\beq
y_{liq}^{HS}(\varphi_{GCP}) = 23.6238 \ ,
\eeq
and the value of $G_0^{HS}$ at the transition is:
\beq
G_0^{HS}(\a^*_{HS}(\varphi_{GCP})) = 0.0170908 \ .
\eeq
Gathering all this we can find the linear behavior of $m^*_{HS}$ near the transition:
\beq \begin{split}
& m^*_{HS}(\varphi) \underset{\varphi \to \varphi_{GCP}}{\sim} \tilde{\m} (\varphi_{GCP}-\varphi) \ , \\
& \tilde{\m} = 20.7487 \ .
\end{split} \eeq
We can also get the scaling of the reduced pressure for $\varphi \to \varphi_{GCP}^-$ by using the expression of the pressure given in 
Eq.~(\ref{pressure_glass}):
\beq
p_{glass} \underset{\varphi \to \varphi_{GCP}}{\sim} \frac{3.03430 \, \varphi_{GCP}}{\varphi_{GCP} - \varphi}.
\eeq
The prefactor is only $1\%$ different from the correct value which is the space dimension $d=3$, predicted by free 
volume theory~\cite{SW62} and by the small cage expansion of Ref.~\cite{PZ10}.

The pressure of the glass is thus found to diverge at the glass close packing point, which is consistent with our postulate that 
the glass close packing is indeed identified, in our mean-field picture, with the jamming transition.

\vspace{1cm}

To sum up, we have detected the glass close packing point, at a density consistent with the numerically observed values.
Upon approaching this point from lower densities at zero temperature, harmonic spheres behave like
hard-spheres, that are in an ideal glassy state, whose reduced pressure is found to diverge at the glass close packing 
point. By studying this point for higher densities, we will confirm that this point is a critical jamming point, as defined in the 
numerical simulations \cite{OSLN03}.

\subsection{Zero temperature, above close packing}
\label{sec:IVB}

For $\varphi > \varphi_{GCP}$, the harmonic spheres have a finite energy at $T = 0$, due to overlaps. Then, because of 
the repulsion the spheres are jammed, and $A^*$ vanishes at $T=0$. It turns out from the numerical calculation of $m^*$ 
and $A^*$, both in the small cage expansion \cite{MP99b} and in the potential approximation (see the previous section) 
that the optimal number of replicas $m^*$ tends to zero when $T \to 0$. One finds that $m = T/\t$ and $A = m \a$ with 
constants $\a$ and $\t$. Therefore for $\varphi > \varphi_{GCP}$ one has to take the $T\to 0$ limit with $\a = A/m$ and 
$\t=T/m$ kept constant. Optimization over $m$ and $A$ is replaced by an optimization over $\a$ and $\t$.

The replicated free entropy simplifies considerably in this limit. We first observe that the $G$ function becomes in the double limit
$A=\a m$ and $T=m \t$:
\beq
G(m,\a ~ m; m ~\t) \underset{m \to 0}{\sim} G_0(\a,\t) = G_0^{HS}\left( \a + \frac{\t}4 \right) - G_0^{HS}\left( \frac \t 4 \right) \ .
\eeq
The free-entropy thus becomes:
\beq
\SS(m,\a ~ m;m ~ \t,\varphi) \underset{m \to 0}{\to} \SS_0(\a,\t;\varphi) = -\frac 32 \left[ \frac{}{}\! \ln(2 \p \a) + 1 \right] + \SS_{liq}(\t,\varphi) 
+ 4 \varphi y_{liq}^{HS}(\varphi) G_0(\a,\t) \ .
\eeq
Optimization over $A$ is again replaced by an optimization over $\a$ by a very similar self-consistent equation than in the HS case:
\beq
 \begin{split}
& J_0^{HS}(\a^*(\t,\varphi)) = \frac{3}{8 \varphi y_{liq}^{HS}(\varphi)} \ , \\
& J_0^{HS}(\a) = \a \frac{d G_0(\a,\t)}{d \a} \ .
\end{split} \eeq
Taking the $\t \to 0$ limit, we see that we recover the hard-sphere results. Thus the jamming limit 
from above ($\varphi > \varphi_{GCP}$) corresponds to sending $\t$ to zero in the previous equations. 
Expanding the equations around $\t = 0$ allows the 
derivation of the scaling behavior of the thermodynamic quantities above the jamming point \cite{BJZ11}.

Expanding the free-entropy we find:
\beq \begin{split}
& \SS_0(\a^*(\t,\varphi),\t;\varphi) = \SS_0^{HS}(\a;\varphi) + \t \SS_1(\a,\varphi) + \OO(\t^2) \ , \\
& \SS_1(\a,\varphi) = \varphi y^{HS}_{liq}(\varphi) \left[ \frac 1\a J_0^{HS}(\a) - 6 \right] \ .
\end{split} \eeq
Evaluating this at the optimal value of $\a$, we get at lowest order in $\t$:
\beq
\SS_0(\t,\varphi) \equiv \SS_0(\a^*(\t,\varphi),\t;\varphi) = \SS_0^{HS}(\a^*_{HS}(\varphi);\varphi) + \t \SS_1(\a^*_{HS}(\varphi),\varphi) + \OO(\t^2) \ .
\eeq

From this we can deduce the complexity and energy of the states by coming back to the reasonings that lead to the expression
of the complexity deduced from the replicated free-energy. We know that:
\beq
\SS(m;T,\varphi) = \Si(f^*(m;T,\varphi)) - m \b f^*(m;T,\varphi) \ .
\eeq
But in the $T \to 0$ limit, the free energy of the states is just their internal energy $e$, thus in the limit where $T=m \t$ we get:
\beq
\SS(m=T/\t;T,\varphi) \underset{T \to 0}{\to} \SS_0(\t,\varphi) = \Si_0(e^*) - e^* / \t \ .
\eeq
From the same reasoning than in Eq.~(\ref{def_ssglass_replicated}) we get that:
\beq 
\left\{ \begin{array}{ll}
& \displaystyle \Si_0(\t,\varphi) = \frac{\partial \left[ \t \SS_0(\t,\varphi) \right]}{\partial \t} \ , \\
& \\
& \displaystyle e(\t,\varphi) = \t^2 \frac{\partial \SS_0(\t,\varphi)}{\partial \t} \ .
\end{array} \right.
\eeq
These relations give us the parametric equation determining the complexity and the energy of the states:
\beq
\left\{ \begin{array}{ll}
& \Si_0(\t,\varphi) = \Si_0^{HS}(\varphi) + \t \SS_1(\varphi) \ , \\
& \\
& e(\t,\varphi) = \t^2 \SS_1(\varphi) \ .
\end{array} \right.
\eeq
This allows to obtain the behavior of the complexity as a function of the energy:
\beq
\Si_0(e,\varphi) = \Si_0^{HS}(\varphi) + 2 \sqrt{e \SS_1(\varphi)} \ .
\eeq
Since $\SS_1$ is finite at $\varphi_{GCP}$ and $\Si_0^{HS}$ vanishes lineraly, we obtain
a quadratic dependance to the ground state energy on $\varphi-\varphi_{GCP}$.
We find:
\beq \begin{split}
& \Si_0^{HS}(\varphi) = 62.9577(\varphi_{GCP}-\varphi) \ , \\
& \SS_1(\varphi_{GCP}) = 3767.51 \ , \\ 
\end{split} \eeq
Using these values we find the scaling of the optimal effective temperature $\t^*$ to scale as:
\beq \begin{split}
& \t^*(\varphi) \underset{\varphi \to \varphi_{GCP}}{\sim} \tilde{\t} ( \varphi-\varphi_{GCP}) \ , \\
& \tilde{\t} = 0.00835535 \ ,
\end{split} \eeq
and the energy of the ground state as:
\beq
e_{GS}(\varphi) \underset{\varphi \to \varphi_{GCP}}{\sim} 0.263017 (\varphi - \varphi_{GCP})^2 \ .
\label{eq:tauGCP} 
\eeq
By a similar procedure, the pressure evaluated at $\varphi_{GCP}$ gives the following scaling:
\begin{equation}
P_{glass}(T=0,\varphi) \underset{\varphi \to \varphi_{GCP}}{\sim} 0.403001 (\varphi-\varphi_{GCP}) \ . 
\label{scalp}
\end{equation}
Clearly the same scaling could have been obtained from the exact relation $P_{glass}(T=0,\varphi) = \frac{6 \varphi^2}\pi \frac{de_{GS}}{d\varphi}$.

\begin{figure}[htb]
\centering
\includegraphics[width=8cm]{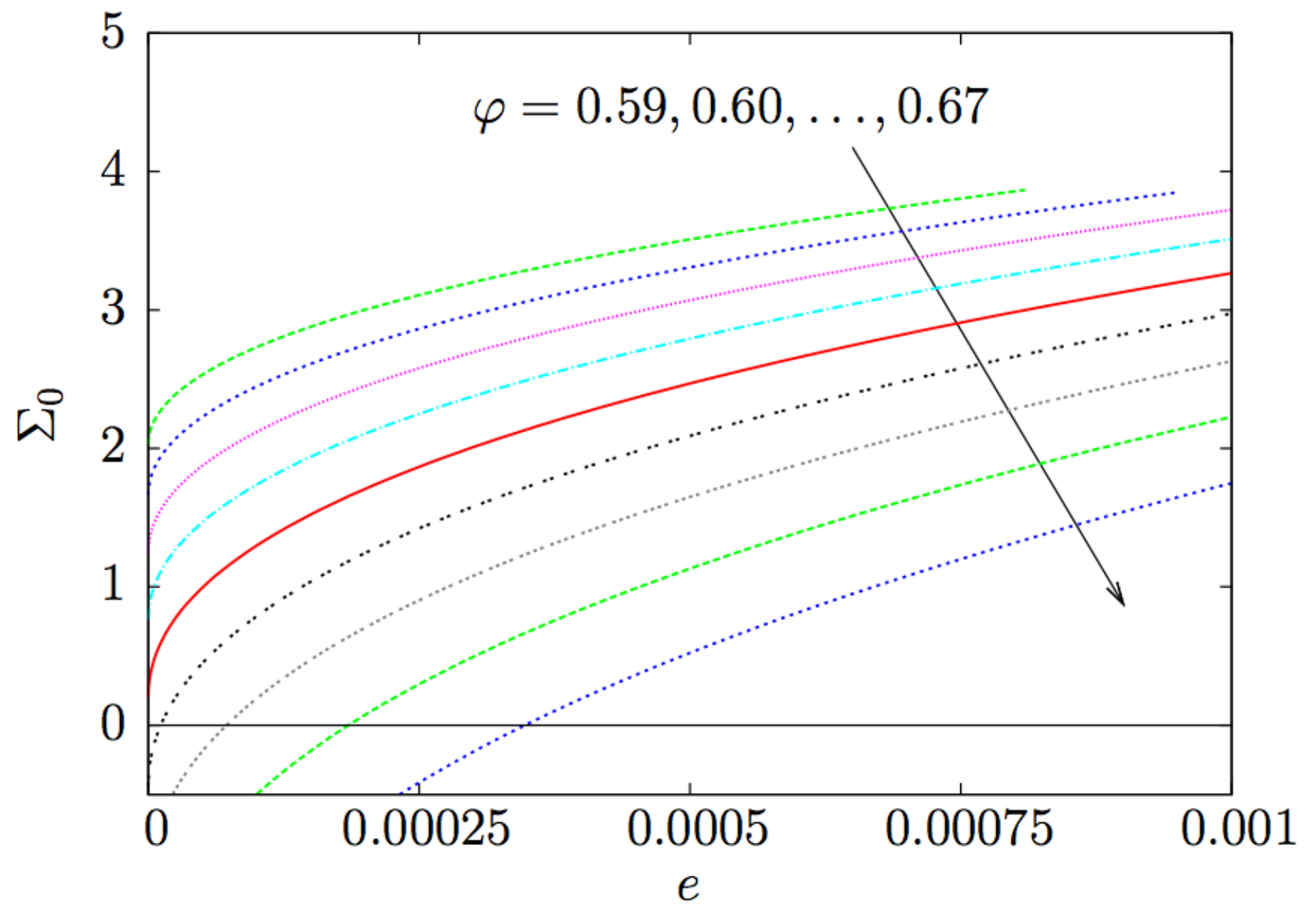}
\caption{
Complexity of the energy minima $\Si_0(e,\varphi)$ as a function of energy for several densities.
}
\label{fig:sigmae}
\end{figure}

We numerically obtain the behavior of the complexity as a function of the energy of the packing, and this is reported in 
Fig.~\ref{fig:sigmae}.

\vspace{1cm}

In this section we have thus derived the scaling 
laws (\ref{eq:tauGCP}) and (\ref{scalp}) first observed numerically
above the jamming transition at zero temperature~\cite{OSLN03}, which 
indicate that solidity emerges continuously at the jamming density. 
These results also confirm the postulated correspondence between the jamming transition observed numerically and the 
glass close packing density defined within our theoretical approach.

\subsection{Scaling around jamming}

We have shown in Secs.~\ref{sec:IVA} and \ref{sec:IVB} 
that all the thermodynamic quantities are singular at $\varphi_{GCP}$
and $T=0$;
for instance $m^*$ is finite below $\varphi_{GCP}$ while it vanishes proportionally to $T$ above $\varphi_{GCP}$.
The reduced pressure is also finite below $\varphi_{GCP}$, while it diverges at $\varphi_{GCP}$ and is formally infinite
above $\varphi_{GCP}$.
Both the energy and the pressure vanish below $\varphi_{GCP}$ while they are finite above $\varphi_{GCP}$.

From this observation, and since all quantities are analytic at finite $T$, 
it follows that all these quantities must satisfy scaling relations if $T$ is small
enough and $\varphi$ is close enough to $\varphi_{GCP}$~\cite{OT07}. 
We discuss explicitly the case of $m^*$ for which we
assume a scaling relation of the form
\beq
m^*(T,\varphi) = T^\g \widetilde m_\pm\left( \frac{|\varphi - \varphi_{GCP}|}{T^\n} \right) \ ,
\eeq
where the two scaling functions correspond to the two sides of the transition.
In the hard sphere limit $T\to 0$ and $\varphi < \varphi_{GCP}$, to recover Eq.~(\ref{eq:mstHS})
we need $\wt m_-(x \to \io) = \wt \mu x$, where $\wt \mu$ is a constant, and $\g = \n$. For $\varphi > \varphi_{GCP}$ instead,
$m^* = T/\t^*(\varphi)$ and to recover Eq.~(\ref{eq:tauGCP}) we need $\wt m_+(x \to \io) = 1/(\wt\t x)$, where $\wt\t$ is a 
constant, and $\g = \n = 1/2$.
Finally, we are led to the scaling
\begin{equation}
\left\{ \begin{array}{ll}
& \displaystyle m^*(T,\varphi) = \sqrt{T} \, \widetilde m_\pm\left( \frac{|\varphi - \varphi_{GCP}|}{\sqrt{T}} \right) \ , \\
& \displaystyle \wt m_-(x \to \io) = \wt \mu x \ , \\
& \displaystyle \wt m_+(x \to \io) = \frac{1}{\wt\t x} \ .
\end{array} \right.
\label{eq:mstarscaling}
\eeq
This scaling is successfully tested in Fig.~\ref{fig:scaling_m}.

\begin{figure}
\centering
\includegraphics[width=8cm]{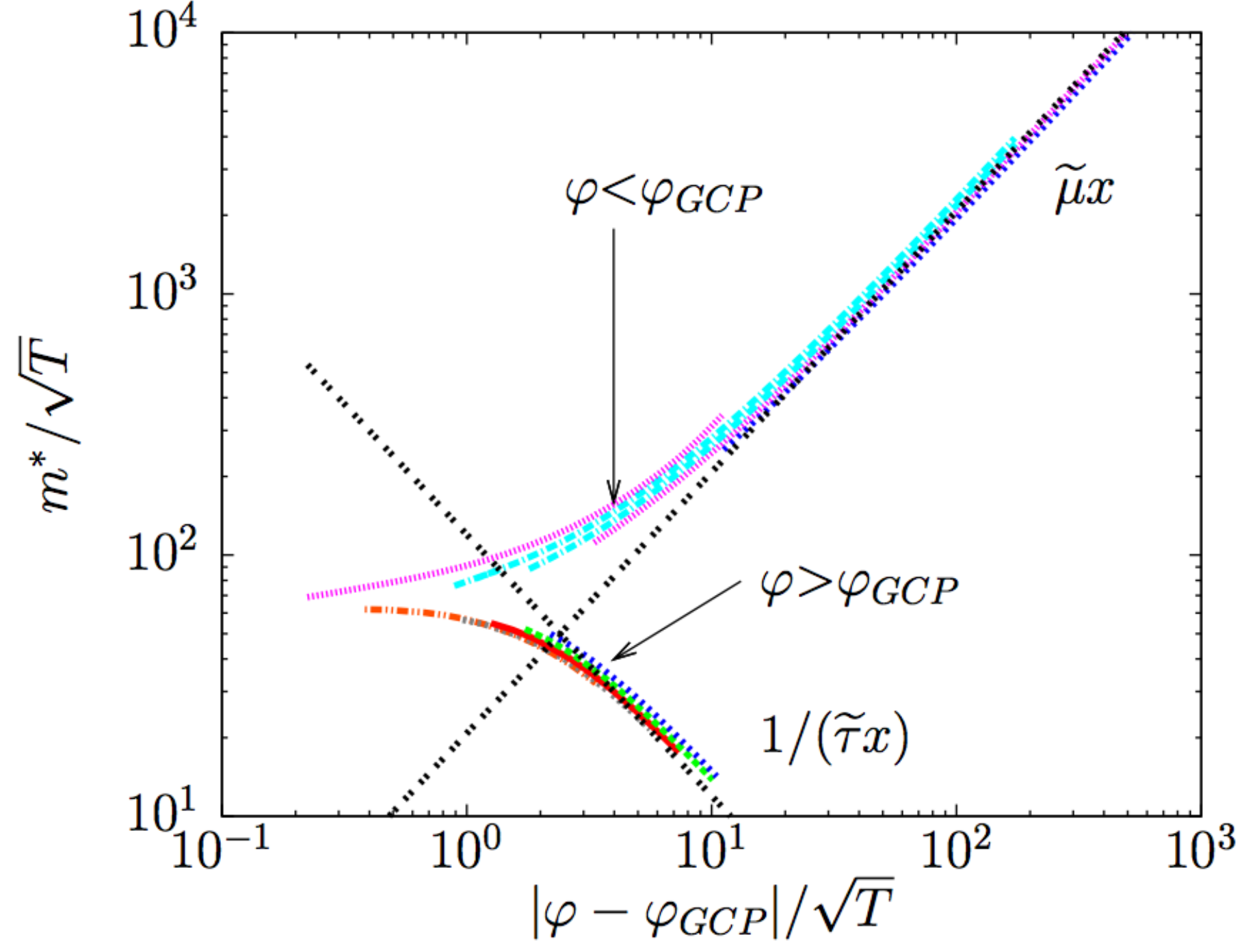}
\caption{Scaling of $m^*(T,\varphi)/\sqrt{T}$ as a function of the rescaled temperature 
$|\varphi-\varphi_{GCP}|/\sqrt{T}$ for $\varphi=0.58 , 0.59 , \ldots , 0.70$. The asymptotic 
forms corresponding to Eq.~(\ref{eq:mstarscaling}) are also reported.
}
\label{fig:scaling_m}
\end{figure}

Qualitatively similar scaling forms (with different exponents) 
apply to thermodynamic quantities such 
as the energy, the pressure, or the complexity. 
Although we do not show the corresponding scaling plots explicitly, 
the energy and the pressure are plotted and
compared with numerical data in the following (Fig.~\ref{fig:ep}).
Note finally that a scaling plot for the complexity 
is related, in the spirit of the Adam-Gibbs relation in 
Eq.~(\ref{adam-gibbs}), to the scaling plot of the 
numerically measured relaxation time 
reported in Ref.~\cite{BW09a}.

\section{The pair correlation function}
\label{correlation}

The pair correlation function is a key quantity that gives a lot of insight into the structure of dense fluids or packings.
In this section we derive an expression for the correlation function of the glass and use it to discuss the scaling
of the contact peak of $g(r)$ close to the jamming transition.

\subsection{General expression of the correlation function of the glass}

It has been shown in \cite[Eqs.(67)-(68)]{PZ10} that one can compute the correlation function of the glass starting from the replicated free energy. 
This is done by adding a small perturbation $u(r)$ to the potential of replica 1 and taking the derivative of the free energy with respect to $u(r)$ 
and finally setting $u$ to zero. The glass correlation function is then calculated as:
\begin{align}
g_{glass}(r) = \frac 2 {\r} \left. \frac{\d \SS}{\d u(r)} \right|_{u=0}.
\end{align}
 We start from Eq.~(\ref{S_of_Q}) and we use approximation (\ref{y_approx_rep}). Since the perturbation acts only on replica 1, we have
\begin{align}
\SS(m,A; & T,\varphi) = \SS_{harm}(m,A) + \SS_{liq}\left[\varphi, e^{m w(r) + u(r)} \right] + \frac{3 \varphi}\pi y^{HS}_{liq}
(\varphi) \int d^3r e^{m w(r) + u(r)}  Q(r),
\end{align}
and $Q(r)$ depends only on the potentials of replicas 2 to $m$, so it is independent of $u$. Taking the derivative with respect to $u(r)$ we have, using again Eq.~(\ref{y_approx_rep}):
\begin{align}
g_{glass}(r)&  = g_{liq}(T/m,\varphi; r) +  y^{HS}_{liq}(\varphi) e^{m w(r)} Q(r) \label{gGlass} \\
& =   y^{HS}_{liq}(\varphi) e^{m w(r)} [1 + Q(r)]  \nonumber \\
& = y^{HS}_{liq}(\varphi) e^{w_{\rm{eff}}(r)} . \nonumber
\end{align}
Therefore the correlation function of the glass is directly related to 
the effective potential within the low temperature 
approximation we used to compute the thermodynamics of the system. Of course, an improved expression for this correlation could be obtained by using 
the HNC approximation to describe the effective liquid. In that case, $g_{glass}(r)$ would be the HNC correlation of the effective liquid. Still, the much 
simpler expression (\ref{gGlass}) is enough to capture the scaling of $g_{glass}(r)$ around jamming, at least close to the first peak. 
In the rest of this section we discuss the scaling form that is obtained starting from Eq.~(\ref{gGlass}).

A very interesting quantity related to $g(r)$ is the number of contacts, defined as the number overlaps per particle.
Within our low temperature approximation it is given by:
\beq
z(T,\varphi) = 24 \varphi \int_0^1 dr \, r^2 g_{glass}(r)  = 24 \varphi y^{HS}_{liq}(\varphi) \int_0^1 dr \, r^2 e^{w_{\rm{eff}}(r)}  \ .
\label{eq:zdef}
\eeq
This expression simplifies into:
\beq
z(T,\varphi) = 24 \varphi y_{liq}^{HS}(\varphi) \int_0^\io du ~ u^2 q(A,T;u)^{m-1} \widehat{q}(A,T;u) \ ,
\eeq
where $\hat{q}$ is the part of the function $q$ that vanishes in the $T=0$ limit:
\beq \begin{split}
\widehat{q}(A,T;r) = & ~ q(A,T;r) - \lim_{T \to 0} q(A,T;r) \ , \\
= & ~ - \sqrt{\frac{A}{\p}} \frac{1}{r(1+4 A \b)} \left( \frac{}{}\! e^{-\frac{(r-1)^2}{4A}} - e^{-\frac{(r+1)^2}{4A}} \right) \\
& ~ + e^{-\frac{(r-1)^2 \b}{1+4 A \b}} \frac{r+4A\b}{2r(1+4A\b)^{3/2}} \left( \erf \left[ \frac{r+4A\b}{2 \sqrt{A(1+4A\b)}} \right]
+ \erf \left[ \frac{1-r}{2 \sqrt{A(1+4A\b)}} \right] \right) \\
& ~ - e^{-\frac{(r+1)^2 \b}{1+4 A \b}} \frac{r-4A\b}{2r(1+4A\b)^{3/2}} \left( \erf \left[ \frac{r-4A\b}{2 \sqrt{A(1+4A\b)}} \right]
- \erf \left[ \frac{1+r}{2 \sqrt{A(1+4A\b)}}\right] \right) \ .
\end{split} \eeq
Similarly the energy of the glass can be computed as:
\beq
U_{glass}(T,\varphi) = 12 \varphi y_{liq}^{HS}(\varphi) \int_0^\io du ~ u^2 q(A,T;u)^{m-1} \left( - \frac{\partial \widehat{q}(A,T;u)}{\partial \b} \right) \ .
\eeq

The starting point of the analysis of the correlation function of the glass are Eq.~(\ref{gGlass}) and
the expression of the effective potential.

\subsection{Scaling of the peak of the pair correlation at finite temperature}

Here we discuss the effect of thermal fluctuations on 
the maximum of $g_{glass}(r)$ near the 
the $T=0$ jamming transition at $\varphi_{GCP}$. This situation was studied 
numerically and experimentally in Refs.~\cite{ZXCYAAHLNY09,JB10}. 
We focus on the scaling of this maximum 
in a region of small $T$ and for $\varphi \sim \varphi_{GCP}$.
In Fig.~\ref{fig:gmax} we show the behavior of the maximum 
of $g_{glass}(r)$, which we call $g_{max}$, 
as a function of the density, for temperatures ranging from $10^{-5}$ to $0$. 
This figure demonstrates that 
we are able to derive analytically all behaviors 
reported in Refs.~\cite{ZXCYAAHLNY09,Ch10a,DTS05a,SLN06}. 
Namely, 
the density at which the pair correlation function reaches its 
maximum shifts towards higher values when the temperature departs 
from $0$ and increases as $\sqrt{T}$, 
while the maximum $g_{max}$ diverges as $|\varphi - \varphi_{GCP}|^{-1}$
on both sides of the transition.

\begin{figure}
\centering
\includegraphics[width=8cm]{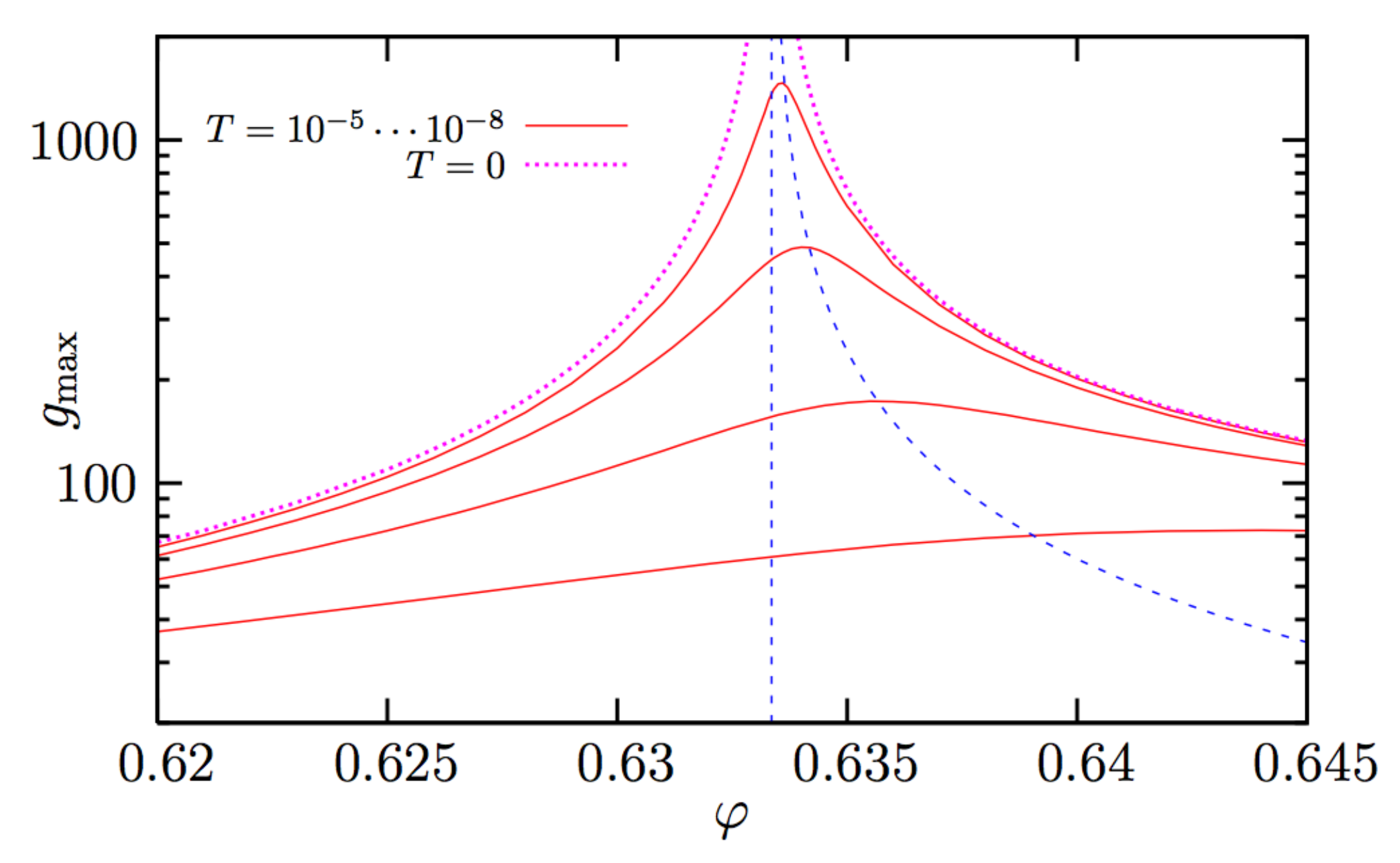}
\caption{Evolution of the maximum of the glass pair correlation function with $T$ and $\varphi$. While $g_{max}$ diverges on both sides of the transition at $T = 0$ as $g_{max} \sim |\varphi - \varphi_{GCP}|^{-1}$, this divergence becomes a smooth maximum at finite $T$ near the transition, whose position shifts with temperature. This behavior compares very well with numerical \cite{ZXCYAAHLNY09,DTS05a,SLN06} and experimental observations \cite{ZXCYAAHLNY09,Ch10a}.
}
\label{fig:gmax}
\end{figure}

As discussed in the previous section, the zero temperature limit is 
very different for $\varphi < \varphi_{GCP}$ 
or $\varphi > \varphi_{GCP}$. Thus we now discuss these two limits separately.

\subsection{Zero temperature below jamming: hard spheres}

As explained above, we send $T \to 0$ before taking the jamming limit $m \to 0$ and $A = \a m$.
By studying the behavior of the effective potential in this limit (starting from the asymptotic behavior of $q$), 
we find that it diverges at $r = 1^+$:
\begin{align}
e^{w_{\rm{eff}}(r)} & =  \frac{1}{m} \D_\a \left( \frac{r-1}{m \sqrt{4 \a}} \right) \ ,
\end{align}
where the scaling function $\D$ is given by:
\begin{equation}
\D_\a(\lam) = 2 \int_0^{1/\sqrt{4 \a}} dx \, x \, (1- x \sqrt{4 \a})^2 \, e^{-  2 \lam x - x^2 } .
\label{DeltaExact}
\end{equation}
For $\varphi \to \varphi_{GCP}^-$ we found that $m \sim \wt \mu \, (\varphi_{GCP} - \varphi)$ and the optimal cage size
\mbox{$A^*=\a^* m$} with a finite $\a^* = \a^*(\varphi_{GCP})$, we thus find the following scaling for the pair
correlation function:
\begin{equation}
\left\{ \begin{array}{ll}
& \displaystyle g_{glass}(r) \, (\varphi_{GCP} - \varphi) =  f\left[ \frac{r-1}{\varphi_{GCP} - \varphi} \right]   , \\
& \\
& \displaystyle f(\lam) =  \frac{y^{HS}_{liq}(\varphi)}{\wt\mu} \D_{\a^*}\left[ \frac{\lam}{\wt \m \sqrt{4 \a^* }  }   \right]  ,
\end{array} \right.
\label{eq:HSscaling}
\end{equation}
The height of the peak scales as:
\beq
g_{glass}(r=1^+,\varphi \to \varphi_{GCP}) \sim \frac{1.09922}{\varphi_{GCP} - \varphi} ,
\eeq 
and its width as $\varphi_{GCP} - \varphi$.

The number of contacts is the integral of the peak at $r = 1^+$ which is given by
\begin{align}
z & \sim 24 \varphi \int_{peak} dr \, g_{glass}(r) \nonumber \\
& \sim 24 \varphi y^{HS}_{liq}(\varphi) \sqrt{4 \a^* } \int_0^\io d\lam \, \D_{\a^*}(\lam)   \label{eq:zHS}
\end{align}
Using the actual values derived in the previous section, we get \mbox{$z = 6.13720$} for 
$\varphi=\varphi_{GCP}$, which is slightly above the expected $z=2d=6$. 

Note that it has been shown in \cite{PZ10} that within a systematic expansion in $\sqrt{\a}$, one gets $z=2d$ at first order. Here instead, in order to match with the soft sphere computation, we are using an approximation scheme which is not a consistent expansion in $\sqrt{\a}$. This could explain the fact that $z \neq 2 d$. It would be interesting to have a full computation of the next term in the small $\a$ expansion to check whether the equality $z=2 d$ survives at this order. Unfortunately this seems to be a quite hard task.

\subsection{Zero temperature above jamming}

For $\varphi > \varphi_{GCP}$, the appropriate limit is to send $T\to 0$ with $m = T/\t$ and $A = \a m$. 
We obtain for the effective potential the following leading behavior:
\begin{align}
e^{w_{eff}(r)} = \theta(r-1) + \theta(1-r) \theta \left( r - \frac{4 \a}{\t+4\a} \right)  \frac{(1+4\a/\t)[1+(r-1)(1+4\a/\t)]^2}{r^2}  e^{-\frac{\t+4\a}{\t^2} (r-1)^2}.
\label{phieffGCPabove}
\end{align}
Close to $\varphi_{GCP}$ this function develops a delta peak close to $r=1$ with height $(\varphi-\varphi_{GCP})^{-1}$ and width $\varphi-\varphi_{GCP}$, in the same way than in the hard-sphere case. The main difference we observe with the hard sphere limit is that, despite the fact that both cases give rise to a delta function at $r=1$, the mechanism is different since for hard spheres the support is concentrated on $r=1^+$ while for soft spheres it is concentrated on $r=1^-$. Another interesting observation is that the theory predicts that $g_{glass}(r)$ should vanish exactly for $r < \frac{4 \a}{\t+4\a}$; this means that no pair of spheres can have an overlap larger than $\frac{\t}{\t+4\a}$. Indeed in the limit $\t \to 0$ we recover the $GCP$ at which no overlaps are present. Unfortunately, a direct numerical test of this prediction is impossible: this is because when $1-r = \frac{\t}{\t+4\a}$, the exponential term in Eq.~(\ref{phieffGCPabove}) is equal to $\exp (-1/(\t + 4 \a))$. Since both $\t$ and $\a$ are already extremely small (of the order of $10^{-3}$ at best), $g_{glass}(r)$ is extremely small in the region where we predict it to vanish, and numerically the signal to noise ratio becomes too large.

Again in the limit $\varphi \to \varphi_{GCP}^+$ we obtain the scaling forms
\begin{equation}
\left\{ \begin{array}{ll}
& \displaystyle g_{glass}(r) \, (\varphi-\varphi_{GCP}) =  f\left[ \frac{1-r}{\varphi - \varphi_{GCP}} \right] \ , \\
& \\
& \displaystyle f(\lam) = \frac{y^{HS}_{liq}(\varphi) 4 \a^*}{\wt\t} \left( 1 - \frac{4 \a^* \lam}{\wt\t} \right)^2 e^{- \frac{4 \a^* \lam^2}{\wt\t^2}} \ ,
\end{array} \right.
\label{eq:SSscaling}
\end{equation}
The number of contacts is easily obtained from Eq.~(\ref{eq:zdef}) and Eq.~(\ref{phieffGCPabove}) and we found:
\beq
z(T=0,\varphi) = z(T=0,\varphi_{GCP}) + C (\varphi - \varphi_{GCP}),
\eeq 
where $z(T=0,\varphi_{GCP})$ is $6.13720$, the same value than in the hard-sphere case.
This result is at odds with the numerical finding of a square root behavior~\cite{DTS05a,SLN06}. We comment further on this 
discrepancy,  in Sec.~\ref{conclusion}. 

Note that this value of $z(T=0,\varphi_{GCP})$ is identical to the one obtained on the hard sphere side (as it can be easily 
proven analytically), despite the fact that the two definitions of $z$ are not equivalent. In the soft sphere case we defined $z$ as the number of overlaps. On the hard sphere side, this number is strictly zero, therefore we defined $z$ as the integral of 
the delta peak at $r=1^+$.

\section{Comparison with numerical data}
\label{comparison}

In this section, we compare the prediction of the theory with numerical data, in particular to test the
scaling in temperature and in $\d \varphi = \varphi - \varphi_{GCP}$ around the glass close packing point. 

\subsection{Details of the numerical procedure}

We investigated
a system of $N=8000$ identical harmonic spheres. The system was prepared using the standard procedure of 
Ref.~\cite{OLLN02}. 
We started from a random configuration at high density; we then minimized the
energy, and then reduced slowly the density, at each step minimizing the energy again, until a jammed
configuration of zero energy and volume fraction $\varphi_j$ was found. This jammed configuration corresponds, according
to the mean-field interpretation of~\cite{PZ10}, to the $T=0$, $p \to \io$ limit of a given hard sphere glass,
or equivalently to the $T\to 0$, $P\to 0$ limit of a corresponding soft sphere glass. 
Starting from that configuration, 
we prepared configurations at 
different $T$ and $\varphi$ by using molecular dynamics simulations
as in Ref.~\cite{BW09a} to equilibrate the 
system inside the glass state selected by the initial jammed configuration. 
Since we always used very small $T$ and $\varphi \sim \varphi_j$, the system is not able to escape from that 
glass state, so it always remains in the vicinity of the original 
jammed configuration. This is a crucial requirement to numerically obtain 
scaling behavior near $\varphi_j$ as microscopic rearrangments would 
directly affect the location of $\varphi_j$ in the simulations~\cite{CBS09}.  
We tested the absence of such an effect in our simulations 
by repeating the energy minimization starting from some equilibrated
configurations at different $T$ and $\varphi$, 
and checking that we always found the same value of $\varphi_j$ (within numerical errors).
For the particular series of run described below, 
we estimated $\varphi_j = 0.643152\pm 0.000020$ by fitting the zero 
temperature energy using $e(\varphi) \propto (\varphi-\varphi_j)^2$.

\subsection{Difficulties in the comparison with the theory}

When comparing the theory with the numerical data, we face two difficulties. First of all, the value
of $\varphi_j$ depends on the numerical protocol and on its particular realization we used, 
and therefore cannot be directly compared with $\varphi_{GCP}$.
Note that $\varphi_{GCP}$ should represent an upper bound for $\varphi_j$, whatever the protocol, but the value
$\varphi_{GCP}=0.633353$ we report seems to contradict this statement.
The reason is simply that our theoretical computations are
based on the HNC equation of state for liquid hard spheres, which (as 
discussed above), underestimates
the value of $\varphi_{GCP}$:
a better result is obtained using the Carnahan-Starling equation of state,
which gives $\varphi_{GCP}=0.683$, consistently with the value of $\varphi_j$ found above~\cite{PZ10}. 
To avoid confusion, here we stick to the use of the HNC equation of state;
the theoretical calculations can be easily repeated for any other equation of state, and the scaling
around $\varphi_{GCP}$ is unaffected by this choice. 
Once again we stress that the absolute values
of density reported here should always 
be taken with this caveat in mind.
Since we are interested here mainly in the relative
distance to jamming, which is $\varphi - \varphi_j$ for the numerical data, and $\varphi - \varphi_{GCP}$ for the theory,
the absolute values of density are not crucial for our purposes. 

The second 
difficulty is the following. For hard spheres, the theory predicts that
the reduced pressure diverges as $p^{HS}_{glass} \sim 3 \varphi_{GCP}/(\varphi_{GCP}-\varphi)$, while
$y^{HS}_{glass}(\varphi) \sim 1.1/(\varphi_{GCP}-\varphi)$ for $\varphi \to \varphi_{GCP}^-$.
As one can easily check, these scalings imply that the theory violates the exact thermodynamic relation 
$p^{HS}_{glass} = 1 + 4 \varphi y^{HS}_{glass}(\varphi)$, as already noticed 
in~\cite{PZ10}. In particular, the divergence of $p^{HS}_{glass}$ is correct, while the
coefficient of the divergence of $y^{HS}_{glass}(\varphi)$ is overestimated by a factor
of $\sim 1.46562$. We will comment on the origin of this problem in Sec.~\ref{conclusion}. This error affects the
prefactors in the scalings of $g_{glass}(r)$ and of $z$ around $\varphi_{GCP}$. 
To correct for this error, we rescaled the $\d \varphi$ of the numerical data in such a way that
the coefficient of the divergence of $y^{HS}_{glass}$ is the same as 
the theoretical one. 

To summarize, when comparing numerical data with the theory, 
we consider $\d \varphi = \varphi -\varphi_{GCP}$ for the theory,
and $\d \varphi = 1.46562 (\varphi - \varphi_j)$ for the numerical data such that
the values of $y^{HS}_{glass}(\d \varphi)$ for theory and simulations are 
exactly the same
close to $\d \varphi = 0$. Note that this adjustment is based solely on 
the analysis of the hard sphere side of the jamming 
transition ($\d \varphi<0$), but we find that it allows
to describe the full scaling also for positive $\d \varphi$. 

If the reader is uncomfortable with the above reasonings, 
another way of thinking is that we need to adjust two free parameters
(the position of the transition and the amplitude of the 
rescaling of $\d\varphi$) 
to obtain the best comparison of theory and numerical simulations.
While the location is $\varphi_j$ was always adjusted in previous
work~\cite{DTS05a,SLN06,OLLN02,Vh10}, we need to adjust also one
prefactor as we seek to compare theoretical predictions to simulations 
in absolute values, and not only at the level of leading 
diverging contributions.

\subsection{Energy, pressure, average coordination}

\begin{figure}[t]
\centering
\includegraphics[width=8cm]{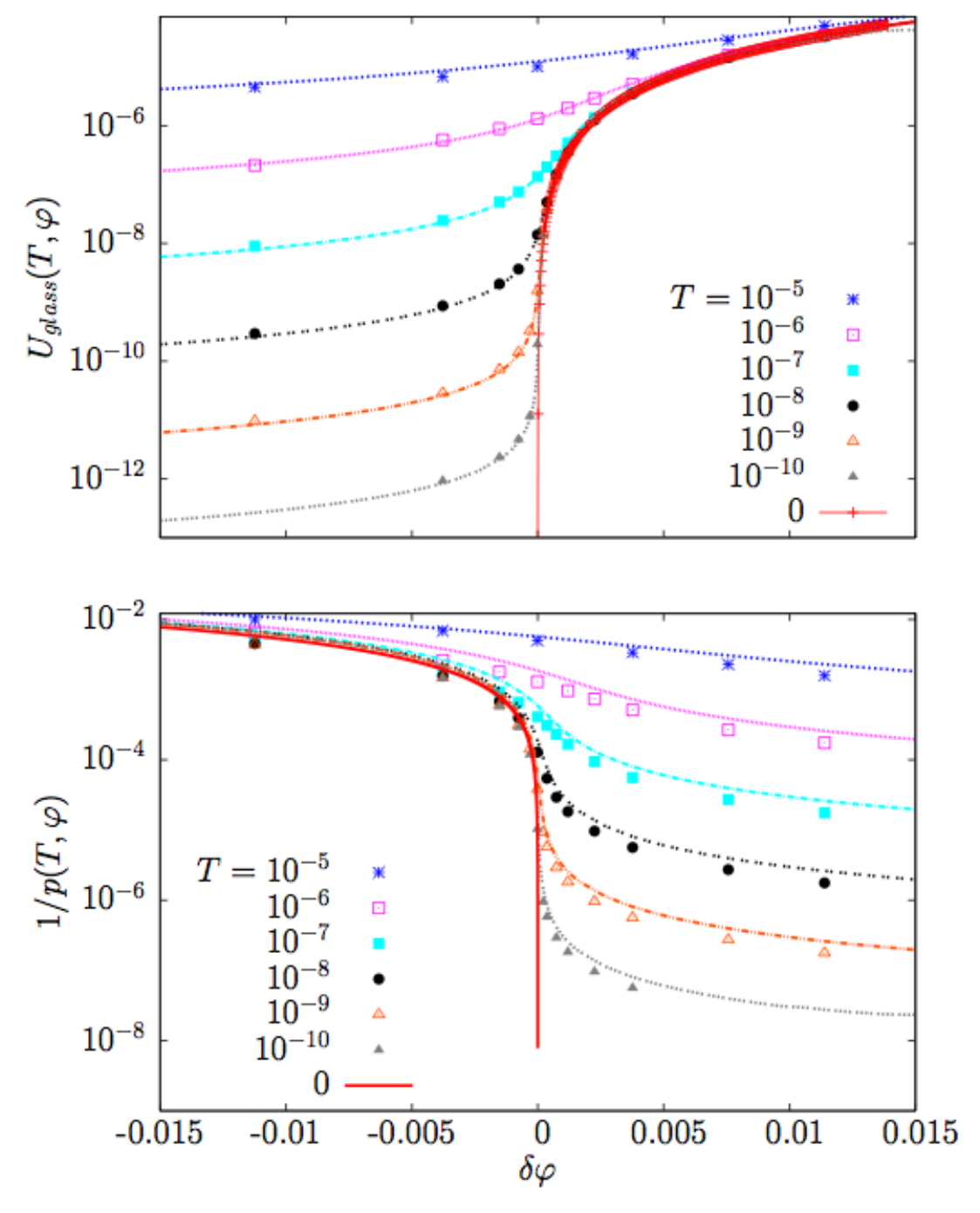}
\caption{
Energy $U_{glass}(T,\varphi)$ (top panel) and inverse reduced pressure $1/p_{glass}(T,\varphi)$ (bottom panel)
as functions of distance from jamming $\d \varphi$, for several temperatures.
We define $\d \varphi = \varphi -\varphi_{GCP}$ for the theory,
and $\d \varphi = 1.46562 (\varphi - \varphi_j)$ for the numerical data.
}
\label{fig:ep}
\end{figure}

We start our discussion by the simplest observables. In Fig.~\ref{fig:ep} we plot the average energy $U_{glass}(T,\varphi)$ 
and the inverse reduced pressure $1/p_{glass}(T,\varphi)$ as functions
of $\d\varphi$ for several temperatures. The agreement between theory and numerical data, with the rescaling of $\d\varphi$ discussed
above, is nearly perfect. The scaling around jamming is clearly visible in the figures. For instance, $U_{glass}(T,\varphi)$ tends
to a finite value for $\d\varphi>0$, while for $\d\varphi < 0$ it goes to zero as a power law, since in this case the system becomes a hard
sphere glass. Similarly, the reduced pressure is finite for $\d\varphi<0$, while it diverges proportionally to $\b$ for $\d\varphi>0$, since
in this case the pressure is finite at zero temperature. 
At finite temperature, the curves interpolate between the two regimes.
Scaling functions similarly to the ones shown in 
Fig.~\ref{fig:scaling_m} for $m^*$ can easily 
be constructed.

\begin{figure}[t]
\centering
\includegraphics[width=8cm]{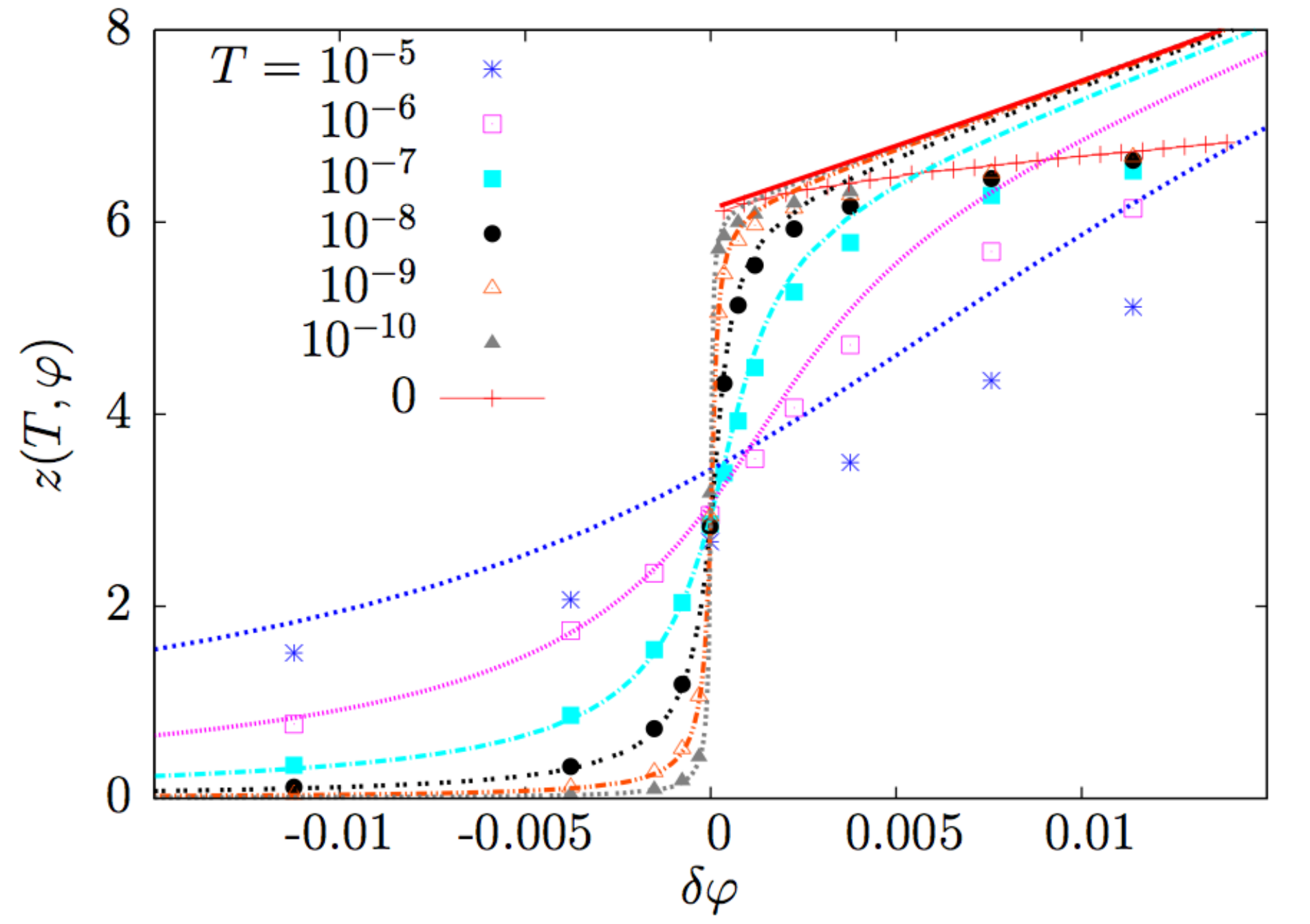}
\caption{
Average number of contacts $z$ as a function of distance from jamming $\d \varphi$, for several temperatures.
We define $\d \varphi = \varphi -\varphi_{GCP}$ for the theory,
and $\d \varphi = 1.46562 (\varphi - \varphi_j)$ for the numerical data.
}
\label{fig:z}
\end{figure}

In Fig.~\ref{fig:z} we report the average coordination number $z(T,\varphi)$, as a function
of $\d\varphi$ for several temperatures. At $T=0$, the average coordination jumps
from 0 to $6$ (for numerical data) or to $6.13720$ (for the theory). 
For $\d\varphi>0$, the average 
coordination grows linearly in $\d\varphi$ for the theory, while it grows 
as $\d\varphi^{1/2}$ for the numerical
data. Therefore, the theory fails to capture the correct
evolution of this structural property at $\d\varphi>0$ and $T=0$. 

A more
detailed description of what happens is obtained by looking to the data 
at finite $T$.
When temperatures become too large, e.g. at $T=10^{-5}$, 
the theory eventually fails because of the low-$T$ approximation we 
made on the liquid theory in
Eq.~(\ref{y_approx_rep}). 
For any fixed temperature $T \leq 10^{-6}$, we observe that the 
theory describes perfectly the numerical data for $\d\varphi < 0$ and also 
for positive $\d\varphi$ up to some crossover value  $\d\varphi_h(T)$. 
Around $\d\varphi_h(T)$ the
theory starts deviating from the numerical data. 
The data in Fig.~\ref{fig:z} suggest that $\d\varphi_h(T)$ is an increasing
function of $T$, \ie that for higher temperatures the theory performs better, if the temperature is low enough
that approximation (\ref{y_approx_rep}) makes sense. Indeed, we 
expect that for $\d\varphi_h(T) \to 0$ for
$T\to 0$, since at $T=0$ the theory fails 
to capture the behavior of the contact number at $\d\varphi>0$. 
We will come back to this point below.

\subsection{Correlation function}

\begin{figure}[t]
\centering
\includegraphics[width=8cm]{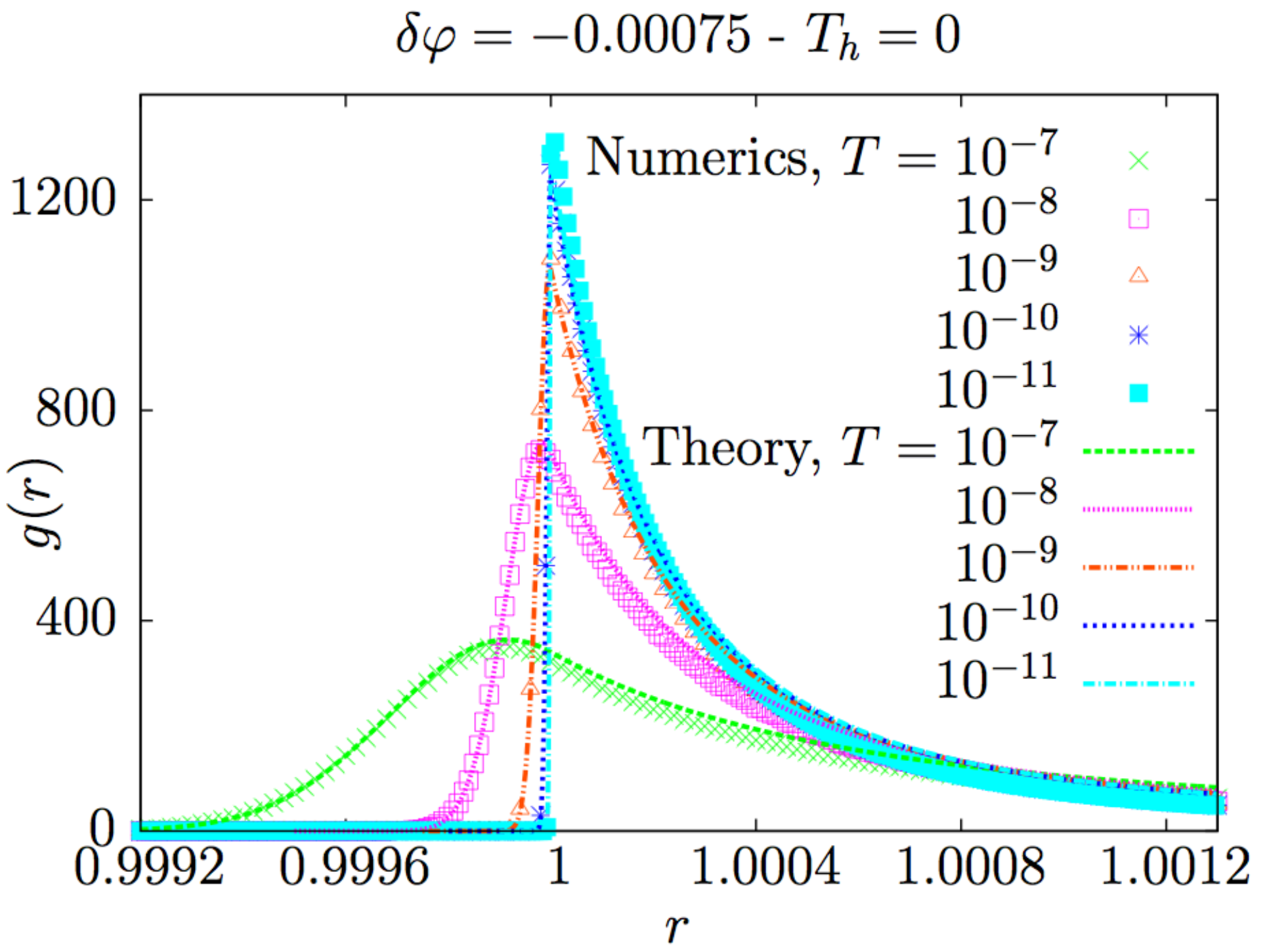}
\caption{
Pair correlation $g_{glass}(r)$ just below jamming predicted by theory 
(full lines) and measured in numerical simulations (symbols).
Here $\d\varphi = -0.00075$.
}
\label{fig:grbelow}
\end{figure}

\begin{figure*}
\centering
\includegraphics[width=11cm]{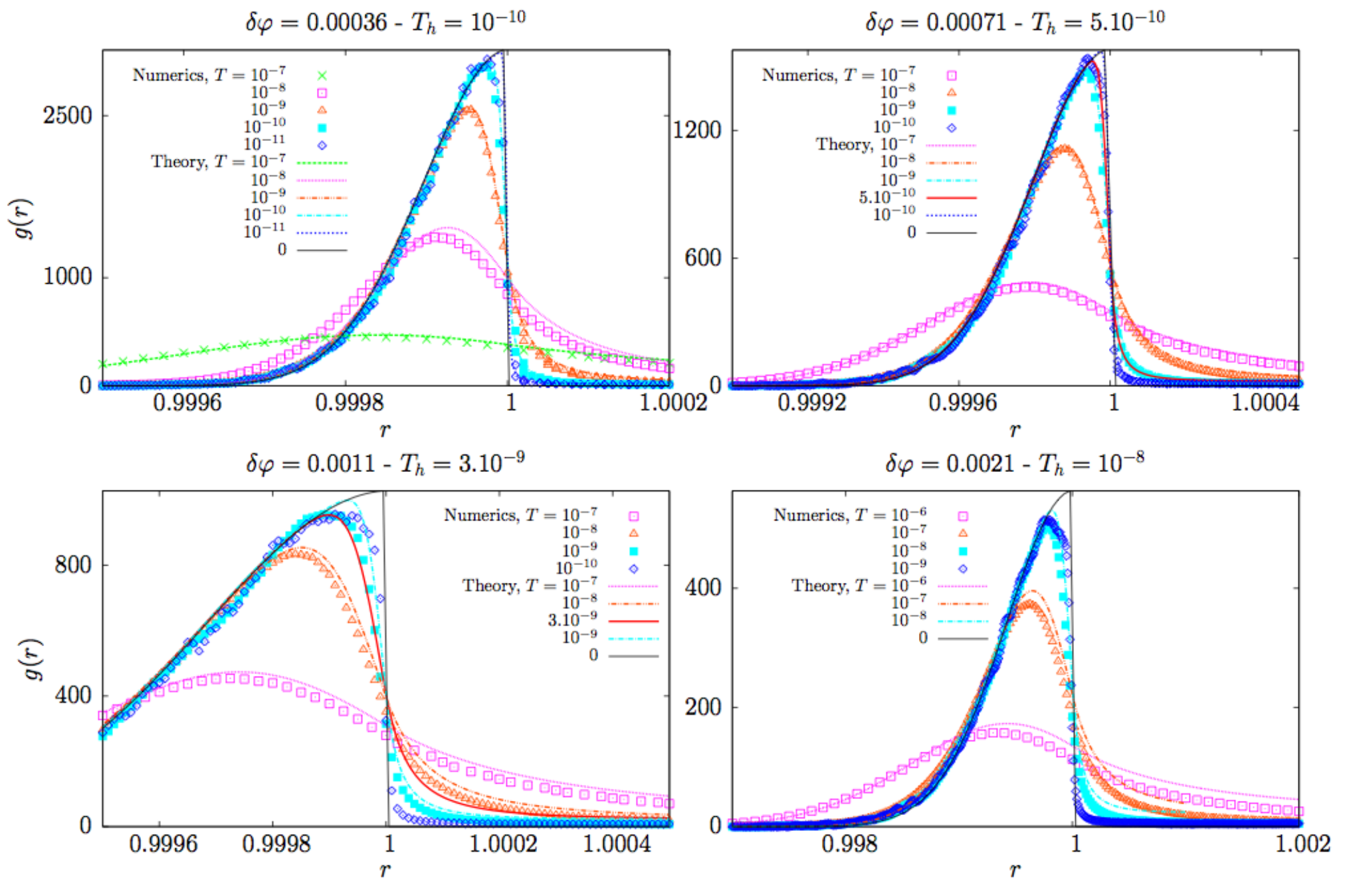}
\caption{
Pair correlations $g_{glass}(r)$ above jamming predicted by theory (full lines) and measured in numerical simulations (symbols).
The agreement is very good for $T > T_h(\d\varphi)$, the values of $T_h(\d\varphi)$ are reported above the figures. Below $T_h(\varphi)$,
the numerical curves do not evolve anymore, while the 
theoretical curves continue to evolve towards the $T=0$ theoretical 
limit.}
\label{fig:grabove}
\end{figure*}

We now report data for the scaling of the contact peak of 
$g_{glass}(r)$ near jamming.
We first consider a fixed $\d\varphi < 0$, 
and change the temperature. In Fig.~\ref{fig:grbelow} we report data for
$\d\varphi = -0.00078$ for several temperatures. In this regime the theory works very well down to $T=0$; this is expected
since we know that the theory works at $T=0$ for hard spheres~\cite{PZ10}.

Next, we consider a fixed $\d\varphi > 0$ and plot $g_{glass}(r)$ for several temperatures. This is done in Fig.~\ref{fig:grabove} for
four different values of $\d\varphi$. Here, we observe that, like for the average coordination (Fig.~\ref{fig:z}), the theory works
very well at moderately large $T$. However, on lowering $T$, at some point we observe that the numerical $g_{glass}(r)$ saturates
to a limiting curve which has a maximum at $r<1$. 
By contrast, the theoretical curves slowly evolve towards the 
$T=0$ theoretical limit, which is a half-Gaussian centered 
in $r=1$, with its $r>1$ part removed.

Therefore, below a certain temperature $T_h(\d\varphi)$ 
(the inverse of the function $\d\varphi_h(T)$ mentioned above), a deviation between
theory and numerical data is observed close to the maximum of $g_{glass}(r)$. Clearly, the deviation between data and theory is a
smooth crossover, so determining $T_h(\d\varphi)$ is not easy.
Here we choose the following procedure. The numerical
data accumulate on a master curve as $T\to 0$. Therefore we define $T_h(\d\varphi)$ as the value of temperature at which the
theoretical curve best fits the $T\to 0$ numerical curve. As an example, in the upper left panel of Fig.~\ref{fig:grabove},
we see that the theoretical curve for $T=10^{-10}$ fits very well the numerical curves for both $T=10^{-10},10^{-11}$, while
the $T=10^{-11}$ theoretical curve is quite different. We fix $T_h=10^{-10}$ for this value of $\d\varphi$. 
Clearly the ambiguity on the precise numerical value of 
$T_h$ is rather large. Despite this, we are able to qualitatively confirm 
that $T_h(\d\varphi)$ is an increasing function of $\d\varphi$, as already 
suggested in the previous section. 
The function $T_h(\d\varphi)$ determined 
from the pair correlation functions is 
reported in Fig.~\ref{fig:Th}. It emerges continuously from zero above 
$\varphi_{GCP}$. 
A comparison with the temperature scale given by the glass 
temperature shows that it is orders of magnitude smaller
than $T_K$. Thus it is only in the small regime of $T<T_h$ 
and $\varphi > \varphi_{GCP}$ that our the effective potential 
approach misses some important physics, as we discuss below 
in Sec.~\ref{subdisc}.

\begin{figure}
\centering
\includegraphics[width=8cm]{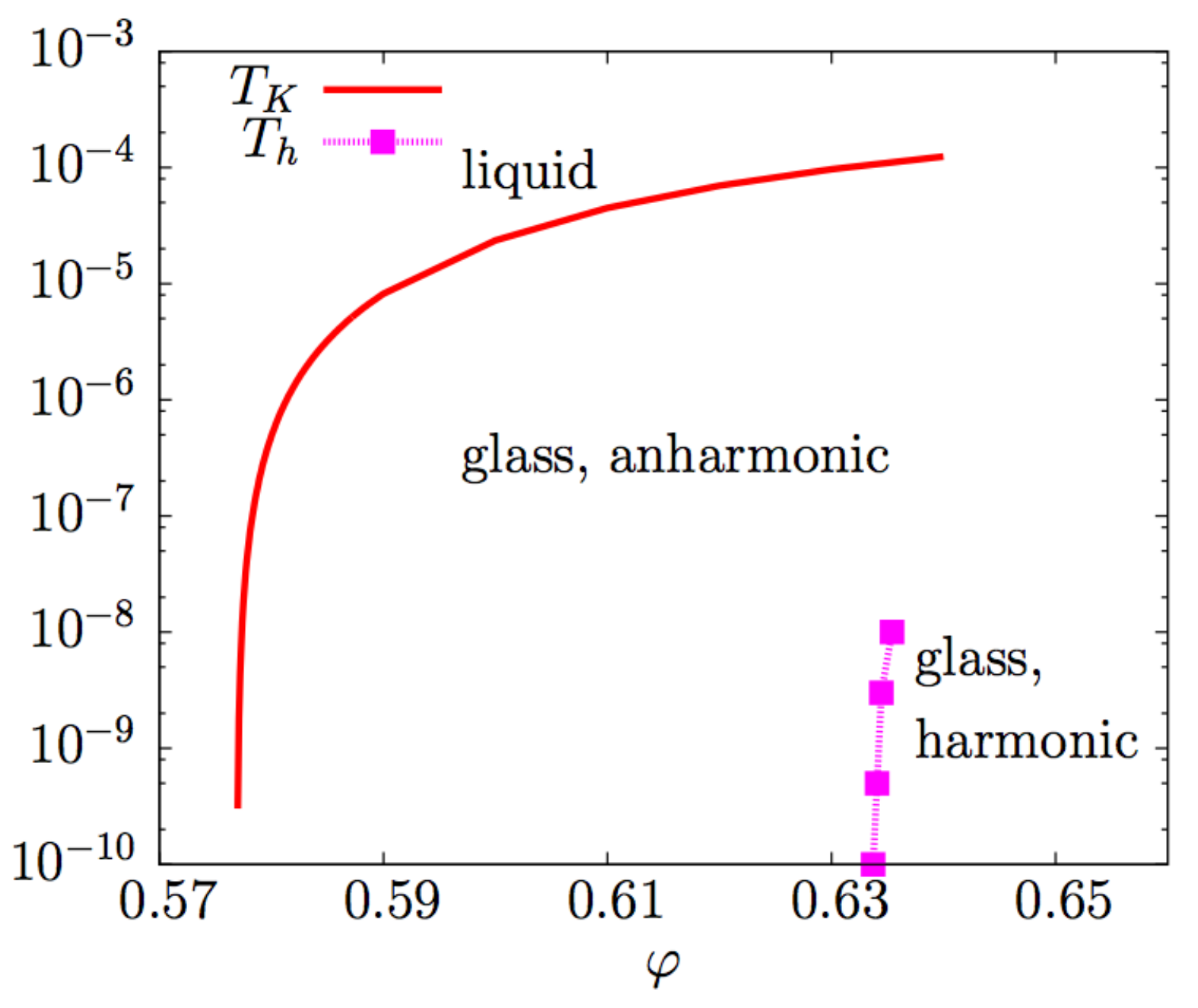}
\caption{
The crossover temperature $T_h(\varphi)$, 
as determined from Fig.~\ref{fig:grabove}, 
and compared with the Kauzmann temperature from Fig.~\ref{fig:sigma_tk}.}
\label{fig:Th}
\end{figure}

\begin{figure*}
\centering
\includegraphics[width=14cm]{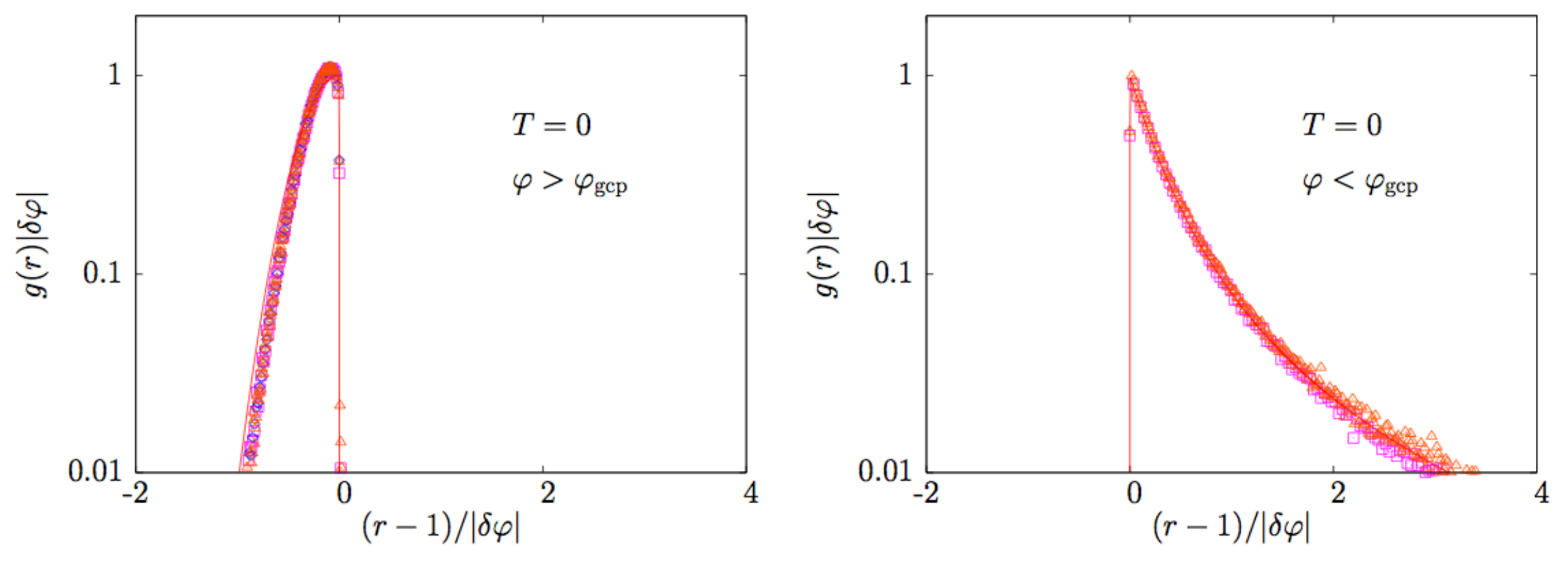}
\caption{
Scaling functions at $T = 0$ above (left) and below (right) the jamming transition showing the convergence of the first peak near $r = 1$ to a delta function with asymmetric scaling functions on both sides of the transition. 
To ease visualization, we show the evolution of $g(r)/g_{max}$, where $g_{max}$ can be read from Fig.~\ref{fig:gmax}.
}
\label{fig:grTzero}
\end{figure*}

Finally, in Fig.~\ref{fig:grTzero} we show that $g_{glass}(r)$, at $T=0$, follows the scaling relations predicted by the theory, Eqs.~(\ref{eq:HSscaling}) and (\ref{eq:SSscaling}), for 
$| \d\varphi | \to 0$, with different scaling functions on the two sides of the transition. While the scaling is perfect for $\d\varphi<0$~\cite{PZ10}, for $\d\varphi>0$ there is a small deviation
around the maximum as already discussed, 
which is however very small in the scaling regime for $\d\varphi \to 0$.

\subsection{Discussion}
\label{subdisc}

The main discrepancy between the theory and numerical data is in the region $\d\varphi>0$ and very small temperature $T<T_h(\d\varphi)$.
Here, the average coordination scaling $z \propto \d\varphi^{1/2}$ at $T=0$ is not obtained. Furthermore, the shape of the peak of $g_{glass}(r)$ is
only partially captured by the theory.
We tentatively attribute these discrepancies to the particular nature of the vibrational modes at $T=0$ and $\d\varphi>0$. It has been
shown that in this regime the low-frequency spectrum is dominated by spatially 
correlated vibrational modes. These modes
cannot be described by our approach, in which vibrations are assumed to 
be Gaussian and only two-body correlations are taken into account.
Therefore it is to be expected that the particular features of 
the jamming transition that are related to these correlated soft modes
are not well reproduced by our theory.

On the hard sphere side, the soft modes induce a square-root divergence of the pair correlation function 
$g_{glass}(r) \propto 1/\sqrt{r-1}$. Yet, for $\d\varphi \to 0^-$, 
this square root singularity is well separated from the contact 
peak~\cite{PZ10};
therefore the shape of the scaling function for the contact peak 
is not affected by the soft 
modes and it is indeed correctly reproduced by the theory (see \cite{PZ10}
and Fig.~\ref{fig:grTzero}).

On the other side of the transition and for small $\d\varphi > 0$ both 
the contact peak and the square root singularity are shifted
 by $\d\varphi$.
In particular the square root contribution should become 
$g_{glass}(r) \sim 1/\sqrt{r-1+ a \d\varphi}$. 
Inserting this form of $g_{glass}(r)$ into Eq.~(\ref{eq:zdef}) yields
the result that $z = z_0 + c \d\varphi^{1/2}$, where $z_0$ is 
the contribution of the contact peak. While $z_0$ is clearly 
unaffected by the shift, the scaling of $z$ with $\delta \varphi$
directly stems from the square root singularity of the pair correlation
function.
Therefore, the scaling of $z$ is dominated 
by the soft modes contribution, which might explain why it is not 
well captured by our theory. 

On the other hand, inserting the same expression in the general formula for the internal energy, Eq.~(\ref{def_U}), one finds
that $e_{GS} \sim \wt e \d\varphi^2 + d \d\varphi^{5/2}$, the first term being the contribution of the contact peak, the second being the one of the square root
singularity. Therefore, for the energy the contribution of the anomalous modes is subdominant, and for this reason we observe that the theory reproduces
well the numerical data in this case. The anomalous correction $\d\varphi^{5/2}$ is of course not reproduced by the theory, that gives an expansion of $e_{GS}$
in integer powers of $\d\varphi$, but these are subdominant terms. 

We also suspect that the shape of $g_{glass}(r)$ near the peak 
is also influenced by a mixing of the contact peak 
with the square root singularity. While both contributions are 
well separated for $\varphi < \varphi_j$, this is not necessarily the case
above jamming, which likely explains the small deviations
observed in Figs.~\ref{fig:grabove} and \ref{fig:grTzero}, but we could not 
find an empirical or scaling argument to disentangle both contributions.

Finally, it is interesting to remark that these deviations
between theory and simulations are washed out by a finite temperature 
$T > T_h(\d\varphi)$. We suspect that $T_h(\d\varphi)$ represents a temperature
scale at which anharmonicity becomes relevant~\cite{XVLN10}, and where 
the zero-temperature vibrational modes become irrelevant.
Above this crossover temperature, vibrations are probably
much less correlated and our approximations become correct. 
Interestingly, since $T_h(\d\varphi) \to 0$ for $\d\varphi \to 0^+$, we 
conclude that the correlations
at $\d\varphi=0$ are extremely fragile, 
such that an arbitrary small temperature is able to 
destroy them and restore the agreement between 
theory and numerical data.

\section{Conclusions}
\label{conclusion}

In this final section we summarize our results and we discuss some 
perspectives for future work.

\subsection{Summary of results}

Understanding the physical properties
of dense packings of soft repulsive particles is a fully 
nonequilibrium problem, because the 
jamming transition occurs in the absence of thermal fluctuations 
deep inside a glassy phase where phase space is dominated by the 
existence of a large number of metastable states. 

Nevertheless, we showed that it can successfully be 
addressed using 
equilibrium statistical mechanics tools.
We have developed a mean-field replica theory of the jamming
transition of soft repulsive spheres which satisfactorily  
derives, from first principles, 
the existence and location of a jamming transition. 
This transition, that happens only at $T=0$, is exactly 
the same as the SAT/UNSAT transition of random constraint satisfaction 
problems~\cite{Mo07,MM09,KZ08}:
it is the point where the Parisi parameter $m^*$ that characterizes 
the 1RSB glass phase goes to zero, the reduced pressure of
the hard sphere glass diverges, or, equivalently, its pressure and 
energy become finite. 
Within our approach, the jamming transition takes place deep into the glassy 
phase, and is thus a phenomenon which is physically distinct from 
the glass transition itself~\cite{MKK09}. 
Finally, we have shown that the average coordination jumps from zero to 
$\sim 6$ at the transition, and we derived scaling functions that
describe well the contact peak of the pair correlation function.

Again, the absolute values of density reported in this study are affected by the fact that we used the HNC equation of state 
for the liquid~\cite{PZ10}, and thus quantitative agreement with simulations cannot be expected. Still, we found that
the scaling around jamming is almost insensitive to the particular equation of state that is chosen for the liquid.
We also investigated the region of small $T$ around the jamming transition. We showed that the theory reproduces quite 
well the scaling numerical data with $T$ and $\d\varphi$, the distance from the jamming transition. On the hard spheres 
side of the transition ($\d\varphi < 0$) the theory works well down to $T=0$. On the soft sphere side ($\d\varphi > 0$) 
we have identified a temperature $T_h(\d\varphi)$ below
which, according to our interpretation, correlated vibrations that are neglected in our theory become relevant for some
observables,and produce a series of interesting scalings that are not captured by our 
theory~\cite{WSNW05,Vh10,LNSWW10}. The most striking of these is the scaling of the number of contacts
which our theory fails to reproduce.

\subsection{Technical aspects}

It is a major open problem to try and include in the theory a correct description of the divergent correlations in the 
vibrational modes that are relevant for $\d\varphi >0$ and $T<T_h(\d\varphi)$. Additionally, there are several more 
technical aspects on which the theory could be improved, even in the region where it works quite well.

Our theory falls within the general framework of the replica method, using the molecular liquid formulation of 
\cite{MP99a,MP99b}.
However, to be able to describe the jamming transition, we had to develop a new approximation scheme, that is based on 
the effective potentials formulation of~\cite{PZ10}, and allows one to view the molecular liquid as a simple liquid with effective 
potentials involving an arbitrary number of particles. Our approximation scheme is based on the following steps:
\begin{enumerate}
\item As in~\cite{PZ10}, we neglect the three (and more)-body interactions, and retain the two-body effective potential only.
\item In addition, we observe that the effective potential is given by the original potential (rescaled by a factor $m$) plus 
a small perturbation, therefore
we use liquid perturbation theory to write the free energy of the replicated liquid as in Eq.~(\ref{S_of_Q}).
\item Finally, we performed the low temperature approximation Eq.~(\ref{y_approx_rep}), in order to simplify the numerical calculations.
\end{enumerate}
The last two approximations have been done mainly for practical reasons. However, we found that they are related to the first one. Indeed, if we get rid of 
approximation 3, we introduce an explicit dependence of $y_{liq}$ on temperature, hence on $\t$ in the limit $T\to 0$ and $\d\varphi>0$. 
It is easy to realize that this dependence affects the scaling close
to jamming. An additional $\sqrt{\t}$ term appears in the small $\t$ expansion, 
which implies that $\t \propto \d\varphi^{2}$ and $e \propto \d\varphi^{3}$. An inspection
of the small $\t$ expansion reveals that if $y_{liq}$ does not depend on $\t$, then a delicate cancellation 
happens to eliminate the $\t^{1/2}$ term and produce
the correct scalings. The crucial observation is that the anomalous term contains the derivative of $y_{liq}$ 
with respect to temperature, which is related to a three-body
correlation function. Therefore we speculate that in a full treatment this term is cancelled by a term 
coming from the three body interaction potential. Therefore, getting
rid of the third step of approximation might require taking into account three-body correlations as well.

This is of course an
extremely interesting direction for future research, mainly because we suspect that getting rid of approximation 2 
should allow us to eliminate the violation of the exact thermodynamic relation 
$p_{glass}^{HS}=1 + 4 \varphi y_{glass}^{HS}(\varphi)$ that
occurs in the theory and forces us to rescale $\d\varphi$ in order to compare with numerical data.
It is also possible that taking into account interactions involving several particles one could capture 
at least partially the long range correlations at jamming.

\subsection{Some general directions for future work}

Our approach is general
enough to be systematically improved and generalized to various
models (\eg Hertzian potentials, or truncated Lennard-Jones potential). 
Moreover, recent work reported on a method to compute the shear modulus within the replica approach~\cite{YM10,SF11}. 
This could allow us to add the external drive to the theory, and obtain
a complete theoretical picture of the jamming transition in the 
three dimensional 
temperature, density, stress jamming phase diagram
originally proposed by Liu and Nagel~\cite{LN98}.

An important points that the theory fails to adress is the 
behavior of pair correlations over a broader range of distances. 
It was shown that pair correlation functions also exhibit singular behavior
beyond the contact peak~\cite{DTS05a,SLN06}, while the 
structure factor~\cite{DTS05b} and isothermal compressibility~\cite{BCCDS11} 
also display intriguing anomalies at large scale. These anomalies are 
postulated to be related to the anomalous scaling of the number of contacts.
Recent simulations at high dimensions \cite{CCPZ12} 
showed that all the previously mentioned anomalies are largely independant of the dimension
The fact that these features are present in all dimensions would suggest that
they can be also described by a mean-field computation like the one we performed.

A possibility would be that such anomalies at large scale are related
to a complex organization of the free-energy minima that dominate the partition
function at very low temperature. This would be reflected in an instability of our
replica symmetric ansatz (in which free-energy minima are supposed to be isolated one from another), 
that would have to be replaced by a full-RSB ansatz.

Another difficulty of the theory comes from the fact that the cage in the glass phase 
was shown by recent simulations to be non-Gaussian \cite{CIPZ12}, which
invalidates, at least in principle, our Gaussian ansatz. The non-Gaussianity of the 
cage should be related to the long-range anomalies found at the transition,
and thus taking properly these two points into account would require a theory that is
able to compute the cage shape from first principles, and that is also able to cope with 
higher levels of replica symmetry breaking. Note that it has been found in \cite{KPZ12} that
in high-dimensions, the free-energy was not affected by the Gaussian ansatz for the cage, 
but this is not expected for our low-dimensional case.

\chapter{Conclusion and perspectives}
\label{chap:conclusion}

In this thesis, we have focused on the theoretical study of the transitions towards an amorphous 
solid state: the glass transition 
and the jamming transition. Our general approach has been to focus 
on finite-dimensional realistic models of structural glasses, and to concentrate on the derivation,  
from the sole knowledge of the pair potential and the evolution equation of the system, of the 
transition from supercooled liquid to glass, or from non-jammed to jammed.

For practical reasons, when an explicit numerical calculation was needed, we specialized to the 
system of harmonic spheres, mainly because of its special importance in the field of the jamming 
transition. The harmonic spheres have bounded interaction strength and range, two properties that we exploited respectively in Chapter \ref{chap:dynamics1} and \ref{chap:jamming}.

\section{Dynamics near the glass transition}

We first studied the dynamics of a system of harmonic spheres and developed a simple perturbative 
approach, that predicts 
the existence of a glass transition in a compact way, giving some hope to the extension of such 
theory to higher-orders. By using the strength of the interaction between particles of the system as 
an expansion parameter, we were able to recover, as the lowest order correction to the ideal gas 
dynamics, a closed equation on the non-ergodicity factor of the system that is very close to the one 
obtained within mode-coupling theory. However, the situation is still unsatisfactory, due to a U(1) 
symmetry that has been broken by hand in our formulation. However, the situation could be 
improved, and several proposals have been put forward in other areas of physics \cite{Ki06,Ki09} 
The mathematical problem has now been formulated in a coherent way, and falls in the general 
class of the 2PI approximations of the free-energy of interacting bosons, which directly connects to
hard condensed matter physics.

In retrospect, the use of the strength of the interaction as an expansion parameter sheds light on 
a set of numerical works by Ikeda and Miyazaki \cite{IM11a,IM11b,IM11c}, who observed that 
mode-coupling predictions became good for the high density Gaussian-core model. This model 
indeed falls in the generic class of bounded repulsive potentials, which behave like ideal gases in 
the high density limit, i. e. the strength of the interaction can be taken as a small parameter in this 
case. Thus the statics become mean-field \cite{LBH00} at high density, and the dynamics start 
obeying closely the mode-coupling phenomenology.

Apart from that case, several theoretical approaches still need to be reconciled. Szamel formulated 
a diagrammatic formulation of the mode-coupling theory for Brownian dynamics \cite{Sz07}, which 
could provide another correct starting point for improvement of the theory. Independently, the 
approach of Kawasaki and Kim \cite{KK08}, which is based on the more traditional Martin-Siggia-
Rose formalism, uses a loop expansion to derive the same result as mode-coupling theory.  
One advantage of Szamel's formulation over the latter is that it contains explicitly the high-order 
correlation functions of the equilibrium liquid, whereas these correlations must be uncovered by 
non-perturbative arguments in the case of standard field-theory. The task is straightforward for the 
two-point correlation function, but we expect it to be much harder in the case of 
higher-order correlation functions. 

We are left with three distinct approach, the three of them being satisfying in one respect: our 
approach provides an expansion parameter, the approach of Szamel takes properly into account 
the static properties of the system, and the one of Kawasaki and Kim respects properly all 
symmetries of the evolution equation. A dynamical approach that could unify these three 
approaches would be a major breakthrough in the field. 

Once a satisfying first-principle derivation of the existence of a glass transition will be derived, it will
be easy to deduce the corresponding expressions for the four-point functions that are currently 
used in numerical works or in experiments, and they will naturally become long range due to the 
sudden appearance of a non-zero value of the non-ergodicity parameter 
\cite{Sz08b,BBBKMR07a,BBBKMR07b}. 

However, it has been argued early that the apparition of a non-zero ergodicity parameter is merely
an artifact of numerical simulations and experiments, that are not able to wait for a sufficient time 
the equilibration of the system. In that case, a cross-over to dynamics that are controlled by the 
static landscape is expected to take place, which, in the replica interpretation at least,  would 
eventually also lead to ergodicity breaking, but with a very different underlying mechanism.

Another possibility would be that a transition of dynamical nature indeed takes place, but is 
described by a more complicated (dynamical) order parameter. This possibility is supported by 
recent numerical simulations \cite{MGC05,PLW11} that uncovered a hidden dynamical transition 
associated with the glass transition. However, the matter of connecting this discovery with observed 
behaviors of the two- or four- point correlation functions, or to thermodynamic quantities is still 
elusive. Moreover the theoretical study of such transitions on realistic models is expected to be 
quite involved.

\section{Statics vs. dynamics}

Independently of dynamical calculations, we also considered the random-first-order transition 
theory, that focuses on the metastable states of the static free-energy landscape. This set of 
theoretical predictions are based on the assumption that the long-time dynamics of the system are 
dominated by the deep minima of the free-energy landscape. By resorting to the replica method, 
and liquid state theory, one is able to predict the appearance of a non-zero non-ergodicity 
parameter. The predictions coming from replica theory and mode-coupling theory for this quantity 
were already noticed to share a similar form \cite{Sz10,PZ10}. Szamel used this similarity to show 
that, in terms of replica theory, the mode-coupling result is obtained with a reasonable assumption 
of vanishing current between different metastable states, and further use of factorization 
approximations much like in the original mode-coupling derivation. We adopted another approach 
and asked, within the static replica theory, for a precise comparison between the dynamical and the 
static self-consistent equations for the non-ergodicity parameter.
The dynamical equation limits itself to a two-mode approximation, so we performed the equivalent 
approximation in the static context.
An immediate observation was that a term containing the three-body static correlation function is 
obtained in both approaches, but the dominant one, that contains two-body correlation, is absent 
from replica theory, and seems to emerge from purely dynamical calculations. 

This calculation allowed to correctly identify the class of 
diagrams that are relevant in the static computation, which can allow for a unified framework for 
replica theory computations. For example, one could make contact with the M\'ezard-Parisi 
\cite{MP99b} and Parisi-Zamponi \cite{PZ10} approximation schemes, or more generally with the 
effective potential 
approximation derived in Chapter \ref{chap:jamming}, by making a Gaussian ansatz on the form 
of the off-diagonal correlation function, and perform cage expansions afterwards.

Once an approximation that is able to detect the glass transition (such as replicated HNC) has 
been found, the calculation that we performed also serves as a basis for setting up a Landau-like 
expansion around this approximation, by using the fact that the bifurcation mechanism with which 
the non-ergodicity appears gives rise to soft modes in the system. 

In a broader perspective, the interplay between the metastable states (i.e. minima of the static free-
energy), that are at the basis of the reasonings in replica theory, and dynamical quantities should 
be investigated. This is by definition impossible in the static context, and has to be 
investigated in a dynamical calculation. More precisely, one has to consider the low lying excited 
states of the evolution operator of the system, and try to put them in correspondence with the 
minima of the static free-energy, in the spirit of \cite{BM98}. An analytical approach that would be 
able to address this question for realistic models of glasses is unfortunately for the moment out of 
reach. However, such a finding would be mandatory if it were found that no dynamical order 
parameter describes the glass transition properly, as suggested above.

Despite this intrinsic limitation of replica theory, it is an important task to push its predictions
as far as possible in order to set the stage for a theory that would be able to incorporate 
both static and dynamical aspects of the glass transition.

\section{Theory of the jamming transition}

Finally, we have developed, based on replica theory, a microscopic theory for the 
jamming transition of harmonic spheres, able to quantitatively recover many observed scalings 
near the jamming transition. Of the three main questions discussed in this thesis, this is where the
current theoretical tools give the most satisfying results. Indeed, the very high density at which 
the jamming transition takes place justifies most of the approximations performed by replica theory:
the out-of-equilibrium procedures used to construct those states place the system directly in an
ergodicity broken picture, where the system is trapped around one given local minima of the 
free-energy. The very high density justifies an expansion in the sizes of the cages seen by the 
particles, which also serves as an order parameter for the transition. 

This successes allowed to clearly separate the glass transition, for which replica theory provides
at worst an upper bound in terms of density, and the jamming transition, which occurs for densities
even larger than this upper bound.

Despite these successes, we have seen that our theory is not able to capture the long-range 
correlations that appear precisely at the jamming transition. A non-linear mechanism that would allow
for the appearance of system-wide correlations could be provided by the inclusion of three-body 
interactions. Another possibility, indicated by recent works in high dimensional systems \cite{KPUZ13},
would be that these long-range anomalies are the product of a complex organisation of the metastable
states near the jamming transition. Indeed, in our 1RSB ansatz, the metastable states are supposed
to be isolated one from the other, and a more intricate organisation of these states could be captured
by a full-RSB ansatz.

Related to this issue is our assumption that the cage shape is Gaussian near the jammgin transition.
It was found in recent numerical simulations \cite{CIPZ12} that the cage is on the contrary non-Gaussian.
Taking these two issues into account would require a theory that is able to derive the cage shape as well 
as cope with more complex ans\"atze for the replica symmetry. The only set-up that has this potential for the
moment is the order-parameter expansion that we set up in Chapter \ref{chap:replicas}, but more powerful 
resummations must be performed in order to establish a link with the effective potential approach that 
we used in Chapter \ref{chap:jamming}.

\pagestyle{empty}

\newpage
\addcontentsline{toc}{chapter}{Bibliography}

\bibliographystyle{plainurl}
\bibliography{full_biblio}

\end{document}